\documentclass[sigplan,nonacm]{acmart}

\usepackage{amsmath}
\usepackage{amsthm}
\usepackage{mathtools}
\usepackage{booktabs}
\usepackage{enumitem}
\usepackage{microtype}
\usepackage{xspace}
\usepackage{hyperref}
\usepackage{multirow}
\usepackage{tcolorbox}
\usepackage{listings}
\tcbuselibrary{breakable,listings,skins}
\usepackage{subcaption}
\usepackage{tablefootnote}
\usepackage{tabularx}

\newcommand{\work}[1]{\textsc{#1}}

\definecolor{promptbackground}{HTML}{F7F8FA}
\definecolor{promptborder}{HTML}{CBD5E1}
\definecolor{promptheading}{HTML}{6D28D9}
\definecolor{promptcode}{HTML}{0F766E}

\lstdefinelanguage{Markdown}{
  sensitive=true,
  morecomment=[l][\color{promptheading}\bfseries]{\#},
  morecomment=[s][\color{promptcode}]{```}{```},
  moredelim=**[s][\color{promptcode}]{`}{`}
}

\tcbset{
  promptlisting/.style={
    enhanced,
    breakable,
    listing only,
    colback=promptbackground,
    colframe=promptborder,
    boxrule=0.4pt,
    arc=1mm,
    left=1.5mm,
    right=1.5mm,
    top=1mm,
    bottom=1mm,
    before skip=0.75em,
    after skip=0.75em,
    listing options={
      language=Markdown,
      basicstyle=\ttfamily\scriptsize,
      columns=fullflexible,
      keepspaces=true,
      breaklines=true,
      breakatwhitespace=true,
      showstringspaces=false,
      upquote=true,
      aboveskip=0pt,
      belowskip=0pt
    }
  }
}

\newcommand{\customsection}[1]{%
  \vspace{0.5em}%
  \noindent\textbf{#1.}\xspace%
}

\usepackage{needspace}
\usepackage{fontawesome5}

\definecolor{solverAccent}{HTML}{224C60}
\definecolor{solverRule}{HTML}{B8C4CA}
\definecolor{solverMuted}{HTML}{59646B}

\title{The Case for Automated Hyperspecialization: \\ Evidence from SAT}

\author{Harrison Green}
\affiliation{%
  \institution{Carnegie Mellon University}
  \city{Pittsburgh}
  \state{PA}
  \country{USA}
}
\email{harrisog@cmu.edu}

\author{Claire Le Goues}
\affiliation{%
  \institution{Carnegie Mellon University}
  \city{Pittsburgh}
  \state{PA}
  \country{USA}
}
\email{clegoues@cmu.edu}

\author{Fraser Brown}
\affiliation{%
  \institution{Carnegie Mellon University}
  \city{Pittsburgh}
  \state{PA}
  \country{USA}
}
\email{fraserb@cmu.edu}

\begin{document}

\begin{abstract}
  The software status quo is to use one system to process
  many different kinds of inputs.
  In contrast, we propose hyperspecialization:
  creating new software that is optimized for a single
  class of inputs.
  Hyperspecializing manually is anywhere from expensive to impossible.
  We conjecture that coding agents make \emph{automated} hyperspecialization
  cheap, effective, and safe for problems with measurable performance
  and checkable output.
  This paper explores one such problem, SAT solving, by synthesizing
  hundreds of workload-specific SAT solvers at an average cost of \$37 each.
  Our specialists outperform their competition-winning, general-purpose cousins
  by 5$\times$ on average,
  and by over $10\times$ on a quarter of benchmark families. %
  A general-purpose solver constructed from
  over a hundred of our prototype
  hyperspecialists won the SAT track at the 2026 SAT Competition. 
\end{abstract}

\maketitle

\section{Introduction}
\label{sec:intro}

\begin{figure}
  \centering
  \resizebox{\linewidth}{!}{\includegraphics{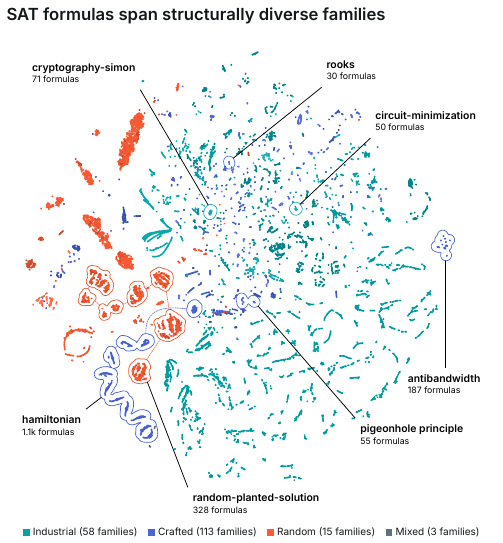}}
  \caption{Each point represents a SAT formula from 189 benchmark families in this paper, projected using t-SNE over 56 structural features. Colors indicate broad benchmark origin, with shades distinguishing families. Seven family clusters are highlighted and labeled.}
  \label{fig:families}
\end{figure}

Different sub-fields of computer science have
staged their own rebellions against generality.
The architecture~\cite{hennessy2019golden,hameed2010understanding},
databases~\cite{stonebraker2005onesize,stonebraker2005cstore},
networking~\cite{clark1990architectural,belay2014ix},
and operating systems~\cite{engler1995exokernel,engler1995exterminate,madhavapeddy2013unikernels}
communities have repeatedly discovered
that domain-specific software or hardware
outperforms its generalist alternative. 
``Exterminate all operating system
abstractions''~\cite{engler1995exterminate}!
The OS should not offer applications foundational primitives
like virtual memory, since doing so requires making
decisions about, for example, page table design---and
``any tradeoff penalizes applications that were not anticipated...by the OS designer''~\cite{engler1995exterminate}.
Instead, the OS should expose almost bare hardware
resources, and the application should use (or avoid)
the relevant ones. 
As a result of the specialist OS, the application
enjoys significant speedups. 

Despite these generality rebellions, not every
specialist has made it from the academic world to 
reality; Linux has not been re-written as a library OS.
This is not because the general- vs. special-purpose
performance gap is overstated, but because
\emph{building specialized software and hardware
is not free}.
Specialists get deployed when
the performance benefit of specialization outweighs
the cost of specializing.

There are domains where the high cost of specialization
is worth it~\cite{dally2020domain}.
It's \emph{extremely} expensive to tape out a new
chip~\cite{khazraee2017moonwalk}---much
more expensive than it is to create new software---yet
the performance benefit of GPUs, amortized across
all graphics problems, justifies the cost of creating
those GPUs in the first place~\cite{lindholm2008nvidia}.
More recently, following a similar cost-benefit analysis,
there has been an explosion of interest in (and startups
about) custom chips for LLM
inference~\cite{jouppi2017datacenter,openai2026jalapeno,techcrunch2026jalapeno}.

Compared to hardware, custom software can justify
a smaller benefit because it's cheaper to produce---and
the cost is now nosediving towards zero.
What once required assembling a team of highly paid
programmers now requires only a credit card and a
horde of Claudes.
The Claudes' results and the programmers' results are not
necessarily the same, though;
while it's \emph{possible} that agents could build specialized
operating systems, such a task involves huge codebases,
informal specifications, and complex evaluations---and there
is no way to check if the resulting specialist is correct.\footnote{In fairness to the machines, all of these facts make OSes hard for people to write, too.}
Thus the animating question of this paper: ``are there
meaningful domains in which (1) specialization improves performance
and (2) \emph{specialization is plausibly automatable}?''

We believe that there may be several such domains
(\S\ref{sec:discussion}), but this paper
presents an existence proof for one: Boolean Satisfiability (SAT)
solving.
A SAT solver takes a formula of Boolean variables, their
negations, and the operators AND and OR, and decides whether
any assignment of true and false to the variables makes the
formula itself true. 
SAT was the first problem shown to be
NP-complete~\cite{cook1971complexity}, meaning that any problem in NP
can be efficiently reduced to SAT, and that there is no
efficient algorithm for solving SAT unless P = NP.
Donald Knuth describes SAT as a ``killer app''
because it can be used to solve so many
(hard!) important
problems~\cite{knuth2015satisfiability}.
Software verification, circuit equivalence checking,
package dependency resolution, combinatorial math problems,
and more all reduce in whole or part to SAT.
These domains have not typically created their own SAT solvers.
Instead, since SAT is a hard problem, they rely on existing tools
that have implemented sixty years of optimizations, including
a long tail of tricks that have been sharpened annually since
the inaugural 2002 SAT Competition~\cite{jarvisalo2012international}.
Rolling your own solver was (usually) not worth it.

We argue that rolling your own solver is now worth it.
First, the SAT community already recognizes SAT as a
domain in which specialization improves
performance.
SAT problems of different origins have very different
structure~\cite{ansotegui2017structure} (\autoref{fig:families}), and
SAT solvers ``may dramatically change their performance
depending on the class of...instances they are trying
to solve''~\cite{ansotegui2017structure}.
There are even theoretical differences between solving
strategies.
For example, most modern solvers use conflict-driven clause
learning (CDCL), which can \emph{require} exponentially large
refutations of XOR constraints~\cite{urquhart1987hard}.
Gauss-Jordan elimination can solve the same
constraints in polynomial time but \emph{cannot}
solve other kinds of constraints, since it 
only reasons about systems of linear
equations~\cite{han2012boolean}.
Empirically, too, no single solver is best on
all workloads: the ``virtual best solver'' (i.e., an oracle
that runs the fastest available solver on a given formula)
would outperform the winning solver in
every competition to date.

Second, SAT solvers are well-suited to being written
by machines.
They have a simple, well-specified interface,
are straightforward to evaluate,
and are
grounded in a vast public literature that spans
algorithms, theory, systems, and formal methods, along with
codebases, system descriptions, benchmarks, and competition results;
the LLMs have been consuming this literature for years.
Most importantly, though, \emph{SAT solvers produce a
checkable result}.
A ``satisfying assignment'' to the formula can be (efficiently)
checked by plugging it in;
an ``unsatisfiable'' answer comes with a proof that a small, formally
verified checker can validate independently. 
Thus, we need never trust the LLM-generated SAT solver, and need
only check its results. 

This paper implements automated hyperspecialization
for SAT.
Our first contribution is presenting hyperspecialization,
a form of program optimization that synthesizes a fresh program
for a given workload, and describing the applications
for which \emph{automated hyperspecialization} is
tractable (\S\ref{sec:design}).
Such problems must handle diverse workloads (so there's something
to specialize for), have measurable objectives
(so there's something to optimize towards), and
have checkable results (so there's no call to
trust the LLM author).

Our second contribution is a framework for SAT
hyperspecialization: we
develop specialist SAT solvers for the 189 ``families''---i.e.,
formulas sharing an origin---in the Global Benchmark
Database (GBD)~\cite{gbd},
the SAT community's benchmark corpus built
from over twenty years of competition submissions.
For each family in the database, we randomly select a training
set of instances, and provide this set to a coding agent.
The agent is prompted to implement (and repeatedly
evaluate when necessary) a specialized
solver from scratch for the entire family.
Many solvers do not use traditional SAT algorithms.
Instead, they recover structure that was lost in the encoding,
and use a different algorithm to solve the structured
problem.
For example, one specialist recognizes a family of rook-placement
  formulas and proves that they're unsatisfiable:
  the solver knows about rooks and chessboards, so it can
  say that satisfying the constraints would require more rooks
  than the board's rows fit.

Our third contribution is an evaluation of specialist
solvers' performance (\S\ref{sec:eval}) that shows
the specialists perform better than their general-purpose cousins,
for low cost.
A prototype of our hyperspecialization approach
won the SAT track at the 2026 SAT Competition.
Specialists match or beat the
per-family virtual best solver, drawn from four past
competition-winning solvers, on 64\% of families, with a geometric
mean $12\times$ improvement (without checking)
and $5\times$ (with checking);
specialists are $\geq100\times$ faster for
27\% of families (without checking)
and 14\% of families (with checking) (\S\ref{sec:rq1}). 
A specialist costs on average \$37 and takes one
and a half hours to produce; the
fastest arrives in twenty-eight minutes for \$11
and a $1.7\times$ improvement (\S\ref{sec:rq2}).

Finally, we discuss the implications of
SAT hyperspecialization for downstream
consumers, and broader opportunities for
hyperspecialization (\S\ref{sec:discussion})---from
SMT solving to compression to compilation and, perhaps,
beyond.

\section{Related work} \label{sec:rel}

In this section, we position our work with respect to prior
literature on software synthesis for optimization.
We first discuss approaches that bound the space of candidate programs so
that correctness follows by search space construction
(\S\ref{sec:superopt}).
We then discuss recent LLM-based approaches, which search a much broader space
of program alternatives but with relaxed correctness guarantees
(\S\ref{sec:llmopt}).
Finally, we discuss prior work on SAT optimization in particular (\S\ref{sec:satopt}).

\subsection{Optimizing to a reference}
\label{sec:superopt}

\customsection{Traditional synthesis techniques}
Program synthesis writ large (see~\cite{gulwani2017program})
has a multi-decade research lineage.
Our work is closest in spirit to \emph{inductive} techniques, which
synthesize programs against partial specifications like
input-output pairs~\cite{polozov2015flashmeta,gulwani2011automating}.
Oracle-guided variants query an oracle that returns counterexamples
to constrain the next
candidate~\cite{jha2010oracle,jha2017theory,solar2006combinatorial,torlak2014lightweight,alur2013syntax}.
Our setting shares the iterative structure: an agent proposes a
candidate, observes how it performs,
and revises.

More closely related is
\emph{superoptimization}, which searches for a fast program equivalent to a given input
program, ranking by a cost function. Equivalence is established by
proof or by testing depending on the technique.
Underlying search techniques include, e.g.,
enumeration~\cite{massalin1987superoptimizer}, deductive
proof search over an axiomatized
machine~\cite{joshi2002denali}, or candidate generation
checked against an oracle~\cite{bansal2006peephole}.
\work{STOKE}~\cite{schkufza2013stochastic} takes the last approach and
comes closest to ours in spirit, sampling whole candidate programs.
It treats correctness as a term in a cost function, with a symbolic
validator checking the candidates that survive.
The reference need not be an existing program; generative libraries
search over equivalent decompositions~\cite{puschel2005spiral} or
schedules~\cite{ragankelley2013halide,chen2018tvm,ikarashi2022exocompilation}
against a fixed mathematical specification.

\customsection{Workload specialization}
Specializing programs to workloads
is longstanding practice across 
computer science.
Architects design domain-specific processors~\cite{hennessy2019golden,hameed2010understanding,dally2020domain,lindholm2008nvidia,jouppi2017datacenter}, while
database researchers argue for engines specialized by workload~\cite{stonebraker2005onesize,stonebraker2005cstore}.

Automating specialist construction has historically meant
\emph{autotuning}, searching a space of implementations laid out in advance
(e.g., for linear algebra
kernels~\cite{whaley1998atlas}, or
signal transforms~\cite{puschel2005spiral}).
\work{FFTW}~\cite{frigo2005fftw3} defers the search to runtime.

Across these systems, the search or design space is
constructed to be semantics-preserving, whether via program
equivalence, specification refinement, or parameterization of an
existing implementation. 
The tradeoff is that a general correctness requirement
constrains the search:
requiring generated SAT solvers to operate correctly on all
formulas rules out, for example, solvers that can
\emph{only} solve rook placement problems (fast!). 

\subsection{Optimizing and synthesizing with LLMs}
\label{sec:llmopt}

Recent LLM-based optimization approaches relax
correctness requirements to search a much larger space of
alternatives.
LLMs have been used to generate whole-system database implementations
customized to a workload.
They have synthesized query engines for a workload
contract~\cite{bespokeolap} or execution code per SQL
template~\cite{gendbdemo}, handling off-nominal inputs by falling
back to a general-purpose engine. 

Most LLM-driven optimization work is considerably more constrained (see~\cite{llm4adsurvey}).
One line of work generates an improved component of a human-written
outer algorithm, such as a scoring function, priority rule, or penalty
term~\cite{funsearch,alphaevolve,eoh,reevo}; another gives the model an
existing implementation and asks for a faster equivalent~\cite{kernelbench}.
\work{SuperCoder}~\cite{supercoder} uses an LLM as a superoptimizer, asking for assembly that runs faster than a compiled reference
while preserving behavior. 

These searches check success against a provided test suite.
Tests are only partial correctness specifications, so
an agent can satisfy the tests without actually solving the problem. Agents exploit evaluation harnesses~\cite{robustkbench}, fail the tests
they were rewarded for passing~\cite{supercoder}, and replace complete
search with approximations that return schema-valid wrong
answers~\cite{heuristictrap}.

This prior work establishes that LLMs can effectively search a
large space of implementations \emph{heuristically}.
We also use LLMs to search for programs,
but limit our search to applications where it's easy
to check programs' results for correctness.

\subsection{Optimization for SAT solvers} \label{sec:satopt}
While SAT is an NP-complete problem---and random SAT instances
are very hard to solve~\cite{chvatal1988many}---formulas
derived from real-world
problems tend to have much faster than average-case
runtime~\cite{ansotegui2019community}.
Modern SAT solvers can solve instances with millions of
clauses and variables, and the SAT Competition shows
huge performance improvements over the years.
The first solver optimizations were
two foundational algorithms,
Davis-Putnam-Logemann-Loveland (DPLL)~\cite{dpll}
and conflict-driven clause learning (CDCL)~\cite{grasp,chaff}.
Improvements since have broadly focused
on engineering~\cite{cadical} or solver heuristics~\cite{liang2016learning}.
LLMs are a recent addition to the solver
optimization literature:
both
SATLUTION~\cite{satlution} and
\work{AE-Kissat-MAB}~\cite{ae-kissat-mab}, which won the
2025 SAT Competition~\cite{satcomp2025results}, use LLMs to optimize
existing solver codebases. 

No prior work builds a fresh solver from scratch for
every workload, but some specialize in a limited way, or manually.
SATenstein~\cite{sateinstein}, which only solves SAT
instances, synthesizes new solvers for a given
workload by combining pieces of existing solvers.
AutoSAT~\cite{autosat} and AutoModSAT~\cite{automodsat}
use LLMs to optimize for several specific families.
They start from a simplified solver codebase,
and only allow the LLM to modify a maximum of
nine heuristics \emph{within} the CDCL implementation.
Prior work also manually specializes solvers to
a specific workload (e.g.,
cryptanalysis~\cite{cryptominisat,nejati2020cdcl}), or
automatically specializes by tuning parameters~\cite{paramils,smac,irace} or choosing the anticipated
best from a portfolio of solvers for a given
instance~\cite{satzilla,isac,algorithm_selection_and_scheduling,hydra}.

\section{Automated hyperspecialization} \label{sec:design}

The software status quo is general-purpose: a single implementation
processes many different workloads.
But these workloads may favor different algorithms,
representations, and heuristics, creating
opportunities to take advantage of a
workload's particular structure.
The goal of automated hyperspecialization is to realize
these opportunities by automatically synthesizing specialist
implementations tailored to each target workload.

In contrast to optimization, the defining feature
of hyperspecialization is that it considers
workload-specific implementations, not
general-purpose ones.
It synthesizes implementations designed to handle, for
example, only SAT problems generated by the
\texttt{Kani}~\cite{kani} verifier, or
only encodings of register allocation problems. %
A particular implementation need only perform
well---or perform at all!---on its target workload. 
Our intuition is that restricting workloads allows
implementations to exploit assumptions about the types
of inputs they will encounter. %
For example, a specialist for register allocation
instances can assume that the problem is graph
coloring, and quickly refute with a counting argument. 

The next section describes the qualities
that make a software
task amenable to hyperspecialization (\S\ref{sec:specable}):
if a task has heterogeneous workloads, measurable objectives,
and a checkable result, it can be iteratively
and safely hyperspecialized by agents. 
This paper focuses on hyperspecialization for SAT
since SAT solvers are critical pieces of
software across a huge range of domains, from
mathematics to software verification to circuit design. 
We describe why SAT is hyperspecializable
(\S\ref{sec:satspec}), including details about workloads,
the specific measurable objective, and how to check the
correctness of SAT solver results. 
Finally, we discuss our realization of hyperspecialization
as a whole-codebase synthesis task, where
agents iteratively produce and measure a SAT solver
implementation tailored to a given workload
(\S\ref{sec:hyper-agent-based}).

\subsection{What kind of software can be hyperspecialized?} \label{sec:specable}

There are three qualities that make a %
task amenable to automated hyperspecialization.
The task must have:
\begin{enumerate}
\item \textbf{Heterogeneous workloads}:
  the task should operate on more than one type of workload
  so that there is plausibly a benefit from building
  workload-tailored specialists.
\item \textbf{Measurable objective(s)}:
  it should be possible to compare one implementation to
another implementation for a given measurable
property (e.g., speed, memory use, etc.).
\item \textbf{Efficiently checkable results}:
  to safely automate software construction---and, as a
  result, deploy untrusted code---implementations'
  results must be checkable.
\end{enumerate}

\subsection{SAT can be hyperspecialized} \label{sec:satspec}

The task of solving SAT formulas satisfies all of our
desiderata: the workload is heterogeneous, because
solvers need to solve
formulas encoding very diverse problems;
faster runtime is the obvious objective;
and we can efficiently check a SAT solver's results.
We discuss SAT and SAT solvers next, then expand
on each criterion in more detail.

\subsubsection{SAT}
Boolean Satisfiability (SAT) asks whether a formula
over Boolean variables and
operators $\land$, $\lor$, and $\neg$ is true
under some assignment to the variables.
Given the space of inputs $\mathcal{X}$ (formulas) and outputs $\mathcal{Y}$ (assignments or proofs), a SAT solver ($\sigma : \mathcal{X} \rightharpoonup \mathcal{Y} \cup \text{unknown}$) is a potentially nondeterministic procedure which, upon receiving input $x \in \mathcal{X}$ either produces an output $y$ (i.e., a satisfying assignment or UNSAT proof), runs indefinitely, or returns \text{unknown}. $\mathcal{R} \subseteq \mathcal{X} \times \mathcal{Y}$ is the set of correct outputs.

\subsubsection{Heterogeneous workloads in SAT}
We use $\mathcal{I}$ to refer to a workload,
a finite set of input instances.
Our implementation of hyperspecialization specializes
by \emph{family}---a set of SAT formulas drawn from the same
task or generator.
Family $k$ has representative workload $\mathcal{I}_k$.
During hyperspecialization, we split this workload into
separate training and validation splits,
denoted $\mathcal{I}_k^{\text{train}}$
and $\mathcal{I}_k^{\text{val}}$. 

\subsubsection{Measuring solver performance}
We care about SAT solvers being fast.
We can measure a particular SAT solver's performance by running it on an input in a fixed environment with a timeout.
Running $\sigma$ on input $x$ with timeout $T$ yields an
output $y$ and runtime measurement $t$
$$
(y,t) \sim \operatorname{Run}_{T}(\sigma, x)
$$
where $y \in \mathcal{Y} \cup \texttt{unknown}$ is the
observed output and $t \in [0, T]$ is the (right-censored)
observed runtime.
Crashes and out-of-memory (OOM) errors are treated
as $(\texttt{unknown},T)$.

\customsection{PAR-2 scores}
Following the SAT Competition's example~\cite{satcomp2020},
we convert runtime measurements into a
\textsf{PAR-2} score in order to encourage both (1) solving instances
faster and (2) solving \emph{more} instances
correctly within the timeout:
$$
\textsf{PAR-2}(x,y,t) :=
\begin{cases}
  t & \text{if } (x,y) \in \mathcal{R}, \\
  2T           & \text{otherwise.}
\end{cases}
$$
A solver's score on a given input $x$ for which it produces $y$
in time $t$ is $t$ if the solver returned a valid solution
within the time budget, and twice the timeout otherwise. 

To compute performance of a solver $\sigma$ on a workload $\mathcal{I}$, denoted $P(\sigma,\mathcal{I})$, we run the solver once on each formula:
$$
(y_x, t_x) \sim \operatorname{Run}_T(\sigma, x) \qquad \text{for each } x \in \mathcal{I}
$$
and compute the average \textsf{PAR-2} score:
$$
P(\sigma, \mathcal{I}) := \frac{1}{|\mathcal{I}|} \sum_{x \in \mathcal{I}} \textsf{PAR-2}(x,y_x,t_x)
$$

\subsubsection{Checking solver results} \label{sec:checksat}
Satisfying assignments are conceptually easy to check by plugging in the values and evaluating the formula.
We use the formally verified \work{gratchk} checker in SAT mode~\cite{grat_checker}.

Checking UNSAT answers, on the other hand, requires augmenting the
solver to emit a proof~\cite{goldberg2003verification,zhang2003validating}.
Our synthesized solvers emit proofs in either \work{DRAT}~\cite{drat_format}, \work{DPR}~\cite{dpr_format}, or \work{VeriPB}~\cite{veripb_format} format. We validate these proofs using the \work{GRAT}~\cite{grat_checker}, \work{DPR}~\cite{dpr_checker}, and \work{VeriPB}~\cite{veripb_checker} proof-checking tool\-chains (as described for the SAT Competition 2025), respectively. Each of these toolchains consists of an \emph{elaborator} which fills in details omitted in the original proof and emits an elaborated certificate which is checked by a formally verified checker.

The choice of proof format and toolchain affects both the reasoning that can be expressed directly and the cost of generating and verifying certificates. For both SAT and UNSAT results, checking takes polynomial time in the (combined) size of the formula and the solver's output.

\subsection{Agent-based hyperspecialization for SAT}
\label{sec:hyper-agent-based}

We treat the problem of SAT solver hyperspecialization as a whole-codebase synthesis task.
Given a target workload $\mathcal{I}_k$, we task a coding agent (GPT-5.6 Sol in Codex) with producing a specialized solver $\sigma_k$ that minimizes $P(\sigma_k, \mathcal{I}_k)$ (smaller \textsf{PAR-2} is better).
To ensure that agents do not simply memorize solutions, we provide the agent 80\% of the instances as training data $\mathcal{I}_k^\text{train}$, and evaluate performance of submitted solvers on the withheld 20\% validation data $\mathcal{I}_k^\text{val}$.

Agents may submit candidate solvers for evaluation.
An external evaluation server runs the candidate solver
on both the training and validation splits, but only
returns training results to the agent as feedback.
We capture the sequence of submissions and their
respective cumulative costs as a trace:
$$
\mathcal{T} :=
\left\langle (\sigma_k^{(i)},C^{(i)}) \right\rangle_{i=1}^{n}
$$
Where $\sigma_k^{(i)}$ is the $i$-th candidate solver and $C^{(i)}$ is
the cumulative cost (in time and money) to produce and evaluate $\sigma_k^{(i)}$.

\begin{figure}
  \centering
  \resizebox{\linewidth}{!}{\includegraphics{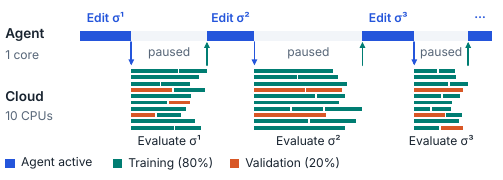}}
  \caption{Our bimodal optimization framework alternates between (a) using a coding agent to construct and optimize candidate solvers and (b) evaluating these solvers on parallelized cloud compute (shown here with 10 CPUs).}
  \label{fig:bimodal-optimization}
\end{figure}

SAT solver development naturally yields a bimodal workload: the agent spends some time
writing, editing, and debugging code (low CPU usage,
inference-bound, largely sequential) and some time running
SAT solver evaluations (high CPU usage, compute-bound,
parallelizable).
To use compute efficiently, we design our hyperspecialization framework
around this division (\autoref{fig:bimodal-optimization}).
We run agents on dedicated machines with bounded compute
(i.e., a single core and limited memory), and instruct them
to submit candidate solvers to an external evaluation server
that runs parallel evaluations in the cloud.

For all experiments, we use GPT-5.6 Sol~\cite{gpt56sol} in
Codex with extra-high reasoning (see \S\ref{sec:agent-choice} for
details on this choice, and \S\ref{app:agent-prompt} for
the prompt).
The agent runs in an isolated Docker container in ``bypass approval''
mode using the default Codex harness.
We run all agents with broad internet access disabled.
While web search could be beneficial in some cases,
we found it too difficult to enable web search without
potentially leaking validation formula data (see
\S\ref{sec:app-alignment} for more).
Each agent run is allocated two hours of \emph{active time}---time
spent synthesizing and testing code, \emph{not} external evaluation
time---and a maximum of ten external evaluations. 

\subsection{When does hyperspecialization pay off?}

Hyperspecialization makes sense when the benefits outweigh the costs.
In our agent-based framework (\S\ref{sec:hyper-agent-based}),
hyperspecialization incurs two costs: money (i.e., LLM token spend)
and time (i.e., waiting for the agent to run). 
The potential benefit is that a faster solver will save time
in the future. 

If we synthesize a specialist that is faster than the general-purpose
baseline, and run that specialist on sufficiently many
instances, we will eventually gain back our upfront
specialization time. 
Running the cost-benefit analysis for monetary cost
is less straightforward, since it requires assigning
monetary value to a speedup.
\emph{How much would you pay for a 2$\times$ faster
solver for your workload?}
For some workloads and companies, the answer is
perhaps ``a lot!''

Our evaluation (\S\ref{sec:eval}) analyzes this cost-benefit
tradeoff in two regimes. 
In the \emph{time-insensitive} regime, specialists are built once and run over an
infinite future workload.
The specialization time is fully amortized, so we only
care about speedups compared to monetary cost.
In the \emph{time-sensitive} regime, specialists will run over a finite future workload,
and our goal is to minimize the total runtime over that workload:
is it faster to hyperspecialize or run an off-the-shelf solver?

\subsubsection{Cost in the time-sensitive regime}

To reason about tradeoffs in the time-sensitive regime,
consider a machine whose goal is to solve as many
instances as possible, and which has compute to devote to either
hyperspecialization or solving (with a generated specialist or an off-the-shelf general-purpose solver).
This machine has $C$ identical cores and an ordered workload $\mathcal{W} := \langle x_1, x_2, \dots, x_N\rangle $
where all $N$ instances are available at time zero.
A deployment policy $\pi$ specifies which computations to run,
when to run them, and how to schedule them on the available cores.
For example, a fixed policy runs the same solver on the next queued
instance whenever a core becomes available.
Other policies may
race multiple solvers or allocate cores to domain-specific
hyperspecialization.

For each policy, we measure its makespan
$M_{\pi}(\mathcal{W},C)$,
the elapsed time until all scheduled tasks have terminated, and its
verified solve count
$Q_{\pi}(\mathcal{W},C)$,
the total number of correct solutions.
Choosing between several policies is inherently a multi-objective problem: we want to both minimize the makespan and maximize the number of solved formulas. The space of policies, therefore, defines a Pareto frontier, where two policies on this frontier are not directly comparable: one is better in one metric and worse in the other.

For our evaluation of the time-sensitive
regime (\S\ref{sec:rq2}),
we make the simplifying observation that the shape of this problem
is analogous to that of comparing individual solver performance.
The best solver should run faster and solve more formulas within the timeout.
Therefore, we lift \textsf{PAR-2} score to the policy level:
$$
P_{\pi}(\mathcal{W},C) = M_{\pi}(\mathcal{W},C) + 2T(N - Q_{\pi}(\mathcal{W},C))
$$
A policy's score is the time it took to run (including hyperspecialization time, when applicable) plus a timeout penalty for every instance it failed to solve.
We discuss a Pareto frontier-based treatment of
optimality in \S\ref{sec:app-pareto-frontier}.

\section{Evaluation} \label{sec:eval}

We address the following broad research questions:

\begin{enumerate}[label=RQ\arabic*:,font=\bfseries]
\item How much does hyperspecialization improve performance compared
  to a general-purpose baseline?
\item When is hyperspecialization economically viable?
\end{enumerate}

We deploy domain-specific hyperspecialization at scale:
we create solvers for the 189 families in the GBD
across three different proof formats.
We evaluate the solvers' cost and performance relative
to a portfolio of four state-of-the-art general-purpose
solvers.
Cumulatively, our evaluation consumed approximately \$25k worth of LLM tokens (at API pricing) and over 5 CPU-years of solver evaluation.
We synthesized over 5,700 unique solvers and ran over 1.8 million solver invocations.

Next, we describe our evaluation setup (\S\ref{sec:eval-setup}),
and the answers to RQ1 (\S\ref{sec:rq1}) and RQ2 (\S\ref{sec:rq2});
\S\ref{sec:casestudies} answers more informal questions
about hyperspecialization.

\subsection{Evaluation setup}
\label{sec:eval-setup}

We describe our dataset (\S\ref{sec:eval-family}),
baseline solvers (\S\ref{sec:eval-baselines}),
the system environment (\S\ref{sec:eval-environment}),
and how we ensured task alignment (\S\ref{sec:cheating});
\S\ref{sec:checksat} already described the three
proof formats we use, with one independent agent per proof
format.

\subsubsection{Dataset}
\label{sec:eval-family}
We draw formula families from the GBD~\cite{gbd},
which consists of\footnote{At evaluation time.}
31,809 CNF formulas split across 189 different families.
These formulas have accumulated over 20 years of SAT
competitions, and vary in size from 2 to
9,842 instances.

\subsubsection{Baseline solvers}
\label{sec:eval-baselines}

We compare against the overall winner of each of the previous four SAT Competitions (\autoref{tbl:baseline-solvers}), which represents
each year's strongest sequential solver across a range of formulas. 
To make the comparison stronger, we also report a \emph{per-family virtual best} solver.
For each family $k$, we pick the baseline with the lowest training \textsf{PAR-2} and evaluate that choice on the validation set.
Given baselines $\Sigma = \{\sigma^{(1)}, \dots, \sigma^{(n)}\}$,
$$
P(\sigma^{\text{VBS}}, \mathcal{I}_k^{\text{val}}) := P(\underset{\sigma \in \Sigma}{\operatorname{argmin}}\ P(\sigma, \mathcal{I}_k^\text{train}), \mathcal{I}_k^{\text{val}})
$$
Across the 189 families in our dataset, each of the four baseline
solvers wins (and is thus selected in the VBS)
between 16.9\% and 32.0\% of the time (\autoref{tbl:baseline-solvers}).

\S\ref{sec:app-extended-rq1} extends our comparison to recent LLM-based SAT solver improvement frameworks: \work{AutoSAT}~\cite{autosat}, \work{AutoModSAT}~\cite{automodsat}, and \work{SolSearch}~\cite{solsearch}.
We do not compare to closed-source SATLUTION~\cite{satlution}.

\begin{table}
  \centering
  \caption{Overall winner from the previous four years of the SAT Competition, and percentage of time it appears in the VBS.}
  \label{tbl:baseline-solvers}
  \begin{tabular}{llr}
    \toprule
    Year & Solver & Baseline wins \\
    \midrule
    2023 & \work{SBVA-CaDiCaL}~\cite{sbva-cadical} & 16.9\% \\
    2024 & \work{Kissat-sc2024}~\cite{kissat-sc2024} & 32.0\% \\
    2025 & \work{AE-Kissat-MAB}~\cite{ae-kissat-mab} & 30.5\% \\
    2026 & \work{satsuma-iter+kissat}\tablefootnote{Proceedings of SAT Competition 2026 not yet available.} & 20.6\% \\
    \bottomrule
  \end{tabular}
\end{table}

\subsubsection{Environment}
\label{sec:eval-environment}
We keep the environment consistent across all experiments.
Solver evaluations use a timeout of 600 seconds,
a memory budget of 8 GB, and a verifier timeout of 6,000 seconds.
They run in Google Cloud Batch using preemptable spot instances
(\texttt{n2-highmem-2} with Intel Ice Lake, to match
the SAT Competition 2026 evaluation environment).
Each task is allocated 1 vCPU and 8 GB RAM, with two
tasks per instance; this is uniform across all solvers.
Coding agents run on an AMD EPYC 9454P
processor with 256 GB of RAM.

\subsubsection{Task alignment}
\label{sec:cheating}
Agents deployed at scale can score well without faithfully solving
the task. We refined the task description and environment beforehand
using \work{Docent}~\cite{docent} to surface avenues for cheating in
pilot runs (\S\ref{sec:app-alignment-pre}).
Afterward, we audited every agent transcript and generated solver for
compliance (\S\ref{sec:app-alignment-post}), found seven invalid runs,
and reran them; a second audit of those seven found no further violations.
All results in this paper refer to valid runs.

\subsection{RQ1: How much does hyperspecialization improve performance?}
\label{sec:rq1}

\begin{figure*}
  \centering
  \includegraphics[width=\textwidth]{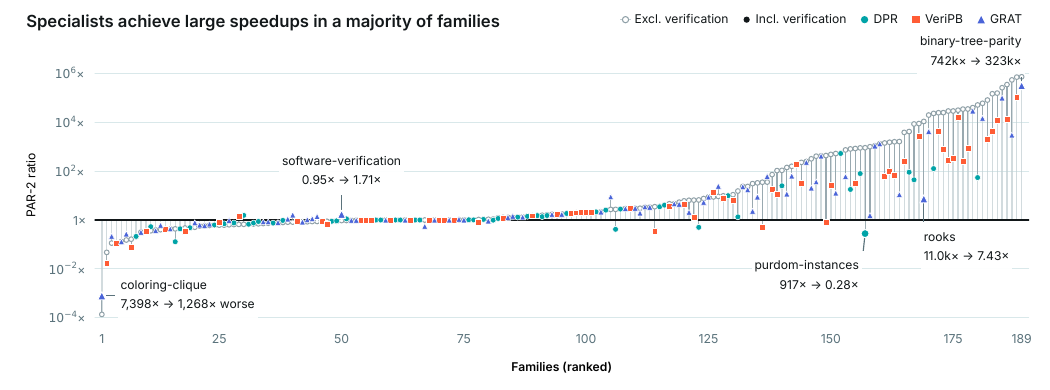}
  \caption{Held-out validation score ratios (baseline / specialist) against the per-family virtual-best of four baseline solvers. Gray stems and open circles exclude verification; colored markers add verification time to both solvers while retaining \textsf{PAR-2} penalties. Both solvers are selected using training \textsf{PAR-2}. Families are ranked by the ratio excluding verification; values above $1\times$ favor specialization. Color and marker shape identify the specialist's proof system. Callouts show speedup excluding $\rightarrow$ including verification time.}
  \label{fig:rq1-solver-speedup}
\end{figure*}

For every family in the dataset, we select the best specialized
solver (across all three proof systems) by evaluating performance
on the training set.
We compute the baseline-to-specialist \textsf{PAR-2} ratio, defined
below, on the validation set, comparing each specialist with the
per-family virtual-best of the four baseline solvers.
\autoref{fig:rq1-solver-speedup} shows this ratio both excluding and including verification time.

\customsection{\textsf{PAR-2} ratio}
We compare the performance of two solvers $\sigma_a$ and $\sigma_b$
using the log \textsf{PAR-2} ratio:
$$
\Delta_{P}(\sigma_a, \sigma_b, \mathcal{I}) := \log_{10} \frac{P(\sigma_b, \mathcal{I})}{P(\sigma_a, \mathcal{I})}
$$
We use a ratio because we care about improvements relative
to $\sigma_b$'s score, rather than absolute improvement
in \textsf{PAR-2}.
The logarithmic scale makes improvements and regressions symmetric
around zero: a positive value $\Delta_{P} > 0$ indicates that
$\sigma_a$ has a better (smaller) \textsf{PAR-2} score, and a negative
value $\Delta_{P} < 0$ indicates that $\sigma_b$ has a better
(smaller) \textsf{PAR-2} score.

\customsection{Speedups}
Agents constructed hyperspecialized solvers that yielded
significant speedups over the baseline solvers.
Excluding verification time, specialization produced a faster
solver in 121/189 families (64.0\%), and produced a solver that
had more than a $1{,}000\times$ reduction in \textsf{PAR-2}
score for 32/189 families (16.9\%).
The geometric mean ratio was a $12.0\times$ reduction in \textsf{PAR-2} score.

Including verification time in both baseline and specialist
scores reduces the geometric mean ratio to $5.03\times$. %
The specialist still outperforms the baseline on 121/189 families:
nine families gain an advantage and nine lose it.
The effect varies substantially by family.
For \texttt{binary-\allowbreak tree-\allowbreak parity},
for example, the ratio remains approximately
$323{,}000\times$.
For \texttt{purdom-\allowbreak instances},
it falls from $917\times$ to $0.278\times$.
In this case, the specialist found a cheap way to generate an expensive-to-check proof.

\customsection{Failure cases}
When agents were not able to discover optimized specialists,
their performance still almost always approached baseline performance.
Specialists had worse than a $10\times$ increase in \textsf{PAR-2} score
on only two families. 
The worst such family is \texttt{coloring-clique},
where the specialist is $10^4\times$ worse.
The GBD includes one shuffled formula---i.e., an instance
with renamed variables and reordered clauses---in
this family, alongside ten structurally distinct
\texttt{coloring-\allowbreak clique} instances.
The shuffled formula was randomly allocated to
the validation set with no counterpart in training;
the agent never observed it, and thus could not build
a solver that accounted for shuffling.
Shuffling itself is not the issue: many other families
on which specialists perform well
contain shuffled variants.
When the shuffled instance is removed from the
\texttt{coloring-clique} validation set, the specialist
outperforms the baseline
($14\times$, $2\times$ accounting for verification time).

\customsection{Evaluating performance transfer}
Specialist performance on a family should still transfer to unseen instances drawn from the same family.
To measure this, we plot the
correlation between solvers' \textsf{PAR-2}
scores on the training set and validation
set.
Highly correlated scores indicate that specialist
performance transfers, while higher \textsf{PAR-2} scores
(i.e., worse performance) on the validation set
(vs. the training set)
imply overfitting to the training set. 

\begin{figure}
  \begin{center}
    \resizebox{\linewidth}{!}{\includegraphics{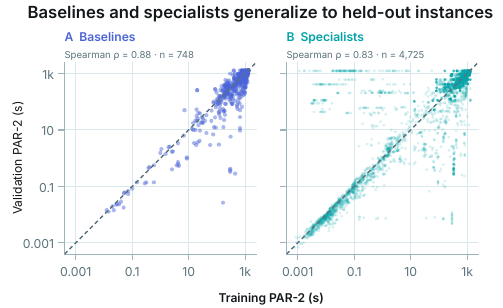}}
  \end{center}
  \caption{Training and validation \textsf{PAR-2} for baseline solvers (A) and all generated specialists (B), with pooled Spearman correlations.}
  \label{fig:rq1-train-val-correlation}
\end{figure}

\autoref{fig:rq1-train-val-correlation} shows
  baseline solver training/validation correlation in the left
  panel (A) as a reference; note that the baseline solvers,
  unlike the specialists, were not actually trained.
  We plot a point at 
$$
(P(\sigma,\mathcal{I}_k^\text{train}),\ P(\sigma,\mathcal{I}_k^{\text{val}}))
$$
for every baseline solver $\sigma \in \Sigma$ and family $k$.
In the right panel (B), we plot the performance of every specialized solver synthesized throughout all of the agent runs.
We plot
$$
(P(\sigma_k^{(i)},\mathcal{I}_k^\text{train}),\ P(\sigma_k^{(i)},\mathcal{I}_k^{\text{val}}))
$$
for every family $k$ and every candidate solver $\sigma_k^{(i)}$ produced during domain-specific hyperspecialization for family $k$.
Training and validation \textsf{PAR-2} were highly rank-correlated both for specialists (Spearman $\rho = 0.83$) and baselines ($\rho = 0.88$).

While specialists were more varied in performance than
baselines (median absolute deviation
of $1.27\times$ vs. $1.16\times$),
they were only \emph{slightly} more biased towards training performance.
The median ratio of validation \textsf{PAR-2}
to training \textsf{PAR-2} was $1.01$ for specialists
(compared to $1.00$ for baselines). 
$6.0\%$ of submissions had a validation/training ratio
above 10, compared to $4.6\%$ of submissions with
validation/training ratio below 0.1.
The near-symmetry of these measures suggests variance
within a family as opposed to overfitting.

\subsection{RQ2: When is hyperspecialization economically viable?}
\label{sec:rq2}

We evaluate specialization in both the \emph{time-insensitive} regime (\S\ref{sec:rq2-time-insensitive}) and the \emph{time-sensitive} regime (\S\ref{sec:rq2-time-sensitive}).

\subsubsection{Time-insensitive regime}
\label{sec:rq2-time-insensitive}
\begin{figure}
  \centering
  \resizebox{\linewidth}{!}{\includegraphics{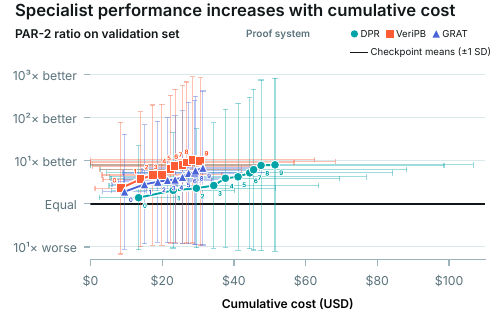}}
  \caption{Validation speedup of specialized solvers relative to cumulative construction cost. Each point shows the average performance of the $i$th submitted solver across all agent runs (separately for each evaluated proof format).}
  \label{fig:rq2-validation-speedup}
\end{figure}

We consider the case where a specialized solver will be reused
many times---or infinitely!---making money the driving cost.
In \autoref{fig:rq2-validation-speedup}, we compare how the improved performance of a candidate specialized solver relates to its cumulative construction cost. For every submission index $i$, we plot the average \textsf{PAR-2} validation ratio over the baselines (same as RQ1) and the cumulative cost (LLM usage + cloud compute) of the candidate solver $\sigma_k^{(i)}$.

Hyperspecialization potential was predominantly influenced by family, not cumulative specialization effort.
On average, even the agents' \emph{first} solver
submissions ($\sigma_k^{(1)}$) were faster than baselines.
The average time to the first submission was just 28.1 minutes at a total cost of \$10.63,
yielding a $1.71\times$ improvement over baseline.
Full runs (limited to 2 hours of active time and 10 evaluations) took on average 1.5 hours at a cost of \$37, yielding a $4.40\times$ improvement.\footnote{This includes submissions from all three proof formats (not just the best one) thus is lower than the average speedup in RQ1.}

The agent accounted for a majority of the total cost (91.4\%), yet a minority of the overall runtime (26.9\%).
Additionally, we observed that among the three proof formats, agents had the most success with \work{VeriPB}, on average yielding slightly more performant solver submissions than with \work{GRAT} at a slightly cheaper average cost. On the other hand, \work{DPR} runs were on average more expensive than both \work{GRAT} and \work{VeriPB} runs for unknown reasons.

\subsubsection{Time-sensitive regime}
\label{sec:rq2-time-sensitive}

\begin{figure*}
  \centering
  \resizebox{\linewidth}{!}{\includegraphics{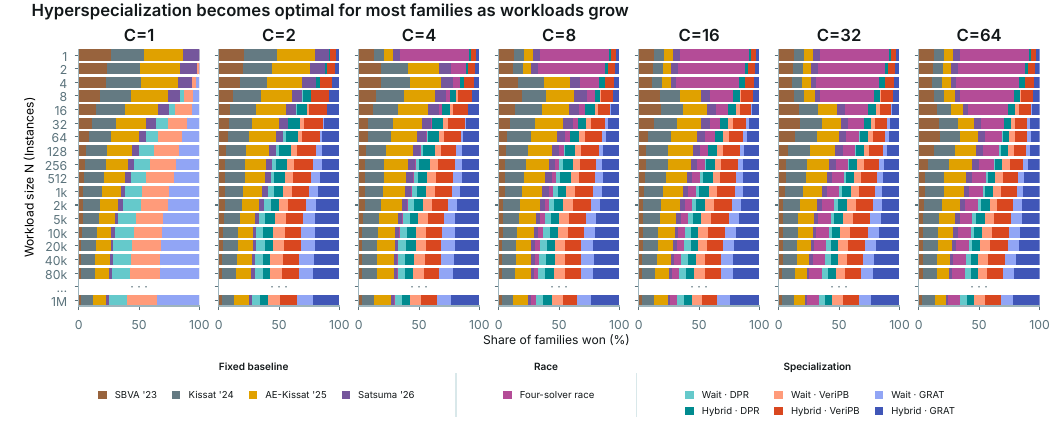}}
  \caption{Share of optimal on-demand policies across simulated workloads. Each horizontal bar specifies a simulated machine with $C$ cores and a workload size of $N$. The bar is colored to represent the share of families for which a given on-demand policy $\pi$ was the optimal policy $\pi^*_{C,N,k}$.}
  \label{fig:on-demand-policy}
\end{figure*}

When there does not yet exist a specialized solver for a particular workload, we must decide whether to attempt
hyperspecialization or just run an existing general-purpose solver.
Here, there is an explicit tradeoff between the amount of compute and time allocated for specialization vs. solving.
We evaluate eleven different scenarios. 

\customsection{Fixed baseline policy (4 instances)} A fixed baseline policy runs the same solver on the next workload instance whenever a core is available. We evaluate a fixed policy separately for each of the four baseline solvers.

\customsection{Four-solver race policy (1 instance)} We evaluate a four-solver race where each of the fixed baselines is deployed in parallel for a single workload instance whenever four cores are free
(which requires $C \geq 4$).
All instances terminate whenever the first solver returns a verified solution.

\customsection{Specialization policy (6 instances)} We evaluate two hyperspecialization deployment policies for each of the three proof formats. In \textbf{wait}, all cores are used for hyperspecialization. We replay the agent and evaluation compute workload based on the available parallelism. Once hyperspecialization completes, the best checkpoint is served on all cores.
In \textbf{hybrid}, hyperspecialization runs until the first solver is submitted. Then, half the cores immediately serve that solver, while the remaining cores continue hyperspecialization. When a new, better solver is produced, the serving cores switch to it. Once specialization completes, all cores serve the best solver.

We performed a
discrete-event simulation of each of these runtime policies. For every core count $C \in \{1,2,4,8,16,32,64\}$, workload size $N \in \{1,2,4,\dots,80\,k,1\,M\}$, and family $k$, we sample $r$ random workloads of size $N$ (with replacement) from the validation set $\mathcal{I}_k^{\text{val}}$, producing $\mathcal{W}_k^{(1)},\dots,\mathcal{W}_k^{(r)}$. We simulate each policy $\pi$ on each of these representative workloads and compute the optimal policy for family $k$ in configuration $(C,N)$ as the policy with the lowest total score:
$$
\pi_{C,N,k}^{*} := \underset{\pi \in \Pi}{\operatorname{argmin}} \sum_{i=1}^r P_{\pi}(\mathcal{W}_k^{(i)},C)
$$

We visualize results in \autoref{fig:on-demand-policy}. Each horizontal bar denotes a particular number of cores $C$ (x-axis) and a particular workload size $N$ (y-axis). The bar is colored based on what share of families a given policy $\pi$ is the optimal policy for the configuration $(C,N)$. This visualization allows us to see how the distribution of optimal strategies changes as we change the machine size ($C$) and the workload size ($N$).

\customsection{Results}
For small workloads, running fixed baseline solvers directly is most frequently the optimal strategy as these solvers incur no upfront cost. In regimes where the workload undersaturates the available compute (i.e. $N < C$), the four-solver race is frequently dominant, as it provides a mechanism to use this ``free'' compute.

Across all machine sizes, as the workload size increases, hyperspecialization increases in viability. On average, \work{GRAT} specialists fare better than both \work{VeriPB} and \work{DPR}. Within each format, the hybrid form of the strategy (where available specialists are immediately deployed) is optimal on more families than the wait form.
Once the workload size exceeds $N=2,000$, some form of hyperspecialization is the optimal strategy on a majority of families across all machine sizes.

\section{Case studies} \label{sec:casestudies}

This section describes four case studies:
classifying the strategies used by generated solvers,
deploying specialists in real systems
that use solvers, measuring how specialists for one
family perform on another, and exploring
how sensitive specialists are to
cardinality encodings (described later). 

\subsection{Solver taxonomy}
\label{sec:solver-taxonomy}

We surveyed the entire set of 567 training-selected specialists across the 189 families,
using LLM-based techniques to analyze and classify the nearly 500,000 lines of
generated solver code.
This section gives broad insights into specialists' strategies, while
\S\ref{sec:app-solver-taxonomy} describes our method
and its limitations; in 
\S\ref{sec:app-gallery}, we provide concise, LLM-generated descriptions of every
family and solver.

\customsection{Results}
Specialists did not compete with generalists by
developing new algorithms; rather, they sampled from the
vast SAT and algorithms literature to solve the problem at hand,
and implemented their solutions with systems-level optimizations. 

Generated solvers were remarkably varied and family-specific.
Our audit found at least forty unique solver strategies:
\work{CDCL}, the backbone of almost all modern solvers,
was the second most common strategy (57.5\%),
after constructing witnesses or deriving proofs directly from problem structure (60.7\%).
Only 16.2\% of solvers deployed
\work{CDCL} on its own,
and 81.1\% of solvers used more than one strategy (\autoref{fig:rq3-algorithms}).

Specialists frequently recovered high-level structure
from input formulas (\autoref{fig:rq3-representation})---for example,
graphs (41.3\%), Bool\-ean circuits (38.6\%), and
cardinality constraints (27.9\%)---and
97.2\% of specialists included some check excluding
formulas that did not fit the expected structure
(\autoref{fig:rq3-assumptions}).
Recovered structure also informed orchestration
for multi-strategy solvers:
93.3\% of solvers selected solving procedures based on
structure, and 60.1\% ran distinct methods sequentially (\autoref{fig:rq3-coordination}).

Specialists used a wide range of known
heuristics and techniques (e.g., XOR reasoning).
They also optimized beyond just
algorithmic and search speedups,
using an array of low-level systems techniques like memory preallocation (96.5\%), specialized parsing (66.0\%), buffered proof logging (45.1\%), going so far as to add branch-prediction hints (1.1\%) and SIMD-accelerated code (1.8\%) (\autoref{fig:rq3-systems}).

\subsection{End-to-end deployment} \label{sec:endtoend}
\begin{figure}
  \centering
  \resizebox{\linewidth}{!}{\includegraphics{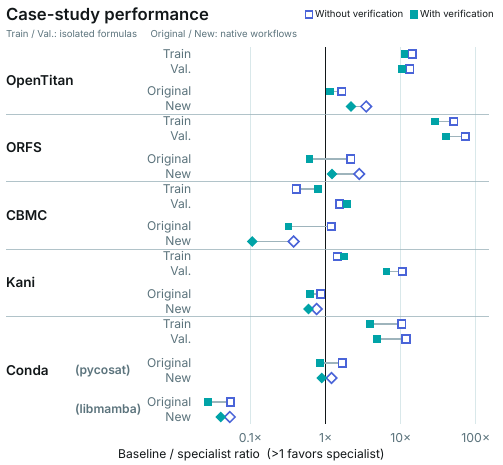}}
  \caption{Specialists' redeployment performance across five real-world systems. Formulas were extracted from a target and used to develop a specialist; Train/Val. show specialist speedup over baselines in isolation on these representative formulas. We then redeploy the specialist in the real-world system and measure end-to-end performance compared to the native tool on the original target for which formulas were extracted (original) and a new target (new). For \work{Conda}, we evaluate two different native modes: \work{pycosat} and \work{libmamba}.}
  \label{fig:end-to-end}
\end{figure}

We ask whether applying hyperspecialization to
real-world systems results in solvers competitive
with native performance. 
We isolated SAT formulas from five real systems, synthesized a specialist, and then
evaluated the specialist in a simulated deployment
on new targets.
\autoref{fig:end-to-end} presents the results compared to
native baselines for circuit equivalence checking
(\texttt{OpenTitan}~\cite{opentitan} and \texttt{ORFS}~\cite{orfs}),
verification (\texttt{CBMC}~\cite{cbmc} and \texttt{Kani}~\cite{kani}),
and package dependency resolution (\texttt{Conda}~\cite{libmamba});
see \S\ref{sec:app-end-to-end} for details.

\customsection{Results}
For two of our five systems (\texttt{OpenTitan} and \texttt{ORFS}),
specialists outperformed native solutions
on novel targets \emph{even when accounting for proof verification time}---something native solutions did not even attempt.

Deploying specialists was challenging for several non-obvious
reasons, though:
(1) real-world systems often implicitly trust solvers and run
them without proof-logging, while our specialists always spend
time producing proofs;
(2) some systems (e.g., \texttt{CBMC} and \texttt{Kani}) optimize
by running native solvers in incremental mode, a mode our
specialists do not yet (but could!) support;
(3) some systems (e.g., \texttt{Conda}) use
custom, higher-level (and thus faster)
solvers as alternatives to SAT.
One potential direction for future work is synthesizing
solvers for a given family and a given application
(e.g., a verification-specific, incremental solver for
\texttt{Kani}).

\subsection{Cross-family sensitivity and performance}

\S\ref{sec:rq1} measures solvers only on their target validation set.
Here, we run the top per-family selected specialist on the
validation set of every \emph{other} family (\autoref{fig:cross-family-matrix}).

\customsection{Results}
Most specialists were very selective, refusing to run (i.e. quickly returning \texttt{UNKNOWN}) on families outside of the target domain: only 27 solvers attempted to run on more than 50\% of other families.
Targeted specialists were also often the \emph{best} solver for their
family: in 134 of the 189 cases (70.9\%), the targeted specialist won its own family.

Specialists were also robust. Only seven source/target combinations resulted in solutions rejected by the verifier.
Six rejections (all on \texttt{multiplier-verification}) were false positives,
caused by a bug in the DPR elaborator \texttt{dpr-trim}\footnote{The elaborator runs before the formally verified part of the checker.}
that mishandled clauses with repeated literals.
We have reported this bug with a patch.
Only one rejection was legitimate.
The \texttt{fpga-routing} solver read clause lines into a fixed 64 KiB buffer with \texttt{fgets}.
It correctly preserved unfinished clauses across multiple
reads, but not unfinished literals: it dropped
minus signs that fell at a buffer boundary,
converting negative literals into positive ones. 
Triggering this bug required a clause line exceeding 64 KiB with a negative
literal precisely at the boundary.

\subsection{Cardinality constraint encodings}

Many high-level problems define cardinality constraints,
or bounds related to counting
(e.g., ``exactly two of these variables must be true'').
These constraints can be encoded into Boolean logic in a variety of ways.
For general-purpose solvers, the specific choice of encoding
is an active area of research~\cite{card_sequential,card_sorting,card_sorting_2,card_network,card_network_2,card_totalizer,card_modulo_totalizer,card_km_totalizer,reeves2025cardinality},
because it plays a significant role in solve time.
Fewer clauses and/or variables, for example,
often correlate with faster solve times.
Here, we test if the same is true under hyperspecialization.
We evaluated baseline and specialist performance across three crafted
benchmarks---chosen because they used many cardinality constraints
and covered both SAT and UNSAT---and up to seven cardinality
encoding methods;
see \S\ref{sec:app-cardinality} for more.

\begin{figure}
  \centering
  \resizebox{\linewidth}{!}{\includegraphics{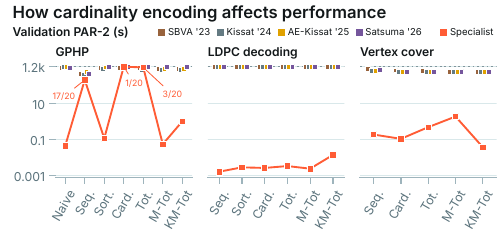}}
  \caption{Average validation \textsf{PAR-2} of baselines and specialists for various cardinality encodings.}
  \label{fig:cardinality-encoding}
\end{figure}

\customsection{Results}
\autoref{fig:cardinality-encoding} shows
the performance of baselines and specialists.\footnote{All using \work{GRAT}, so specialists must construct resolution-based proofs and cannot simply map the cardinality constraints into algebraic proofs.}
Unlike general-purpose solvers,\footnote{The general-purpose encoding benefit is more visible in \S\ref{sec:app-cardinality}.}
specialists do not benefit from more compact encodings;
in fact, for one of the benchmarks, the naive (pairwise) cardinality encoding yielded the \emph{fastest} specialist (yet the worst
baseline performance).
We hypothesize that it is more important for a
formula to be encoded in a conceptually \emph{clear} way
for the hyperspecialist, as opposed to a \emph{compact}---but
perhaps convoluted---way.
Compactness helps general-purpose solvers because it plays nicely
with those solvers' fixed internal heuristics.
In contrast, the specialist chooses its \emph{own} internal heuristics---in many
cases extracting and reasoning about cardinality constraints at a high
level.
An optimized encoding can stymie the specialist who's trying to reconstruct
a higher-level problem.

\section{Discussion} \label{sec:discussion}

This section considers the ramifications of hyperspecialization
for SAT and beyond.

\customsection{Hyperspecialization and SAT}
Hyperspecialization (we hope!) makes it easier for SAT research to have practical impact.
In the past, a technique benefiting a small set of formulas
was difficult to justify maintaining in a mainline solver.  %
Now, if the technique is useful for a specific workload,
an agent can discover and implement it in a specialist.

Hyperspecialization may also open up new research avenues.
For example, a CDCL solver's reasoning maps cleanly to
DRAT proofs; it doesn't need the richness
of formats like VeriPB.
Specialist solvers \emph{can} take advantage of
more expressive reasoning, though---so the 
bottleneck becomes proof verification time. 
Can we design formats that speed up verification
with specialist-authored hints?
What does a proof format designed for hyperspecialists
look like? 

Finally, we may consider the SAT solver within
the context of the embedding application.
\S\ref{sec:endtoend} describes how synthesizing a specialist
from workload examples alone may not be enough to
beat native speeds.
The synthesis step could improve performance by
accounting for information about the embedding application.

\customsection{Hyperspecialization beyond SAT}
The most obvious hyperspecialization domains beyond SAT
are its close relatives that also support verification: SMT~\cite{cert_smt}, MaxSAT~\cite{cert_maxsat}, pseudo-Boolean optimization~\cite{cert_pb},
finite-domain constraint solving~\cite{cert_cp}, QBF~\cite{cert_qbf}, certified model counting~\cite{cert_counting},
and more.
Hyperspecialization should work more generally, though:
we claim that it applies to any problem with a given shape
(\S\ref{sec:specable}).
We now provide three possibilities in rough order of audacity.

A lossless compressor sees many types of inputs
(e.g., text or audio), and produces a small archive that
decompresses back to the original.
There already exist specialized compressors that
use domain knowledge to achieve better compression ratios
than general-purpose compressors~\cite{wei2021logreducer,pelkonen2015gorilla,afroozeh2023alp,chandak2019spring,horn2017lepton}.
Automated hyperspecialization may allow us to tailor
compressors to other workloads---or even individual inputs!
The correctness check is simply decompressing the output
and comparing the result with the input.

Compilers compile many types of programs, and understanding
program semantics often improves performance of the
generated binary.
Could we build hyperspecialized compilers for certain classes
of programs, using translation
validation~\cite{pnueli1998translation,alive2} to check output?
What about web browsers? They serve many different web pages, and
today, behemoth applications wring new drips of
performance out of JavaScript engines and renderers with
each new release. 
What would per-workload browsers look like, and how might
we check them?
More broadly, in a world where untrusted authors can specialize
(or ``specialize'') anything, how might we re-imagine
checking?

\section*{Generative AI disclosure}

We used generative AI to assist with various aspects of this research. We used coding agents (primarily GPT-5.6 Sol in Codex, but also other models in Cursor and Claude Code) to help with the development of the SAT hyperspecialization framework, evaluation pipelines, and case studies.
Our core experiment tests how well coding agents can synthesize SAT solvers and thus naturally uses generative AI.
During evaluation, we used LLMs (GPT-6 Astra and GPT-5.6 Luna) to build taxonomies of solver behavior (detailed in \S\ref{sec:app-solver-taxonomy}).
In the preparation of this manuscript, we used coding agents (GPT-5.6 Sol / GPT-6 Astra) to help create graphs and figures.
Except for three clearly labeled sections in the appendix (\S\ref{sec:app-gallery}, \S\ref{sec:app-full-taxonomy}, and \S\ref{sec:app-end-to-end-case-studies}),
the words in this paper were written by humans.

\bibliographystyle{ACM-Reference-Format}
\bibliography{references}

\appendix

\newpage
\onecolumn
\section{Agent prompt}
\label{app:agent-prompt}

Each agent is prompted with the core task prompt (\S\ref{sec:app-prompt-task}) plus a proof instruction extension depending on the selected proof format: \work{DPR}: \S\ref{sec:app-prompt-dpr}, \work{GRAT}: \S\ref{sec:app-prompt-grat}, or \work{VeriPB}: \S\ref{sec:app-prompt-veripb}. In addition, we also provide condensed documentation for each of the formats, mounted in the container at \texttt{/opt/autospec/docs} (not shown here).

\subsection{Task definition}
\label{sec:app-prompt-task}
\tcbinputlisting{
  promptlisting,
  listing file={prompts/system.md}
}

\subsection{DPR extension}
\label{sec:app-prompt-dpr}
\tcbinputlisting{
  promptlisting,
  listing file={prompts/dpr.md}
}

\subsection{GRAT extension}
\label{sec:app-prompt-grat}
\tcbinputlisting{
  promptlisting,
  listing file={prompts/grat.md}
}

\subsection{VeriPB extension}
\label{sec:app-prompt-veripb}
\tcbinputlisting{
  promptlisting,
  listing file={prompts/veripb.md}
}

\newpage
\twocolumn

\section{Extended performance results}
\label{sec:app-extended-rq1}

In the main body of RQ1 (\S\ref{sec:rq1}) we evaluate the top specialist against the per-family VBS of the four core baselines. This serves as a sort of lower bound on the potential of hyperspecialization: \emph{How effective are the best specialists compared to a strong baseline?}

\begin{figure*}
  \centering
  \resizebox{\linewidth}{!}{\includegraphics{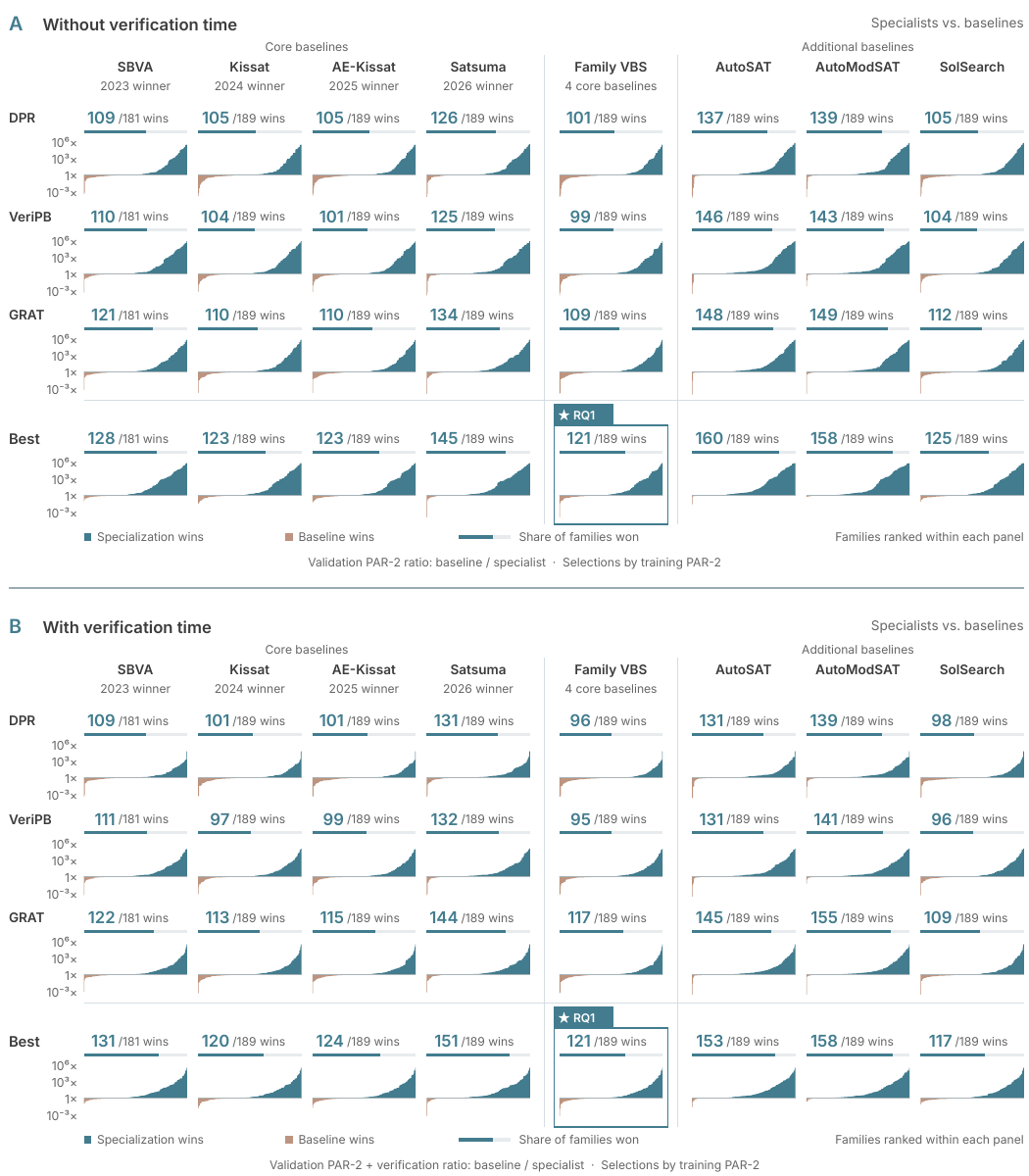}}
  \caption{Full head-to-head per-family performance between specialists (rows) and baselines (columns). (A) Validation PAR-2 without verification time. (B) Validation PAR-2 including verification time, using the same training-selected solvers. Each subplot ranks families by the baseline/specialist PAR-2 ratio. Labels denote families where the specialist outperforms the baseline. The two specific subplots marked RQ1 denote the data highlighted in the main body of the paper.}
  \label{fig:app-full-data}
\end{figure*}

Here we include a complete enumeration of head-to-head results, visualized in a similar format. \autoref{fig:app-full-data} shows the per-family head-to-head performance of specialists constructed for every format (\work{DPR}, \work{VeriPB}, \work{GRAT}) as well as the best specialist compared to: each of the four core baselines individually, the per-family virtual-best of the four core baselines, and three additional baselines from related work in LLM-based solver synthesis: \work{AutoSAT}~\cite{autosat}, \work{AutoModSAT}~\cite{automodsat}, and \work{SolSearch}~\cite{solsearch}.

\clearpage

\section{Pareto frontier of on-demand specialization}
\label{sec:app-pareto-frontier}

Developing an optimal runtime policy is inherently a multi-objective optimization problem. We want to both minimize the makespan ($M_\pi$) and maximize the number of solved formulas ($Q_\pi$). Picking an optimal policy therefore requires determining the relationship between these two metrics: \emph{Do we care more about solving quickly or solving every formula?}

In RQ2 (\S\ref{sec:rq2}), we define optimality by a combined metric $P_\pi$ where an unsolved formula is equivalent to a penalty of $2T$ seconds. This formula is not unreasonable (indeed it mirrors \textsf{PAR-2} which is widely used to measure SAT solver performance), but it is \emph{arbitrary} in the sense that we could have decided an unsolved formula was worth more or less.

\begin{figure*}
  \centering
  \resizebox{\linewidth}{!}{\includegraphics{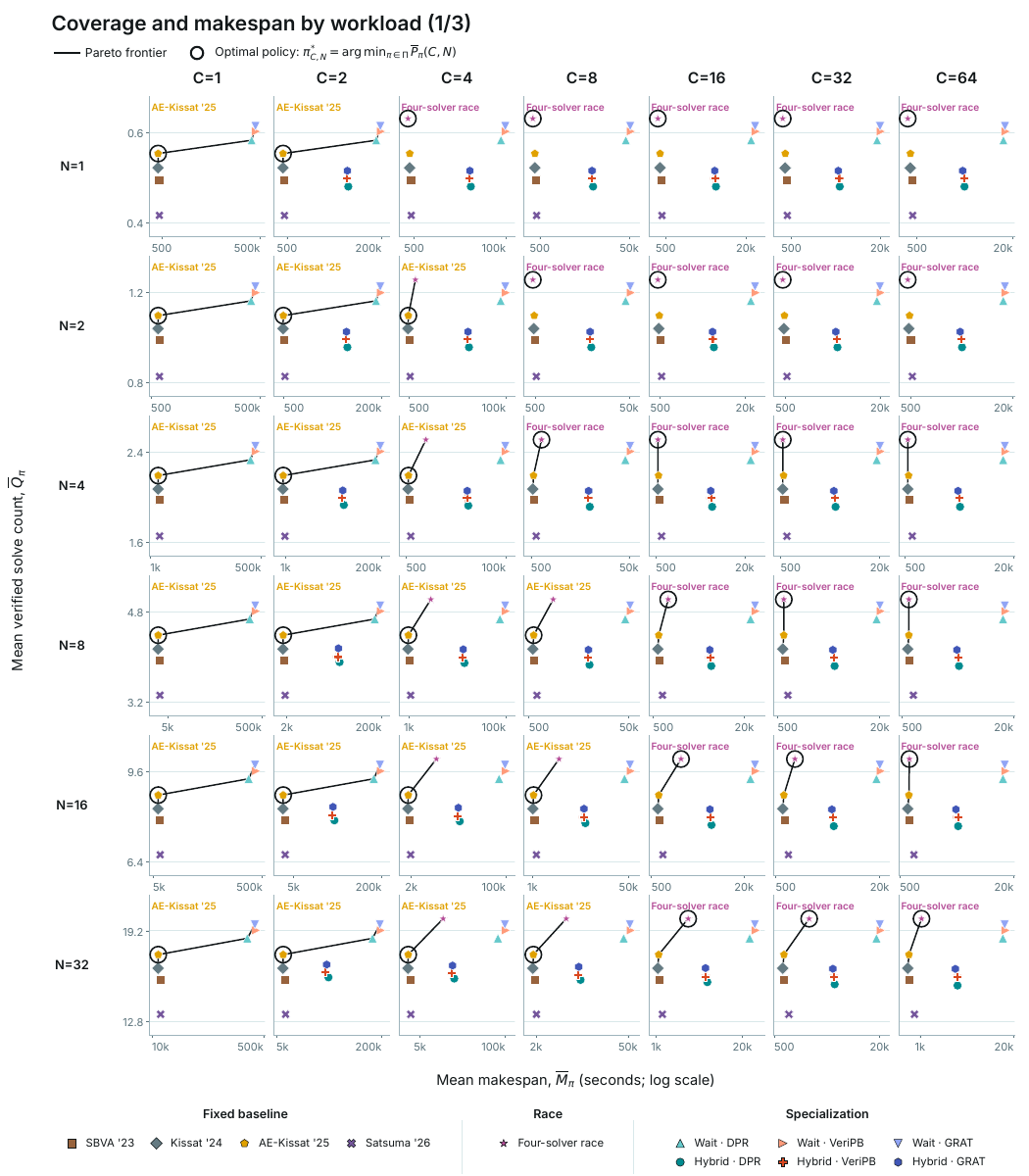}}
  \caption{Pareto frontier of on-demand runtime strategies. N=1 to N=32.}
  \label{fig:app-pareto-1}
\end{figure*}

\begin{figure*}
  \centering
  \resizebox{\linewidth}{!}{\includegraphics{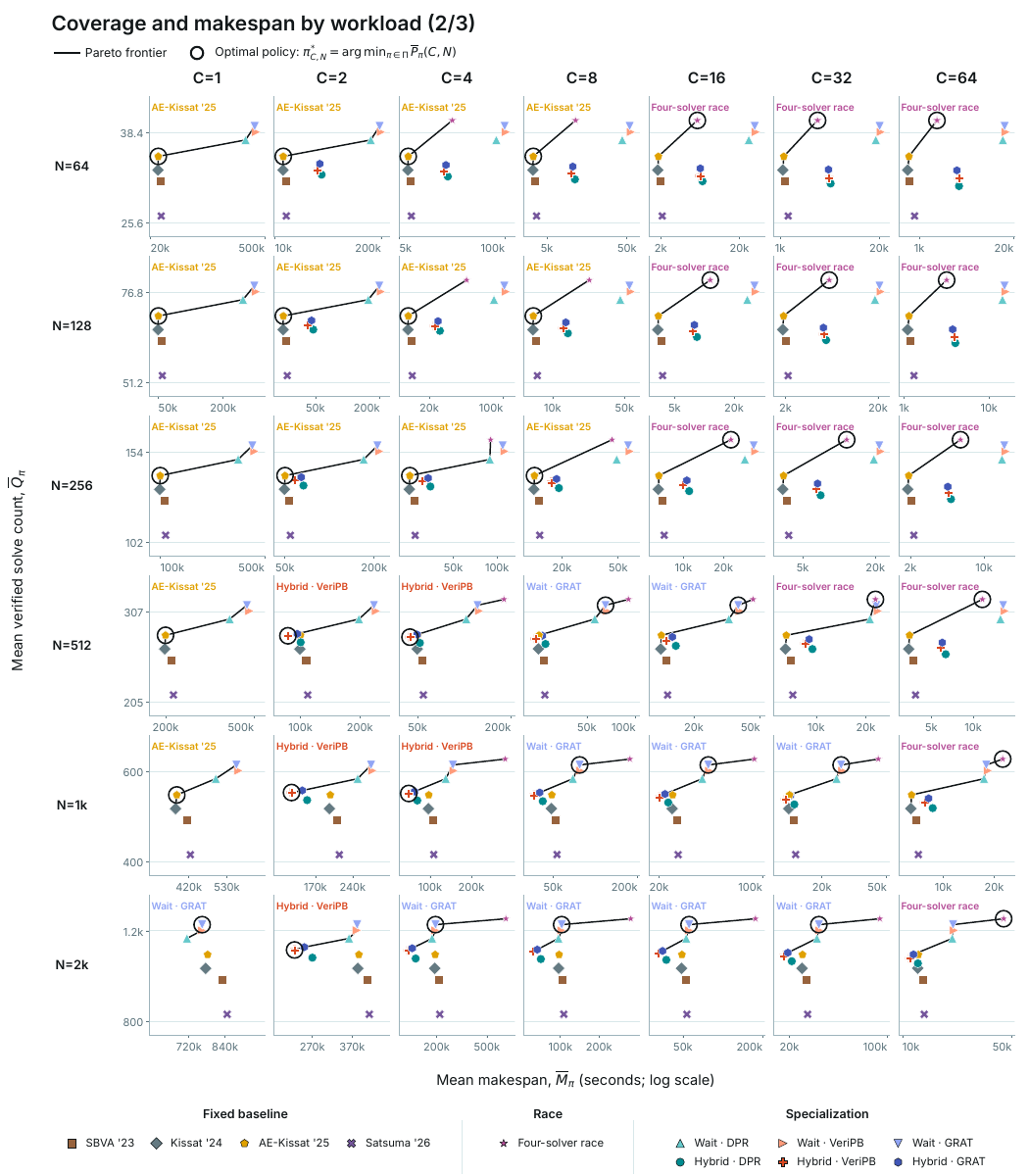}}
  \caption{Pareto frontier of on-demand runtime strategies. N=64 to N=2k.}
  \label{fig:app-pareto-2}
\end{figure*}

\begin{figure*}
  \centering
  \resizebox{\linewidth}{!}{\includegraphics{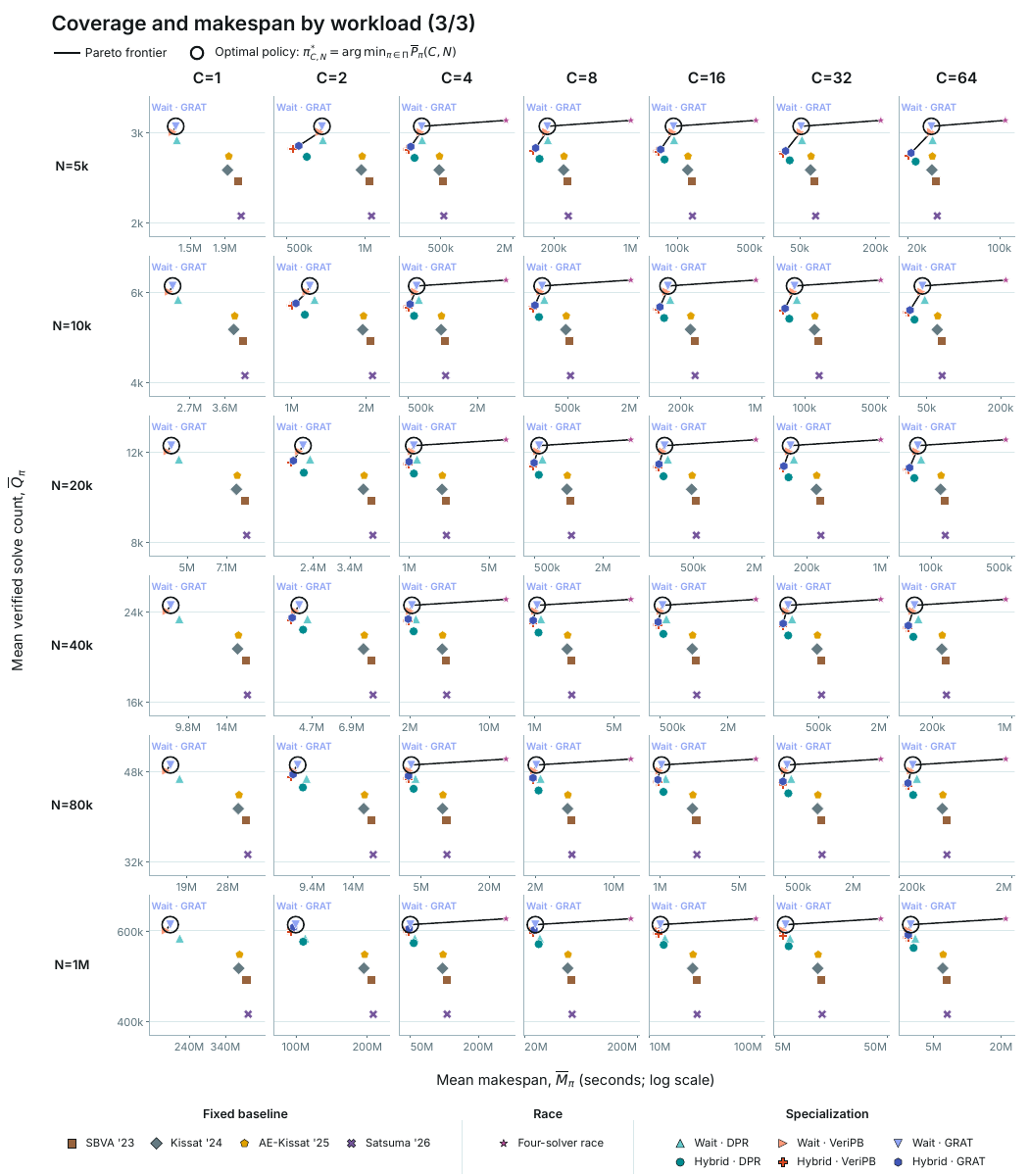}}
  \caption{Pareto frontier of on-demand runtime strategies. N=5k to N=1M.}
  \label{fig:app-pareto-3}
\end{figure*}

Here, we visualize these on-demand runtime policy results without imposing a combined metric. Instead, for each machine size ($C$) and workload size ($N$), we plot the Pareto frontier of makespan and coverage tradeoffs.
Each subplot in \autoref{fig:app-pareto-1}, \autoref{fig:app-pareto-2}, and \autoref{fig:app-pareto-3} plots, for a particular $(C,N)$ configuration, the average makespan $\overline{M}_\pi$ and average coverage $\overline{Q}_\pi$ for each runtime policy $\pi$ computed over all family workloads of size $N$:
\begin{align*}
  \overline{M}_\pi &:= \frac{1}{r|K|} \sum_{k \in K} \sum_{i=1}^{r} M_\pi(\mathcal{W}_k^{(i)}, C) \\
  \overline{Q}_\pi &:= \frac{1}{r|K|} \sum_{k \in K} \sum_{i=1}^{r} Q_\pi(\mathcal{W}_k^{(i)}, C) \\
\end{align*}

In each subplot, we draw a line between runtime policies that sit on the Pareto frontier (i.e. no other policy has both a smaller makespan and larger coverage). Additionally, we circle the point with the optimal average \textsf{PAR-2} score computed as:
$$
\overline{P}_\pi := \overline{M}_\pi + 2T(N - \overline{Q}_\pi)
$$

Note that since results here are averaged over every family, the optimal policy selected here is not necessarily the same as the most frequently optimal policy selected on a per-family basis (as visualized in \autoref{fig:on-demand-policy}).

\clearpage

\section{Extended: Methods and techniques}
\label{sec:app-rq3}

\begin{figure}
  \centering
  \resizebox{\linewidth}{!}{\includegraphics{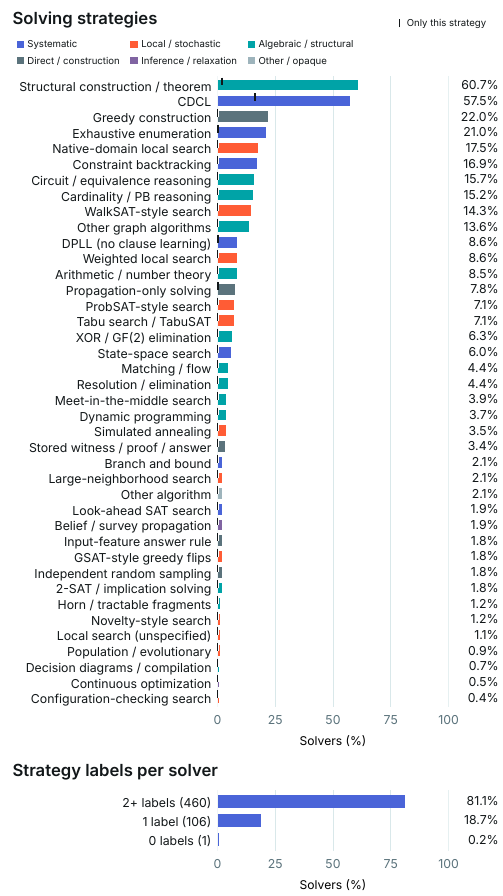}}
  \caption{What algorithms do specialists use?}
  \label{fig:rq3-algorithms}
\end{figure}

\begin{figure}
  \centering
  \resizebox{\linewidth}{!}{\includegraphics{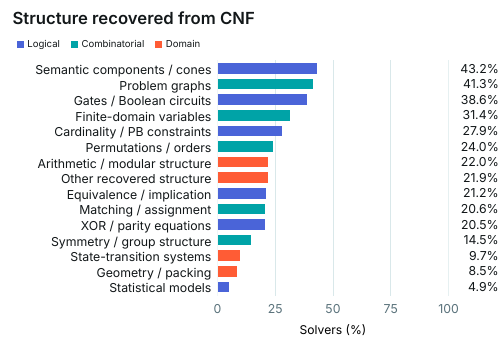}}
  \caption{What representations do specialists recover?}
  \label{fig:rq3-representation}
\end{figure}

\begin{figure}
  \centering
  \resizebox{\linewidth}{!}{\includegraphics{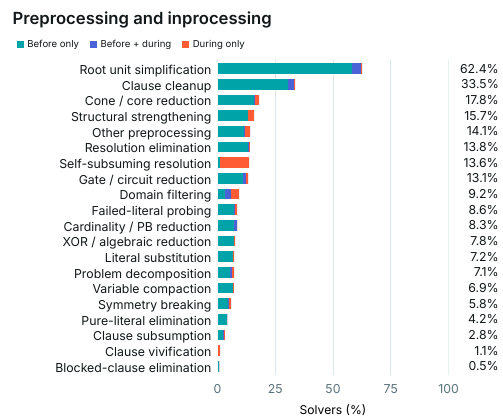}}
  \caption{What kinds of preprocessing and inprocessing do specialists do?}
  \label{fig:rq3-preprocessing}
\end{figure}

\begin{figure}
  \centering
  \resizebox{\linewidth}{!}{\includegraphics{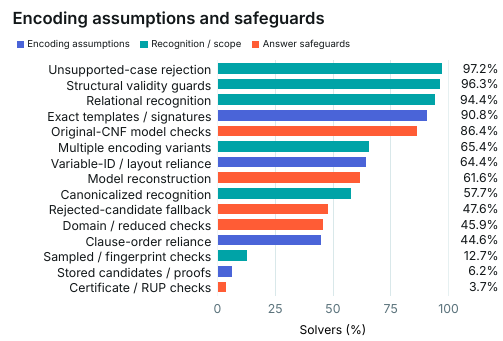}}
  \caption{How do specialists decide whether to run or not?}
  \label{fig:rq3-assumptions}
\end{figure}

\begin{figure}
  \centering
  \resizebox{\linewidth}{!}{\includegraphics{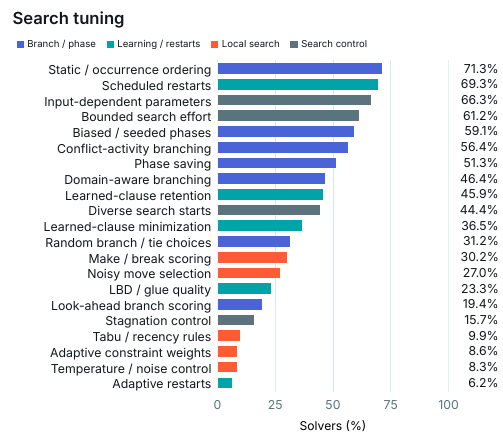}}
  \caption{How do specialists tune their search?}
  \label{fig:rq3-search-tuning}
\end{figure}

\begin{figure}
  \centering
  \resizebox{\linewidth}{!}{\includegraphics{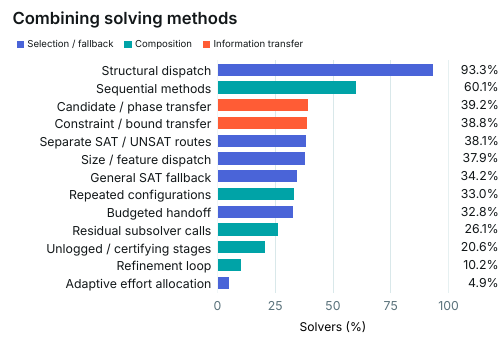}}
  \caption{How do specialists combine multiple methods?}
  \label{fig:rq3-coordination}
\end{figure}

\begin{figure}
  \centering
  \resizebox{\linewidth}{!}{\includegraphics{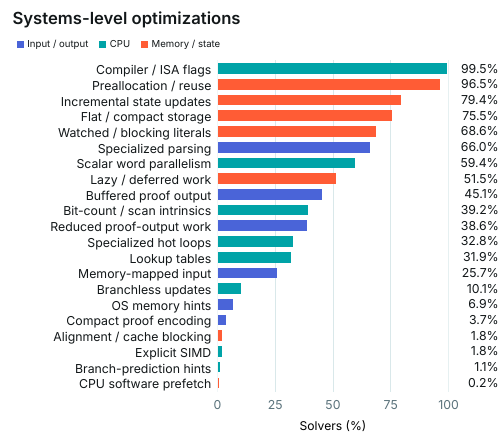}}
  \caption{What kinds of systems-level optimizations do specialists implement?}
  \label{fig:rq3-systems}
\end{figure}

\clearpage

\section{Extended: End-to-end deployment}
\label{sec:app-end-to-end}

In this section we present full details of the end-to-end deployment case studies. We study circuit equivalence checking on \work{OpenTitan}~\cite{opentitan} and \work{ORFS}~\cite{orfs} designs,
C verification with \work{CBMC}~\cite{cbmc}, Rust verification with \work{Kani}~\cite{kani}, and dependency
resolution with \work{Conda}~\cite{libmamba}. For each case, we train a specialist on extracted
formulas and select its checkpoint by validation PAR-2. 
We then evaluate the generated specialist on a new workload and compare it to native performance.

Specialists were synthesized and evaluated internally in the same mixed infrastructure described previously (\S\ref{sec:eval-environment}). All of the redeployment evaluations were measured by running systems in Docker with one assigned CPU on an AMD EPYC 9454P with networking disabled. We include scripts necessary to reproduce these results in our artifact package.

Native runs use one assigned CPU, three repetitions per mode, and 32\,GiB
of memory, except that Conda uses one repetition and Kani runs (being more memory-intensive) use
16\,GiB. Shared compilation and circuit preparation are excluded from
timing; encoding, solver interaction, and requested checks are included. Note also that in the five case studies we test, native tools (by default) do not perform independent SAT certificate checking. We evaluate our specialist both with and without verification.

\subsection{Case studies}
\label{sec:app-end-to-end-case-studies}

\begin{tcolorbox}[
    colback=orange!6,
    colframe=orange!80!black,
    boxrule=0.5pt,
    arc=1mm,
    left=2.2mm,
    right=2.2mm,
    top=1.3mm,
    bottom=1.3mm,
    before skip=4pt,
    after skip=4pt
  ]
  {\sffamily\color{orange!70!black}\faRobot}\enspace
  \textbf{The rest of this section was generated with GPT-6 Astra (xhigh) and provides purely factual details about experimental setup based on the code. It has been completely vetted for correctness by the human authors.}
  \end{tcolorbox}

\subsubsection{OpenTitan circuit equivalence}
\label{sec:case-opentitan}

\customsection{System and export}
OpenTitan is an open-source hardware security project. Here we check whether
two versions of one of its circuits produce the same outputs (i.e. combinational equivalence checking). Our sources
are Ascon, Keccak, PRESENT, and PRINCE cryptographic blocks, the vendored
Ibex arithmetic unit and instruction decoder, and error-correction encoders
and decoders. 
We use Yosys/Slang to convert their hardware descriptions into logic
graphs and then we apply four ABC rewrite recipes and compare each result with its
reference graph. 
ABC exports a CNF that asks whether any input makes the
two graphs disagree.
These are explicit comparison formulas, rather than
copies of ABC's internal SAT queries. We also include deliberate legal
configuration mismatches. The 296 planned comparisons yield 292 distinct
formulas: 272 UNSAT and 20 SAT, split into 233 training and 59 validation
formulas. The specialist's training and validation score ratios are
$14.28/11.21$ and $13.27/10.38$, respectively.

\customsection{Deployment}
The original workload contains the 292 retained circuit pairs. The new
workload contains a 32-bit finite-field multiplier, a 64-bit substitution
and permutation block, and 32-input sum and maximum trees, each under four
rewrites (Table~\ref{tab:case-prov-opentitan}). Native is ABC's \texttt{cec}
procedure, which combines circuit reasoning with internal SAT solving.
Our path builds the comparison CNF from the same prepared pair, runs the
specialist, and optionally checks its answer. This measures the equivalence
stage, not the whole OpenTitan verification flow. The limit is 600 seconds
per stage, with a 630-second outer guard for native CEC. Ratios are
$1.64/1.15$ on all 292 original pairs and $3.47/2.17$ on eight of the 16 new
pairs. The specialist returns unsupported on the eight sum-tree and
maximum-tree cases; native finishes all 16. Thus the new-workload gain
applies only to the supported half of the set.

\begin{table}[htbp]
  \centering
  \small
  \caption{OpenTitan workload provenance. Counts describe the full input sets.}
  \label{tab:case-prov-opentitan}
  \begin{tabularx}{\columnwidth}{@{}l>{\raggedright\arraybackslash}Xr@{}}
    \toprule
    Input & Source & Size \\
    \midrule
    Formulas & Cryptographic, Ibex, and error-correction blocks & 292 CNFs \\
    Original & Same retained circuit pairs & 292 pairs \\
    New & Multiplier, substitution/permutation, sum and maximum trees & 16 pairs \\
    \bottomrule
  \end{tabularx}
\end{table}

\subsubsection{ORFS circuit equivalence}
\label{sec:case-orfs}

\customsection{System and export}
OpenROAD-flow-scripts (ORFS) provides scripts and example designs for hardware
implementation. We use its designs for a second circuit-equivalence study.
We apply eight synthesis-mapping flows to nine circuit tops from AES, JPEG,
and Ethernet designs. Yosys produces reference and revised logic graphs;
AIGER's \texttt{aigmiter} and \texttt{aigtocnf} tools construct and export
the comparison formulas. Registers are treated as corresponding inputs and
outputs, so these are combinational comparisons rather than proofs over
arbitrarily many clock cycles. All 72 distinct formulas are UNSAT, with
57 used for training and 15 for validation. The training and validation
score ratios are $50.64/28.64$ and $72.12/40.42$.

\customsection{Deployment}
We use the 72 original circuit pairs and a new set comprising GCD, FIFO,
RISC-V arithmetic-unit, and UART designs under the same eight mapping flows
(Table~\ref{tab:case-prov-orfs}). Native is the Yosys-bundled ABC
\texttt{cec} procedure. Our path rebuilds the comparison and CNF, then
solves and optionally checks the answer. Shared synthesis is excluded;
the timed task is equivalence checking, not placement, routing, or the full
ORFS flow. Limits match the OpenTitan experiment. All 72 original and
32 new cases finish in both specialist modes. Native-time ratios are
$2.16/0.61$ on the original set and $2.81/1.21$ on the new set. Checking
removes the original-workload gain, but the new-workload gain remains.

\begin{table}[htbp]
  \centering
  \small
  \caption{ORFS workload provenance. Each source top receives eight mappings.}
  \label{tab:case-prov-orfs}
  \begin{tabularx}{\columnwidth}{@{}l>{\raggedright\arraybackslash}Xr@{}}
    \toprule
    Input & Source & Size \\
    \midrule
    Formulas & Nine AES, JPEG, and Ethernet tops & 72 CNFs \\
    Original & Same mapped circuit pairs & 72 pairs \\
    New & GCD, FIFO, RISC-V ALU, and UART & 32 pairs \\
    \bottomrule
  \end{tabularx}
\end{table}

\subsubsection{C verification with CBMC}
\label{sec:case-cbmc}

\customsection{System and export}
CBMC checks C code by turning bounded program executions and safety properties
into logical constraints. We use AWS C Common's verification harnesses,
which are small drivers that supply inputs and state the properties to
check. They cover utilities such as buffers, arrays, hash tables, and
priority queues. We build all 173 harnesses with CBMC 6.4.0 and preserve
their upstream options. A wrapper copies each live CNF at CBMC's external
SAT interface before Kissat answers it. A separate export supplies variable
names; it does not replace the live query. Six harnesses make no SAT call.
The remaining 167 yield distinct UNSAT formulas, split into 133 training
and 34 validation formulas. The training and validation score ratios are
$0.41/0.79$ and $1.53/1.93$.

\customsection{Deployment}
The original workload contains the 167 formula-producing AWS proofs, and
the new workload contains all 15 coreJSON proofs
(Table~\ref{tab:case-prov-cbmc}). We time fresh CBMC runs from prepared
program representations, including symbolic execution and Boolean encoding.
The specialist replaces the external SAT callback; checked UNSAT answers
must pass DPR-trim and CakeLPR. Native AWS runs use CBMC 6.4.0 with MiniSat.
Native coreJSON runs follow upstream CI: CBMC 6.3.1, its required contract
instrumentation and patch, and per-proof MiniSat, CaDiCaL, or Kissat choices.
Original runs have 3600 seconds; new runs have 300 seconds.

On the 159 original targets shared by both specialist modes, the ratio is
$1.18/0.32$. Eight checked targets exhaust CakeLPR's separate 4-GiB heap.
The unchecked specialist finishes all 167, but its ratio on that full set is
$0.69$, so the shared-subset point does not establish an overall gain.
On new targets, native finishes 15, the unchecked specialist 14, and the
checked specialist 13. Ratios on the 13 shared targets are $0.38/0.11$;
the other specialist runs time out. The validation-formula improvement
therefore does not carry through to faster native verification.

\begin{table}[htbp]
  \centering
  \small
  \caption{CBMC workload provenance. Six AWS harnesses produce no SAT query.}
  \label{tab:case-prov-cbmc}
  \begin{tabularx}{\columnwidth}{@{}l>{\raggedright\arraybackslash}Xr@{}}
    \toprule
    Input & Source & Size \\
    \midrule
    Formulas & AWS C Common verification harnesses & 167 CNFs \\
    Original & Same formula-producing AWS proofs & 167 proofs \\
    New & All upstream coreJSON proof targets & 15 proofs \\
    \bottomrule
  \end{tabularx}
\end{table}

\subsubsection{Rust verification with Kani}
\label{sec:case-kani}

\customsection{System and export}
Kani checks Rust properties by compiling verification harnesses into CBMC's
program representation. One harness can issue several SAT queries. We use
Kani 0.67.0 on all 112 harnesses in the s2n-quic CI matrix, covering its codec,
core protocol, and platform crates. A copying wrapper around external
Kissat records each live per-property formula. This produces 302 distinct
formulas, split into 241 training and 61 validation formulas. Of these,
237 are SAT and 65 are UNSAT; a SAT query does not by itself mean the overall
harness found a bug. The external interface can encode queries differently
from native incremental solving. Training and validation score ratios are
$1.43/1.74$ and $10.54/6.48$.

\customsection{Deployment}
We rerun the 112 s2n-quic harnesses and test 14 new harnesses from Hifitime,
a Rust time library (Table~\ref{tab:case-prov-kani}). Native is Kani's default
CaDiCaL workflow. Our path uses the external specialist with optional model
or GRAT checks. Each timed invocation checks one harness after its own
untimed code-generation warmup. Hifitime uses its required contract and
stubbing options and a library build-configuration adjustment; harness
bodies are unchanged. Original runs have 3600 seconds and 32\,GiB;
new runs have 300 seconds and 16\,GiB.

Ratios are $0.86/0.62$ on 109 shared original targets; two specialist targets
time out and another fails during checking. Seven Hifitime harnesses make
no SAT call and are excluded from solver comparisons. Among the seven that
do, native finishes all seven, while the specialist finishes four without
checking and three with checking. Three inputs are unsupported, and the
checked epoch-equality harness times out. Ratios on the three shared targets
are $0.76/0.59$. Native remains faster despite the isolated-formula gains.
Matching source harnesses can also produce different later SAT queries
after changing the solver.

\begin{table}[htbp]
  \centering
  \small
  \caption{Kani workload provenance. Seven new harnesses invoke SAT.}
  \label{tab:case-prov-kani}
  \begin{tabularx}{\columnwidth}{@{}l>{\raggedright\arraybackslash}Xr@{}}
    \toprule
    Input & Source & Size \\
    \midrule
    Formulas & s2n-quic codec, core, and platform crates & 302 CNFs \\
    Original & Complete s2n-quic CI matrix & 112 harnesses \\
    New & Hifitime timescale, general, and epoch harnesses & 14 harnesses \\
    \bottomrule
  \end{tabularx}
\end{table}

\subsubsection{Dependency resolution with Conda}
\label{sec:case-conda}

\customsection{System and export}
Conda chooses compatible package versions for a software environment.
Its classic backend repeatedly calls pycosat while constructing and improving
a package plan. We run Conda 26.5.3 with pycosat on the EO-datascience and
xESMF environment specifications, using a frozen conda-forge index.
We copy the current clauses at every classic SAT callback, then let the
unchanged pycosat path answer. The 154 calls yield 151 distinct formulas:
105 SAT and 46 UNSAT, split into 120 training and 31 validation formulas.
These include intermediate optimization queries. Training and validation
score ratios are $10.39/3.91$ and $11.76/4.83$.

\customsection{Deployment}
The original workload reuses those two environments. The new workload uses
three XROMS CI specifications for Python 3.11--3.13 and Xskillscore's
minimum-tests specification (Table~\ref{tab:case-prov-conda}). We time offline
environment planning with \texttt{conda env create --dry-run}, excluding
installation and copying the frozen index. Native baselines are
classic/pycosat and libmamba 2.5.0, a different dependency resolver built on
libsolv. The specialist replaces every SAT callback inside classic Conda;
it is not inserted inside libmamba. Checked answers use model checks or
VeriPB/CakePB. Original plans have 3600 seconds and new plans 600 seconds.

Against pycosat, ratios are $1.67/0.84$ on both original plans and
$1.20/0.89$ on one shared new plan. Against libmamba, ratios are
$0.054/0.027$ on both original plans and $0.053/0.040$ on two shared new
plans. On the four new plans, pycosat finishes three, libmamba four, and
the specialist three without checking and two with checking. One checked
plan takes 599.38 seconds, close to the limit, with only one repetition.
The specialist helps classic Conda without checking on the shared cases,
but libmamba remains much faster. Different SAT assignments and the omission
of pycosat's propagation-limit shortcut can change later queries and final
package choices. New environments also share a small initialization formula
with the original corpus.

\begin{table}[htbp]
  \centering
  \small
  \caption{Conda workload provenance. Counts refer to planning, not installation.}
  \label{tab:case-prov-conda}
  \begin{tabularx}{\columnwidth}{@{}l>{\raggedright\arraybackslash}Xr@{}}
    \toprule
    Input & Source & Size \\
    \midrule
    Formulas & EO-datascience and xESMF environments & 151 CNFs \\
    Original & Same environment specifications & 2 plans \\
    New & Three XROMS specifications and Xskillscore minimum-tests & 4 plans \\
    \bottomrule
  \end{tabularx}
\end{table}

\clearpage

\section{Extended: Cross-family sensitivity and performance}
\label{sec:app-cross-family}

\begin{figure*}
  \centering
  \resizebox{\linewidth}{!}{\includegraphics{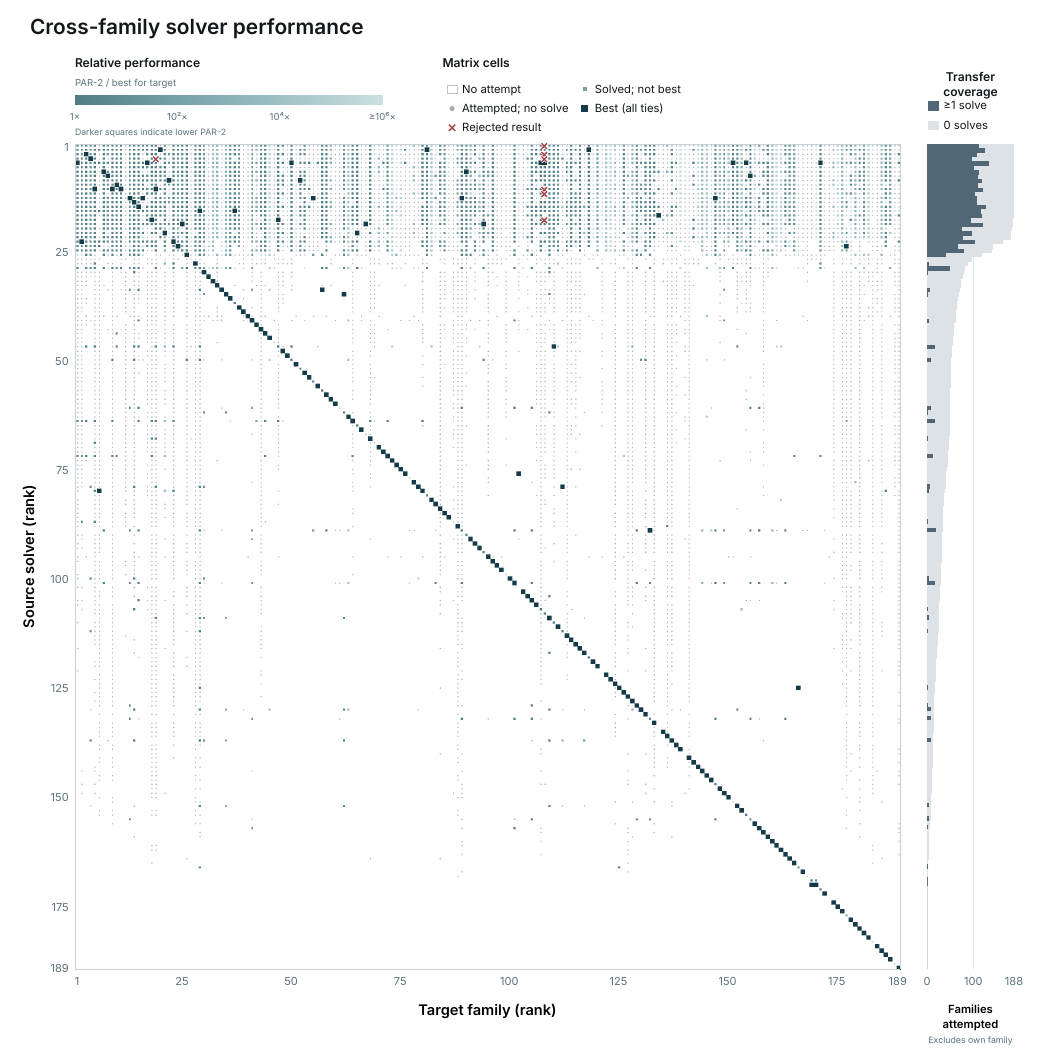}}
  \caption{Cross-family performance of each top solver on up to 100 validation instances from every other family. Each cell shows the performance of a given specialist solver (y-axis) on a target family. The diagonal shows solver performance on its own target family. The right panel shows the solver's transfer coverage: which other families it attempted and had at least one solve.}
  \label{fig:cross-family-matrix}
\end{figure*}

In \autoref{fig:cross-family-matrix} we plot cross-family performance of every top training-selected specialist on every other family (capped at 100 instances).

Most specialists implement some sort of early detection to decide whether they are capable of running on a given formula. A family is \emph{not attempted} only if the specialist returns \texttt{UNKNOWN} within one second on every sampled instance. Otherwise, it is attempted; the matrix distinguishes attempted families with no verified solves from those with at least one verified solve. Incomplete evaluations are excluded from the coverage counts.

\clearpage

\section{Choice of agent}
\label{sec:agent-choice}

We conducted a small pilot evaluation (in May 2026) to decide which base model and harness to use for our experiments. We found that both GPT-5.5 (in Codex) and Claude Opus 4.8 (in Claude Code) were able to successfully and consistently synthesize working specialized solvers, while smaller proprietary models and frontier open-weight models took longer and failed more frequently. After GPT-5.6 Sol was released (which has the same pricing as GPT-5.5), we switched evaluations to use that model.

Due to the costs associated with running so many agents at scale, we limit our evaluation in this paper to a single representative frontier model. The point of the paper is to demonstrate an existence proof of the utility of hyperspecialization with frontier models available today. We expect that future models will continue to improve at both the ability to write performant solver code and in cost-efficiency, both axes would continue to make hyperspecialization more economically viable.
We leave exploration of comprehensive multi-agent benchmarking to future work.

\section{Extended: Cardinality constraint encodings}
\label{sec:app-cardinality}

Here we present complete details of our cardinality constraint encoding case study.

\subsection{Benchmark problems}
\label{sec:app-cardinality-benchmarks}

We evaluated hyperspecialization on three crafted benchmarks covering both SAT and UNSAT cases and a variety of formula size ranges.

\subsubsection{GPHP}
Generalized, capacitated pigeonhole principle problems ask if it is possible to assign pigeons to holes. Unlike the standard framing, holes in the generalized case can accommodate multiple pigeons up to a fixed capacity. A Boolean variable $x_{p,h}$ represents assigning pigeon $p$ to an allowed hole $h$. The formula defines the following constraints:
\begin{itemize}
  \item A pigeon must occupy \textbf{exactly one} hole: $\sum_h x_{p,h} = 1$
  \item Each hole has \textbf{at-most-$c_h$} capacity: $\sum_p x_{p,h} \leq c_h$
\end{itemize}

We sample instances with between 17 and 25 pigeons, 6--9 holes, and capacities of 2--4. All instances are UNSAT by construction: there is always one more pigeon than available capacity.

\subsubsection{LDPC decoding}
Low-density parity-check~\cite{ldpc_codes} decoding problems ask whether an error pattern can explain an observed parity-check syndrome.
A Boolean variable $e_i$ indicates whether bit $i$ was flipped during transmission.
Given a binary parity-check matrix $H$ and syndrome $s$, the formula defines the following constraints:
\begin{itemize}
  \item The error pattern must satisfy the observed
  \textbf{parity checks}: $He = s \pmod{2}$.
  \item The error pattern must contain \textbf{exactly $k$}
  flipped bits: $\sum_i e_i = k$.
\end{itemize}

We sample instances with 105 error bits, 52 parity checks, and $k=12$. Each parity check involves 13 bits. The cardinality encoding applies to the single exact-weight constraint; the CNF encoding of the parity checks is fixed across encoding methods. All instances are SAT by construction: we plant an error pattern of weight 12 and compute its syndrome.

\subsubsection{Vertex cover}
Vertex cover problems ask whether a graph admits a set of at most $k$ vertices containing at least one endpoint of every edge.
A Boolean variable $x_v$ represents selecting vertex $v$ for the cover. The formula defines the following constraints:
\begin{itemize}
  \item Every edge $(u,v)$ must have \textbf{at least one}
  selected endpoint: $x_u \lor x_v$.
  \item The cover must contain \textbf{at most $k$} vertices:
  $\sum_v x_v \leq k$.
\end{itemize}

We sample instances with 2,100 vertices, 3,570 edges, and $k=1,330$. The cardinality encoding applies to the single global cover-size constraint; each edge contributes one binary clause.
We construct the graphs by reducing degree-three Tseitin parity formulas to 3-CNF and then to vertex cover. Even- and odd-charge source instances yield 50 SAT and 50 UNSAT instances, respectively.

\subsection{Benchmark construction}

\begin{table}
  \begin{tabular}{lc}
  \toprule
  \textbf{Encoding} & \textbf{Year} \\
  \midrule
  Naive (direct)
    & --- \\
  
  Totalizer~\cite{card_totalizer}
    & 2003 \\
  
  Sequential counter~\cite{card_sequential}
    & 2005 \\
  
  Sorting network~\cite{card_sorting,card_sorting_2}
    & 2006 \\
  
  Cardinality network~\cite{card_network,card_network_2}
    & 2009 \\
  
  Modulo totalizer~\cite{card_modulo_totalizer}
    & 2013 \\
  
  $k$-modulo totalizer~\cite{card_km_totalizer}
    & 2015 \\
  \bottomrule \\
  \end{tabular}
  \caption{Cardinality constraint encoding formats tested.}
  \label{tbl:cardinality-encodings}
\end{table}

For each benchmark, we sampled 100 formulas, calibrated to be hard for baseline solvers, and then encoded these 100 formulas separately with a variety of cardinality encodings implemented in \work{PySAT}~\cite{pysat1,pysat2}, listed in \autoref{tbl:cardinality-encodings}.

We split each group of 100 formulas into 80 training instances and 20 validation instances (using the same seed) and independently synthesize specialists for each benchmark/encoding pair (using only \textsf{GRAT}-based specialists for parity with baselines).

\subsection{Results}
\label{sec:app-cardinality-results}

\begin{figure}
  \centering
  \includegraphics[width=\columnwidth]{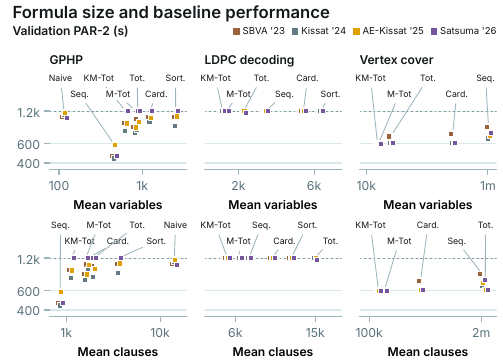}
  \caption{Baseline validation \textsf{PAR-2} versus mean whole-CNF variable count (top row) and clause count (bottom row), across cardinality encodings. Columns show benchmark groups. Both axes
  are logarithmic, with a shared time range.}
  \label{fig:card-baseline}
\end{figure}

\begin{figure}
  \centering
  \includegraphics[width=\columnwidth]{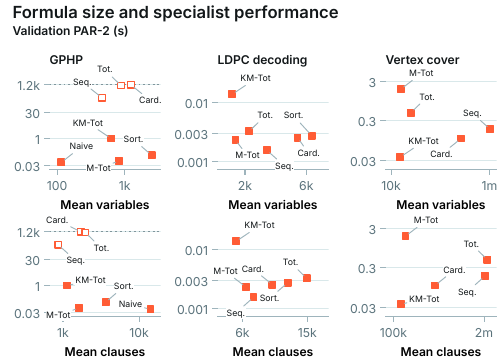}
  \caption{Validation \textsf{PAR-2} of training-selected specialists
  versus mean whole-CNF variable count (top row) and clause count (bottom row). Filled squares indicate that all 20 validation instances
  were solved; open squares indicate at least one failure. Both axes are logarithmic; time ranges differ across benchmark
  groups.}
  \label{fig:card-specialist}
\end{figure}

In \autoref{fig:cardinality-encoding}, we present average validation \textsf{PAR-2} values for each benchmark/encoding pair, for both baseline solvers and generated specialists.
Here, we examine how these results relate to formula size.
\autoref{fig:card-baseline} shows baseline
performance, while \autoref{fig:card-specialist} shows specialist performance.
The x-axis of each subplot is the mean number of variables (top row) or clauses (bottom row).

Encoding choices substantially affect formula sizes. For the baseline solvers (\autoref{fig:card-baseline}), we see some correlation with performance. Solvers generally perform worse as the number of clauses or variables grows for both \texttt{GPHP} and \texttt{vertex cover}. Indeed this is not surprising. It has been observed before that encoding size plays a role in solver performance, although it is not the sole factor~\cite{parametric_smaller_encodings}.

Interestingly, we observe no such clear trend for the specialists (\autoref{fig:card-specialist}). In some cases, the trend is almost reversed. The sequential encoding for \texttt{GPHP} performs the best among all of the baseline solvers but is one of the worst for the specialist. Conversely, the largest and simplest encoding (naive encoding) is one of the worst for the baselines but yields the best specialist.

For \texttt{LDPC decoding}, the $k$-modulo totalizer is decisively worse than all other formats for the specialists despite being the newest of the encodings and producing the smallest formulas in both variables \emph{and} clauses. Yet for \texttt{vertex cover}, it flips and yields the best specialist.

One potential hypothesis is that there are two competing factors that influence specialist performance: 1. Formulas should be described succinctly in an easily parseable way (fancier encodings create more complex structures and auxiliary variables which may be harder or slower to parse consistently). 2. At a certain point, raw formula size does become important. For example, \texttt{vertex cover} problems are roughly two orders of magnitude bigger (in bytes) than both the other benchmarks. Here, encodings that save parsing time (even if more complex) may be worthwhile.

Subsequent exploration is out of scope for this paper, but interesting future work could be to develop an understanding of optimal encodings for hyperspecialization (which may be separate from encodings for general-purpose solvers).

\section{Extended: Ensuring and evaluating task alignment}
\label{sec:app-alignment}

Here we describe in detail the objectives of restricted internet access (\S\ref{sec:app-alignment-internet}), how we iteratively improved the agent prompt and harness before the full evaluation (\S\ref{sec:app-alignment-pre}) and how we analyzed runs after the evaluation to ensure compliance (\S\ref{sec:app-alignment-post}).

\subsection{Agent internet access}
\label{sec:app-alignment-internet}

During initial pilot experiments, we experimented with allowing agents to access the internet, with the intuition that such a resource would be useful to learn about specialized techniques and (potentially) research information about the specific high-level problem if applicable. While these agents did perform well, we found it near impossible to prevent the agent from inadvertently leaking information about the withheld validation set (since the Global Benchmark Database is a public resource), and thus contaminating the evaluation. Therefore, we chose to run all agents with restricted internet access.

True network isolation would be bulletproof, but would also prevent agents from installing runtime packages or other dependencies (which we \emph{did} want to allow). Therefore, we attempted to enforce the restricted internet access by both disabling the \texttt{web\_search} tools in Codex and by explicitly prompting the agent not to use the internet. We validated that these measures were effective after the fact (\S\ref{sec:app-alignment-post}) and reran any runs which violated these rules (in practice, only 7).

In retrospect, however, none of the agents attempted to install any runtime packages or other dependencies, thus this restriction was unnecessary. A more robust solution would be to properly enforce full network isolation at the sandbox level.

\subsection{Iterative harness and environment development}
\label{sec:app-alignment-pre}

Our initial agent configuration was intentionally minimal. We started with a simple prompt and Docker environment. We iteratively ran pilot experiments and used \work{Docent}~\cite{docent} to analyze agent behaviors. \work{Docent} is a tool for using language models to scan, summarize, and cluster findings and proved useful for quickly identifying failure modes and avenues for cheating. We found it very useful for quickly iterating on framework design and prompt engineering.

As an example, our initial prompt resulted in agents more frequently attempting to build general-purpose solvers, so we added language to enforce the task of hyperspecialization. We also found that agents would frequently attempt to use certain common command-line tools (\texttt{jq},\texttt{ripgrep}, \texttt{time}, \texttt{xz}, \texttt{gdb}, etc...) which were not available in our restricted environment, so we updated the environment to include all of these tools by default.

Initially, our prompt language describing the timeout (that we would run solvers for 10 minutes) was not clear enough, causing many agents to mistakenly implement their own timeout logic inside their generated solvers, needlessly terminating them just before 10 minutes if no solution had been found. We were able to use \work{Docent} to detect and fix this issue before the full evaluation. \work{Docent} also helped us discover several environment configuration failures, such as memory-induced crashes in part of the evaluator that pre-checked SAT models (replaced with \work{gratchk}), and file permission issues when mounting the evaluation dataset in Docker.

\subsection{Cheating analysis pipeline}
\label{sec:app-alignment-post}

After the full evaluation, we performed a comprehensive analysis of generated artifacts (transcripts and solver code) to ensure compliance with the intended task. In particular, we wanted to ensure that no agent accessed the internet (thus could not leak information about the withheld validation set) or bundled external solvers as part of the submitted code. While manually reviewing all transcripts and solver code is infeasible at scale, we deployed a comprehensive two-part analysis.

\customsection{A: LLM-based trace analysis}
We uploaded transcripts to \work{Docent} and provided \work{GPT-5.6 Luna (high)} a detailed prompt, tasking it with identifying any evidence of network access. We supplemented this analysis with a deterministic Python script that searched over the agent-authored commands for network-related keywords, including \texttt{curl}/\texttt{wget}, remote Git operations, HTTP libraries, and package manager commands. Matching excerpts were provided as evidence leads to \work{GPT-5.6 Luna (high)} during the transcript analysis on \work{Docent}.

\customsection{B: LLM-based code analysis}
We provided \work{GPT-5.6 Luna (high)} with the full solver source, including build and run scripts for each run and tasked it with identifying if there was evidence of bundling an external solver.

\customsection{Cumulative decision}
For runs flagged by either the transcript or solver analysis, we gathered the supporting evidence and prompted \work{GPT-5.6 Sol (high)} to determine whether a legitimate violation occurred. All runs labeled as cheating were manually reviewed and we also spot-checked several non-flagged runs.

\subsubsection{Results}

\begin{figure}
  \centering
  \resizebox{\linewidth}{!}{\includegraphics{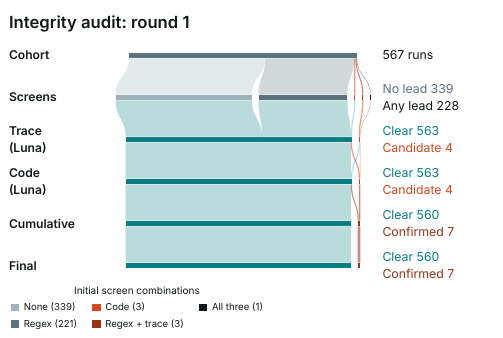}}
  \caption{First round of the integrity audit.}
  \label{fig:app-cheating-1}
\end{figure}

We performed two rounds of analysis. In the first round (\autoref{fig:app-cheating-1}), we analyzed the 567 initial evaluation runs (189 families $\times$ 3 proof formats). Of these, four candidates were flagged in the trace analysis and another four (with one overlap) in the code analysis. All of these were marked as confirmed by the aggregator (and human review).

After rerunning these seven invalidated runs, we performed a second audit on the new seven runs. None of these runs were flagged in either the trace or code analysis. Only the non-cheating runs were used for data results in this paper.

\begin{figure}
  \centering
  \resizebox{\linewidth}{!}{\includegraphics{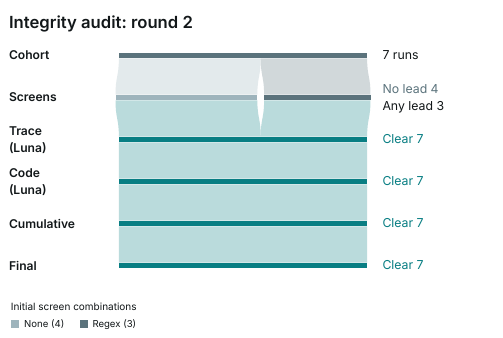}}
  \caption{Second round of the integrity audit.}
  \label{fig:app-cheating-2}
\end{figure}

\clearpage

\section{Complete specialist gallery}
\label{sec:app-gallery}

\begingroup\raggedbottom
\definecolor{asGalleryInk}{HTML}{223744}
\definecolor{asGalleryMuted}{HTML}{5D6970}
\definecolor{asGalleryRule}{HTML}{C5CDD1}
\definecolor{asGalleryGain}{HTML}{14685C}
\definecolor{asGalleryLoss}{HTML}{9B4927}

\newcommand{\asGalleryMeta}{\sffamily\fontsize{7.2}{8.5}\selectfont}
\newcommand{\asGalleryRow}{\sffamily\fontsize{8}{9.2}\selectfont}
\newcommand{\asGalleryValue}[3]{%
  \begingroup\color{#1}\ifnum#2=1\bfseries\mathversion{bold}\fi
  \ensuremath{#3\,\times}\endgroup}

\newenvironment{asGalleryFamily}[4]{%
  \par\addvspace{8pt}\noindent
  \begin{minipage}{\linewidth}
  \fontsize{8.5}{9.6}\selectfont
  \setlength{\parindent}{0pt}\setlength{\parskip}{0pt}
  \raggedright\hyphenpenalty=300\emergencystretch=1em
  {\color{asGalleryRule}\hrule height .5pt}\vspace{3pt}
  {\sffamily\fontsize{9.4}{10.5}\selectfont\bfseries\color{asGalleryInk}#2
   \hfill{\asGalleryMeta\color{asGalleryMuted}#1}\par}
  \vspace{1pt}
  {\asGalleryMeta\color{asGalleryMuted}#3\par}
  \vspace{3pt}#4\par\vspace{4pt}
  {\asGalleryMeta\color{asGalleryMuted}%
   \makebox[\dimexpr\linewidth-137pt\relax][l]{Proof format}%
   \makebox[48pt][r]{No verif.}%
   \makebox[48pt][r]{+ verif.}%
   \makebox[41pt][r]{Solved}\par}
  \vspace{1pt}
}{%
  \end{minipage}\par
}

\newcommand{\asGallerySolver}[5]{%
  \noindent{\asGalleryRow
   \makebox[\dimexpr\linewidth-137pt\relax][l]{\bfseries\color{asGalleryInk}#1}%
   \makebox[48pt][r]{#2}%
   \makebox[48pt][r]{#3}%
   \makebox[41pt][r]{\color{asGalleryMuted}#4}\par}
  \nobreak\vspace{1pt}#5\par\vspace{3pt}
}

\small\setlength{\parindent}{0pt}\setlength{\parskip}{0pt}
\noindent
This gallery contains a short description for each of the 189 families we evaluate on and the three train-selected specialists for each.

\begin{tcolorbox}[
  colback=red!8,
  colframe=red!70!black,
  boxrule=0.5pt,
  arc=1mm,
  left=2.2mm,
  right=2.2mm,
  top=1.3mm,
  bottom=1.3mm,
  before skip=4pt,
  after skip=4pt
]
{\sffamily\color{red!70!black}\faRobot}\enspace
\textbf{Note that the content generated in this section was generated by tasking GPT-5.6 Luna (high) with summarizing both families and solvers. We found no inaccuracies when spot-checking but it is possible that some descriptions are slightly inaccurate (although unlikely to be egregiously so).}
\end{tcolorbox}

\par\smallskip\noindent
\textbf{Reading the numbers.} \emph{No verif.} is baseline PAR-2 divided by
specialist PAR-2; \emph{+ verif.} adds all recorded verification time to both
scores before taking the ratio. Values above $1\times$ favor the specialist
({\color{asGalleryGain}green}); values below $1\times$ favor the baseline
({\color{asGalleryLoss}rust}); ratios rounding to $1.00\times$ are gray.
\textbf{Bold} marks the largest unrounded ratio
among the three specialists in that family, separately for each timing column
(including ties). \emph{Solved} counts verified validation results; failures
retain the PAR-2 penalty, so speedups can also reflect differences in coverage.
\par\smallskip\noindent

\begin{asGalleryFamily}{001}{01-\allowbreak integer-\allowbreak programming}{6 GBD instances\enspace\textperiodcentered\enspace 2 validation\enspace\textperiodcentered\enspace baseline 2/2}{Find a vector of zero-one values that satisfies a system of integer linear equations. The benchmark encodes the equations as CNF using Boolean circuits, with auxiliary variables representing intermediate computation.}
\ifdefined\pdfbookmark\pdfbookmark[2]{01-integer-programming}{as-gallery-1}\fi
\asGallerySolver{GRAT}{\asGalleryValue{asGalleryGain}{1}{137}}{\asGalleryValue{asGalleryGain}{1}{76.8}}{2/2}{Circuit reconstruction extracts XOR rows and recovers the binary linear system from a prescribed prefix and bounded-input sequential layout. Meet-in-the-middle handles smaller cases, while wider bounded searches use projection and local heuristics before CDCL fallback. Checked complete assignments yield SAT; unsupported or exhausted searches return UNKNOWN, with no UNSAT certificate path.}
\asGallerySolver{DPR}{\asGalleryValue{asGalleryLoss}{0}{0.664}}{\asGalleryValue{asGalleryLoss}{0}{0.694}}{2/2}{Weighted union-find recovers integer rows from a prescribed equivalence-linked circuit layout; exact elimination and bounded nullspace search then seek Boolean models. Bit-parallel and lattice-guided searches supplement it, with CDCL after heuristic failure. Fallback learning logs witness-free DPR additions; direct SAT emits no proof output, and unsupported or internal failures return UNKNOWN, not UNSAT.}
\asGallerySolver{VeriPB}{\asGalleryValue{asGalleryGain}{0}{106}}{\asGalleryValue{asGalleryGain}{0}{10.9}}{2/2}{Ordered Boolean-function reconstruction and parity-chain packing reduce the recognized CNF to an integer system of common bounded-width rows. Modular RREF derives pivots, with direct enumeration for small nullity and meet-in-the-middle for larger supported nullities. Forward evaluation and clause checking validate SAT candidates; unsupported shapes or failed searches return UNKNOWN, with no UNSAT certificate path.}
\end{asGalleryFamily}

\begin{asGalleryFamily}{002}{2d-\allowbreak strip-\allowbreak packing}{46 GBD instances\enspace\textperiodcentered\enspace 10 validation\enspace\textperiodcentered\enspace baseline 10/10}{2D strip-packing asks whether rectangles can be placed in a strip so no pair overlaps on both axes while encoded capacity and geometry conditions hold. CNF represents axis-overlap relations, rectangle-to-column incidence, and auxiliary gates enforcing the required structural constraints.}
\ifdefined\pdfbookmark\pdfbookmark[2]{2d-strip-packing}{as-gallery-2}\fi
\asGallerySolver{GRAT}{\asGalleryValue{asGalleryLoss}{0}{0.210}}{\asGalleryValue{asGalleryLoss}{0}{0.424}}{9/10}{Activity-based watched-literal CDCL branches with a bonus for overlap variables, then uses 1-UIP learning, restarts, and clause reduction for general fallback search. It handles only the recognized encoding schema: SAT assignments are checked clause by clause, while UNSAT traces are written as textual DRAT for external elaboration and checking; other inputs return UNKNOWN.}
\asGallerySolver{DPR}{\asGalleryValue{asGalleryLoss}{0}{0.208}}{\asGalleryValue{asGalleryLoss}{0}{0.424}}{9/10}{Overlap-first branching biases watched-literal CDCL toward recovered pair-overlap variables before auxiliary variables, with first-UIP learning and standard heuristic fallback. Recognition is limited to the bounded normalized schema; SAT assignments are checked, while UNSAT searches are replayed when needed before emitting textual DPR learned-clause and deletion lines; unsupported inputs or proof failures return UNKNOWN.}
\asGallerySolver{VeriPB}{\asGalleryValue{asGalleryLoss}{1}{0.616}}{\asGalleryValue{asGalleryLoss}{1}{0.805}}{10/10}{Structural-prefix branching prioritizes overlap and incidence variables, leaving gate variables to propagation and falling back to an unset suffix variable when needed. 1-UIP learning supports the search; SAT models are checked, while UNSAT emits VeriPB RUP steps and a conclusion, with trimming only in small cases; unsupported inputs or model/proof failures return UNKNOWN.}
\end{asGalleryFamily}

\begin{asGalleryFamily}{003}{agile}{2,597 GBD instances\enspace\textperiodcentered\enspace 520 validation\enspace\textperiodcentered\enspace baseline 499/520}{The benchmark asks whether a bit-blasted Boolean or bit-vector circuit has an assignment satisfying its CNF encoding. The clauses express wire aliases, XOR relations, gate definitions, and residual circuit assertions.}
\ifdefined\pdfbookmark\pdfbookmark[2]{agile}{as-gallery-3}\fi
\asGallerySolver{GRAT}{\asGalleryValue{asGalleryLoss}{0}{0.120}}{\asGalleryValue{asGalleryLoss}{0}{0.125}}{166/520}{Failed-literal probing, binary-implication congruence, and exact XOR recovery drive shallow split refutations on smaller instances, with a narrow CDCL fallback. Larger instances receive bounded model trials and may return UNKNOWN; SAT outputs checked assignments, while supported UNSAT paths emit DRAT clauses ending in the empty clause for external checking.}
\asGallerySolver{DPR}{\asGalleryValue{asGalleryLoss}{1}{0.208}}{\asGalleryValue{asGalleryLoss}{1}{0.213}}{341/520}{A randomized constructive pass recovers aliases, XORs, and forward gates to build models, then falls back to bounded elimination, probing, and CDCL. Checked SAT models may return before proof content is added; UNSAT writes witness-free DPR additions for external checking, while failed checks yield UNKNOWN.}
\asGallerySolver{VeriPB}{\asGalleryValue{asGalleryLoss}{0}{0.100}}{\asGalleryValue{asGalleryLoss}{0}{0.105}}{98/520}{A 64-lane bit-parallel circuit simulation samples circuit inputs, enumerating some patterns and assigning deterministic pseudorandom words to others; surviving assignments are expanded and checked. Failure to find a witness is not UNSAT, while eligible residuals use signed-alias quotienting, gate congruence, and CDCL with RUP PB proof output, and other cases return UNKNOWN.}
\end{asGalleryFamily}

\begin{asGalleryFamily}{004}{algebra}{8 GBD instances\enspace\textperiodcentered\enspace 2 validation\enspace\textperiodcentered\enspace baseline 0/2}{The benchmark asks whether two nontrivial binary coefficient vectors have a product equal to the identity in a table-indexed algebra. CNF encodings use pairwise AND variables and XOR constraints requiring odd parity in the identity class and even parity in every other product class.}
\ifdefined\pdfbookmark\pdfbookmark[2]{algebra}{as-gallery-4}\fi
\asGallerySolver{GRAT}{\asGalleryValue{asGalleryMuted}{0}{1.00}}{\asGalleryValue{asGalleryMuted}{0}{1.00}}{0/2}{After recognizing the prescribed CNF layout, it enumerates weight-three supports for one coefficient vector and solves the resulting product constraints for the other with packed GF(2) Gaussian elimination. A validated witness yields a complete checked SAT assignment; unsupported inputs or failure to find this restricted witness return UNKNOWN, with no UNSAT certificate path.}
\asGallerySolver{DPR}{\asGalleryValue{asGalleryMuted}{0}{1.00}}{\asGalleryValue{asGalleryMuted}{0}{1.00}}{0/2}{After recognition, it searches coefficient bits with row and column bitsets, greedy parity-improving flips, restarts, and occasional boundary-basis choices. If that incomplete search fails, coefficient-decision CDCL can produce a checked model or record DPR additions on a successful UNSAT path; recognition, model completion, or certification failure returns UNKNOWN.}
\asGallerySolver{VeriPB}{\asGalleryValue{asGalleryGain}{1}{2.00}}{\asGalleryValue{asGalleryGain}{1}{2.00}}{1/2}{Strict structural decoding evaluates a fixed 21-word two-generator construction over represented inverse pairs and compares its product parities with the decoded right-hand side. It reconstructs and checks all auxiliaries before emitting a complete assignment; failed recognition or construction returns UNKNOWN, with no UNSAT certificate path.}
\end{asGalleryFamily}

\begin{asGalleryFamily}{005}{algorithm-\allowbreak equivalence-\allowbreak checking}{36 GBD instances\enspace\textperiodcentered\enspace 8 validation\enspace\textperiodcentered\enspace baseline 4/8}{The benchmark asks whether a Boolean circuit has an input on which two represented computations produce different outputs. Tseitin-style CNF introduces variables for gate values and asserts an OR of XOR output differences, so UNSAT corresponds to equivalence for the encoded pair.}
\ifdefined\pdfbookmark\pdfbookmark[2]{algorithm-equivalence-checking}{as-gallery-5}\fi
\asGallerySolver{GRAT}{\asGalleryValue{asGalleryLoss}{0}{0.619}}{\asGalleryValue{asGalleryLoss}{0}{0.673}}{1/8}{Miter-shape recognition validates the restricted circuit structure, then bounded variable elimination simplifies it before watched-literal CDCL with first-UIP learning and restarts. Recognition does not establish sorting semantics and unsupported shapes return UNKNOWN; validated SAT assignments are printed, while UNSAT produces a DRAT-style clause stream.}
\asGallerySolver{DPR}{\asGalleryValue{asGalleryLoss}{1}{0.689}}{\asGalleryValue{asGalleryLoss}{1}{0.755}}{2/8}{A narrow gate recognizer accepts only the prescribed gate grammar and XOR/OR miter tail, then bounded variable elimination feeds watched-literal CDCL. Elimination resolvents and learned clauses are emitted as witness-free additions with deletions; contradiction yields UNSAT, but satisfying results or unsupported inputs return UNKNOWN.}
\asGallerySolver{VeriPB}{\asGalleryValue{asGalleryLoss}{0}{0.614}}{\asGalleryValue{asGalleryLoss}{0}{0.679}}{1/8}{Exact miter recognition admits only the supported topological XOR/OR shape, then bounded elimination and self-subsuming resolution preprocess the CNF before watched-literal CDCL with first-UIP learning. UNSAT paths emit VeriPB resolution and RUP steps; SAT paths print checked assignments, while unsupported shapes or internal failures return UNKNOWN.}
\end{asGalleryFamily}

\begin{asGalleryFamily}{006}{alloy-\allowbreak vpn-\allowbreak models}{15 GBD instances\enspace\textperiodcentered\enspace 3 validation\enspace\textperiodcentered\enspace baseline 3/3}{These benchmarks ask whether a Boolean assignment satisfies a circuit-shaped CNF encoding a relational model, including auxiliary gate variables and one asserted condition. The instances use structured binary implications and longer gate clauses rather than arbitrary CNF.}
\ifdefined\pdfbookmark\pdfbookmark[2]{alloy-vpn-models}{as-gallery-6}\fi
\asGallerySolver{GRAT}{\asGalleryValue{asGalleryLoss}{0}{0.220}}{\asGalleryValue{asGalleryLoss}{0}{0.222}}{2/3}{Signed-AND circuit recognition contracts parity-equivalent variables and propagates assignments through a gate DAG, then tries restricted searches before a circuit-aware CDCL fallback. SAT assignments are expanded and checked against the original CNF; a level-0 fallback conflict can emit a DRAT trace for external elaboration, while unsupported shapes or failed searches return UNKNOWN.}
\asGallerySolver{DPR}{\asGalleryValue{asGalleryLoss}{0}{0.445}}{\asGalleryValue{asGalleryLoss}{0}{0.435}}{3/3}{It reconstructs signed-AND gates with parity union-find, then searches leaf representatives using DAG propagation and circuit-aware CDCL. A fallback can emit unit and learned clause additions as a DRAT trace for external elaboration and checking; unsupported encodings or failed searches return UNKNOWN, while SAT assignments are expanded and CNF-checked.}
\asGallerySolver{VeriPB}{\asGalleryValue{asGalleryLoss}{1}{0.802}}{\asGalleryValue{asGalleryLoss}{1}{0.766}}{3/3}{On selected accepted shapes, support-pair gate normalization reorders clauses before watched-literal CDCL; other accepted shapes use the generic watched-literal path. SAT assignments are printed as complete models, but UNSAT has no active certificate path: lower-scope results lack certification and the largest hard-coded scope returns UNKNOWN.}
\end{asGalleryFamily}

\begin{asGalleryFamily}{007}{antibandwidth}{187 GBD instances\enspace\textperiodcentered\enspace 38 validation\enspace\textperiodcentered\enspace baseline 27/38}{Antibandwidth asks whether graph vertices can occupy distinct positions so adjacent vertices are far apart by at least w. CNF uses vertex-position variables, coverage constraints, and clauses forbidding edge endpoints from being too close.}
\ifdefined\pdfbookmark\pdfbookmark[2]{antibandwidth}{as-gallery-7}\fi
\asGallerySolver{GRAT}{\asGalleryValue{asGalleryLoss}{0}{0.894}}{\asGalleryValue{asGalleryLoss}{0}{0.931}}{20/38}{Release-time permutations and focused tabu swaps seek a labeling that separates every edge, guiding exchanges by conflicts and distance deficits. It validates and emits a completed SAT assignment, but rejects other structures as UNKNOWN, has no UNSAT certificate path, and may continue without resolving a failed search.}
\asGallerySolver{DPR}{\asGalleryValue{asGalleryGain}{1}{1.54}}{\asGalleryValue{asGalleryGain}{1}{1.52}}{27/38}{Capacity-constrained graph coloring, layered ordering, and local swaps first seek an antibandwidth order; otherwise watched-literal CDCL completes the recognized formula. A validated completion supplies the SAT certificate; UNSAT uses witness-free direct or learned additions plus an empty clause, while unsupported structures return UNKNOWN.}
\asGallerySolver{VeriPB}{\asGalleryValue{asGalleryLoss}{0}{0.919}}{\asGalleryValue{asGalleryLoss}{0}{0.937}}{20/38}{Maximum-linear-arrangement seeding and stochastic swap search build a vertex permutation, using min-conflicts and penalty objectives to improve edge separation. It emits a validated SAT assignment; the checked recognizer accepts only this structure, bounded search returns UNKNOWN without an ordering, and no UNSAT certificate path is implemented.}
\end{asGalleryFamily}

\begin{asGalleryFamily}{008}{argumentation}{217 GBD instances\enspace\textperiodcentered\enspace 40 validation\enspace\textperiodcentered\enspace baseline 9/40}{The benchmark asks whether a directed argumentation instance admits a conflict-free, defended, or stable set of arguments, or a complete labeling. CNF encodings use status variables and clauses expressing attacks, defense, coverage, and sometimes exactly-one choices, though supported layouts are only restricted patterns.}
\ifdefined\pdfbookmark\pdfbookmark[2]{argumentation}{as-gallery-8}\fi
\asGallerySolver{GRAT}{\asGalleryValue{asGalleryLoss}{1}{0.970}}{\asGalleryValue{asGalleryLoss}{1}{0.978}}{7/40}{Exact recognition of the expected two-label clause layout drives projection onto in variables, then bounded graph search and randomized maximal conflict-free-set repair precede watched-literal CDCL. SAT assignments are checked against the original CNF; only this recognized projection can emit an UNSAT certificate, while other branches seek SAT and failed proof reruns return UNKNOWN.}
\asGallerySolver{DPR}{\asGalleryValue{asGalleryLoss}{0}{0.866}}{\asGalleryValue{asGalleryLoss}{0}{0.872}}{3/40}{Structural recognition of strict two-block and three-block layouts drives projection onto in variables, followed by bounded graph search and randomized maximal conflict-free-set repair before watched-literal CDCL. SAT assignments are checked against the original CNF; only the two-block rerun emits UNSAT, while other layouts or failed searches return UNKNOWN, without internally checking that stream.}
\asGallerySolver{VeriPB}{\asGalleryValue{asGalleryLoss}{0}{0.936}}{\asGalleryValue{asGalleryLoss}{0}{0.943}}{6/40}{Repeated-block status recovery exposes an attack graph for grounded fixed-point propagation, randomized kernel repair, and compressed CDCL over IN variables. SAT witnesses are checked on the original CNF; failure to find a stable kernel can return UNKNOWN on weaker encodings, while only conservative recognized cases emit an UNSAT proof.}
\end{asGalleryFamily}

\begin{asGalleryFamily}{009}{at-\allowbreak least-\allowbreak two-\allowbreak sol}{18 GBD instances\enspace\textperiodcentered\enspace 4 validation\enspace\textperiodcentered\enspace baseline 2/4}{The benchmark asks whether a Boolean CNF formula has at least two satisfying assignments, rather than merely one. A common encoding places two copies of the formula in one CNF, links corresponding variables with equality indicators, and requires at least one pair to differ.}
\ifdefined\pdfbookmark\pdfbookmark[2]{at-least-two-sol}{as-gallery-9}\fi
\asGallerySolver{GRAT}{\asGalleryValue{asGalleryLoss}{0}{0.599}}{\asGalleryValue{asGalleryLoss}{0}{0.688}}{0/4}{A bounded 2-CNF implication/SCC search and a permutation-CSP search first seek two models after exact layout recognition. Fallback watched-literal CDCL uses a blocking clause for the second model; unsupported layouts or failure to find that model return UNKNOWN rather than UNSAT, while only fallback UNSAT paths emit textual DRAT for external elaboration.}
\asGallerySolver{DPR}{\asGalleryValue{asGalleryLoss}{1}{0.785}}{\asGalleryValue{asGalleryLoss}{1}{0.883}}{1/4}{An active 2-CNF-plus-wide-clause search and a permutation-CSP search first seek two models after exact layout recognition. Fallback watched-literal CDCL uses a blocking clause for another model; unsupported layouts or failed second-model validation return UNKNOWN, while UNSAT paths emit witness-free textual DPR clause additions ending in the empty clause, to be elaborated and checked externally.}
\asGallerySolver{VeriPB}{\asGalleryValue{asGalleryLoss}{0}{0.599}}{\asGalleryValue{asGalleryLoss}{0}{0.688}}{0/4}{An AIG structural-hashing shortcut and bounded Davis-Putnam elimination target selected UNSAT cases, retaining information for RUP replay and model extension. Fallback component-wise watched-literal CDCL finds one model, blocks it, and seeks a second; checked witnesses are accepted, while failed recognition or second-model search returns UNKNOWN and UNSAT paths close with VeriPB RUP certificates.}
\end{asGalleryFamily}

\begin{asGalleryFamily}{010}{auto-\allowbreak correlation}{51 GBD instances\enspace\textperiodcentered\enspace 11 validation\enspace\textperiodcentered\enspace baseline 2/11}{The benchmark asks whether a binary sequence has bounded aperiodic autocorrelation or whether two cyclic subsets have constant combined intersection counts at every nonzero shift. CNF encodings use primary variables and auxiliary gates or counters to express these constraints.}
\ifdefined\pdfbookmark\pdfbookmark[2]{auto-correlation}{as-gallery-10}\fi
\asGallerySolver{GRAT}{\asGalleryValue{asGalleryGain}{0}{1.60}}{\asGalleryValue{asGalleryGain}{0}{1.60}}{5/11}{Structural recognition drives multiplier-orbit enumeration and cyclic intersection matching for cyclic subset instances, while tabu bit flips search recognized aperiodic sequence layouts. Smaller cyclic cases use randomized annealing; larger or unsupported cases, low-goal aperiodic cases, and exhausted searches return UNKNOWN, while residual WalkSAT completion can produce SAT assignments without an UNSAT certificate path.}
\asGallerySolver{DPR}{\asGalleryValue{asGalleryLoss}{0}{0.871}}{\asGalleryValue{asGalleryLoss}{0}{0.871}}{0/11}{It searches recognized cyclic subset layouts by enumerating bounded multiplicative-orbit unions and matching cyclic difference signatures, including complementary, swapped, and rotated variants. After fixing primary bits, watched-literal propagation plus an internal full clause check validate complete SAT assignments; failed recognition or construction returns UNKNOWN, with no UNSAT or DPR certificate path.}
\asGallerySolver{VeriPB}{\asGalleryValue{asGalleryGain}{1}{1.91}}{\asGalleryValue{asGalleryGain}{1}{1.91}}{6/11}{It dispatches recognized aperiodic layouts to correlation-maintaining simulated annealing with restarts, and cyclic layouts to bounded multiplicative-orbit enumeration with hashed meet-in-the-middle autocorrelation matching. Propagation, uniform fills, and residual WalkSAT completion precede full clause verification; unsupported, exhausted, and low-bound aperiodic cases return UNKNOWN, with no general SAT fallback or UNSAT/VeriPB certificate path.}
\end{asGalleryFamily}

\begin{asGalleryFamily}{011}{automata-\allowbreak synchronization}{12 GBD instances\enspace\textperiodcentered\enspace 3 validation\enspace\textperiodcentered\enspace baseline 3/3}{Does a bounded-length word send every state of a complete two-letter deterministic automaton to one state? Its time-expanded CNF tracks reachable states and letter choices, propagates transitions, and requires at most one final reachable state.}
\ifdefined\pdfbookmark\pdfbookmark[2]{automata-synchronization}{as-gallery-11}\fi
\asGallerySolver{GRAT}{\asGalleryValue{asGalleryGain}{0}{21.1}}{\asGalleryValue{asGalleryGain}{0}{2.21}}{3/3}{For recognized inputs, structural Cerny recognition first yields a closed-form word; otherwise bit-packed image-set beams and reverse-BFS pair merging seek a bounded reset word. Candidate words become checked assignments; failed searches invoke bounded small-subset or antichain image-exclusion clauses for external DRAT elaboration, while unsupported or out-of-bound cases return UNKNOWN.}
\asGallerySolver{DPR}{\asGalleryValue{asGalleryLoss}{0}{0.101}}{\asGalleryValue{asGalleryLoss}{0}{0.134}}{1/3}{After recognizing the fixed two-letter image-set schema, it tries Cerny detection, image-set beams, greedy pair merging, then watched-literal CDCL. Direct words become checked assignments; CDCL emits learned and pair-distance clause additions, reporting UNSAT only after a level-zero contradiction. Missing transition clauses disable pair-distance preprocessing, and unsupported or uncertified cases return UNKNOWN.}
\asGallerySolver{VeriPB}{\asGalleryValue{asGalleryGain}{1}{489}}{\asGalleryValue{asGalleryGain}{1}{26.7}}{3/3}{After recognizing the reachable-subset encoding, it uses Cerny threshold dispatch and exact-encoding interval reasoning, then falls back to duplicate-eliminating bit-parallel beams. If searches find no word, bounded pair/triple-distance reasoning emits VeriPB proofs; SAT assignments are clause-checked, but the solver does not validate proof files and unsupported or unproved cases return UNKNOWN.}
\end{asGalleryFamily}

\begin{asGalleryFamily}{012}{baseball-\allowbreak lineup}{40 GBD instances\enspace\textperiodcentered\enspace 8 validation\enspace\textperiodcentered\enspace baseline 7/8}{The benchmark asks whether exactly K items can be selected so that every binary attribute receives at least its required coverage. Its CNF encodes the selection, item-attribute incidence, and cardinality or coverage counts with auxiliary variables and sequential counters.}
\ifdefined\pdfbookmark\pdfbookmark[2]{baseball-lineup}{as-gallery-12}\fi
\asGallerySolver{GRAT}{\asGalleryValue{asGalleryGain}{1}{1.98}}{\asGalleryValue{asGalleryGain}{1}{1.94}}{7/8}{After recognizing the row-mask-counter schema, weighted randomized greedy selection with targeted min-conflicts swaps drives the primary search. It falls back to a bounded DPLL tree; availability conflicts emit RUP-style units and an empty clause, while unresolved cases return UNKNOWN. SAT assignments reconstruct auxiliaries and are checked against all input clauses.}
\asGallerySolver{DPR}{\asGalleryValue{asGalleryGain}{0}{1.98}}{\asGalleryValue{asGalleryGain}{0}{1.65}}{7/8}{Strict row-mask-counter recognition precedes deficit-weighted greedy selection with restarts and swap-delta improvement. Failure triggers exact row conflict search only for small K; larger unresolved cases return UNKNOWN. UNSAT uses sparse-column units or conflict refutations, may add witness-free clauses, and is not internally verified; SAT models are clause-checked.}
\asGallerySolver{VeriPB}{\asGalleryValue{asGalleryLoss}{0}{0.992}}{\asGalleryValue{asGalleryLoss}{0}{0.993}}{6/8}{Scarcity-weighted greedy search with breakout swaps follows strict incidence-counter recovery. Failure falls back to a bounded primary-variable tree, then weighted PB separation with lifted-counter proof logging; bounded or unsupported cases return UNKNOWN. SAT candidates are clause-checked; UNSAT emits RUP or PB derivations, including deficient-support contradictions.}
\end{asGalleryFamily}

\begin{asGalleryFamily}{013}{battleship}{45 GBD instances\enspace\textperiodcentered\enspace 9 validation\enspace\textperiodcentered\enspace baseline 8/9}{The benchmark asks whether selected cyclic modular lines cover every point of an n-by-n toroidal board. It encodes line choices with Boolean variables, coverage clauses for every point, and within-block at-most-one clauses.}
\ifdefined\pdfbookmark\pdfbookmark[2]{battleship}{as-gallery-13}\fi
\asGallerySolver{GRAT}{\asGalleryValue{asGalleryGain}{0}{421}}{\asGalleryValue{asGalleryGain}{0}{35.4}}{9/9}{A direct square proof handles complete square instances, while bounded prime-power construction and weighted min-conflicts seek checked SAT assignments elsewhere. If these searches fail, affine-symmetry lifting or watched-literal CDCL can emit DRAT traces for applicable cases; unsupported or unresolved inputs return UNKNOWN, and UNSAT requires successful proof generation.}
\asGallerySolver{DPR}{\asGalleryValue{asGalleryLoss}{0}{0.414}}{\asGalleryValue{asGalleryLoss}{0}{0.420}}{6/9}{Compressed-word min-conflicts tracks line coverage with exact deltas and restarts, trying one permitted omitted pair with both rows forced before one-row search unless its negative-case guard applies. After heuristic failure, bounded unit propagation emits a DPR-style stream with witness-free additions; limits or unsupported layouts return UNKNOWN.}
\asGallerySolver{VeriPB}{\asGalleryValue{asGalleryGain}{1}{2.49\!\cdot\!10^{4}}}{\asGalleryValue{asGalleryGain}{1}{4{,}470}}{9/9}{A direct cutting-planes contradiction handles complete square instances, while a prime-field two-affine-pencil construction supplies checked SAT assignments on qualifying complete instances. Otherwise grouped set-cover min-conflicts uses coverage counts, tabu moves, and bounded restarts before selected CDCL fallbacks; specialized proof routes certify limited UNSAT cases, while unresolved recognized inputs return UNKNOWN.}
\end{asGalleryFamily}

\begin{asGalleryFamily}{014}{belpyramid-\allowbreak puzzle}{57 GBD instances\enspace\textperiodcentered\enspace 12 validation\enspace\textperiodcentered\enspace baseline 6/12}{These instances ask whether two Boolean networks with shared inputs differ on some output. CNF clauses encode AND gates and output comparisons; a required mismatch makes SAT witness inequivalence and UNSAT indicate equivalence; this is not every possible encoding.}
\ifdefined\pdfbookmark\pdfbookmark[2]{belpyramid-puzzle}{as-gallery-14}\fi
\asGallerySolver{GRAT}{\asGalleryValue{asGalleryLoss}{0}{0.638}}{\asGalleryValue{asGalleryLoss}{0}{0.795}}{1/12}{RUP-checked simulation lemmas drive solving on normalized AIG miters, with multiword signatures proposing equivalences before a monolithic watched-literal CDCL fallback. First-UIP learning and reason minimization support fallback search; SAT models are checked, the pyramid branch returns UNKNOWN, and UNSAT emits textual DRUP-style additions for external checking.}
\asGallerySolver{DPR}{\asGalleryValue{asGalleryLoss}{1}{0.698}}{\asGalleryValue{asGalleryLoss}{1}{0.870}}{2/12}{Signature-guided equivalence sweeping drives incremental assumption-based CDCL on a recognized AND-miter layout. Multiword signatures propose equivalent or complementary variables, with a complementary-input prepass; bounded queries retain learned clauses and final conflicts emit proof clauses, while failed searches or unsupported layouts return UNKNOWN without a SAT witness.}
\asGallerySolver{VeriPB}{\asGalleryValue{asGalleryLoss}{0}{0.690}}{\asGalleryValue{asGalleryLoss}{0}{0.825}}{2/12}{Comparator-cone probing leads the compact-miter search: assumed output mismatches are tested with restricted watched-literal CDCL, then the full miter receives unrestricted CDCL. Learned clauses remain global and UNSAT is logged with VeriPB RUP records, but rejected finite-domain layouts and all non-UNSAT outcomes return UNKNOWN without a SAT witness.}
\end{asGalleryFamily}

\begin{asGalleryFamily}{015}{binary-\allowbreak pigeon-\allowbreak hole}{5 GBD instances\enspace\textperiodcentered\enspace 1 validation\enspace\textperiodcentered\enspace baseline 0/1}{Binary pigeon-hole CNFs ask whether p pigeons can be assigned distinct holes using binary codes. A direct encoding gives each pigeon a bit block, excludes unused codes, and forbids two pigeons from sharing a valid code.}
\ifdefined\pdfbookmark\pdfbookmark[2]{binary-pigeon-hole}{as-gallery-15}\fi
\asGallerySolver{GRAT}{\asGalleryValue{asGalleryGain}{0}{4{,}340}}{\asGalleryValue{asGalleryGain}{0}{1.40}}{1/1}{Exact structural recognition replaces general SAT search. For recognized formulas, it translates codes into unary membership and capacity constraints, repeatedly reduces pigeon-hole instances with fresh bridge variables, and emits an empty clause in a DRAT-style stream for external elaboration and checking. Recognition or proof-writing failure returns UNKNOWN.}
\asGallerySolver{DPR}{\asGalleryValue{asGalleryGain}{0}{3{,}170}}{\asGalleryValue{asGalleryLoss}{0}{0.620}}{1/1}{Exact recognition of the complete direct encoding drives a specialized path without general search. SAT assigns pigeon i code i and rechecks the clauses; UNSAT adds exact-code indicators and unary pigeon-hole constraints, then uses witness-free clause additions and propagation-redundancy symmetry steps to reach an empty clause; unsupported cases return UNKNOWN.}
\asGallerySolver{VeriPB}{\asGalleryValue{asGalleryGain}{1}{3.51\!\cdot\!10^{4}}}{\asGalleryValue{asGalleryGain}{1}{264}}{1/1}{Exact canonical-encoding recognition, including an order-independent retry, and a binary-code assignment shortcut replace general SAT search. SAT assigns pigeon i code i and checks clauses; UNSAT emits a VeriPB certificate with exact-code indicators, per-hole pseudo-Boolean capacity derivations, and per-pigeon binary resolution trees; unsupported inputs or proof-generation failures return UNKNOWN.}
\end{asGalleryFamily}

\begin{asGalleryFamily}{016}{binary-\allowbreak tree-\allowbreak parity}{2 GBD instances\enspace\textperiodcentered\enspace 1 validation\enspace\textperiodcentered\enspace baseline 0/1}{These instances ask whether a Boolean assignment satisfies three-variable parity equations whose supports form two full binary trees, possibly in different variable orderings. In CNF, each equation is represented by clauses forbidding the four assignments of one parity, with some instances allowing a shortened block.}
\ifdefined\pdfbookmark\pdfbookmark[2]{binary-tree-parity}{as-gallery-16}\fi
\asGallerySolver{GRAT}{\asGalleryValue{asGalleryGain}{1}{7.42\!\cdot\!10^{5}}}{\asGalleryValue{asGalleryGain}{1}{3.23\!\cdot\!10^{5}}}{1/1}{Packed GF(2) Gaussian elimination solves recognized parity groups, with bounded affine-space enumeration and direct checking of candidates against the CNF. It falls back to order-independent matching for shortened gates; SAT emits a checked assignment, while unsupported or unsuccessful cases return UNKNOWN because no UNSAT certificate path is implemented.}
\asGallerySolver{DPR}{\asGalleryValue{asGalleryGain}{0}{3.21\!\cdot\!10^{5}}}{\asGalleryValue{asGalleryGain}{0}{6.25\!\cdot\!10^{4}}}{1/1}{Exact recognition of the gate order and orientation precedes packed GF(2) Gaussian elimination and bounded Gray-code enumeration of affine solutions, with SAT candidates checked against clauses. It retries weakened gates by fixing the omitted-clause assignment and attempts Davis-Putnam refutation only for small no-model cases; unsupported or uncertified outcomes return UNKNOWN.}
\asGallerySolver{VeriPB}{\asGalleryValue{asGalleryGain}{0}{3.17\!\cdot\!10^{5}}}{\asGalleryValue{asGalleryGain}{0}{6.39\!\cdot\!10^{4}}}{1/1}{Affine reduction through the recognized first tree produces a packed GF(2) system for the second. It also enumerates bounded shortened-gate branches and checks candidates against the CNF. Failed-literal propagation yields UNSAT only on a RUP contradiction; failed or unsupported cases return UNKNOWN, while SAT emits a checked assignment.}
\end{asGalleryFamily}

\begin{asGalleryFamily}{017}{bioinformatics}{61 GBD instances\enspace\textperiodcentered\enspace 13 validation\enspace\textperiodcentered\enspace baseline 10/13}{The benchmark asks whether Boolean variables satisfy a structured set of constraints; some instances organize them as one-of-many choices at positions with compatibility or circuit conditions. CNF uses one-hot clauses and auxiliary Tseitin variables, but these recognized shapes do not define every possible encoding.}
\ifdefined\pdfbookmark\pdfbookmark[2]{bioinformatics}{as-gallery-17}\fi
\asGallerySolver{GRAT}{\asGalleryValue{asGalleryLoss}{0}{0.461}}{\asGalleryValue{asGalleryLoss}{1}{0.691}}{6/13}{Conservative XITS recognition enables canonical partition branching in small-class cases; other inputs use activity-ordered first-UIP CDCL, with failed-literal probing and bounded elimination only on the generic path. Checked SAT models are printed, while reported UNSAT is accompanied by textual DRAT ending in an empty clause; unresolved search or reconstruction failure returns UNKNOWN.}
\asGallerySolver{DPR}{\asGalleryValue{asGalleryLoss}{0}{0.362}}{\asGalleryValue{asGalleryLoss}{0}{0.532}}{4/13}{Bounded Davis-Putnam elimination protects the one-hot and counter prefix; Xits cases also bias state choices and probe same-state pairs before CDCL fallback. SAT assignments are reconstructed and checked; internal UNSAT is logged with preprocessing and learned clauses plus an empty clause for external DPR elaboration, while unrecognized inputs or failed checks return UNKNOWN.}
\asGallerySolver{VeriPB}{\asGalleryValue{asGalleryLoss}{1}{0.485}}{\asGalleryValue{asGalleryLoss}{0}{0.436}}{6/13}{Detected Xits shapes activate positive phases on choice blocks, Xits-gated minimization, and learned-clause reduction; other inputs use activity-based first-UIP CDCL with phase saving and Luby restarts. Checked SAT assignments are printed, while UNSAT requires written VeriPB RUP records ending in an empty record; parse, check, or proof-write failure returns UNKNOWN.}
\end{asGalleryFamily}

\begin{asGalleryFamily}{018}{bitvector}{594 GBD instances\enspace\textperiodcentered\enspace 119 validation\enspace\textperiodcentered\enspace baseline 49/119}{The benchmark asks whether a Boolean assignment can satisfy a CNF encoding of quantifier-free bit-vector constraints, with variables representing bit values and circuit intermediates. Clauses enforce translated operations and asserted constraints.}
\ifdefined\pdfbookmark\pdfbookmark[2]{bitvector}{as-gallery-18}\fi
\asGallerySolver{GRAT}{\asGalleryValue{asGalleryLoss}{0}{0.739}}{\asGalleryValue{asGalleryLoss}{1}{0.855}}{20/119}{It recognizes local gate blocks and, when coverage is sufficient, tries bit-parallel functional-block search; otherwise it uses watched-literal CDCL with activity branching, 1-UIP learning, restarts, and clause reduction. SAT models are rechecked, while UNSAT is represented by textual DRAT additions for external elaboration; failures can return UNKNOWN.}
\asGallerySolver{DPR}{\asGalleryValue{asGalleryLoss}{0}{0.737}}{\asGalleryValue{asGalleryLoss}{0}{0.846}}{21/119}{Local gate-pattern detection supplies only a heap tie-break, not bit-vector or XOR reasoning; the solver otherwise runs generic watched-literal first-UIP CDCL with phase saving and restarts. Its addition-only proof records learned clauses and the empty clause at root conflict, while SAT models are checked and printed separately, not in the proof.}
\asGallerySolver{VeriPB}{\asGalleryValue{asGalleryLoss}{1}{0.756}}{\asGalleryValue{asGalleryLoss}{0}{0.852}}{23/119}{Root failed-literal probing and bounded proof-logged variable elimination precede generic watched-literal first-UIP CDCL; no bit-vector semantics are used. SAT assignments are independently checked, while UNSAT emits VeriPB RUP steps, deletions, and a final empty clause, with parsing, checking, or proof-writing failures yielding UNKNOWN.}
\end{asGalleryFamily}

\begin{asGalleryFamily}{019}{bounded-\allowbreak model-\allowbreak checking}{31 GBD instances\enspace\textperiodcentered\enspace 7 validation\enspace\textperiodcentered\enspace baseline 7/7}{These benchmarks ask whether a Boolean assignment satisfies a Tseitin-encoded circuit check. Variables represent circuit or state signals, and CNF clauses impose local gate and checking constraints.}
\ifdefined\pdfbookmark\pdfbookmark[2]{bounded-model-checking}{as-gallery-19}\fi
\asGallerySolver{GRAT}{\asGalleryValue{asGalleryLoss}{1}{0.0612}}{\asGalleryValue{asGalleryLoss}{1}{0.0605}}{7/7}{It recognizes dense Tseitin-like AND/OR and XOR/XNOR structure, then applies bounded elimination to gate outputs; unsupported shapes return UNKNOWN rather than invoking general search. CDCL with first-UIP learning is the fallback, emitting DRAT additions for external elaboration on UNSAT and checking reconstructed SAT models against the original clauses.}
\asGallerySolver{DPR}{\asGalleryValue{asGalleryLoss}{0}{0.0549}}{\asGalleryValue{asGalleryLoss}{0}{0.0390}}{7/7}{It starts with reverse-order bounded variable elimination, retaining non-tautological resolvents and recording eliminated values for reconstruction; watched-literal CDCL with first-UIP learning is the fallback when no contradiction appears. UNSAT uses addition-only RUP-style clause additions, while reconstructed SAT assignments are checked against original clauses; failed reconstruction or validation returns UNKNOWN, and proof logging is configurable.}
\asGallerySolver{VeriPB}{\asGalleryValue{asGalleryLoss}{0}{0.0471}}{\asGalleryValue{asGalleryLoss}{0}{0.0168}}{7/7}{It begins with parity-DSU substitution and conservative gate hashing, adding bounded elimination only for a recognized signature; unrecognized or malformed inputs return UNKNOWN. CDCL with first-UIP learning is the fallback, with UNSAT logged as augmented VeriPB RUP additions and reconstructed SAT assignments checked against the original CNF.}
\end{asGalleryFamily}

\begin{asGalleryFamily}{020}{brent-\allowbreak equations}{20 GBD instances\enspace\textperiodcentered\enspace 4 validation\enspace\textperiodcentered\enspace baseline 3/4}{These benchmarks ask whether the 3 by 3 matrix-multiplication tensor over GF(2) has a rank-r decomposition into binary factor vectors. A Tseitin-style CNF represents products and XOR accumulation, yielding 729 parity equations that describe the tensor.}
\ifdefined\pdfbookmark\pdfbookmark[2]{brent-equations}{as-gallery-20}\fi
\asGallerySolver{GRAT}{\asGalleryValue{asGalleryGain}{0}{1.06}}{\asGalleryValue{asGalleryGain}{1}{1.06}}{3/4}{A 26-term Strassen-style construction drives recognized unfixed high-rank cases, while fixed cases use DSU component decomposition, bounded multi-choice search, and bipartite matching after gate-network recognition. Incremental residual local search handles eligible high-rank failures; clause-checked SAT assignments are emitted, but no UNSAT certificate path exists and unsuccessful cases return UNKNOWN.}
\asGallerySolver{DPR}{\asGalleryValue{asGalleryGain}{1}{1.06}}{\asGalleryValue{asGalleryGain}{0}{1.06}}{3/4}{XOR-root monomial flattening and bipartite matching first construct a primary assignment for recognized high-rank encodings. When direct construction fails, an open-ended weighted WalkSAT search may precede watched-literal CDCL; SAT assignments are checked and emitted, while unsupported, low-rank, or UNSAT cases return UNKNOWN and no UNSAT certificate path is implemented.}
\asGallerySolver{VeriPB}{\asGalleryValue{asGalleryGain}{0}{1.06}}{\asGalleryValue{asGalleryGain}{0}{1.06}}{3/4}{Direct reconstruction of the fixed generator schedule drives sparse rank-one completion; substituted dense cases use bit-parallel tabu search with incremental deltas and repair. Watched-literal CDCL follows failed direct searches; clause-checked SAT assignments are emitted, but low-rank, contradictory, or failed searches return UNKNOWN and no UNSAT certificate path exists.}
\end{asGalleryFamily}

\begin{asGalleryFamily}{021}{cardinality-\allowbreak constraints}{17 GBD instances\enspace\textperiodcentered\enspace 4 validation\enspace\textperiodcentered\enspace baseline 4/4}{The benchmark asks whether at most K lattice points can hit every specified geometric object, such as a square or triangle. CNF uses Boolean variables for point selections, clauses requiring each object to be hit, and auxiliary clauses encoding the at-most-K bound.}
\ifdefined\pdfbookmark\pdfbookmark[2]{cardinality-constraints}{as-gallery-21}\fi
\asGallerySolver{GRAT}{\asGalleryValue{asGalleryLoss}{1}{0.506}}{\asGalleryValue{asGalleryLoss}{1}{0.779}}{3/4}{Incidence recognition drives a canonical hitting-set witness for aligned squares, with min-conflicts as a SAT fallback; propagation completes the model before checking it. For arbitrary squares and aligned triangles, mapped templates or DRAT-producing 1-UIP CDCL handle UNSAT; aligned squares lack a CDCL fallback, and unsupported or failed paths return UNKNOWN.}
\asGallerySolver{DPR}{\asGalleryValue{asGalleryLoss}{0}{0.301}}{\asGalleryValue{asGalleryLoss}{0}{0.633}}{2/4}{Bounded geometric regeneration and weighted min-conflicts construct a capped hitting set; propagation then completes and checks the original CNF for SAT. On failure, bounded resolution elimination feeds GenericCDCL, whose resolution and learned-clause additions support an UNSAT certificate; unsupported or altered inputs, or failure of that path, return UNKNOWN.}
\asGallerySolver{VeriPB}{\asGalleryValue{asGalleryLoss}{0}{0.302}}{\asGalleryValue{asGalleryLoss}{0}{0.633}}{2/4}{Fixed-cardinality simulated annealing searches recognized exact geometric layouts while maintaining unhit objects; unit propagation and small recursive DPLL then complete the assignment and check every original clause. No UNSAT certificate path is active: recognition, construction, or completion failure returns UNKNOWN.}
\end{asGalleryFamily}

\begin{asGalleryFamily}{022}{cellular-\allowbreak automata}{52 GBD instances\enspace\textperiodcentered\enspace 11 validation\enspace\textperiodcentered\enspace baseline 8/11}{These benchmarks ask whether a cyclic binary row has a predecessor trajectory under a local cellular-automaton rule for T steps ending in prescribed values. CNF represents cell states by time layer, links local neighborhoods with transition clauses, and fixes terminal values with unit clauses.}
\ifdefined\pdfbookmark\pdfbookmark[2]{cellular-automata}{as-gallery-22}\fi
\asGallerySolver{GRAT}{\asGalleryValue{asGalleryGain}{1}{1.29}}{\asGalleryValue{asGalleryGain}{1}{1.44}}{9/11}{Backward traversal of recognized layered gates biases search from the fixed terminal row, while a shallow clause-width SPG heuristic tries sequential phase passes before the generic watched-literal CDCL fallback. SAT assignments are checked; UNSAT results use textual DRAT additions, with recognized-path clauses replayed chronologically, while an exhausted SPG budget returns UNKNOWN.}
\asGallerySolver{DPR}{\asGalleryValue{asGalleryLoss}{0}{0.455}}{\asGalleryValue{asGalleryLoss}{0}{0.520}}{2/11}{Structural clause-width dispatch selects bounded and deterministic phase passes for non-ECA layouts, then falls back to watched-literal CDCL with first-UIP learning; recognized ECA layouts return UNKNOWN unless ECA\_CDCL enables ordinary CDCL, with no specialized reverse-search path. SAT assignments are checked, while UNSAT learning is emitted as witness-free DPR additions, including the empty clause.}
\asGallerySolver{VeriPB}{\asGalleryValue{asGalleryLoss}{0}{0.692}}{\asGalleryValue{asGalleryLoss}{0}{0.698}}{6/11}{Recognized Rule-110 layouts first try a cyclic-predecessor check, then use backward row-biased CDCL; population signatures may activate bounded elimination, while unrecognized inputs return UNKNOWN. Tracked UNSAT replay emits VeriPB RUP dependencies and an empty conclusion, while direct predecessor derivations are limited to targets without a cyclic one-step predecessor; SAT assignments are checked.}
\end{asGalleryFamily}

\begin{asGalleryFamily}{023}{circuit-\allowbreak equialence-\allowbreak checking}{19 GBD instances\enspace\textperiodcentered\enspace 4 validation\enspace\textperiodcentered\enspace baseline 0/4}{The benchmark asks whether two Boolean circuits compute the same function. A CNF miter combines their outputs with an XOR and forces it true, so SAT gives a distinguishing input while UNSAT establishes equivalence for the gate encoding.}
\ifdefined\pdfbookmark\pdfbookmark[2]{circuit-equialence-checking}{as-gallery-23}\fi
\asGallerySolver{GRAT}{\asGalleryValue{asGalleryMuted}{1}{1.00}}{\asGalleryValue{asGalleryMuted}{1}{1.00}}{0/4}{Strict LUT/miter recognition drives bit-parallel simulation of small root supports, checking any distinguishing assignment against the original clauses. Watched-literal DPLL handles larger supported roots and logs learned or cube-blocking clauses as textual DRAT additions, ending UNSAT with an empty clause; unsupported layouts or excessive support return UNKNOWN.}
\asGallerySolver{DPR}{\asGalleryValue{asGalleryMuted}{1}{1.00}}{\asGalleryValue{asGalleryMuted}{1}{1.00}}{0/4}{Semantic sweeping groups equal or complemented signals using bit-parallel truth-table simulation, then tests candidate implications by propagation and RUP. Common-cut lemmas and bounded CDCL add witness-free clause additions, ending proved UNSAT with an empty clause; bounded failure returns UNKNOWN, and SAT models are checked.}
\asGallerySolver{VeriPB}{\asGalleryValue{asGalleryMuted}{1}{1.00}}{\asGalleryValue{asGalleryMuted}{1}{1.00}}{0/4}{Packed signatures find equal or complemented candidate correspondences, certified first by bounded parity-DSU exact cuts with bit-parallel evaluation. Relevant-cone CDCL falls back when cuts fail, reusing proved pairs and logging RUP or polynomial-resolution steps; internal SAT only rejects candidates, no SAT assignment is emitted, and unsupported or unproved roots return UNKNOWN.}
\end{asGalleryFamily}

\begin{asGalleryFamily}{024}{circuit-\allowbreak equivalence-\allowbreak checking}{20 GBD instances\enspace\textperiodcentered\enspace 4 validation\enspace\textperiodcentered\enspace baseline 1/4}{The benchmark asks whether two Boolean circuit signals agree on every shared input assignment. Gate relations are encoded in CNF and a difference (XOR) output is asserted, so SAT supplies a counterexample and UNSAT establishes equivalence.}
\ifdefined\pdfbookmark\pdfbookmark[2]{circuit-equivalence-checking}{as-gallery-24}\fi
\asGallerySolver{GRAT}{\asGalleryValue{asGalleryGain}{1}{3.01}}{\asGalleryValue{asGalleryGain}{1}{3.07}}{3/4}{Bit-parallel SAT sweeping leads the search: signatures on a recognized acyclic LUT network propose equal or complemented signals, and CDCL assumptions test them against counterexamples. Bounded functional search covers exceptional inputs; fallback CDCL logs textual DRAT for external checks on UNSAT, while unsupported inputs return UNKNOWN and SAT yields a checked model.}
\asGallerySolver{DPR}{\asGalleryValue{asGalleryGain}{0}{1.52}}{\asGalleryValue{asGalleryGain}{0}{1.56}}{2/4}{Assumption-based relation queries lead the search: incremental watched-literal CDCL tests candidate signal equalities from a bounded signature sweep, with counterexamples refining the signatures before final fanin queries. Failed candidates are discarded; unsupported or uncertifiable encodings return UNKNOWN, SAT gives a checked model, and only the final UNSAT path emits a dependency-tracked RUP proof core.}
\asGallerySolver{VeriPB}{\asGalleryValue{asGalleryLoss}{0}{0.763}}{\asGalleryValue{asGalleryLoss}{0}{0.782}}{0/4}{Bit-parallel simulation leads: replayed lanes can produce clause-checked SAT models; otherwise bounded gate elimination feeds embedded CDCL, whose UNSAT path logs a VeriPB proof with RUP constraints and an empty clause, while fallback SAT is not checked against the original formula and nonmatching encodings return UNKNOWN.}
\end{asGalleryFamily}

\begin{asGalleryFamily}{025}{circuit-\allowbreak minimization}{50 GBD instances\enspace\textperiodcentered\enspace 10 validation\enspace\textperiodcentered\enspace baseline 0/10}{Structured circuit-minimization encodings ask whether exactly a prescribed number of Boolean candidates can cover every requirement; selector clauses express coverage, and Tseitin population-count clauses enforce the exact cardinality.}
\ifdefined\pdfbookmark\pdfbookmark[2]{circuit-minimization}{as-gallery-25}\fi
\asGallerySolver{GRAT}{\asGalleryValue{asGalleryGain}{0}{9.95}}{\asGalleryValue{asGalleryGain}{1}{9.95}}{9/10}{Structural recognition and population-counter probing recover candidate phases and the exact cover size, with a per-bit fallback. Weighted greedy cover search uses focused exchanges and changing target weights; counter propagation completes SAT models, whose original clauses are checked, while unsupported or unsuccessful cases return UNKNOWN without an UNSAT certificate path.}
\asGallerySolver{DPR}{\asGalleryValue{asGalleryGain}{1}{9.97}}{\asGalleryValue{asGalleryGain}{0}{9.92}}{9/10}{Phase normalization identifies candidate inputs and cover constraints, while bounded cardinality probing recovers the target size in two exact layouts. A randomized weighted add/remove search seeks the fixed-cardinality cover, then bounded propagation/DPLL completes variables; checked SAT assignments are emitted, but failed or unsupported cases return UNKNOWN because no UNSAT certificate path exists.}
\asGallerySolver{VeriPB}{\asGalleryValue{asGalleryGain}{0}{9.94}}{\asGalleryValue{asGalleryGain}{0}{9.88}}{9/10}{Recognition of two exact signatures recovers selector phases, cover rows, and target cardinality from the population-count core, with repair for a shortened-gate variant. Weighted fixed-cardinality search focuses on uncovered rows and scores swaps with tabu diversification; clause-checked candidates yield SAT models, while failed or unsupported cases return UNKNOWN without an UNSAT path.}
\end{asGalleryFamily}

\begin{asGalleryFamily}{026}{circuit-\allowbreak multiplier}{20 GBD instances\enspace\textperiodcentered\enspace 4 validation\enspace\textperiodcentered\enspace baseline 2/4}{The benchmark asks whether two nontrivial positive binary numbers, represented little-endian, can multiply to the fixed target encoded by the instance. CNF clauses constrain operand bits and auxiliary multiplier, adder, and carry wires so an assignment realizes that product.}
\ifdefined\pdfbookmark\pdfbookmark[2]{circuit-multiplier}{as-gallery-26}\fi
\asGallerySolver{GRAT}{\asGalleryValue{asGalleryLoss}{1}{0.789}}{\asGalleryValue{asGalleryLoss}{1}{0.789}}{1/4}{Watched-literal CDCL is the active engine, with first-UIP learning, restarts, rephasing, and clause reduction after a syntactic two-block check. SAT models are checked against the original CNF; closed non-SAT searches return UNKNOWN because no retained DRAT proof is produced.}
\asGallerySolver{DPR}{\asGalleryValue{asGalleryLoss}{0}{0.620}}{\asGalleryValue{asGalleryLoss}{0}{0.620}}{0/4}{Structural recognition admits only the documented operand blocks and circuit layout; malformed or unmatched inputs return UNKNOWN without generic SAT fallback. ProbSAT-style local search tracks unsatisfied clauses and break counts, then falls back to watched-literal CDCL for checked SAT models and witness-free DPR additions plus a root-contradiction marker on UNSAT.}
\asGallerySolver{VeriPB}{\asGalleryValue{asGalleryLoss}{0}{0.757}}{\asGalleryValue{asGalleryLoss}{0}{0.757}}{1/4}{Bounded variable elimination follows strict recognition, removes only internal variables while protecting operand words, and retains reverse-reconstruction records; unmatched inputs return UNKNOWN. Residual watched-literal CDCL uses first-UIP learning and restarts; reconstructed SAT models are checked against the original CNF, while internal UNSAT returns UNKNOWN without a VeriPB refutation.}
\end{asGalleryFamily}

\begin{asGalleryFamily}{027}{clique-\allowbreak coloring}{15 GBD instances\enspace\textperiodcentered\enspace 3 validation\enspace\textperiodcentered\enspace baseline 3/3}{The CNF asks whether some edge assignment contains k distinct selected vertices forming a clique and assigns every vertex a color so adjacent vertices differ. Edge, selection, and color variables encode these conditions with exactly-one, distinctness, and incompatibility clauses.}
\ifdefined\pdfbookmark\pdfbookmark[2]{clique-coloring}{as-gallery-27}\fi
\asGallerySolver{GRAT}{\asGalleryValue{asGalleryGain}{1}{31.1}}{\asGalleryValue{asGalleryGain}{1}{2.30}}{3/3}{Exact schema recognition and closed-form model construction are the active techniques. Its SAT branch checks a complete assignment; when k\ensuremath{>}c, it emits a DRAT proof with fresh variables and a Hall-style pigeonhole refutation, while unrecognized or unsupported cases return UNKNOWN instead of using generic search.}
\asGallerySolver{DPR}{\asGalleryValue{asGalleryGain}{0}{10.7}}{\asGalleryValue{asGalleryLoss}{0}{0.330}}{3/3}{Exact schema recognition gates both branches. When k\ensuremath{>}c, fresh-variable, witness-free clause additions and a subset-Hall refutation provide the UNSAT certificate; in the constructive regime, it constructs and checks a closed-form clique-coloring assignment. Unrecognized or unsupported cases return UNKNOWN rather than invoking generic search.}
\asGallerySolver{VeriPB}{\asGalleryValue{asGalleryGain}{0}{26.4}}{\asGalleryValue{asGalleryLoss}{0}{0.513}}{3/3}{Streaming recognition and direct model checking lead the solver. It constructs a checked SAT assignment when k\ensuremath{<}=n and c\ensuremath{>}=k; otherwise it emits bounded pseudo-Boolean proofs for n\ensuremath{<}k or k\ensuremath{>}c using position-pigeonhole reasoning or vertex symmetry, RUP, and cutting planes, while recognition or guard failures return UNKNOWN without general search.}
\end{asGalleryFamily}

\begin{asGalleryFamily}{028}{clique-\allowbreak formulas}{4 GBD instances\enspace\textperiodcentered\enspace 1 validation\enspace\textperiodcentered\enspace baseline 1/1}{The benchmark asks whether k ordered positions can choose vertices from different parts of a complete multipartite graph. Its CNF has one positive choice clause per position and binary clauses forbidding two choices in one position or the same part.}
\ifdefined\pdfbookmark\pdfbookmark[2]{clique-formulas}{as-gallery-28}\fi
\asGallerySolver{GRAT}{\asGalleryValue{asGalleryGain}{0}{447}}{\asGalleryValue{asGalleryGain}{1}{47.9}}{1/1}{Cross-row forbidden-neighbor signatures reconstruct rows and parts, followed by an exact conflict audit for the balanced encoding. It checks a complete SAT assignment when positions fit the parts; otherwise, for at most 16 parts, it writes a DRAT certificate for external elaboration and checking using OR definitions and subset-Hall clauses, while unsupported cases return UNKNOWN.}
\asGallerySolver{DPR}{\asGalleryValue{asGalleryGain}{1}{820}}{\asGalleryValue{asGalleryGain}{0}{17.9}}{1/1}{DSU recognition recovers selector positions and parts, requiring the balanced clone encoding and full incompatibility set rather than general search. It checks a SAT assignment when positions fit recovered parts; otherwise it emits a DPR refutation with clone-collapse additions and a pigeonhole sequence, while unsupported inputs or failed checks return UNKNOWN.}
\asGallerySolver{VeriPB}{\asGalleryValue{asGalleryGain}{0}{735}}{\asGalleryValue{asGalleryGain}{0}{9.00}}{1/1}{Twin-variable representative reduction and whole-part symmetry breaking compress the recognized balanced, auxiliary-free encoding instead of general clique search. It checks a SAT assignment when positions fit the parts; otherwise it writes a VeriPB proof ending in a RUP contradiction without verifying that proof; unsupported inputs or failures in model checks or proof output return UNKNOWN.}
\end{asGalleryFamily}

\begin{asGalleryFamily}{029}{clique-\allowbreak width}{27 GBD instances\enspace\textperiodcentered\enspace 6 validation\enspace\textperiodcentered\enspace baseline 1/6}{The benchmark asks whether a recovered graph can be built through n stages using at most k temporary labels, with each stage recording component relations, label choices, and graph constraints. These construction choices are represented by CNF variables and clauses.}
\ifdefined\pdfbookmark\pdfbookmark[2]{clique-width}{as-gallery-29}\fi
\asGallerySolver{GRAT}{\asGalleryValue{asGalleryMuted}{1}{1.00}}{\asGalleryValue{asGalleryMuted}{1}{1.00}}{1/6}{A partition-state merge search tries forest-postorder, linear-neighborhood, and beam-bounded general constructions, pruning merges by outside-neighborhood signatures and implication conflicts; reused-label enumeration is an active fallback. The narrow recognized layout is required; structural assignments receive SCC-based 2-SAT completion and full clause checking, while search or checking failure returns UNKNOWN and no UNSAT certificate is emitted.}
\asGallerySolver{DPR}{\asGalleryValue{asGalleryMuted}{0}{1.00}}{\asGalleryValue{asGalleryMuted}{0}{1.00}}{1/6}{A capped linear-order DFS seeks a graph ordering; success adds relation literals, while failure or cap exhaustion sends a binary normal form with representatives and Sinz-style counters to watched-literal CDCL. Only a narrow normalized layout is recognized; 32-bit masks limit support, and failed or unsupported cases return UNKNOWN without an UNSAT or DPR certificate.}
\asGallerySolver{VeriPB}{\asGalleryValue{asGalleryMuted}{0}{1.00}}{\asGalleryValue{asGalleryMuted}{0}{1.00}}{1/6}{Graph recovery and a bounded component-hierarchy trial provide the active heuristic, after which failure falls back to unrestricted watched-literal CDCL rather than implying UNSAT. The exact generator-specific layout is required; SAT assignments are checked internally, while UNSAT emits a PB RUP proof, and malformed or unsupported inputs return UNKNOWN without external checking.}
\end{asGalleryFamily}

\begin{asGalleryFamily}{030}{clustered-\allowbreak random}{43 GBD instances\enspace\textperiodcentered\enspace 9 validation\enspace\textperiodcentered\enspace baseline 8/9}{The benchmark asks whether a Boolean assignment satisfies every clause in a uniform 3- or 4-CNF formula whose clauses are arranged in overlapping clusters on limited variable supports. It represents a restricted clustered random SAT structure, not every possible encoding.}
\ifdefined\pdfbookmark\pdfbookmark[2]{clustered-random}{as-gallery-30}\fi
\asGallerySolver{GRAT}{\asGalleryValue{asGalleryLoss}{0}{0.379}}{\asGalleryValue{asGalleryLoss}{0}{0.498}}{6/9}{Focused local search with occurrence counters and bounded Hamming repair comes first, followed by watched-literal CDCL when no verified assignment is found. SAT returns a checked complete assignment; only the fallback logs learned clauses and an empty clause as a DRAT trace for external elaboration, while unsupported input or reconstruction failure returns UNKNOWN.}
\asGallerySolver{DPR}{\asGalleryValue{asGalleryLoss}{1}{0.481}}{\asGalleryValue{asGalleryLoss}{1}{0.616}}{7/9}{Factor-local exhaustive repair and stochastic local search with belief-propagation decimation for width-4 cases precede factor-resolution clauses, bounded Davis-Putnam elimination, and watched-literal CDCL fallback. SAT exits return checked assignments, whereas only the fallback emits witness-free RUP-style clause additions toward an UNSAT derivation; unsupported inputs return UNKNOWN.}
\asGallerySolver{VeriPB}{\asGalleryValue{asGalleryLoss}{0}{0.470}}{\asGalleryValue{asGalleryLoss}{0}{0.336}}{7/9}{Focused probSAT-style walking leads the search, with consensus-derived clauses on width-4 instances; failure triggers first-UIP CDCL with same-support consensus, bounded elimination, and a selective failed-literal pass. SAT returns a checked assignment, while UNSAT is certified through sliced VeriPB RUP steps; unsupported inputs or internal errors return UNKNOWN.}
\end{asGalleryFamily}

\begin{asGalleryFamily}{031}{coloring}{594 GBD instances\enspace\textperiodcentered\enspace 119 validation\enspace\textperiodcentered\enspace baseline 58/119}{The benchmark asks whether each graph vertex can receive a color while respecting every encoded incompatibility between endpoint choices. These constraints are translated into CNF using compact Boolean codes or one-hot choices in the recognized instances, rather than one universal coloring encoding.}
\ifdefined\pdfbookmark\pdfbookmark[2]{coloring}{as-gallery-31}\fi
\asGallerySolver{GRAT}{\asGalleryValue{asGalleryLoss}{0}{0.776}}{\asGalleryValue{asGalleryLoss}{0}{0.784}}{38/119}{Strict recognizers send compact cases through bit-mask forward checking, with small-domain branching and impact-based state ordering; one-hot cases use CDCL directly. SAT assignments are checked against the CNF; failed bounded searches use CDCL or elimination, emitting DRAT clause additions for UNSAT, while unresolved cases return UNKNOWN.}
\asGallerySolver{DPR}{\asGalleryValue{asGalleryLoss}{0}{0.845}}{\asGalleryValue{asGalleryLoss}{0}{0.854}}{44/119}{Incidence-signature recognition reduces supported compact encodings to finite-domain masks and binary support tables, followed by arc consistency and tabu/min-conflicts search. Validated assignments provide SAT witnesses; eligible failures fall back to watched-literal CDCL, emitting DRAT additions including an empty clause for UNSAT, while larger failed compact cases return UNKNOWN by default.}
\asGallerySolver{VeriPB}{\asGalleryValue{asGalleryLoss}{1}{0.879}}{\asGalleryValue{asGalleryLoss}{1}{0.888}}{47/119}{Recognized encodings become finite-domain CSPs: min-conflicts repair is followed by propagation and bounded DSATUR-style search. SAT assignments are checked against the original CNF; exact rope signatures may use triangle-cycle dynamic programming, while CDCL or rope DP emit VeriPB proof steps, including RUP steps, for UNSAT; inconclusive one-hot or dense-choice searches return UNKNOWN.}
\end{asGalleryFamily}

\begin{asGalleryFamily}{032}{coloring-\allowbreak clique}{11 GBD instances\enspace\textperiodcentered\enspace 3 validation\enspace\textperiodcentered\enspace baseline 3/3}{The benchmark asks whether an existentially chosen graph can contain a k-vertex clique while admitting a proper coloring with c colors. The CNF uses edge, clique-position, and color variables for graph choices, vertex selection, distinctness, and coloring conflicts.}
\ifdefined\pdfbookmark\pdfbookmark[2]{coloring-clique}{as-gallery-32}\fi
\asGallerySolver{GRAT}{\asGalleryValue{asGalleryLoss}{0}{1.35\!\cdot\!10^{-4}}}{\asGalleryValue{asGalleryLoss}{1}{7.88\!\cdot\!10^{-4}}}{2/3}{Exact structural recognition accepts only the expected numbering and complete clause set; malformed or unrecognized formulas return UNKNOWN. Recognized instances receive a clause-checked SAT assignment when k \ensuremath{<}= c; when k \ensuremath{>} c, the solver introduces position-color variables and emits a textual DRAT pigeonhole refutation, with generation failures returning UNKNOWN.}
\asGallerySolver{DPR}{\asGalleryValue{asGalleryLoss}{1}{1.35\!\cdot\!10^{-4}}}{\asGalleryValue{asGalleryLoss}{0}{7.88\!\cdot\!10^{-4}}}{2/3}{Exact clause-set recognition accepts only the recovered k = c+1 encoding; equivalent encodings return UNKNOWN. For matches, it skips SAT search, adds selected-color and derived-color clauses, reduces the contradiction to pigeonhole, and emits a DPR certificate ending in the empty clause; proof failures return UNKNOWN.}
\asGallerySolver{VeriPB}{\asGalleryValue{asGalleryLoss}{0}{1.35\!\cdot\!10^{-4}}}{\asGalleryValue{asGalleryLoss}{0}{7.87\!\cdot\!10^{-4}}}{2/3}{Symmetry-based paired vertex swaps canonicalize selected vertices after accepting the exact contiguous layout and complete templates; alternate layouts or unsupported formulas return UNKNOWN. For c \ensuremath{>}= k, it checks a SAT assignment; for c \ensuremath{<} k, RUP exclusions and cutting planes derive a VeriPB pigeonhole contradiction, with generation failures returning UNKNOWN.}
\end{asGalleryFamily}

\begin{asGalleryFamily}{033}{coloring-\allowbreak mycielski-\allowbreak graph}{19 GBD instances\enspace\textperiodcentered\enspace 4 validation\enspace\textperiodcentered\enspace baseline 0/4}{The benchmark asks whether a recursively constructed Mycielski graph has a proper coloring with one fewer color than its level. CNF encodings use vertex-color variables, exactly-one constraints, and edge clauses forbidding equal colors, with supported compact and permuted variants.}
\ifdefined\pdfbookmark\pdfbookmark[2]{coloring-mycielski-graph}{as-gallery-33}\fi
\asGallerySolver{GRAT}{\asGalleryValue{asGalleryGain}{0}{1.33}}{\asGalleryValue{asGalleryLoss}{0}{0.786}}{1/4}{One-hot recognition drives an extension-variable recoloring proof that repeatedly reduces a Mycielski coloring instance to fewer colors. Shadow copies handle hinted levels, bit encodings first define color indicators, and an opt-in CDCL fallback can verify SAT assignments; without it, unsupported inputs return UNKNOWN, while refutations are emitted as DRAT text for external elaboration.}
\asGallerySolver{DPR}{\asGalleryValue{asGalleryGain}{0}{1.01\!\cdot\!10^{4}}}{\asGalleryValue{asGalleryGain}{0}{120}}{4/4}{It recognizes supported compact, permuted, and canonical layouts, reconstructing graph, color, and hint structure rather than running general SAT search. Complete hints drive RUP shadow-edge derivations and a PHP refutation; incomplete hints use fresh-variable one-hot recursion only at bounded ranks, while rejected inputs return UNKNOWN and no SAT-witness path exists.}
\asGallerySolver{VeriPB}{\asGalleryValue{asGalleryGain}{1}{2.94\!\cdot\!10^{4}}}{\asGalleryValue{asGalleryGain}{1}{271}}{4/4}{It recognizes canonical and recovered bit, one-hot, and XOR layouts, then uses color symmetry to fix the newest root and recursively derive a coloring instance with one fewer color. Proof-only indicators and recoloring steps support the induction to a final contradiction; failed recognition or certificate generation returns UNKNOWN, with no SAT-witness path.}
\end{asGalleryFamily}

\begin{asGalleryFamily}{034}{core-\allowbreak based-\allowbreak generator}{20 GBD instances\enspace\textperiodcentered\enspace 4 validation\enspace\textperiodcentered\enspace baseline 3/4}{The benchmark asks whether a Boolean CNF is satisfiable when a residual core is surrounded by repeated three-row gadgets over signed variables. Each gadget encodes all but one sign pattern with chain auxiliaries and links the omitted pattern to the core.}
\ifdefined\pdfbookmark\pdfbookmark[2]{core-based-generator}{as-gallery-34}\fi
\asGallerySolver{GRAT}{\asGalleryValue{asGalleryGain}{1}{35.9}}{\asGalleryValue{asGalleryGain}{1}{61.6}}{4/4}{Missing-sign decoding fixes gadget variables and reduces recognized inputs to output units plus a residual core for watched-literal CDCL with first-UIP learning. SAT assignments are checked; UNSAT emits a textual DRAT trace with gadget clauses, learned residual clauses, and a final contradiction, while rejected structures or failures return UNKNOWN.}
\asGallerySolver{DPR}{\asGalleryValue{asGalleryGain}{0}{1.05}}{\asGalleryValue{asGalleryGain}{0}{1.01}}{3/4}{Strict structural extraction and preprocessing remove recognized padding, pure variables, and low-product occurrence pairs while recording resolvents, leaving a reduced core for watched-literal CDCL. SAT models are reconstructed and checked; UNSAT writes a textual DPR trace that can include witness-free clause additions, while unsupported structures or proof failures return UNKNOWN.}
\asGallerySolver{VeriPB}{\asGalleryValue{asGalleryGain}{0}{1.05}}{\asGalleryValue{asGalleryGain}{0}{3.22}}{3/4}{Local search first attacks the bounded, structurally extracted core; if it fails, its best phase seeds watched-literal CDCL. SAT models are extended through the gadgets and checked; UNSAT emits VeriPB cutting-planes and RUP derivations, while unsupported structures or internal failures return UNKNOWN.}
\end{asGalleryFamily}

\begin{asGalleryFamily}{035}{cover}{18 GBD instances\enspace\textperiodcentered\enspace 4 validation\enspace\textperiodcentered\enspace baseline 1/4}{The benchmark asks whether an n-point Steiner triple system has a cap of at least k points with no complete triple. CNF uses point variables, triple clauses forbidding all three selected, and auxiliary counters for the size threshold.}
\ifdefined\pdfbookmark\pdfbookmark[2]{cover}{as-gallery-35}\fi
\asGallerySolver{GRAT}{\asGalleryValue{asGalleryMuted}{0}{1.00}}{\asGalleryValue{asGalleryMuted}{0}{1.00}}{1/4}{Recognized layouts first use randomized cap search with maximal-cap growth, local exchanges, tabu or annealing moves, and a special projective quotient search. Selected targets use watched-literal CDCL; checked SAT models are printed, while UNSAT emits learned clauses and the empty clause as DRAT. Unrecognized inputs or noncertifying failures return UNKNOWN.}
\asGallerySolver{DPR}{\asGalleryValue{asGalleryMuted}{1}{1.00}}{\asGalleryValue{asGalleryMuted}{1}{1.00}}{1/4}{Recognized Steiner/BDD layouts use quasigroup cap search: completion conflicts trigger endpoint removal, refill, and periodic ruin-and-recreate. Selected cases use a propagation tree or watched-literal CDCL, emitting clause additions, some witness-free, that end in the empty clause; unsupported layouts and proof failures return UNKNOWN, while remaining regimes lack an UNSAT path.}
\asGallerySolver{VeriPB}{\asGalleryValue{asGalleryMuted}{0}{1.00}}{\asGalleryValue{asGalleryMuted}{0}{1.00}}{1/4}{Embedded affine and product cap constructions lead, with bit-mask branch-and-bound and selected tabu-style repair on recognized layouts. Selected cases then use proof-logging CDCL or a tree proof; direct SAT models are checked and printed, but CDCL SAT results are not printed as models, and unsupported or noncertifying cases return UNKNOWN.}
\end{asGalleryFamily}

\begin{asGalleryFamily}{036}{crafted-\allowbreak cec}{20 GBD instances\enspace\textperiodcentered\enspace 4 validation\enspace\textperiodcentered\enspace baseline 4/4}{The benchmark asks whether a primary-input assignment can make an asserted output of a signed AND circuit true. Each topologically ordered gate is represented by one wide clause plus binary clauses, and the final output is asserted by a unit clause.}
\ifdefined\pdfbookmark\pdfbookmark[2]{crafted-cec}{as-gallery-36}\fi
\asGallerySolver{GRAT}{\asGalleryValue{asGalleryGain}{1}{23.7}}{\asGalleryValue{asGalleryGain}{1}{23.5}}{4/4}{Packed genetic search simulates the recognized circuit in bit lanes, mutating assignments to seek a 12-input model and rechecking clauses. The SAT path has no UNSAT route; a narrow two-input branch uses equivalence checks and RUP case proofs with CDCL fallback, emitting a proof stream on closure, while unsupported layouts or unfinished proofs return UNKNOWN.}
\asGallerySolver{DPR}{\asGalleryValue{asGalleryGain}{0}{21.1}}{\asGalleryValue{asGalleryGain}{0}{18.7}}{4/4}{Packed bit-parallel evaluation combines an incumbent with one-bit neighbors; coordinate ascent and sparse mutations seek the asserted root, with large inputs falling back to random assignments. After validation it emits a SAT model but no DPR certificate; heuristic failure has no UNSAT or timeout result, while binary-root and unrecognized layouts return UNKNOWN.}
\asGallerySolver{VeriPB}{\asGalleryValue{asGalleryGain}{0}{9.89}}{\asGalleryValue{asGalleryGain}{0}{9.44}}{4/4}{Weighted breakout search scores packed assignments and one-bit neighbors, raises weights on false root targets, and uses random moves to seek a satisfying 12-way output without handling a negative root. It reconstructs gates and checks clauses before emitting the model; no UNSAT certificate path is active, and unsupported shapes or other roots return UNKNOWN.}
\end{asGalleryFamily}

\begin{asGalleryFamily}{037}{cril-\allowbreak misc}{62 GBD instances\enspace\textperiodcentered\enspace 13 validation\enspace\textperiodcentered\enspace baseline 4/13}{The benchmark asks whether Boolean signal and gate-output variables can satisfy a CNF encoding of a circuit or netlist. Clauses encode local gate relations, which may include parity-like constraints, gate definitions, or equivalences.}
\ifdefined\pdfbookmark\pdfbookmark[2]{cril-misc}{as-gallery-37}\fi
\asGallerySolver{GRAT}{\asGalleryValue{asGalleryLoss}{1}{0.848}}{\asGalleryValue{asGalleryGain}{1}{1.04}}{1/13}{Structural recognition limits this solver to a large, clause-dense circuit-CNF envelope; parity-like blocks remain ordinary clauses, not GF(2)-eliminated. Watched-literal first-UIP CDCL handles recognized inputs; unsupported or unresolved cases return UNKNOWN, SAT assignments are checked, and UNSAT learned clauses are emitted as raw DRAT additions for external elaboration.}
\asGallerySolver{DPR}{\asGalleryValue{asGalleryLoss}{0}{0.799}}{\asGalleryValue{asGalleryLoss}{0}{0.977}}{0/13}{It recognizes ordered ternary parity blocks and NAND definitions, then uses bipartite matching to orient parity equations before branching with structure-informed preferences in watched-literal CDCL. Unsupported or unmatched cases return UNKNOWN; SAT models are checked, while UNSAT emits witness-free DPR clause additions, including an empty clause, but the source does not establish proof correctness.}
\asGallerySolver{VeriPB}{\asGalleryValue{asGalleryLoss}{0}{0.799}}{\asGalleryValue{asGalleryLoss}{0}{0.977}}{0/13}{Order-sensitive XOR/NAND template recognition selects probe variables, followed by hidden-unit extraction and failed-literal probing before watched-literal first-UIP CDCL. UNSAT traces use RUP steps and a VeriPB conclusion, SAT models are fully assigned and rechecked, and input or post-check failures return UNKNOWN.}
\end{asGalleryFamily}

\begin{asGalleryFamily}{038}{cryptography}{860 GBD instances\enspace\textperiodcentered\enspace 172 validation\enspace\textperiodcentered\enspace baseline 84/172}{These benchmarks ask whether a chosen Boolean input block can make a one-block cryptographic hash or compression circuit meet specified output constraints. The circuit becomes CNF with variables for input bits and intermediate values, plus clauses fixing constants, selected inputs, and outputs.}
\ifdefined\pdfbookmark\pdfbookmark[2]{cryptography}{as-gallery-38}\fi
\asGallerySolver{GRAT}{\asGalleryValue{asGalleryLoss}{0}{0.961}}{\asGalleryValue{asGalleryLoss}{0}{0.961}}{74/172}{Recognized hash layouts enumerate a bounded free message prefix, evaluate the hash, and use CDCL to complete auxiliary variables; other formulas use watched-literal CDCL with 1-UIP learning and restarts. Models are checked and returned; direct search lacks an UNSAT certificate and reports UNKNOWN on exhaustion or completion failure, while fallback logging supplies DRAT for elaboration.}
\asGallerySolver{DPR}{\asGalleryValue{asGalleryGain}{1}{1.01}}{\asGalleryValue{asGalleryGain}{1}{1.01}}{80/172}{Exact structural recognition drives free-message enumeration and native hash evaluation, followed by CDCL completion of auxiliary variables; unrecognized inputs use watched-literal CDCL with 1-UIP learning and restarts. Verified SAT assignments are returned, while only the fallback can log learned DPR/RUP-style additions and a final empty clause; direct-search failure or unavailable logging yields UNKNOWN.}
\asGallerySolver{VeriPB}{\asGalleryValue{asGalleryLoss}{0}{0.896}}{\asGalleryValue{asGalleryLoss}{0}{0.896}}{66/172}{Functional recovery of ordered gate blocks and 64-way bit-sliced input search drives the specialized path, filtering derived values against fixed clauses. SAT assignments are checked against original clauses; only compact CDCL emits a VeriPB RUP proof for UNSAT, while unsupported structures, fast-path exhaustion, and non-SAT sampled cases return UNKNOWN.}
\end{asGalleryFamily}

\begin{asGalleryFamily}{039}{cryptography-\allowbreak ascon}{26 GBD instances\enspace\textperiodcentered\enspace 6 validation\enspace\textperiodcentered\enspace baseline 5/6}{These formulas ask whether two unknown message bytes can make a fixed Ascon-Hash v1.2 digest through a bit-blasted Boolean circuit. The benchmark uses a specific optimized, topologically ordered CNF encoding with auxiliary variables and fixed constraints, rather than arbitrary cryptographic formulas.}
\ifdefined\pdfbookmark\pdfbookmark[2]{cryptography-ascon}{as-gallery-39}\fi
\asGallerySolver{GRAT}{\asGalleryValue{asGalleryGain}{1}{1{,}010}}{\asGalleryValue{asGalleryGain}{1}{1.55}}{6/6}{Bit-parallel enumeration evaluates all 16-bit source assignments through the recognized circuit and intersects the target constraints. A surviving assignment is completed and checked clause by clause; exhaustion emits textual DRAT additions for external elaboration, including generated shortcuts and a resolution tree, while unsupported shapes or failed checks return UNKNOWN.}
\asGallerySolver{DPR}{\asGalleryValue{asGalleryGain}{0}{747}}{\asGalleryValue{asGalleryGain}{0}{1.50}}{6/6}{Bit-sliced truth-table evaluation exhaustively tests 16-bit source assignments, using small gates and frontier summaries to filter target constraints. A surviving input is completed and checked against every clause; otherwise the solver emits witness-free assignment-blocking clauses and a resolution sequence for a DPR certificate, with recognition or certificate failures returning UNKNOWN.}
\asGallerySolver{VeriPB}{\asGalleryValue{asGalleryGain}{0}{1.17}}{\asGalleryValue{asGalleryLoss}{0}{0.400}}{4/6}{Incidence decomposition and bit-parallel compiled-gate evaluation enumerate the 16-bit source assignments while separating four small side components. A surviving lane is reconstructed with recursive DPLL and fully clause-checked; with none, the solver emits assignment-blocking RUP leaves and a balanced VeriPB polynomial-elimination tree, while unsupported structures or proof failures return UNKNOWN.}
\end{asGalleryFamily}

\begin{asGalleryFamily}{040}{cryptography-\allowbreak cbmc}{13 GBD instances\enspace\textperiodcentered\enspace 3 validation\enspace\textperiodcentered\enspace baseline 3/3}{These benchmarks ask whether Boolean constraints from a bounded cryptographic circuit check, typically a CBMC AES encoding, have a satisfying assignment. Circuit inputs and intermediate values become Boolean variables, while gate equations, bookkeeping conditions, and an assertion or miter obligation become CNF clauses.}
\ifdefined\pdfbookmark\pdfbookmark[2]{cryptography-cbmc}{as-gallery-40}\fi
\asGallerySolver{GRAT}{\asGalleryValue{asGalleryGain}{1}{2.72}}{\asGalleryValue{asGalleryGain}{1}{8.95}}{3/3}{Root propagation, binary-equivalence substitution, bounded variable elimination, and failed-literal probing simplify recognized CNFs before activity-guided CDCL with restarts and clause reduction. It checks SAT assignments against the original clauses and logs preprocessing and search additions for externally checked UNSAT proofs; unsupported inputs and some cases after elimination return UNKNOWN.}
\asGallerySolver{DPR}{\asGalleryValue{asGalleryLoss}{0}{0.419}}{\asGalleryValue{asGalleryGain}{0}{1.36}}{3/3}{Signature filtering identifies Tseitin-like inputs, after which root simplification, implication-SCC substitution, bounded elimination, and failed-literal probing precede focused output-assumption search for a unique width-16, 56, or 184 candidate, with generic CDCL as fallback. Logged witness-free clause additions and empty-clause exits support UNSAT, while SAT assignments are reported; unsupported inputs or unresolved post-elimination searches return UNKNOWN.}
\asGallerySolver{VeriPB}{\asGalleryValue{asGalleryLoss}{0}{0.159}}{\asGalleryValue{asGalleryLoss}{0}{0.654}}{2/3}{Lookup-map congruence, signed AND/NAND matching, and binary-implication equivalences drive clause rewriting, bounded elimination, and failed-literal probing before watched-literal CDCL with 1-UIP learning. UNSAT receives VeriPB RUP records ending in an empty constraint; SAT is reported only when no variables were eliminated, while malformed inputs, setup failures, or post-elimination SAT cases return UNKNOWN.}
\end{asGalleryFamily}

\begin{asGalleryFamily}{041}{cryptography-\allowbreak simon}{71 GBD instances\enspace\textperiodcentered\enspace 15 validation\enspace\textperiodcentered\enspace baseline 5/15}{These instances ask whether a 32-bit key maps fixed plaintext to ciphertext through Simon-style Feistel rounds, with a related key obtained by bit permutation. The CNF bit-blasts keys, round states, and gate intermediates, and fixes both endpoint words.}
\ifdefined\pdfbookmark\pdfbookmark[2]{cryptography-simon}{as-gallery-41}\fi
\asGallerySolver{GRAT}{\asGalleryValue{asGalleryGain}{1}{1{,}110}}{\asGalleryValue{asGalleryGain}{1}{1{,}100}}{15/15}{Structured Hamming-weight-ordered enumeration exploits the related-key bit permutation after exact canonical recognition, with SIMD batches and an exhaustive key-search fallback. A found key is reconstructed and checked against every input clause before a SAT assignment is emitted; unsupported, exhausted, or failed-validation cases return UNKNOWN, with no UNSAT certificate path.}
\asGallerySolver{DPR}{\asGalleryValue{asGalleryGain}{0}{624}}{\asGalleryValue{asGalleryGain}{0}{596}}{15/15}{AVX-512 exhaustive enumeration of the first key, with a last-round inverse filter, follows fixed-key probes and exact clause-multiset recognition. Each found key is expanded and checked against every clause before a SAT assignment is emitted; unsupported, failed-validation, or exhausted cases return UNKNOWN because no UNSAT certificate path is implemented.}
\asGallerySolver{VeriPB}{\asGalleryValue{asGalleryGain}{0}{716}}{\asGalleryValue{asGalleryGain}{0}{679}}{15/15}{Strict ordered-template recognition is followed by word-level endpoint screening and exhaustive vectorized key enumeration, including a copied-half test before the final nonlinear round. Each surviving key is expanded and clauses are validated before a SAT assignment is emitted; rejected templates or keyless instances return UNKNOWN, with no UNSAT certificate path.}
\end{asGalleryFamily}

\begin{asGalleryFamily}{042}{design-\allowbreak debugging}{119 GBD instances\enspace\textperiodcentered\enspace 24 validation\enspace\textperiodcentered\enspace baseline 24/24}{These benchmarks ask whether fixed stimuli and, when present, state or observed values are consistent with a combinational or unrolled sequential circuit. Tseitin-style CNF clauses constrain Boolean signals, while unit clauses fix selected values; SAT means a completion exists and UNSAT means inconsistency.}
\ifdefined\pdfbookmark\pdfbookmark[2]{design-debugging}{as-gallery-42}\fi
\asGallerySolver{GRAT}{\asGalleryValue{asGalleryGain}{1}{3.04}}{\asGalleryValue{asGalleryGain}{1}{1.93}}{24/24}{Uniform-phase completion precedes search, assigning DIMACS variables in fixed order and accepting SAT only after checking every original clause. Failed trials invoke first-UIP CDCL with activity-based branching and phase saving, logging learned clauses as a DRAT refutation; SAT paths return checked models, and malformed or inconclusive cases return UNKNOWN.}
\asGallerySolver{DPR}{\asGalleryValue{asGalleryLoss}{0}{4.43\!\cdot\!10^{-3}}}{\asGalleryValue{asGalleryLoss}{0}{0.0379}}{22/24}{Gate-pattern congruence closure recovers supported AND, XOR, and ITE relations and merges them with a DSU quotient after propagation and phase trials. Failed-literal probing can emit clauses toward a DPR contradiction, but unsupported residual encodings or inconclusive search return UNKNOWN; no general DPLL or CDCL fallback exists.}
\asGallerySolver{VeriPB}{\asGalleryValue{asGalleryLoss}{0}{8.85\!\cdot\!10^{-3}}}{\asGalleryValue{asGalleryLoss}{0}{0.0704}}{23/24}{Occurrence-counter propagation scans CSR literal occurrences to decrement residual counts, with CDCL fallback using 1-UIP learning, phase saving, and restarts. UNSAT produces a VeriPB RUP proof, or a conditional compact pol chain for qualifying root conflicts; SAT assignments are clause-checked, while malformed or unresolved cases return UNKNOWN.}
\end{asGalleryFamily}

\begin{asGalleryFamily}{043}{diagnosis}{130 GBD instances\enspace\textperiodcentered\enspace 26 validation\enspace\textperiodcentered\enspace baseline 19/26}{The benchmark asks whether a finite-domain diagnosis model has a consistent assignment, with each object selecting one of six states across repeated scenarios. Observations and transition or relation constraints are compiled into CNF using one-hot choices and auxiliary variables.}
\ifdefined\pdfbookmark\pdfbookmark[2]{diagnosis}{as-gallery-43}\fi
\asGallerySolver{GRAT}{\asGalleryValue{asGalleryLoss}{0}{0.586}}{\asGalleryValue{asGalleryGain}{1}{1.40}}{12/26}{It recognizes the six-state exact-one structure, performs binary-implication SCC substitution and bounded variable elimination, then solves accepted formulas with watched-literal CDCL and first-UIP learning. Unsupported inputs return UNKNOWN; SAT outputs are checked assignments, while UNSAT uses textual DRAT for external elaboration; preprocessing conflicts can lack a final empty clause.}
\asGallerySolver{DPR}{\asGalleryValue{asGalleryLoss}{0}{0.520}}{\asGalleryValue{asGalleryGain}{0}{1.22}}{10/26}{It dispatches on recognized disjoint six-state exact-one groups and uses watched-literal CDCL with first-UIP learning, implication-graph minimization, activity branching, and restarts. SAT returns a checked assignment; UNSAT uses logged additions and deletions ending in an empty clause, with recovery after the serialization guard, while unsupported or certificate failures return UNKNOWN.}
\asGallerySolver{VeriPB}{\asGalleryValue{asGalleryLoss}{1}{0.661}}{\asGalleryValue{asGalleryGain}{0}{1.39}}{14/26}{It uses binary-clause parity union-find to quotient variables and validate a SAT model from the reduced instance; otherwise inputs use watched-literal CDCL with first-UIP learning. Unsupported or unresolved cases return UNKNOWN; UNSAT emits VeriPB RUP additions with hints, deletions, and an empty-clause conclusion, but shortcut paths have no UNSAT certificate.}
\end{asGalleryFamily}

\begin{asGalleryFamily}{044}{dimacs-\allowbreak sorter}{32 GBD instances\enspace\textperiodcentered\enspace 7 validation\enspace\textperiodcentered\enspace baseline 7/7}{These benchmarks ask whether one option can be selected from each finite-domain block, with auxiliary Boolean choices, so that all remaining constraints are satisfied. In CNF, each block uses an at-least-one clause and pairwise at-most-one clauses, followed by residual circuit or row constraints.}
\ifdefined\pdfbookmark\pdfbookmark[2]{dimacs-sorter}{as-gallery-44}\fi
\asGallerySolver{GRAT}{\asGalleryValue{asGalleryLoss}{0}{0.0591}}{\asGalleryValue{asGalleryLoss}{0}{0.0591}}{3/7}{Ordered one-hot and gate-DAG recognition is the distinguishing technique: recognized circuits are evaluated over domain choices and Boolean controls, while shuffled inputs receive only one-hot recognition. After an initial population pass, watched-literal CDCL supplies the fallback; failed search or checking returns UNKNOWN, with no UNSAT certificate path.}
\asGallerySolver{DPR}{\asGalleryValue{asGalleryGain}{0}{1.31}}{\asGalleryValue{asGalleryGain}{0}{1.31}}{7/7}{Bit-parallel local search is the distinguishing technique: after recovering signed one-hot blocks and functional gates, it evaluates many candidate states at once and applies tabu reassignment with randomized restarts. Recognition is bounded; search is incomplete. SAT models are printed, while failures return UNKNOWN; no UNSAT or DPR certificate is emitted.}
\asGallerySolver{VeriPB}{\asGalleryValue{asGalleryGain}{1}{2.50}}{\asGalleryValue{asGalleryGain}{1}{2.49}}{7/7}{Row abstraction and propagation distinguish this solver: recognized sorter rows become source-count constraints, narrowed by fixpoint intervals and bounded affine checks before branching on the row with fewest patterns. Leaves reconstruct gates and validate every clause; unsupported layouts or exhausted search return UNKNOWN, with no VeriPB UNSAT certificate path.}
\end{asGalleryFamily}

\begin{asGalleryFamily}{045}{dining-\allowbreak philosophers}{37 GBD instances\enspace\textperiodcentered\enspace 8 validation\enspace\textperiodcentered\enspace baseline 8/8}{These benchmarks ask whether a bounded dining-philosophers transition system can reach a deadlocked ring at some time frame. CNF encodes Boolean state and action signals, transition circuitry, and deadlock indicators, with a clause requiring one indicator to be true.}
\ifdefined\pdfbookmark\pdfbookmark[2]{dining-philosophers}{as-gallery-45}\fi
\asGallerySolver{GRAT}{\asGalleryValue{asGalleryLoss}{1}{0.573}}{\asGalleryValue{asGalleryLoss}{1}{0.595}}{8/8}{After recognizing its gate/XOR structure, the solver orders branches on deadlock outputs by AND-cone size; a size heuristic changes only that order. It then uses watched-literal CDCL, checks SAT assignments, and logs UNSAT clauses as textual DRAT additions for external checking; unsupported inputs return UNKNOWN.}
\asGallerySolver{DPR}{\asGalleryValue{asGalleryLoss}{0}{0.255}}{\asGalleryValue{asGalleryLoss}{0}{0.177}}{8/8}{It recovers paths from Tseitin structure, prioritizes temporal goals, and for the exact UNSAT shape adds action-coverage clauses before bounded Davis-Putnam elimination, followed by CDCL. Non-special recognized inputs use CDCL; SAT models are checked, while UNSAT logging uses witness-free DPR additions with preludes outside live RUP checks, and unrecognized inputs or search/check failures return UNKNOWN.}
\asGallerySolver{VeriPB}{\asGalleryValue{asGalleryLoss}{0}{0.115}}{\asGalleryValue{asGalleryLoss}{0}{0.0576}}{7/8}{Bounded Davis-Putnam elimination first shrinks recognized formulas while recording data for reverse model reconstruction, then watched-literal CDCL handles the reduced instance. SAT assignments are reconstructed and checked against all original clauses; UNSAT preprocessing and learned clauses are logged as VeriPB RUP constraints with an empty constraint; unsupported inputs, reconstruction failures, or validation failures return UNKNOWN.}
\end{asGalleryFamily}

\begin{asGalleryFamily}{046}{discrete-\allowbreak logarithm}{20 GBD instances\enspace\textperiodcentered\enspace 4 validation\enspace\textperiodcentered\enspace baseline 1/4}{The benchmark asks whether a bounded exponent x satisfies a\textasciicircum{}x mod n = r. CNF encodings represent exponent bits and staged accumulator, conditional-product, and modular-reduction words, with clauses fixing the final residue; implementations recognize particular layouts.}
\ifdefined\pdfbookmark\pdfbookmark[2]{discrete-logarithm}{as-gallery-46}\fi
\asGallerySolver{GRAT}{\asGalleryValue{asGalleryGain}{0}{2.85\!\cdot\!10^{4}}}{\asGalleryValue{asGalleryGain}{1}{7{,}850}}{4/4}{Prefix CDCL probes recover intermediate residues; GCD-based reconstruction recovers the modulus before baby-step/giant-step search finds an exponent. That exponent and the circuit state are assigned as roots, and occurrence-indexed propagation completes and checks clauses; unsupported or ambiguous recovery, search or completion failures return UNKNOWN, with no UNSAT certificate path implemented.}
\asGallerySolver{DPR}{\asGalleryValue{asGalleryGain}{1}{4.09\!\cdot\!10^{4}}}{\asGalleryValue{asGalleryGain}{0}{1{,}070}}{4/4}{Baby-step/giant-step search directly recovers a bounded exponent after structural recognition, then pins exponent, accumulator, and product words to reduce circuit completion to propagation. A watched-literal pass is followed by an active CDCL fallback and clause checking; unsupported layouts, noninvertible giant steps, and search or completion failures return UNKNOWN, with no UNSAT or DPR certificate path.}
\asGallerySolver{VeriPB}{\asGalleryValue{asGalleryGain}{0}{3.55\!\cdot\!10^{4}}}{\asGalleryValue{asGalleryGain}{0}{885}}{4/4}{Extended baby-step/giant-step search peels gcd factors and uses GCD-based modulus recovery to solve the bounded logarithm. Stage residues and quotient hints pin exponent and state bits; alternating forward and reverse propagation completes auxiliaries and checks clauses, while unsupported layouts or search/completion failures return UNKNOWN, with no general SAT fallback or UNSAT/VeriPB certificate path implemented.}
\end{asGalleryFamily}

\begin{asGalleryFamily}{047}{edge-\allowbreak matching}{97 GBD instances\enspace\textperiodcentered\enspace 20 validation\enspace\textperiodcentered\enspace baseline 8/20}{The benchmark asks whether a square grid can be filled using each tile once while colors on shared edges agree between neighbors. CNF uses one-hot placement and internal-edge groups, with clauses enforcing legal local orientations; recognized layouts are specific encodings, not the whole edge-matching family.}
\ifdefined\pdfbookmark\pdfbookmark[2]{edge-matching}{as-gallery-47}\fi
\asGallerySolver{GRAT}{\asGalleryValue{asGalleryGain}{1}{2.17}}{\asGalleryValue{asGalleryGain}{1}{2.16}}{14/20}{Structural recognition targets compact and unshuffled historical layouts, using grid embedding, arc consistency, and depth-first edge synchronization, with watched propagation and false-first completion on the historical path. SAT models undergo full clause checking; only a root-level propagation conflict emits a one-step RUP/DRAT empty-clause addition, while unrecognized layouts and other failures return UNKNOWN.}
\asGallerySolver{DPR}{\asGalleryValue{asGalleryGain}{0}{1.10}}{\asGalleryValue{asGalleryGain}{0}{1.10}}{8/20}{Compact tile-major recognition and local orientation enumeration drive a min-conflict constructor, followed by connected-frontier backtracking with edge propagation, MRV choices, capacity checks, and symmetry breaking. The fallback forcing and watched-literal search prints SAT only after complete clause validation; unsupported, unsatisfiable, or exhausted cases return UNKNOWN because no UNSAT certificate path is implemented.}
\asGallerySolver{VeriPB}{\asGalleryValue{asGalleryGain}{0}{1.64}}{\asGalleryValue{asGalleryGain}{0}{1.61}}{12/20}{Generalized exact-cover search uses colored secondary edge columns, bounded local edge tuples, and minimum-active-row branching after recognizing the compact placement-and-color structure. An alternate path requires a recovered contiguous square gate layout, forces identity placement, and uses watched-literal propagation; complete models are clause-checked, while failures return UNKNOWN because no UNSAT certificate path is implemented.}
\end{asGalleryFamily}

\begin{asGalleryFamily}{048}{edit-\allowbreak distance}{20 GBD instances\enspace\textperiodcentered\enspace 4 validation\enspace\textperiodcentered\enspace baseline 0/4}{The benchmark asks whether a signed complete graph can be partitioned into clusters so that disagreements with target edge signs stay within an encoded edit budget. Pair-equality variables represent co-membership, transitivity clauses enforce consistent clusters, and additional bounded-width clauses constrain the allowed edit count.}
\ifdefined\pdfbookmark\pdfbookmark[2]{edit-distance}{as-gallery-48}\fi
\asGallerySolver{GRAT}{\asGalleryValue{asGalleryGain}{0}{2.00}}{\asGalleryValue{asGalleryGain}{1}{2.00}}{2/4}{The solver first uses partition local search with singleton, all-in-one, and graph-derived starts, then vertex relocations and cluster merges. It fixes pair variables, uses watched-literal CDCL to complete and validate a SAT model, and returns UNKNOWN on missed partitions, unsupported encodings, or failed completion; no UNSAT certificate path is implemented.}
\asGallerySolver{DPR}{\asGalleryValue{asGalleryGain}{0}{2.00}}{\asGalleryValue{asGalleryGain}{0}{2.00}}{2/4}{Positive-gain agglomeration, single-vertex descent, and correlation-clustering restarts generate fixed pair assignments, then watched-literal CDCL completes and checks a SAT model. On failure, induced-P3 packing and bounded minimum-fill Davis-Putnam elimination feed residual CDCL and emit proof records for external checking; recognition mismatches return UNKNOWN, while UNSAT requires empty-clause or conflict evidence, not heuristic failure.}
\asGallerySolver{VeriPB}{\asGalleryValue{asGalleryGain}{1}{2.00}}{\asGalleryValue{asGalleryGain}{0}{2.00}}{2/4}{Multi-start cluster-editing search uses graph-based and random partitions, greedy vertex moves, merges, exchanges, and perturbations to find a budget-feasible partition. It fixes pair variables and runs watched-literal CDCL only on the budget suffix, checks every original clause, and returns UNKNOWN on unsupported encodings or failed search/completion; no UNSAT certificate path is implemented.}
\end{asGalleryFamily}

\begin{asGalleryFamily}{049}{ensemble-\allowbreak computation}{13 GBD instances\enspace\textperiodcentered\enspace 3 validation\enspace\textperiodcentered\enspace baseline 3/3}{These benchmarks ask whether a bounded collection of set-union operations can generate required target subsets from basic elements. They encode operation choices and resulting memberships as structured Boolean circuits, often expressed as CNF with auxiliary gate variables.}
\ifdefined\pdfbookmark\pdfbookmark[2]{ensemble-computation}{as-gallery-49}\fi
\asGallerySolver{GRAT}{\asGalleryValue{asGalleryLoss}{0}{0.152}}{\asGalleryValue{asGalleryLoss}{1}{0.271}}{1/3}{Template recognition identifies bounded direct or complement layouts and encodes candidate subsets as integer masks. Randomized disjoint-union and min-conflicts searches propose programs; assumptions on result bits are propagated and clause-checked, while unsupported inputs or failed search return UNKNOWN and no active UNSAT or DRAT emission path exists.}
\asGallerySolver{DPR}{\asGalleryValue{asGalleryLoss}{0}{0.101}}{\asGalleryValue{asGalleryLoss}{0}{0.181}}{0/3}{Bounded structural recognition filters for the six-row, k-column indicator layout and uses the recovered block to seed branching phases and order. Watched-literal CDCL with first-UIP learning searches the CNF; SAT models are completed and checked, while unsupported inputs or unproved UNSAT outcomes return UNKNOWN because proof logging is disabled.}
\asGallerySolver{VeriPB}{\asGalleryValue{asGalleryLoss}{1}{0.152}}{\asGalleryValue{asGalleryLoss}{0}{0.271}}{1/3}{Acyclic XOR/AND recognition groups rows into masks and extracts targets. A construction handles one-one cases; randomized common-subexpression elimination and OR closure repair handle others, with CDCL fallback on an abstract subset closure. Checked SAT witnesses are emitted; failures or unsupported cases return UNKNOWN, with no UNSAT or VeriPB certificate path.}
\end{asGalleryFamily}

\begin{asGalleryFamily}{050}{equivalence-\allowbreak chain}{19 GBD instances\enspace\textperiodcentered\enspace 4 validation\enspace\textperiodcentered\enspace baseline 0/4}{This benchmark asks whether a Boolean assignment satisfies a structured CNF. Groups of four ternary clauses encode parity relations over three variables, while binary and other ternary clauses impose residual constraints, sometimes using auxiliary variables.}
\ifdefined\pdfbookmark\pdfbookmark[2]{equivalence-chain}{as-gallery-50}\fi
\asGallerySolver{GRAT}{\asGalleryValue{asGalleryMuted}{1}{1.00}}{\asGalleryValue{asGalleryMuted}{1}{1.00}}{0/4}{Affine parity extraction and bounded-weight meet-in-the-middle decoding drive this solver: it recognizes four-clause ternary parity groups, eliminates them over GF(2), and searches for a low-weight parity-check solution. It propagates candidates through residual constraints and checks SAT assignments; without general fallback or UNSAT certification, structural, decoding, or completion failure returns UNKNOWN.}
\asGallerySolver{DPR}{\asGalleryValue{asGalleryMuted}{1}{1.00}}{\asGalleryValue{asGalleryMuted}{1}{1.00}}{0/4}{Bounded-weight syndrome decoding is its core: it reduces recognized XOR structure to a system and uses meet-in-the-middle search for support of weight at most seven. It reconstructs variables and applies DPLL to residuals; checked SAT assignments succeed, while unsupported structure or failed decoding/completion returns UNKNOWN without an UNSAT path.}
\asGallerySolver{VeriPB}{\asGalleryValue{asGalleryMuted}{1}{1.00}}{\asGalleryValue{asGalleryMuted}{1}{1.00}}{0/4}{Affine propagation and randomized conflict-core learning drive its search: Gaussian elimination expresses variables as affine forms, then residual-clause decisions yield dependency-based affine nogoods. It checks reconstructed SAT assignments; unsupported structures or failed search return UNKNOWN, with no UNSAT refutation implemented.}
\end{asGalleryFamily}

\begin{asGalleryFamily}{051}{equivalence-\allowbreak chain-\allowbreak principle}{4 GBD instances\enspace\textperiodcentered\enspace 1 validation\enspace\textperiodcentered\enspace baseline 0/1}{These CNFs ask whether a chain of Boolean tables can satisfy fixed endpoint assignments while auxiliary matching choices transport truth values between adjacent tables. In the benchmark encoding, CNF clauses express exact-one matching choices and guarded implications between corresponding entries.}
\ifdefined\pdfbookmark\pdfbookmark[2]{equivalence-chain-principle}{as-gallery-51}\fi
\asGallerySolver{GRAT}{\asGalleryValue{asGalleryMuted}{1}{1.00}}{\asGalleryValue{asGalleryMuted}{1}{1.00}}{0/1}{Bounded structural recognition activates a fiber-independence invariant, propagating selector-controlled equivalences across the chain and using a backward dependency pass from a contradictory endpoint fiber. For recognized instances, it emits a textual DRAT clause-addition stream ending in the empty clause for external elaboration; unsupported, malformed, or proof-generation failures return UNKNOWN, with no general SAT fallback.}
\asGallerySolver{DPR}{\asGalleryValue{asGalleryMuted}{1}{1.00}}{\asGalleryValue{asGalleryMuted}{1}{1.00}}{0/1}{Bounded recognition handles k=3 patterns by recursively transporting guarded existential-row clauses and branching on one matching coordinate; it does not require every pairwise-negative exact-one clause. Recognized k=4 patterns invoke watched-literal CDCL with 1-UIP learning and log UNSAT clause additions; checked SAT assignments are supported, while unsupported cases or solver failures return UNKNOWN.}
\asGallerySolver{VeriPB}{\asGalleryValue{asGalleryMuted}{1}{1.00}}{\asGalleryValue{asGalleryMuted}{1}{1.00}}{0/1}{Structural recognition dispatches accepted chains to either a pseudo-Boolean equivalence proof, using matching-tree cardinality inequalities that telescope across layers, or a guarded implication proof that propagates nonempty fibers through an n-ary path tree. It emits a VeriPB proof rather than performing SAT search; inputs outside the reconstructed layout, proof-generation failures, and output failures return UNKNOWN.}
\end{asGalleryFamily}

\begin{asGalleryFamily}{052}{equivalence-\allowbreak checking}{28 GBD instances\enspace\textperiodcentered\enspace 6 validation\enspace\textperiodcentered\enspace baseline 0/6}{These benchmarks ask whether two acyclic Boolean networks produce the same output for every primary-input assignment. A Tseitin CNF introduces variables for internal signals and asserts a final mismatch output, so a satisfying assignment supplies a distinguishing input and unsatisfiability establishes equivalence.}
\ifdefined\pdfbookmark\pdfbookmark[2]{equivalence-checking}{as-gallery-52}\fi
\asGallerySolver{GRAT}{\asGalleryValue{asGalleryMuted}{1}{1.00}}{\asGalleryValue{asGalleryMuted}{1}{1.00}}{0/6}{Bit-parallel simulation proposes equal or complementary signals, which the solver verifies while decomposing negated mismatch goals through recognized gate polarities. An incremental watched-literal CDCL fallback handles unresolved goals and logs learned clauses plus an empty clause for external UNSAT checking; unsupported or incomplete searches return UNKNOWN, and no SAT-model output path is implemented.}
\asGallerySolver{DPR}{\asGalleryValue{asGalleryMuted}{1}{1.00}}{\asGalleryValue{asGalleryMuted}{1}{1.00}}{0/6}{Recursive OR decomposition and LUT shortcuts try to force final-miter literals false before cone-local watched-literal CDCL handles the remainder. Its three-input four-clause check is generic rather than mux-specific; on UNSAT, local clauses are mapped back and logged with an empty clause for external checking, while SAT completion is checked and failures return UNKNOWN.}
\asGallerySolver{VeriPB}{\asGalleryValue{asGalleryMuted}{1}{1.00}}{\asGalleryValue{asGalleryMuted}{1}{1.00}}{0/6}{Hash-consed complemented-edge DAG reconstruction turns recognized Tseitin gates into simplified Boolean structure, while simulation signatures only nominate candidate equivalences and constants. Watched-literal RUP checks and bounded branching elaborate these claims; targeted CDCL emits RUP proof steps for closed UNSAT cases, but unresolved residues return UNKNOWN.}
\end{asGalleryFamily}

\begin{asGalleryFamily}{053}{erdos-\allowbreak discrepancy}{21 GBD instances\enspace\textperiodcentered\enspace 5 validation\enspace\textperiodcentered\enspace baseline 1/5}{The benchmark asks whether a finite sequence of positive or negative ones can keep every prefix sum along each homogeneous progression d, 2d, and so on within absolute value three. Its CNF uses sign variables plus auxiliary state variables to encode these bounded-prefix conditions.}
\ifdefined\pdfbookmark\pdfbookmark[2]{erdos-discrepancy}{as-gallery-53}\fi
\asGallerySolver{GRAT}{\asGalleryValue{asGalleryGain}{0}{1.01}}{\asGalleryValue{asGalleryGain}{0}{1.01}}{1/5}{Primary-boundary recognition guides activity and phase initialization for the restricted three-state encoding, after which watched-literal CDCL performs the active search with learning, restarts, and clause reduction. SAT assignments receive full clause verification; rejected inputs, search or verification failure, and an UNSAT result return UNKNOWN, with no UNSAT certificate path.}
\asGallerySolver{DPR}{\asGalleryValue{asGalleryGain}{1}{905}}{\asGalleryValue{asGalleryGain}{1}{80.2}}{5/5}{Mod-3 phase seeding, with ladder reconstruction for shuffled instances, targets recognized encodings before watched-literal CDCL completes or repairs assignments through learning and restarts. It verifies clauses for SAT; unsupported inputs, recognition failure, search failure, and an UNSAT result all return UNKNOWN, with no DPR proof path for UNSAT.}
\asGallerySolver{VeriPB}{\asGalleryValue{asGalleryLoss}{0}{0.815}}{\asGalleryValue{asGalleryLoss}{0}{0.815}}{0/5}{A conditional family-specific phase warm start seeds sequence variables from a generated instance, then watched-literal CDCL searches recognized inputs with learning and restarts; only phases, not learned state, are carried over. It rebuilds states and rechecks clauses before SAT; recognition, reconstruction, or search failure and root contradiction return UNKNOWN, with no UNSAT certificate path.}
\end{asGalleryFamily}

\begin{asGalleryFamily}{054}{fdmus}{1,000 GBD instances\enspace\textperiodcentered\enspace 200 validation\enspace\textperiodcentered\enspace baseline 200/200}{Given a Boolean circuit, the benchmark asks whether some input makes its output 1 and makes it 0 after either of two coordinate flips. CNF uses three circuit copies, sharing inputs except at those flips, with gate and output constraints.}
\ifdefined\pdfbookmark\pdfbookmark[2]{fdmus}{as-gallery-54}\fi
\asGallerySolver{GRAT}{\asGalleryValue{asGalleryGain}{0}{3.22}}{\asGalleryValue{asGalleryGain}{1}{2.08}}{200/200}{Transitive-fanout equivalences collapse unaffected gates onto first-copy representatives and restrict decisions to changed inputs and affected cones. Watched-literal CDCL handles the reduced instance and SAT returns a complete assignment; exact-layout rejection or solver/proof failures return UNKNOWN, while UNSAT retains a textual DRAT/RUP-style certificate.}
\asGallerySolver{DPR}{\asGalleryValue{asGalleryGain}{0}{3.03}}{\asGalleryValue{asGalleryLoss}{0}{0.794}}{200/200}{Congruence propagation equates gates across copies; quotienting and bounded elimination simplify affected cones. Residual watched-literal CDCL searches the remainder and emits textual DPR additions for preprocessing and learning, with an empty clause on UNSAT. SAT reconstruction checks the full original assignment; unsupported layouts or solver, proof, or reconstruction failures return UNKNOWN.}
\asGallerySolver{VeriPB}{\asGalleryValue{asGalleryGain}{1}{3.51}}{\asGalleryValue{asGalleryLoss}{0}{0.352}}{200/200}{Structural hashing in lockstep with disjoint-set equalities derives gate congruences across the three copies, followed by root-level unit propagation. A two-watched-literal CDCL fallback handles remaining cases and logs equivalences and learned clauses as VeriPB RUP steps; SAT returns a checked full assignment, while unsupported layouts or solver/proof-writing failures return UNKNOWN.}
\end{asGalleryFamily}

\begin{asGalleryFamily}{055}{fermat}{8 GBD instances\enspace\textperiodcentered\enspace 2 validation\enspace\textperiodcentered\enspace baseline 2/2}{This family asks whether two bounded, equal-width integers p and q satisfy p squared minus q squared equals N, usually with ordering and range conditions. A CNF encodes the squaring and subtraction circuits with auxiliary variables.}
\ifdefined\pdfbookmark\pdfbookmark[2]{fermat}{as-gallery-55}\fi
\asGallerySolver{GRAT}{\asGalleryValue{asGalleryGain}{1}{6.07\!\cdot\!10^{4}}}{\asGalleryValue{asGalleryGain}{1}{1.48\!\cdot\!10^{4}}}{2/2}{Template recognition recovers the odd target N rather than searching arbitrary CNF. It factors N, selects a balanced divisor pair for p and q, propagates through the circuit, and checks every clause before emitting a SAT assignment; unsupported or unsuccessful cases return UNKNOWN, with no UNSAT certificate path.}
\asGallerySolver{DPR}{\asGalleryValue{asGalleryGain}{0}{4.25\!\cdot\!10^{4}}}{\asGalleryValue{asGalleryGain}{0}{1{,}430}}{2/2}{Strict recognition of 8-65-bit primary vectors and a ripple-subtractor layout extracts the odd target N. Pollard-rho factorization and divisor splits construct p and q, after which ordered propagation and full clause checks validate a SAT assignment; unsupported or failed cases return UNKNOWN, with no UNSAT or DPR path.}
\asGallerySolver{VeriPB}{\asGalleryValue{asGalleryGain}{0}{4.79\!\cdot\!10^{4}}}{\asGalleryValue{asGalleryGain}{0}{1{,}300}}{2/2}{Parity and carry-chain recognition extracts an odd target from the narrow CNF layout rather than solving arbitrary formulas. It factors the target, selects a closest nontrivial factor pair, builds a full model by forward propagation, and checks every clause; failures return UNKNOWN, with no UNSAT or VeriPB proof path.}
\end{asGalleryFamily}

\begin{asGalleryFamily}{056}{finite-\allowbreak state-\allowbreak machines}{12 GBD instances\enspace\textperiodcentered\enspace 3 validation\enspace\textperiodcentered\enspace baseline 3/3}{These benchmarks ask whether a bounded, unrolled finite-state transition system has an assignment satisfying its transition constraints and a stated property condition. State, signal, and auxiliary gate equations are converted to CNF, typically through Tseitin clauses.}
\ifdefined\pdfbookmark\pdfbookmark[2]{finite-state-machines}{as-gallery-56}\fi
\asGallerySolver{GRAT}{\asGalleryValue{asGalleryLoss}{0}{0.822}}{\asGalleryValue{asGalleryGain}{1}{1.62}}{3/3}{Bounded variable elimination resolves pivots only when pair, width, and growth limits permit, retaining resolvents for reconstruction and logging. It then uses watched-literal CDCL with 1-UIP learning; SAT reverses elimination and checks a model, unsupported layouts return UNKNOWN, and UNSAT is written as DRAT additions for external elaboration and checking.}
\asGallerySolver{DPR}{\asGalleryValue{asGalleryLoss}{0}{0.478}}{\asGalleryValue{asGalleryLoss}{0}{0.971}}{3/3}{It recognizes only TIP/PICO layouts; TIP prioritizes frames containing bad-state goals. On PICO, bounded Davis-Putnam elimination uses gate definitions and capped pairwise resolution before residual CDCL; SAT reverses elimination and checks a model, unsupported or setup-failing cases return UNKNOWN, while UNSAT ends with witness-free clause additions and an empty clause.}
\asGallerySolver{VeriPB}{\asGalleryValue{asGalleryGain}{1}{1.24}}{\asGalleryValue{asGalleryLoss}{0}{0.318}}{3/3}{Binary-implication SCC preprocessing merges equivalent literals, selecting bounded elimination regimes before a watched-literal CDCL fallback. SAT restores eliminated variables and checks a model; UNSAT logs preprocessing and learned clauses as VeriPB RUP constraints, ending with an empty RUP and UNSAT command, while malformed or failed cases return UNKNOWN.}
\end{asGalleryFamily}

\begin{asGalleryFamily}{057}{fixed-\allowbreak shape-\allowbreak random}{30 GBD instances\enspace\textperiodcentered\enspace 6 validation\enspace\textperiodcentered\enspace baseline 2/6}{This benchmark asks whether shared Boolean base variables can satisfy all components of a bounded, fixed-shape CNF encoding. Each component uses center or root auxiliaries and selectors to express compatible local supports, and the question becomes whether one assignment satisfies every original clause.}
\ifdefined\pdfbookmark\pdfbookmark[2]{fixed-shape-random}{as-gallery-57}\fi
\asGallerySolver{GRAT}{\asGalleryValue{asGalleryGain}{1}{4.64}}{\asGalleryValue{asGalleryGain}{1}{4.64}}{5/6}{On recognized inputs, noisy minimum-break local search works on the sparse CNF and checks candidates before reporting SAT. A failed walk does not imply UNSAT; watched-literal first-UIP CDCL is the fallback, logging learned clauses and a final empty clause as textual DRAT for external GRAT elaboration, while unsupported inputs return UNKNOWN.}
\asGallerySolver{DPR}{\asGalleryValue{asGalleryGain}{0}{4.55}}{\asGalleryValue{asGalleryGain}{0}{4.55}}{5/6}{Recognized inputs use focused probSAT on compact CNF, then support-level min-conflicts over shared bases. Base models get exhaustive auxiliary completion; if these searches fail, watched-literal CDCL logs learned clauses and the empty clause as witness-free DPR additions, and heuristic failure is not UNSAT. Unsupported shapes return UNKNOWN; no elaboration or kernel validation is supplied.}
\asGallerySolver{VeriPB}{\asGalleryValue{asGalleryGain}{0}{4.62}}{\asGalleryValue{asGalleryGain}{0}{4.62}}{5/6}{Plan projection first reduces each recognized gadget to minimal signed-primary supports, then weighted min-conflicts flips all false literals of selected plans while updating affected gadgets incrementally. Exhaustive auxiliary reconstruction and full CNF checking certify SAT; no UNSAT or VeriPB certificate path exists, so unsupported inputs return UNKNOWN and unsatisfiable inputs may continue until externally stopped.}
\end{asGalleryFamily}

\begin{asGalleryFamily}{058}{floodit-\allowbreak puzzle}{40 GBD instances\enspace\textperiodcentered\enspace 8 validation\enspace\textperiodcentered\enspace baseline 8/8}{These instances ask whether a fixed-source Flood-It board can be completely flooded within at most m moves by choosing colors that absorb adjacent monochromatic regions. The CNF uses one-hot color choices and time-indexed flooded-state variables.}
\ifdefined\pdfbookmark\pdfbookmark[2]{floodit-puzzle}{as-gallery-58}\fi
\asGallerySolver{GRAT}{\asGalleryValue{asGalleryGain}{1}{745}}{\asGalleryValue{asGalleryGain}{1}{61.2}}{8/8}{Disjoint-set contraction and bitset state tracking support multi-heuristic greedy construction, with seeded noisy trials providing a fallback for finding a plan within the move bound. It recognizes only ten-color square grids; found plans are expanded into move and state variables and clause-checked before SAT, while construction failure returns UNKNOWN and no UNSAT certificate path exists.}
\asGallerySolver{DPR}{\asGalleryValue{asGalleryGain}{0}{328}}{\asGalleryValue{asGalleryGain}{0}{5.71}}{8/8}{Monochromatic-region contraction supports greedy region-graph color selection with randomized trials as a fallback for finding a bounded move sequence. It reconstructs and clause-checks the full assignment before SAT; this narrow recognized encoding has no UNSAT certificate path, so recognition, search, or checking failure returns UNKNOWN.}
\asGallerySolver{VeriPB}{\asGalleryValue{asGalleryGain}{0}{582}}{\asGalleryValue{asGalleryGain}{0}{6.51}}{8/8}{Randomized greedy rollouts on a contracted region graph lead into bounded beam search over flooded-component states. It recognizes only a ten-color layout, expands a found plan into all action and auxiliary variables, and rechecks every CNF clause before SAT; missed plans or failures return UNKNOWN, with no UNSAT certificate path.}
\end{asGalleryFamily}

\begin{asGalleryFamily}{059}{fpga-\allowbreak routing}{97 GBD instances\enspace\textperiodcentered\enspace 20 validation\enspace\textperiodcentered\enspace baseline 16/20}{These CNFs ask whether each route or vertex can select one of k channels so conflicting pairs receive different channels. They encode choices with group clauses and matched binary conflicts, with some instances using equality-expanded or binary representations.}
\ifdefined\pdfbookmark\pdfbookmark[2]{fpga-routing}{as-gallery-59}\fi
\asGallerySolver{GRAT}{\asGalleryValue{asGalleryGain}{0}{1.29}}{\asGalleryValue{asGalleryGain}{0}{1.04}}{17/20}{The active strategy reconstructs conflict graphs, uses bit-parallel clique search, and tries chordal or tabu coloring before finite-domain search and CDCL fallback. SAT assignments are checked against the CNF; UNSAT proof output supplies learned clauses or finite-domain nogoods for external DRAT elaboration, while unsupported or unfinished cases return UNKNOWN.}
\asGallerySolver{DPR}{\asGalleryValue{asGalleryGain}{0}{4.07}}{\asGalleryValue{asGalleryGain}{0}{4.05}}{19/20}{The active strategy finds bitset (k+1)-cliques, then tries bounded coloring, binary equality chains, and tabu search before CDCL fallback. SAT assignments are checked against input clauses; UNSAT uses PR/PHP clique proofs or proof-logged CDCL with normalized/original fallback and witness-free additions. Unsupported inputs or unresolved searches return UNKNOWN; verification is external.}
\asGallerySolver{VeriPB}{\asGalleryValue{asGalleryGain}{1}{4.09}}{\asGalleryValue{asGalleryGain}{1}{4.08}}{19/20}{The active strategy reconstructs route-conflict graphs and searches for bounded oversized cliques, using cutting-planes pigeonhole proofs when a clique is found; dense bit encodings can use one-hot extensions. Otherwise watched-literal CDCL supplies RUP logging, while SAT models are checked against original clauses and unsupported or failed cases return UNKNOWN.}
\end{asGalleryFamily}

\begin{asGalleryFamily}{060}{generic-\allowbreak csp}{83 GBD instances\enspace\textperiodcentered\enspace 17 validation\enspace\textperiodcentered\enspace baseline 8/17}{Each instance asks whether a ternary constraint problem has one of four values for every variable while satisfying all relations. Its CNF encoding uses four-literal domain clauses and three-literal clauses forbidding disallowed value triples; the recognized subset has unique extension of any two values.}
\ifdefined\pdfbookmark\pdfbookmark[2]{generic-csp}{as-gallery-60}\fi
\asGallerySolver{GRAT}{\asGalleryValue{asGalleryLoss}{0}{0.702}}{\asGalleryValue{asGalleryLoss}{0}{0.793}}{3/17}{Bounded MAC first applies tuple-mask arc consistency, then the active CDCL search adds domain and functional clauses and uses conflict learning to seek a checked SAT assignment. UNSAT emits DRAT proof text for external elaboration and checking; unrecognized formulas or proof-output failure return UNKNOWN, and the local-search path is inactive.}
\asGallerySolver{DPR}{\asGalleryValue{asGalleryLoss}{1}{0.892}}{\asGalleryValue{asGalleryLoss}{1}{0.928}}{6/17}{Unique-extension propagation builds bootstrap seeds, then four-way DFS seeks checked SAT assignments and records rejected cubes. UNSAT is limited to at most 11 seeds and a complete bounded search with an externally checked clause/deletion log; otherwise unbounded local search is used, while unrecognized inputs return UNKNOWN.}
\asGallerySolver{VeriPB}{\asGalleryValue{asGalleryLoss}{0}{0.820}}{\asGalleryValue{asGalleryLoss}{0}{0.678}}{6/17}{Support filtering and bounded CSP search branch on domains, then fall back to watched-literal CDCL with learned clauses when that search fails; SAT assignments are checked against the original CNF. UNSAT uses VeriPB text with red witnesses and RUP steps for external checking; recognition or proof setup/finalization failure returns UNKNOWN.}
\end{asGalleryFamily}

\begin{asGalleryFamily}{061}{genurq}{12 GBD instances\enspace\textperiodcentered\enspace 3 validation\enspace\textperiodcentered\enspace baseline 3/3}{Instances use small clause blocks with graph-like variable sharing; one form encodes parity constraints, while another combines positive binary clauses with an exact sequential encoding of an at-most-k constraint. The task is to determine whether all clauses can be satisfied for these restricted encodings.}
\ifdefined\pdfbookmark\pdfbookmark[2]{genurq}{as-gallery-61}\fi
\asGallerySolver{GRAT}{\asGalleryValue{asGalleryGain}{0}{64.9}}{\asGalleryValue{asGalleryGain}{1}{41.2}}{3/3}{Recognizes bounded-width parity blocks as XOR equations, solves their incidence graph by spanning forests, and enumerates one possible exceptional block. It verifies resulting assignments against the CNF before SAT, but has no UNSAT certificate path; unsupported or unresolved inputs return UNKNOWN instead of using general SAT search.}
\asGallerySolver{DPR}{\asGalleryValue{asGalleryGain}{0}{56.4}}{\asGalleryValue{asGalleryGain}{0}{18.1}}{3/3}{Combines XOR block recognition and incidence-graph parity solving with a fallback for positive binary clauses and a sequential at-most-k counter. The base branch emits a checked SAT assignment; only bounded line-graph reconstruction can produce an UNSAT DPR proof, reducing the matching contradiction to PHP with PR swaps, while failed searches and unsupported shapes return UNKNOWN.}
\asGallerySolver{VeriPB}{\asGalleryValue{asGalleryGain}{1}{72.7}}{\asGalleryValue{asGalleryGain}{0}{17.6}}{3/3}{Recognizes bounded XOR/table blocks, enumerates exceptional assignments, and solves incidence components by spanning trees. Compact SAT models are checked, while supported contradictions use VeriPB red/rup proofs or, after bounded line-graph reconstruction, pol proofs using a sequential counter; the debug route has no SAT branch and failures return UNKNOWN.}
\end{asGalleryFamily}

\begin{asGalleryFamily}{062}{glassy-\allowbreak gen}{140 GBD instances\enspace\textperiodcentered\enspace 28 validation\enspace\textperiodcentered\enspace baseline 9/28}{The benchmark asks whether a Boolean assignment satisfies every clause in a sparse near-14-regular CNF formula. It uses mostly three-literal clauses, at most one two-literal clause, and variables occurring 13 or 14 times, defining a structural envelope rather than the whole family.}
\ifdefined\pdfbookmark\pdfbookmark[2]{glassy-gen}{as-gallery-62}\fi
\asGallerySolver{GRAT}{\asGalleryValue{asGalleryGain}{1}{1{,}340}}{\asGalleryValue{asGalleryGain}{1}{1{,}330}}{28/28}{Bit-packed GF(2) elimination treats the three-literal clauses as favored odd-parity equations, then recursive implication-closure repair tests the candidate; failure falls back to replica-exchange Metropolis search with local CNF repair and reseeding. Only clause-checked SAT assignments are emitted; unsupported inputs return UNKNOWN and no active UNSAT certificate path exists.}
\asGallerySolver{DPR}{\asGalleryValue{asGalleryGain}{0}{4.35}}{\asGalleryValue{asGalleryGain}{0}{4.35}}{24/28}{Ordered planted-prefix reconstruction leads the search, followed by GF(2) parity elimination that tests both a solution and its complement; failures fall back to watched-literal 1-UIP CDCL and then open-ended probSAT repair. Only verified SAT assignments are emitted, while unsupported or unverified cases have no UNSAT/DPR path and return UNKNOWN.}
\asGallerySolver{VeriPB}{\asGalleryValue{asGalleryGain}{0}{60.0}}{\asGalleryValue{asGalleryGain}{0}{60.0}}{28/28}{Ordered selector reconstruction is tried first, with candidate enumeration and CNF verification, followed by packed GF(2) elimination of odd-parity equations; remaining recognized cases use XOR annealing and focused probSAT-style search with Hamming-ball repair. Only verified SAT models are emitted; unsupported or unsuccessful searches return UNKNOWN, with no UNSAT/VeriPB certificate path exists.}
\end{asGalleryFamily}

\begin{asGalleryFamily}{063}{graceful-\allowbreak production}{20 GBD instances\enspace\textperiodcentered\enspace 4 validation\enspace\textperiodcentered\enspace baseline 4/4}{The benchmark asks whether a structured CNF encoding of finite-domain order and equality constraints is satisfiable. Boolean variables represent semantic choices and auxiliary comparison, arithmetic, cardinality, or gate relations, while clauses enforce consistency in a restricted ordered construction.}
\ifdefined\pdfbookmark\pdfbookmark[2]{graceful-production}{as-gallery-63}\fi
\asGallerySolver{GRAT}{\asGalleryValue{asGalleryGain}{1}{6.44}}{\asGalleryValue{asGalleryGain}{1}{2.08}}{4/4}{Two-watched-literal propagation drives a positive, increasing-variable completion after root units, avoiding branching or backtracking. It recognizes only a narrow normalized DIMACS signature; an independent clause scan verifies the selected-literal SAT witness, while root conflicts alone emit a minimal proof artifact and completion conflicts or unsupported inputs return UNKNOWN.}
\asGallerySolver{DPR}{\asGalleryValue{asGalleryLoss}{0}{0.0179}}{\asGalleryValue{asGalleryLoss}{0}{0.0418}}{2/4}{Positive increasing-index construction, driven by watched-literal propagation, is the active technique rather than general SAT search or a graph-level decoder. It independently checks every original clause and emits a clause-covering SAT witness; conflicts, unsupported inputs, or oversized outputs return UNKNOWN, and no UNSAT or DPR certificate path is implemented.}
\asGallerySolver{VeriPB}{\asGalleryValue{asGalleryLoss}{0}{0.0179}}{\asGalleryValue{asGalleryLoss}{0}{0.0410}}{2/4}{Parameterized opening-signature recognition selects the supported ordered CNF subset, followed by unit propagation and increasing positive assignment without DPLL or fallback search. A verified essential-variable witness is emitted for SAT; structurally different, conflicting, or oversized inputs return UNKNOWN, with no UNSAT or VeriPB certificate path.}
\end{asGalleryFamily}

\begin{asGalleryFamily}{064}{grandtour-\allowbreak puzzle}{20 GBD instances\enspace\textperiodcentered\enspace 4 validation\enspace\textperiodcentered\enspace baseline 4/4}{The benchmark asks whether choices can cover every required source while each destination is used at most once. CNF encodings use coverage clauses and binary incompatibility clauses; supported instances have one more source than destination, producing a pigeonhole contradiction.}
\ifdefined\pdfbookmark\pdfbookmark[2]{grandtour-puzzle}{as-gallery-64}\fi
\asGallerySolver{GRAT}{\asGalleryValue{asGalleryGain}{0}{2.50\!\cdot\!10^{4}}}{\asGalleryValue{asGalleryGain}{0}{3{,}200}}{4/4}{Bounded dual-rail recovery and line-graph inversion reconstruct a bipartite incidence graph, accepting only a checked connected component with one more source than destination. A swap-and-remove reduction then emits ordered DRAT additions, including extensions, to the empty clause; recognition, parsing, or proof-output failure returns UNKNOWN, with no SAT search.}
\asGallerySolver{DPR}{\asGalleryValue{asGalleryGain}{0}{1.17\!\cdot\!10^{4}}}{\asGalleryValue{asGalleryGain}{0}{2.99}}{4/4}{Signed union-find normalization, propagation, and clique checks recognize a one-over-capacity pigeonhole core, while matching exposes a Hall circuit. For accepted cores, it adds fresh variables and emits deletion and witness-free clause additions completing a polynomial pigeonhole proof. It does not run SAT search, and unsupported or failed cases return UNKNOWN.}
\asGallerySolver{VeriPB}{\asGalleryValue{asGalleryGain}{1}{3.17\!\cdot\!10^{4}}}{\asGalleryValue{asGalleryGain}{1}{1.56\!\cdot\!10^{4}}}{4/4}{Parity union-find collapses equality and complement classes, then line-graph reconstruction recovers endpoint cliques and the one-unit source surplus. The solver emits RUP clauses and incremental pseudo-Boolean at-most-one derivations for a counting proof, but performs no SAT search or external proof check; unsupported or failed cases return UNKNOWN.}
\end{asGalleryFamily}

\begin{asGalleryFamily}{065}{graph-\allowbreak isomorphism}{92 GBD instances\enspace\textperiodcentered\enspace 19 validation\enspace\textperiodcentered\enspace baseline 14/19}{The benchmark asks whether two colored graphs have a vertex bijection that preserves the relevant adjacency relations. In the supported CNF form, long clauses require candidate mapping choices, while binary clauses forbid incompatible choices.}
\ifdefined\pdfbookmark\pdfbookmark[2]{graph-isomorphism}{as-gallery-65}\fi
\asGallerySolver{GRAT}{\asGalleryValue{asGalleryGain}{1}{1.79}}{\asGalleryValue{asGalleryGain}{1}{1.71}}{17/19}{After recognizing the restricted candidate-mapping grammar, it tries a bounded grid-factor shortcut with graph-isomorphism search, then falls back to CDCL with binary-conflict propagation and clause learning. SAT assignments are checked against the original CNF; UNSAT is emitted as textual DRAT for external elaboration, while unsupported inputs or failed checks return UNKNOWN.}
\asGallerySolver{DPR}{\asGalleryValue{asGalleryLoss}{0}{0.476}}{\asGalleryValue{asGalleryLoss}{0}{0.721}}{0/19}{It reconstructs recognized graphs, using paired color refinement and individualization search, while a bounded CFI branch solves induced GF(2) constraints and falls back on this search. SAT mappings are CNF-checked; narrow UNSAT proof search uses Hall/pigeonhole branching and RUP/PR only for full instances without phase flips or repair cells, while other nonisomorphic cases return UNKNOWN.}
\asGallerySolver{VeriPB}{\asGalleryValue{asGalleryLoss}{0}{0.476}}{\asGalleryValue{asGalleryLoss}{0}{0.721}}{0/19}{It solves the recognized incidence structure as an independent-transversal problem, using bitset arc consistency and minimum-domain DFS with conflict-based support checks. After checking complete assignments against the CNF, it emits SAT models; unrecognized shapes and failed searches return UNKNOWN, with no active UNSAT certificate path.}
\end{asGalleryFamily}

\begin{asGalleryFamily}{066}{gray\_\allowbreak codes}{35 GBD instances\enspace\textperiodcentered\enspace 7 validation\enspace\textperiodcentered\enspace baseline 6/7}{This benchmark asks whether a cycle of w distinct binary words of width b exists, with each pair of cyclically adjacent words differing in exactly one bit. The CNF uses a binary matrix and auxiliary variables for shifted tracks, XOR relations, and one-bit transitions.}
\ifdefined\pdfbookmark\pdfbookmark[2]{gray\_codes}{as-gallery-66}\fi
\asGallerySolver{GRAT}{\asGalleryValue{asGalleryLoss}{0}{0.683}}{\asGalleryValue{asGalleryLoss}{0}{0.866}}{5/7}{A bounded 64-bit exact-cover search over cyclic translates is the SAT technique, constructing a shifted-track cycle and checking the resulting assignment against all clauses. If construction fails, watched-literal CDCL is used; unsupported cases may return UNKNOWN, and UNSAT certificates are emitted as DRAT streams ending in empty clause.}
\asGallerySolver{DPR}{\asGalleryValue{asGalleryLoss}{0}{0.673}}{\asGalleryValue{asGalleryLoss}{0}{0.849}}{5/7}{A bounded residue construction is the primary SAT technique: it chooses periodic transition representatives and track offsets, then checks the assignment against every clause. If construction fails, CDCL is the fallback; unsupported cases return UNKNOWN, and root-conflict UNSAT logging records only witness-free learned-clause additions, ending in empty clause without PR witnesses.}
\asGallerySolver{VeriPB}{\asGalleryValue{asGalleryGain}{1}{1{,}540}}{\asGalleryValue{asGalleryGain}{1}{100}}{7/7}{Bounded canonical cyclic-factor search is the active SAT technique: it enumerates derivative choices and shift lifts, builds words, and checks adjacency, uniqueness, and all clauses. Failure invokes no CDCL; counting and parity arguments can emit VeriPB proofs in eligible nondivisible or odd-quotient cases, while other recognized cases return UNKNOWN.}
\end{asGalleryFamily}

\begin{asGalleryFamily}{067}{greentao}{12 GBD instances\enspace\textperiodcentered\enspace 3 validation\enspace\textperiodcentered\enspace baseline 0/3}{Green-Tao-style instances ask whether prime arithmetic-progression edges can be two-colored so every positive edge contains color 1 and every negative edge contains color 0. In CNF, these two requirements are monotone positive and negative clauses.}
\ifdefined\pdfbookmark\pdfbookmark[2]{greentao}{as-gallery-67}\fi
\asGallerySolver{GRAT}{\asGalleryValue{asGalleryMuted}{1}{1.00}}{\asGalleryValue{asGalleryMuted}{1}{1.00}}{0/3}{Stochastic incremental local search repairs violated clauses with occurrence-local satisfaction counts, weighted break/make scores, restarts, and noisy flips, followed by a longer fallback. Checked assignments are certificates and produce SAT output; the inactive exact proof path leaves failed searches as UNKNOWN, with no UNSAT certificate.}
\asGallerySolver{DPR}{\asGalleryValue{asGalleryMuted}{1}{1.00}}{\asGalleryValue{asGalleryMuted}{1}{1.00}}{0/3}{Min-conflicts search leads with negative-edge covers, balanced starts, and focused flips; failure guides an exact fallback using blocker reduction, counter propagation, watched clauses, and gain branching. No cutoff applies; it emits witness-free clause additions and an empty-clause UNSAT certificate. SAT has no proof file; unsupported inputs return UNKNOWN.}
\asGallerySolver{VeriPB}{\asGalleryValue{asGalleryMuted}{1}{1.00}}{\asGalleryValue{asGalleryMuted}{1}{1.00}}{0/3}{With shorter positive clauses, asymmetric greedy hitting-set initialization covers negative edges while avoiding complete positive edges; stalled cases use Novelty+ local search with make-break scores, restarts, and incumbent retention. Checked assignments are printed as SAT certificates; no UNSAT or VeriPB certificate path exists, so unsupported and unsatisfiable inputs return UNKNOWN.}
\end{asGalleryFamily}

\begin{asGalleryFamily}{068}{grs-\allowbreak fp-\allowbreak comm}{64 GBD instances\enspace\textperiodcentered\enspace 13 validation\enspace\textperiodcentered\enspace baseline 1/13}{The benchmark asks whether a bit-blasted Boolean circuit for floating-point commutativity has a satisfying assignment, representing a counterexample. Its CNF uses unit constraints and ordered local clauses for two- or three-input gates; recognized layouts are restricted and do not independently verify arithmetic semantics.}
\ifdefined\pdfbookmark\pdfbookmark[2]{grs-fp-comm}{as-gallery-68}\fi
\asGallerySolver{GRAT}{\asGalleryValue{asGalleryLoss}{1}{0.955}}{\asGalleryValue{asGalleryGain}{1}{1.12}}{0/13}{Gate-aware root propagation and bounded variable elimination, with defined outputs avoided as decisions, simplify accepted streams before watched-literal first-UIP CDCL on the residual formula. For UNSAT, it emits plain clause additions; unsupported inputs or a conflict-free completion are reported as UNKNOWN rather than producing a SAT witness.}
\asGallerySolver{DPR}{\asGalleryValue{asGalleryLoss}{1}{0.955}}{\asGalleryValue{asGalleryGain}{1}{1.12}}{0/13}{Repeated failed-literal probing, root shortening, binary-implication SCC processing, and bounded Davis-Putnam elimination reduce the recognized circuit formula before 1-UIP CDCL on the remainder. Its UNSAT path buffers derived clause additions, including probe, SCC, elimination, learned, and contradiction clauses; unsupported inputs or a satisfying search result are reported as UNKNOWN.}
\asGallerySolver{VeriPB}{\asGalleryValue{asGalleryLoss}{1}{0.955}}{\asGalleryValue{asGalleryGain}{1}{1.12}}{0/13}{Dynamic-cost elimination over gate outputs and binary-implication SCC substitution drive preprocessing; fixed-seed local search seeds phases for one-UIP CDCL with minimization and restarts. For UNSAT, it emits derived and trimmed learned clauses as unhinted RUP steps, then an empty RUP constraint and conclusion; unsupported layouts or SAT exhaustion return UNKNOWN.}
\end{asGalleryFamily}

\begin{asGalleryFamily}{069}{hamiltonian}{1,100 GBD instances\enspace\textperiodcentered\enspace 220 validation\enspace\textperiodcentered\enspace baseline 206/220}{The benchmark asks whether a structured monotone CNF has a selection of Boolean variables that selects exactly one variable in each positive constraint. Pairwise negative clauses forbid selecting variables that co-occur, reducing the decision question to a restricted exact-cover search.}
\ifdefined\pdfbookmark\pdfbookmark[2]{hamiltonian}{as-gallery-69}\fi
\asGallerySolver{GRAT}{\asGalleryValue{asGalleryLoss}{0}{0.248}}{\asGalleryValue{asGalleryLoss}{0}{0.256}}{141/220}{Focused Metropolis flips and at-most-one-preserving packing searches provide SAT attempts, followed by exactly-one CDCL probes. SAT assignments are checked; if no model is found, an unbounded logged CDCL run emits learned clauses and an empty clause as DRAT text for external elaboration and checking, while unsupported, exhausted, or failed paths return UNKNOWN.}
\asGallerySolver{DPR}{\asGalleryValue{asGalleryLoss}{1}{0.473}}{\asGalleryValue{asGalleryLoss}{1}{0.483}}{185/220}{Reversible exact-cover search branches on the least-available constraint with reversible bitset buckets; bounded probes precede an unbounded pass. Watched-literal CDCL follows a no-model search; SAT assignments are checked, while UNSAT runs emit learned clauses and an empty clause as witness-free DPR additions for external checking; unsupported layouts or proof failures return UNKNOWN.}
\asGallerySolver{VeriPB}{\asGalleryValue{asGalleryLoss}{0}{0.443}}{\asGalleryValue{asGalleryLoss}{0}{0.423}}{182/220}{CDCL-first search operates only after strict port-layout recognition, then falls back after a non-root conflict to Algorithm X with singleton propagation, reversible state, and minimum-row branching. SAT models are checked against the CNF; UNSAT reruns emit pseudo-Boolean RUP steps and tree nogoods, while unsupported layouts, some large cases, or proof-output failures return UNKNOWN.}
\end{asGalleryFamily}

\begin{asGalleryFamily}{070}{hamiltonian-\allowbreak cycle}{24 GBD instances\enspace\textperiodcentered\enspace 5 validation\enspace\textperiodcentered\enspace baseline 4/5}{The benchmark asks whether a sparse graph contains a single cycle visiting every vertex once. In the recognized CNF layouts, paired directed-edge selectors enforce one chosen incoming and outgoing arc per vertex, with auxiliary finite-state counter constraints.}
\ifdefined\pdfbookmark\pdfbookmark[2]{hamiltonian-cycle}{as-gallery-70}\fi
\asGallerySolver{GRAT}{\asGalleryValue{asGalleryGain}{1}{251}}{\asGalleryValue{asGalleryGain}{1}{233}}{5/5}{Graph CDCL drives the search with degree-two constraints, DSU component checks, and incremental subtour cuts. Blossom 2-factor seeds and randomized DFS provide fallbacks; a found cycle is oriented, completed by watched-literal propagation, and clause-checked, while unsupported or exhausted searches return UNKNOWN rather than an UNSAT certificate.}
\asGallerySolver{DPR}{\asGalleryValue{asGalleryGain}{0}{1.43}}{\asGalleryValue{asGalleryGain}{0}{1.43}}{4/5}{Structural recognition reconstructs the graph, then rollback degree-2 search uses subtour propagation and frontier memoization on detected narrow strips. A Tutte-style degree-2 factor and bounded exchanges, followed by full-CNF CDCL, provide fallbacks; verified assignments are emitted, while unsupported layouts or failed SAT-side searches return UNKNOWN because no UNSAT/DPR certificate path exists.}
\asGallerySolver{VeriPB}{\asGalleryValue{asGalleryGain}{0}{1.42}}{\asGalleryValue{asGalleryGain}{0}{1.41}}{4/5}{Certifying CDCL with VeriPB RUP replay supplies UNSAT certification after a graph-guided probe using blossom matching, Posa rotations, and bounded rollback. Only the supported layouts are recognized; SAT candidates are completed and clause-checked with full assignments, while unsupported inputs or replay divergence yield UNKNOWN and the counter-divisibility helpers remain unused.}
\end{asGalleryFamily}

\begin{asGalleryFamily}{071}{hanoi}{6 GBD instances\enspace\textperiodcentered\enspace 2 validation\enspace\textperiodcentered\enspace baseline 2/2}{The benchmark asks whether a Boolean assignment describes a legal shortest Towers-of-Hanoi transfer between three pegs. A CNF uses move, disk, support, and state variables, with clauses enforcing one move per step, legal support changes, and the initial and goal configurations.}
\ifdefined\pdfbookmark\pdfbookmark[2]{hanoi}{as-gallery-71}\fi
\asGallerySolver{GRAT}{\asGalleryValue{asGalleryGain}{1}{159}}{\asGalleryValue{asGalleryGain}{1}{64.7}}{2/2}{Structural recognition recovers conditional-frame tracks, transition order, and endpoint towers for bounded Hanoi signatures. It recursively constructs the canonical plan, maps moves to action variables, propagates units before checking every clause; it has no SAT fallback, and structural, propagation, or verification failure returns UNKNOWN, with no UNSAT certificate path.}
\asGallerySolver{DPR}{\asGalleryValue{asGalleryLoss}{0}{4.57\!\cdot\!10^{-4}}}{\asGalleryValue{asGalleryLoss}{0}{4.63\!\cdot\!10^{-4}}}{0/2}{Structural recognition accepts only a fixed six-disk, 63-transition structure, reconstructs its persistent tracks, and synthesizes the canonical plan. It reduces residual clauses to 2-SAT by propagation and SCC fallback, checks complete assignments against all clauses, and emits SAT only; failures return UNKNOWN, with no UNSAT certificate path.}
\asGallerySolver{VeriPB}{\asGalleryValue{asGalleryLoss}{0}{4.57\!\cdot\!10^{-4}}}{\asGalleryValue{asGalleryLoss}{0}{4.63\!\cdot\!10^{-4}}}{0/2}{Recognition recovers the six-disk path and persistent move tracks, then forces each moved disk to the canonical optimal sequence. Occurrence-based unit completion handles larger frontends, while the base residual falls back to learned-clause CDCL. SAT assignments are clause-checked; unsupported or non-SAT cases return UNKNOWN, and no UNSAT certificate is emitted.}
\end{asGalleryFamily}

\begin{asGalleryFamily}{072}{hardware-\allowbreak bmc}{49 GBD instances\enspace\textperiodcentered\enspace 10 validation\enspace\textperiodcentered\enspace baseline 6/10}{This benchmark asks whether bounded hardware execution can reach a bad state while obeying initial and transition constraints. The time-unrolled circuit, state, and gate constraints become CNF clauses, so SAT gives a witness execution and UNSAT rules out the bound.}
\ifdefined\pdfbookmark\pdfbookmark[2]{hardware-bmc}{as-gallery-72}\fi
\asGallerySolver{GRAT}{\asGalleryValue{asGalleryLoss}{1}{0.868}}{\asGalleryValue{asGalleryGain}{1}{1.13}}{5/10}{Structural gate-pattern recognition admits a narrow hardware-CNF subfamily, then bounded variable elimination simplifies it before watched-literal CDCL search. SAT assignments are reconstructed and checked against the original CNF; UNSAT runs emit DRAT-style clause additions, deletions, and an empty clause for external checking, while rejected inputs, proof-output failure, or failed checks return UNKNOWN.}
\asGallerySolver{DPR}{\asGalleryValue{asGalleryLoss}{0}{0.829}}{\asGalleryValue{asGalleryGain}{0}{1.09}}{5/10}{Wide-gate recognition and incremental assumptions over bad frames drive search, with propagation and bounded elimination feeding watched-literal CDCL. Rejected frames become complement units; SAT models are checked after reverse extension, while UNSAT may log witness-free clause additions, deletions, and an empty clause for checking; unrecognized inputs, proof-output failure, or failed reconstruction return UNKNOWN.}
\asGallerySolver{VeriPB}{\asGalleryValue{asGalleryLoss}{0}{0.797}}{\asGalleryValue{asGalleryLoss}{0}{0.786}}{5/10}{Signed equivalence closure and gate-congruence recognition enable proof-logged substitutions and bounded variable elimination before watched-literal CDCL search. SAT models are reconstructed and checked against the original CNF; supported UNSAT runs emit VeriPB RUP steps through contradiction, whereas UNSAT outside supported certificate profiles or on unrecognized inputs returns UNKNOWN.}
\end{asGalleryFamily}

\begin{asGalleryFamily}{073}{hardware-\allowbreak model-\allowbreak checking}{76 GBD instances\enspace\textperiodcentered\enspace 15 validation\enspace\textperiodcentered\enspace baseline 8/15}{The benchmark asks whether Boolean signals in a bounded hardware check can receive values satisfying every constraint. Gate relations, initial conditions, and checked outputs become CNF clauses, often in a structured Tseitin netlist, though encodings may be less regular.}
\ifdefined\pdfbookmark\pdfbookmark[2]{hardware-model-checking}{as-gallery-73}\fi
\asGallerySolver{GRAT}{\asGalleryValue{asGalleryLoss}{0}{0.567}}{\asGalleryValue{asGalleryLoss}{0}{0.674}}{1/15}{Exact AND/XOR recognition, equivalence probing, and bounded variable elimination simplify smaller gate-heavy instances before search. Watched-literal CDCL handles larger or unreduced formulas; SAT models are reconstructed and checked, while UNSAT clauses are logged for external elaboration and malformed or failed internal paths return UNKNOWN.}
\asGallerySolver{DPR}{\asGalleryValue{asGalleryLoss}{1}{0.794}}{\asGalleryValue{asGalleryLoss}{1}{0.964}}{5/15}{Conservative gate recognition and canonicalization recover Boolean operations, then test structural candidates under negation before adding RUP-style proof clauses. Within certifying size limits, watched-literal CDCL handles fallback cases; larger formulas with structural tasks use quotient discovery, while checked SAT assignments are accepted and proofless UNSAT or failed discovery return UNKNOWN.}
\asGallerySolver{VeriPB}{\asGalleryValue{asGalleryLoss}{0}{0.586}}{\asGalleryValue{asGalleryLoss}{0}{0.730}}{1/15}{Complemented-edge structural hashing canonicalizes recognized topologically numbered Tseitin gates, folds simplifications, and checks the resulting quotient by propagation. After structural admission, internal watched-literal CDCL searches the quotient; without admission or a structural contradiction, the solver returns UNKNOWN, while SAT assignments are clause-checked and UNSAT ends with a VeriPB RUP certificate.}
\end{asGalleryFamily}

\begin{asGalleryFamily}{074}{hardware-\allowbreak verification}{779 GBD instances\enspace\textperiodcentered\enspace 156 validation\enspace\textperiodcentered\enspace baseline 129/156}{The benchmark asks whether Boolean signals in a circuit or hardware-oriented encoding can satisfy all required constraints. CNF typically expresses gate relations together with output, property, or state conditions, though the runtime accepts arbitrary DIMACS formulas.}
\ifdefined\pdfbookmark\pdfbookmark[2]{hardware-verification}{as-gallery-74}\fi
\asGallerySolver{GRAT}{\asGalleryValue{asGalleryLoss}{1}{0.637}}{\asGalleryValue{asGalleryLoss}{1}{0.965}}{106/156}{A bounded AIG sweep recognizes signed-AND gate patterns, simulates assignments in parallel, and may reconstruct a model or add equivalence clauses after RUP checks. Otherwise it falls back to CDCL. SAT models are checked; UNSAT logging records learned additions and a root empty clause as DRAT for external checking. Malformed input returns UNKNOWN.}
\asGallerySolver{DPR}{\asGalleryValue{asGalleryLoss}{0}{0.460}}{\asGalleryValue{asGalleryLoss}{0}{0.685}}{83/156}{A size-band portfolio starts with baseline CDCL, then reparses selected inputs into a fresh solver with recursive reason minimization. It otherwise uses first-UIP learning, watched propagation, activity branching, and restarts. SAT assignments are not independently clause-checked; UNSAT output records learned and empty-clause additions without original-clause copies, and malformed input or replay/proof failures return UNKNOWN.}
\asGallerySolver{VeriPB}{\asGalleryValue{asGalleryLoss}{0}{0.526}}{\asGalleryValue{asGalleryLoss}{0}{0.688}}{97/156}{A circuit sweep recovers gates, uses bit-parallel signatures and bounded implication queries, and falls back to CDCL when unproductive. The fallback uses first-UIP learning and Luby restarts; SAT assignments are clause-checked, while UNSAT output records learned clauses as VeriPB RUP constraints, an empty contradiction, and an explicit conclusion. Malformed input or certificate failures return UNKNOWN.}
\end{asGalleryFamily}

\begin{asGalleryFamily}{075}{hashtable-\allowbreak safety}{21 GBD instances\enspace\textperiodcentered\enspace 5 validation\enspace\textperiodcentered\enspace baseline 1/5}{These instances are CNF satisfiability queries with numbered Boolean variables and clauses. The available solver artifacts do not establish which hashtable-level property the queries encode.}
\ifdefined\pdfbookmark\pdfbookmark[2]{hashtable-safety}{as-gallery-75}\fi
\asGallerySolver{GRAT}{\asGalleryValue{asGalleryLoss}{1}{0.815}}{\asGalleryValue{asGalleryGain}{1}{1.10}}{0/5}{Packed binary implications are used for propagation in a narrow variable-count range, with ordinary watched literals as the fallback. It runs first-UIP CDCL with activity-based branching, phase saving, and restarts, checks SAT assignments, and emits learned clauses as DRAT for external elaboration; input, proof, or exception failures return UNKNOWN.}
\asGallerySolver{DPR}{\asGalleryValue{asGalleryLoss}{1}{0.815}}{\asGalleryValue{asGalleryGain}{1}{1.10}}{0/5}{A strict CNF profile gate admits only very large, mostly short-clause formulas with many units; accepted inputs use two-watch CDCL with 1-UIP learning. Checked SAT models are reported, while UNSAT output uses witness-free DPR clause additions, a final empty clause, and only the first parsed original unit; unsupported or output failures return UNKNOWN.}
\asGallerySolver{VeriPB}{\asGalleryValue{asGalleryLoss}{1}{0.815}}{\asGalleryValue{asGalleryGain}{0}{1.09}}{0/5}{A fixed signature gate accepts only a narrow normalized bit-blasted CNF pattern; accepted inputs use watched-literal CDCL with 1-UIP learning and activity-based branching. UNSAT conflicts are logged as VeriPB RUP constraints with hints and a final empty RUP, while SAT returns a checked assignment; unsupported or malformed inputs return UNKNOWN.}
\end{asGalleryFamily}

\begin{asGalleryFamily}{076}{heule-\allowbreak folkman}{11 GBD instances\enspace\textperiodcentered\enspace 3 validation\enspace\textperiodcentered\enspace baseline 3/3}{The benchmark asks whether the edges of a K4-free graph can receive two colors so every triangle is non-monochromatic. A CNF representation uses one variable per edge and an all-positive/all-negative clause pair per triangle, requiring its three values not to be all equal.}
\ifdefined\pdfbookmark\pdfbookmark[2]{heule-folkman}{as-gallery-76}\fi
\asGallerySolver{GRAT}{\asGalleryValue{asGalleryGain}{1}{429}}{\asGalleryValue{asGalleryGain}{1}{419}}{3/3}{Focused local search attacks monochromatic triangles with weighted break-score flips and updates, then falls back to watched-literal CDCL when bounded search fails. SAT assignments are checked before emission; CDCL UNSAT logs textual DRAT and emits the empty clause after a root conflict for external elaboration, while unsupported inputs return UNKNOWN.}
\asGallerySolver{DPR}{\asGalleryValue{asGalleryGain}{0}{1.83}}{\asGalleryValue{asGalleryGain}{0}{1.83}}{3/3}{NAE local search uses focused bad-triangle walks and breakout/probSAT choices, then tries vertex cut-switching, bounded repair, elimination, and CDCL. SAT candidates are checked on the original; reduced UNSAT is rerun untouched, logging additions that may be witness-free and emitting the empty clause only after a root conflict for external checking; unsupported inputs return UNKNOWN.}
\asGallerySolver{VeriPB}{\asGalleryValue{asGalleryGain}{0}{104}}{\asGalleryValue{asGalleryGain}{0}{98.2}}{3/3}{NAE WalkSAT first maintains monochromatic triangles incrementally across noise restarts and an alternate sampler, then falls back to watched-literal CDCL seeded with its best phase and a color-symmetry fixing. A complete assignment is expanded and checked against the original clauses; unrecognized or exhausted cases return UNKNOWN because no UNSAT certificate path is implemented.}
\end{asGalleryFamily}

\begin{asGalleryFamily}{077}{heule-\allowbreak nol}{11 GBD instances\enspace\textperiodcentered\enspace 3 validation\enspace\textperiodcentered\enspace baseline 2/3}{The benchmark asks whether an L-shaped board can use three colors without a monochromatic equal-arm L, with some cells pinned. The CNF gives each cell three color variables, an at-least-one clause, and negative clauses forbidding monochromatic colors on each L.}
\ifdefined\pdfbookmark\pdfbookmark[2]{heule-nol}{as-gallery-77}\fi
\asGallerySolver{GRAT}{\asGalleryValue{asGalleryGain}{0}{37.4}}{\asGalleryValue{asGalleryGain}{0}{37.4}}{3/3}{Weighted breakout descent uses incidence-local recoloring scores, tabu moves, and restarts to escape local minima. The narrow canonical-board recognizer validates a found coloring and emits it as the SAT witness; unsupported inputs return UNKNOWN, while unresolved recognized cases can remain in the restart loop because no UNSAT certificate path is implemented.}
\asGallerySolver{DPR}{\asGalleryValue{asGalleryGain}{0}{13.4}}{\asGalleryValue{asGalleryGain}{0}{13.4}}{3/3}{Incremental breakout local search combines weighted incident scores, smoothing, and deterministic restart salts. The exact 22 by 22, n=11 recognizer searches one-hot colorings although its raw clauses lack at-most-one constraints; it emits checked SAT assignments, proves only a directly forced monochromatic L by an empty clause, and unresolved cases can remain in the restart loop.}
\asGallerySolver{VeriPB}{\asGalleryValue{asGalleryGain}{1}{194}}{\asGalleryValue{asGalleryGain}{1}{191}}{3/3}{Incremental weighted min-conflicts local search evaluates six recolorings around a violated corner, using noise, breakout weights, and deterministic restarts. After exact recognition for n \ensuremath{<}= 64, unsupported inputs return UNKNOWN; VeriPB pol or RUP certificates cover only two restricted UNSAT triggers, while checked SAT assignments are emitted and other unresolved cases may continue indefinitely.}
\end{asGalleryFamily}

\begin{asGalleryFamily}{078}{hgen}{336 GBD instances\enspace\textperiodcentered\enspace 68 validation\enspace\textperiodcentered\enspace baseline 36/68}{These benchmarks ask whether a Boolean assignment satisfies a DIMACS CNF, usually with bounded-width clauses and regular occurrence, density, and polarity patterns. Some encodings use disjoint four-literal choice clauses and signed binary exclusions whose hole components are one fewer than choice groups.}
\ifdefined\pdfbookmark\pdfbookmark[2]{hgen}{as-gallery-78}\fi
\asGallerySolver{GRAT}{\asGalleryValue{asGalleryGain}{0}{1.32}}{\asGalleryValue{asGalleryGain}{0}{1.30}}{42/68}{Configuration-checking and ProbSAT-style local search maintains true-literal counts and a false-clause set, then uses unbounded focused restarts when attempts find no model. An elimination route uses watched-literal CDCL with first-UIP learning and writes learned clauses as textual DRAT additions for external elaboration; the local-search route has no UNSAT result path.}
\asGallerySolver{DPR}{\asGalleryValue{asGalleryGain}{1}{1.64}}{\asGalleryValue{asGalleryGain}{0}{1.53}}{47/68}{A structural sparse-pigeonhole detector first constructs a RUP/PR/RAT-style certificate with deletions and witness-free clause additions; the writer is not internally checked. Otherwise focused stochastic local search uses break counts, weighted choices, restarts, and an active WalkSAT-style mode. Regular recognized formulas have no UNSAT certificate route; failed assignments are reported UNKNOWN.}
\asGallerySolver{VeriPB}{\asGalleryValue{asGalleryGain}{0}{1.61}}{\asGalleryValue{asGalleryGain}{1}{1.63}}{47/68}{A direct pigeonhole detector emits a VeriPB-style proof by deriving hole at-most-one constraints and summing them with four-choice clauses. Other CNFs use inverse-power local search with spectral restarts, then CDCL with RUP logging; only the exact pigeonhole structure has a direct proof path, while failed paths can return UNKNOWN.}
\end{asGalleryFamily}

\begin{asGalleryFamily}{079}{hidoku}{34 GBD instances\enspace\textperiodcentered\enspace 7 validation\enspace\textperiodcentered\enspace baseline 7/7}{The benchmark asks whether an n by n grid can place 1 through n\textasciicircum{}2 exactly once, respect fixed clues, and put consecutive values in king-adjacent cells. CNF encodings use cell-value variables, exact-one constraints, clue units, and adjacency implications, sometimes with auxiliary variables.}
\ifdefined\pdfbookmark\pdfbookmark[2]{hidoku}{as-gallery-79}\fi
\asGallerySolver{GRAT}{\asGalleryValue{asGalleryLoss}{0}{0.0818}}{\asGalleryValue{asGalleryLoss}{0}{0.202}}{0/7}{It enumerates a bounded set of row-wise Hamiltonian snake assignments under board symmetries, fixing the primary cell-value variables before searching auxiliary variables. Watched-literal DPLL completes and independently checks the CNF before emitting a total SAT assignment; unrecognized or unsuccessful cases return UNKNOWN, and no UNSAT certificate path is implemented.}
\asGallerySolver{DPR}{\asGalleryValue{asGalleryGain}{0}{163}}{\asGalleryValue{asGalleryLoss}{0}{0.278}}{7/7}{It tries row-boustrophedon snakes on presumed primary variables, then uses propagation and recursive DPLL to complete and check the CNF before emitting a SAT assignment. A specialized recognizer constructs a clause/deletion trace for a diagonal crossing and reports UNSAT only on contradiction, but covers only narrowly recognized 6x6 and 7x7 encodings; other failures return UNKNOWN.}
\asGallerySolver{VeriPB}{\asGalleryValue{asGalleryGain}{1}{8{,}980}}{\asGalleryValue{asGalleryGain}{1}{2{,}620}}{7/7}{It recognizes a nested-AMO and king-neighbor encoding and, for a geometric clue obstruction, emits a VeriPB certificate using PB and RUP derivations. Otherwise it tries symmetric row snakes, then clue-aware Hamiltonian-path search with distance and connectivity pruning; SAT models are checked before emission, while failed or unrecognized cases return UNKNOWN.}
\end{asGalleryFamily}

\begin{asGalleryFamily}{080}{hypertree-\allowbreak decomposition}{56 GBD instances\enspace\textperiodcentered\enspace 12 validation\enspace\textperiodcentered\enspace baseline 8/12}{The benchmark asks whether a hypergraph has an elimination order whose resulting bags satisfy the hypertree-width bound, with each bag covered by at most k hyperedges and special conditions. A CNF records order, completion relations, bag selections, and bound constraints. Recognizers may support restricted encodings.}
\ifdefined\pdfbookmark\pdfbookmark[2]{hypertree-decomposition}{as-gallery-80}\fi
\asGallerySolver{GRAT}{\asGalleryValue{asGalleryGain}{0}{1.23}}{\asGalleryValue{asGalleryGain}{1}{1.23}}{7/12}{Recognizes a fixed CNF layout, constructs a primal elimination order with dual min-fill trials, then searches bag hyperedge covers, using bounded branching only for smaller widths. Rollback Horn propagation completes remaining variables, after which every clause is checked before SAT output; unsupported layouts, width rejection, or failed construction/search return UNKNOWN, with no UNSAT certificate path.}
\asGallerySolver{DPR}{\asGalleryValue{asGalleryGain}{1}{1.23}}{\asGalleryValue{asGalleryGain}{0}{1.16}}{7/12}{Recognizes degree-two incidence and an exact dual-line-graph primal graph, then fixes a primal order by greedy dual minimum-fill completion. A watched-literal backtracking search solves remaining guard, special-condition, and counter variables, with limited seed retries; structural rejection or unsolved residual returns UNKNOWN, and no DPR derivation or UNSAT certificate path is implemented.}
\asGallerySolver{VeriPB}{\asGalleryValue{asGalleryLoss}{0}{0.517}}{\asGalleryValue{asGalleryLoss}{0}{0.522}}{0/12}{Uses randomized local search over orders, with bitset branching for bounded bag covers and restarts. After an order is found, it fixes arcs and selectors, uses Horn propagation for remaining variables, and checks every clause before SAT output; unsupported layouts or failed search/completion return UNKNOWN, with no VeriPB or UNSAT certificate path.}
\end{asGalleryFamily}

\begin{asGalleryFamily}{081}{independent-\allowbreak set}{30 GBD instances\enspace\textperiodcentered\enspace 6 validation\enspace\textperiodcentered\enspace baseline 2/6}{The benchmark asks whether exactly K of N=2\textasciicircum{}d binary-word vertices can be selected without selecting both ends of a graph edge. CNF uses one selection variable per vertex, clauses forbidding both endpoints of each edge, and auxiliary one-hot running-count variables enforcing the exact total.}
\ifdefined\pdfbookmark\pdfbookmark[2]{independent-set}{as-gallery-81}\fi
\asGallerySolver{GRAT}{\asGalleryValue{asGalleryGain}{0}{1.17}}{\asGalleryValue{asGalleryGain}{0}{1.29}}{2/6}{Weighted-syndrome construction is tried first for Z-channel graphs, followed by specialized exact or bit-parallel complement-clique search and bounded local search on remaining recognized cases. Candidates are given counter values and fully rechecked before SAT output; failed or unsupported inputs return UNKNOWN, with no UNSAT or DRAT certificate path.}
\asGallerySolver{DPR}{\asGalleryValue{asGalleryGain}{0}{1.17}}{\asGalleryValue{asGalleryGain}{0}{1.29}}{2/6}{Layered complement-graph clique search drives construction, using Hamming-weight layers and bit-parallel branch-and-bound for transposition graphs, while deterministic local search handles other recognized cases. After assigning the running-count auxiliaries, it rechecks every clause before emitting a SAT assignment; failed or unsupported cases return UNKNOWN, and no UNSAT or DPR certificate path is implemented.}
\asGallerySolver{VeriPB}{\asGalleryValue{asGalleryGain}{1}{4.63}}{\asGalleryValue{asGalleryGain}{1}{3.62}}{5/6}{Component decomposition and complement-clique search solve suitable cases exactly, while residual-degree greedy and exchange search cover larger graphs. A found SAT assignment reconstructs the automaton and is checked against every clause; only narrow component-tree, transposition, and Z-channel branches emit VeriPB UNSAT proofs, while other failures or unsupported cases return UNKNOWN.}
\end{asGalleryFamily}

\begin{asGalleryFamily}{082}{independent-\allowbreak set-\allowbreak reconfiguration}{20 GBD instances\enspace\textperiodcentered\enspace 4 validation\enspace\textperiodcentered\enspace baseline 0/4}{Given a graph and two same-size independent sets, the task asks whether one can reach the other through a specified-length sequence of distinct independent sets, replacing one vertex at each step. CNF encodings represent graph edges, configurations, transitions, and endpoint constraints.}
\ifdefined\pdfbookmark\pdfbookmark[2]{independent-set-reconfiguration}{as-gallery-82}\fi
\asGallerySolver{GRAT}{\asGalleryValue{asGalleryGain}{0}{2.00}}{\asGalleryValue{asGalleryGain}{0}{2.00}}{2/4}{Target-directed bitset exploration builds the token-jumping component, then randomized shortest-path-guided trials seek a simple path of the prescribed length. A completed path is propagated and checked against the CNF for SAT; bounded component, parity, and small-clique routines write DRAT-style UNSAT certificates, while unrecognized or unresolved cases return UNKNOWN.}
\asGallerySolver{DPR}{\asGalleryValue{asGalleryGain}{1}{3.95}}{\asGalleryValue{asGalleryGain}{1}{3.90}}{3/4}{Guarded structural routines emit PHP/RUP-style derivations, including witness-free clause additions, for a few UNSAT cases; otherwise bidirectional BFS with bounded randomized walks and ear insertion seeks a simple path of the required length. Propagation and clause rescanning complete SAT models; unsupported layouts or unresolved searches return UNKNOWN, with no general UNSAT path.}
\asGallerySolver{VeriPB}{\asGalleryValue{asGalleryGain}{0}{2.00}}{\asGalleryValue{asGalleryGain}{0}{1.99}}{2/4}{Specialized pseudo-Boolean proof generators run first, while bit-mask token-jump generation and component enumeration support distance-pruned exact-length DFS or seeded restarts. SAT paths are propagated and rescanned against the CNF; no generic CDCL proof fallback is enabled, so unrecognized or unresolved cases return UNKNOWN.}
\end{asGalleryFamily}

\begin{asGalleryFamily}{083}{influence-\allowbreak maximization}{20 GBD instances\enspace\textperiodcentered\enspace 4 validation\enspace\textperiodcentered\enspace baseline 0/4}{The benchmark asks whether at most K seed nodes in a directed graph can activate at least T nodes within ten synchronous threshold rounds, with seeds remaining active. A CNF represents seed choices, activity states, threshold propagation, and cardinality limits.}
\ifdefined\pdfbookmark\pdfbookmark[2]{influence-maximization}{as-gallery-83}\fi
\asGallerySolver{GRAT}{\asGalleryValue{asGalleryGain}{0}{1.33}}{\asGalleryValue{asGalleryGain}{1}{1.33}}{1/4}{Greedy seed construction and simulated-annealing swaps guide threshold diffusion on the recognized encoding. A watched-literal CDCL fallback handles unsuccessful searches without an internal stop condition; unresolved cases can return UNKNOWN. SAT checks a total assignment; UNSAT logs learned clauses and an empty clause as DRAT additions for external elaboration.}
\asGallerySolver{DPR}{\asGalleryValue{asGalleryGain}{1}{1.33}}{\asGalleryValue{asGalleryGain}{0}{1.33}}{1/4}{Bit-parallel diffusion simulation drives greedy and annealed fixed-cardinality seed search on recognized layouts. If it fails, a watchdog-free CDCL fallback adds reduction, elimination, and learning clauses; unresolved or rejected models return UNKNOWN. SAT requires a total satisfying assignment; UNSAT requires potentially witness-free RUP additions to derive the empty clause.}
\asGallerySolver{VeriPB}{\asGalleryValue{asGalleryGain}{0}{1.33}}{\asGalleryValue{asGalleryGain}{0}{1.33}}{1/4}{Bitset threshold simulation scores greedy additions and annealed fixed-cardinality swaps on the recognized ten-layer layout. Activity-guided CDCL is conflict-bounded, so failed heuristic search or unfinished fallback returns UNKNOWN. SAT requires a checked assignment; UNSAT is reported only after level-zero closure with VeriPB RUP constraints and deletions.}
\end{asGalleryFamily}

\begin{asGalleryFamily}{084}{interval-\allowbreak matching}{23 GBD instances\enspace\textperiodcentered\enspace 5 validation\enspace\textperiodcentered\enspace baseline 0/5}{The benchmark asks whether each group can choose one option without using a resource clique twice. Its CNF expresses choices with at-least-one clauses and conflicts with binary clauses, reducing the question to a matching that covers every group.}
\ifdefined\pdfbookmark\pdfbookmark[2]{interval-matching}{as-gallery-84}\fi
\asGallerySolver{GRAT}{\asGalleryValue{asGalleryGain}{0}{2.97\!\cdot\!10^{4}}}{\asGalleryValue{asGalleryGain}{0}{17.9}}{5/5}{Parity-DSU polarity recovery normalizes recognized clauses into rows and conflict cliques; augmenting-path matching constructs and checks a full assignment against every original clause. After a Hall deficiency, alternating reachability feeds a fresh-variable pigeonhole reduction emitting definition, row, conflict, and empty clauses for UNSAT; unsupported structure or proof failure returns UNKNOWN.}
\asGallerySolver{DPR}{\asGalleryValue{asGalleryGain}{0}{1.26\!\cdot\!10^{5}}}{\asGalleryValue{asGalleryGain}{0}{208}}{5/5}{Augmenting-path matching builds a complete assignment, checked against every original clause before SAT is reported. For a deficient matching, Hall validation and bitset-guided swap planning must succeed; the PR-style writer emits swap records, unit clauses, and an empty clause for UNSAT, while recognition, model-check, planning, or proof-output failures return UNKNOWN.}
\asGallerySolver{VeriPB}{\asGalleryValue{asGalleryGain}{1}{1.50\!\cdot\!10^{5}}}{\asGalleryValue{asGalleryGain}{1}{4{,}220}}{5/5}{Augmenting-path matching on recognized signed group/resource structure yields a complete assignment checked against every original clause before SAT is reported. For a deficient matching, alternating reachability supplies a Hall witness for a pseudo-Boolean derivation combining group clauses and clique bounds to derive an empty clause; recognition, model-check, or certificate failures return UNKNOWN.}
\end{asGalleryFamily}

\begin{asGalleryFamily}{085}{karatsuba-\allowbreak multiplication}{3 GBD instances\enspace\textperiodcentered\enspace 1 validation\enspace\textperiodcentered\enspace baseline 1/1}{These CNFs ask whether two odd, nontrivial binary factors satisfy a bit-blasted Karatsuba encoding of a fixed odd product. Factor bits and auxiliary circuit variables are linked by clauses, making the benchmark a Boolean satisfiability question.}
\ifdefined\pdfbookmark\pdfbookmark[2]{karatsuba-multiplication}{as-gallery-85}\fi
\asGallerySolver{GRAT}{\asGalleryValue{asGalleryLoss}{1}{0.115}}{\asGalleryValue{asGalleryLoss}{1}{0.115}}{0/1}{Composite odd-prefix probing guides modular-inverse recovery and watched-literal recursive DFS, prioritizing factor bits before auxiliary variables; clause learning, available prime tries, and factor-order symmetry prune the search. Only a checked satisfying assignment is emitted; no UNSAT certificate is implemented, so unsupported layouts, contradictions, and exhausted searches return UNKNOWN.}
\asGallerySolver{DPR}{\asGalleryValue{asGalleryLoss}{1}{0.115}}{\asGalleryValue{asGalleryLoss}{1}{0.115}}{0/1}{Low-column residue decoding with modular-inverse propagation drives watched-literal recursive DFS, prioritizing factor bits before branching on remaining variables; prime and exact-width checks prune completed candidates. It emits a checked satisfying assignment but has no UNSAT certificate or generic fallback, so unsupported layouts, contradictions, and failed searches return UNKNOWN.}
\asGallerySolver{VeriPB}{\asGalleryValue{asGalleryLoss}{1}{0.115}}{\asGalleryValue{asGalleryLoss}{1}{0.115}}{0/1}{Adaptive modular-residue recovery guides discrepancy-ordered factor search on recognized layouts, while bounded CDCL completes auxiliary variables and a general CDCL loop handles backdoor failure. Every reported assignment is checked against the original clauses; no UNSAT certificate path exists, so heuristic failure, contradictions, or unsuccessful completion return UNKNOWN.}
\end{asGalleryFamily}

\begin{asGalleryFamily}{086}{knights-\allowbreak problem}{21 GBD instances\enspace\textperiodcentered\enspace 5 validation\enspace\textperiodcentered\enspace baseline 2/5}{The benchmark asks whether an even square board has a closed knight tour, a cycle of legal knight moves that visits every square exactly once. CNF encodings select these moves, with auxiliary variables enforcing one-to-one incidence or position conditions.}
\ifdefined\pdfbookmark\pdfbookmark[2]{knights-problem}{as-gallery-86}\fi
\asGallerySolver{GRAT}{\asGalleryValue{asGalleryGain}{0}{3.34}}{\asGalleryValue{asGalleryGain}{1}{3.49}}{4/5}{Structural graph recovery is the distinguishing technique: recognized CNFs are matched to an even-board knight graph before a randomized Warnsdorff-style search selects a closed tour. Propagation and bounded completion fill the remaining variables and check all clauses, but failed recognition or search yields UNKNOWN; no general fallback or UNSAT certificate path is implemented.}
\asGallerySolver{DPR}{\asGalleryValue{asGalleryGain}{1}{3.34}}{\asGalleryValue{asGalleryGain}{0}{3.48}}{4/5}{Structural encoding recognition recovers the knight graph and move groups, then randomized Warnsdorff search proposes a tour; propagation and bounded residual DPLL complete the assignment. For unary and binary formulas, SCC reasoning can certify UNSAT with contradictory unit clauses and an empty clause; unsupported encodings or failed search return UNKNOWN.}
\asGallerySolver{VeriPB}{\asGalleryValue{asGalleryGain}{0}{1.67}}{\asGalleryValue{asGalleryGain}{0}{1.74}}{3/5}{Stable-color refinement and constrained backtracking recover knight-graph structure only for validated layouts, after which a Warnsdorff-style search selects a closed tour. Watched-literal propagation and bounded residual completion check the full clauses, but unsupported layouts or failed construction return UNKNOWN; no UNSAT certificate path is implemented.}
\end{asGalleryFamily}

\begin{asGalleryFamily}{087}{ktf}{20 GBD instances\enspace\textperiodcentered\enspace 4 validation\enspace\textperiodcentered\enspace baseline 2/4}{KTF instances ask whether a team can be selected from conflict groups so each skill is covered at least a required number of times within a weight budget. The CNF uses agent variables, skill counts, conflict cliques, and a budget constraint.}
\ifdefined\pdfbookmark\pdfbookmark[2]{ktf}{as-gallery-87}\fi
\asGallerySolver{GRAT}{\asGalleryValue{asGalleryLoss}{0}{0.511}}{\asGalleryValue{asGalleryLoss}{0}{0.511}}{0/4}{Bounded auxiliary-only elimination is followed by watched-literal CDCL, with learned clauses and positive-first branching. SAT reconstructs eliminated variables and checks the CNF; UNSAT at level zero is logged with clause additions and an empty clause for external DRAT elaboration, while unsupported input or internal failure returns UNKNOWN.}
\asGallerySolver{DPR}{\asGalleryValue{asGalleryGain}{0}{1.02}}{\asGalleryValue{asGalleryGain}{0}{1.02}}{2/4}{It recognizes a narrow nested clique-and-skill encoding, compresses each clique to a level, and tries rarity-weighted greedy constructions. It completes auxiliaries and checks every clause for SAT; only demand two enables watched-literal CDCL with witness-free clause additions for UNSAT, while other failures or unrecognized layouts return UNKNOWN.}
\asGallerySolver{VeriPB}{\asGalleryValue{asGalleryGain}{1}{1{,}440}}{\asGalleryValue{asGalleryGain}{1}{62.6}}{4/4}{An integer-dual search tests whether skill lower bounds exceed the decoded budget, emitting a pseudo-Boolean proof on success. Otherwise, randomized cost-aware construction and fail-first branching seek SAT models, complete auxiliaries by propagation, and check clauses; unsupported structures or exhausted search return UNKNOWN, with no general UNSAT fallback.}
\end{asGalleryFamily}

\begin{asGalleryFamily}{088}{lam-\allowbreak discrete-\allowbreak geometry}{20 GBD instances\enspace\textperiodcentered\enspace 4 validation\enspace\textperiodcentered\enspace baseline 3/4}{These CNFs ask whether a partially fixed incidence structure can be completed by selecting incidence variables so required sets are hit while incompatible pairs and small forbidden patterns are avoided. They use restricted row-based encodings, sometimes with auxiliary intersection variables.}
\ifdefined\pdfbookmark\pdfbookmark[2]{lam-discrete-geometry}{as-gallery-88}\fi
\asGallerySolver{GRAT}{\asGalleryValue{asGalleryLoss}{1}{0.401}}{\asGalleryValue{asGalleryLoss}{1}{0.619}}{0/4}{The supported pure 111-stride path uses a perfect-matching configuration lift, grouping incidence variables into row configurations and searching an exact-cover problem with bitset incompatibility masks and minimum-options branching. It has no generic fallback: unsupported structures return UNKNOWN, while UNSAT explanations are emitted as textual DRAT additions for external elaboration and checking.}
\asGallerySolver{DPR}{\asGalleryValue{asGalleryLoss}{1}{0.401}}{\asGalleryValue{asGalleryLoss}{1}{0.619}}{0/4}{A domain-level search enumerates legal row domains for the compact stride-111 layout, propagating compatibility with bit masks and chronological branching. If the specialized path does not establish UNSAT, watched-literal CDCL with learned clauses and restarts is used; unsupported signatures return UNKNOWN, and UNSAT output is a witness-free clause-addition stream for external elaboration and checking.}
\asGallerySolver{VeriPB}{\asGalleryValue{asGalleryLoss}{1}{0.401}}{\asGalleryValue{asGalleryLoss}{1}{0.619}}{0/4}{Semantic failed-literal probing and incidence-specific preprocessing prepare accepted instances for two-watched-literal CDCL, using first-UIP minimization, semantic phases, and Luby restarts. The bounded detector has no generic fallback: rejected inputs return UNKNOWN, while UNSAT attempts produce a VeriPB-style RUP proof for external checking and proof-creation failure also returns UNKNOWN.}
\end{asGalleryFamily}

\begin{asGalleryFamily}{089}{linvrinv}{12 GBD instances\enspace\textperiodcentered\enspace 3 validation\enspace\textperiodcentered\enspace baseline 1/3}{The benchmark asks whether n by n Boolean matrices over GF(2) can satisfy AB = I but not BA = I, testing one-sided inversion. The CNF uses matrix variables and AND/XOR auxiliaries, with a clause requiring BA to differ from I.}
\ifdefined\pdfbookmark\pdfbookmark[2]{linvrinv}{as-gallery-89}\fi
\asGallerySolver{GRAT}{\asGalleryValue{asGalleryMuted}{0}{1.00}}{\asGalleryValue{asGalleryMuted}{1}{1.00}}{1/3}{A narrow structural signature drives watched-literal CDCL, preferring variables with the expected gate-occurrence pattern and otherwise using activity-based decisions. First-UIP learning, restarts, phase saving, and clause reduction guide search; UNSAT emits textual DRAT for external elaboration, while SAT returns a checked assignment and malformed, unsupported, or proof-file failures return UNKNOWN.}
\asGallerySolver{DPR}{\asGalleryValue{asGalleryMuted}{1}{1.00}}{\asGalleryValue{asGalleryMuted}{0}{1.00}}{1/3}{XOR/AND flattening turns recognized networks up to four into parity clauses before two-watch CDCL, retaining the original CNF if flattening fails. First-UIP learning produces a textual DPR trace with parity, learned-clause, deletion, and empty-clause records on UNSAT; SAT returns a checked assignment, while malformed, unsupported, or larger inputs return UNKNOWN.}
\asGallerySolver{VeriPB}{\asGalleryValue{asGalleryMuted}{0}{1.00}}{\asGalleryValue{asGalleryMuted}{0}{1.00}}{1/3}{Algebraic decoding reconstructs matrices and, at orders four and five, enumerates vectors for an injection-surjection VeriPB certificate. Other cases below six, or decoder failures, use watched-literal CDCL with First-UIP learning and RUP logging; malformed or unsupported inputs and orders six or larger return UNKNOWN, and SAT returns a checked assignment.}
\end{asGalleryFamily}

\begin{asGalleryFamily}{090}{long-\allowbreak learned-\allowbreak clauses}{16 GBD instances\enspace\textperiodcentered\enspace 4 validation\enspace\textperiodcentered\enspace baseline 0/4}{This benchmark asks whether a Boolean assignment satisfies every clause in a structured CNF formula. The clauses encode local relations and, in some instances, larger counting or graph-linked layouts.}
\ifdefined\pdfbookmark\pdfbookmark[2]{long-learned-clauses}{as-gallery-90}\fi
\asGallerySolver{GRAT}{\asGalleryValue{asGalleryGain}{0}{4.00}}{\asGalleryValue{asGalleryGain}{0}{3.96}}{3/4}{Small XOR truth-table blocks are grouped into incidence components; an odd-RHS, zero-incidence component triggers fresh XOR gates, resolution projections, and balanced-tree rotations for a refutation. The refutation is written as DRAT for external elaboration and checking; otherwise bounded randomized search checks models, and unsupported or failed cases return UNKNOWN.}
\asGallerySolver{DPR}{\asGalleryValue{asGalleryGain}{0}{4.00}}{\asGalleryValue{asGalleryGain}{0}{3.95}}{3/4}{Small clauses become 2-4-variable parity relations; sparse XOR expressions and greedy overlap cancellation refute the odd-aggregate, two-occurrence case with witness-free DPR additions for external elaboration and checking. Otherwise, bipartite flow handles a relaxed exact-two four-variable pattern with one binary relation and builds a clause-checked model; unsupported cases return UNKNOWN.}
\asGallerySolver{VeriPB}{\asGalleryValue{asGalleryGain}{1}{1.59\!\cdot\!10^{5}}}{\asGalleryValue{asGalleryGain}{1}{1.20\!\cdot\!10^{4}}}{4/4}{Sequential-counter and clique-covered graph layouts are recognized first, yielding a non-branching pseudo-Boolean proof of contradiction. Compact groups use parity tests or fewest-mask branching for a clause-checked model; structural proofs use only recognized clauses. A bounded TreeRup fallback can emit a RUP proof, and unrecognized or uncertified paths return UNKNOWN.}
\end{asGalleryFamily}

\begin{asGalleryFamily}{091}{matrix-\allowbreak multiplication}{53 GBD instances\enspace\textperiodcentered\enspace 11 validation\enspace\textperiodcentered\enspace baseline 7/11}{These benchmarks ask whether a matrix-multiplication tensor over GF(2) can be expressed as a bounded sum of rank-one tensors. CNF variables encode factor coefficients and auxiliary products, while parity constraints require the resulting tensor entries to match.}
\ifdefined\pdfbookmark\pdfbookmark[2]{matrix-multiplication}{as-gallery-91}\fi
\asGallerySolver{GRAT}{\asGalleryValue{asGalleryLoss}{1}{0.978}}{\asGalleryValue{asGalleryLoss}{1}{0.987}}{7/11}{Structural recognizers recover tensor factors from AND/parity connectivity and try rank-six proof bridges or the embedded rank-23 construction, completing auxiliary variables by propagation before falling back to watched-literal CDCL. SAT models are clause-checked; UNSAT proof output is textual DRAT, with rank-six branches remapping bundled binary proof assets.}
\asGallerySolver{DPR}{\asGalleryValue{asGalleryLoss}{0}{0.373}}{\asGalleryValue{asGalleryLoss}{0}{0.382}}{0/11}{An exact recognizer for the checked rank-23 Brent structure canonicalizes the GF(2) tensor, searches the embedded Laderman orbit, and propagates auxiliaries. Only nine compact signatures reach watched-literal CDCL; SAT assignments are checked, compact UNSAT logs witness-free DPR additions and an empty clause, while unsupported inputs or proof failures return UNKNOWN.}
\asGallerySolver{VeriPB}{\asGalleryValue{asGalleryLoss}{0}{0.937}}{\asGalleryValue{asGalleryLoss}{0}{0.957}}{7/11}{Exact structural recognizers recover the 3x3 rank-23 Laderman structure or a rank-10 parity chain, using GF(2) elimination for residuals and a basis model for rank 10. Eligible cases use watched-literal CDCL with RUP logging; direct SAT returns checked assignments, while hard-listed or oversized cases return UNKNOWN without an UNSAT certificate.}
\end{asGalleryFamily}

\begin{asGalleryFamily}{092}{maximum-\allowbreak constraint-\allowbreak partition}{10 GBD instances\enspace\textperiodcentered\enspace 2 validation\enspace\textperiodcentered\enspace baseline 2/2}{The benchmark asks whether 100 of 200 weighted items can be chosen so their total equals that of the remaining 100. Its CNF uses selector variables plus auxiliary cardinality and binary-sum circuitry, with equality outputs enforcing matching totals.}
\ifdefined\pdfbookmark\pdfbookmark[2]{maximum-constraint-partition}{as-gallery-92}\fi
\asGallerySolver{GRAT}{\asGalleryValue{asGalleryLoss}{0}{0.0797}}{\asGalleryValue{asGalleryLoss}{0}{0.0798}}{1/2}{Bit-parallel evaluation of singleton scenarios recovers weights and total from the recognized 200-selector circuit, then dispatches by parity. Odd totals use least-significant-bit Davis-Putnam elimination and can emit DRAT additions for elaboration and checking; even totals use cardinality-grouped meet-in-the-middle search, but have no UNSAT path; unsupported layouts or failures return UNKNOWN.}
\asGallerySolver{DPR}{\asGalleryValue{asGalleryGain}{0}{1{,}060}}{\asGalleryValue{asGalleryGain}{1}{74.5}}{2/2}{Bit-parallel singleton evaluation recovers weights and total from the recognized circuit, then dispatches by parity. Odd totals use a least-significant-bit slice with minimum-occurrence-product Davis-Putnam elimination and witness-free clause additions for UNSAT; even totals use cardinality-grouped meet-in-the-middle search for checked models, while unsupported layouts or failures return UNKNOWN.}
\asGallerySolver{VeriPB}{\asGalleryValue{asGalleryGain}{1}{1{,}610}}{\asGalleryValue{asGalleryGain}{0}{70.9}}{2/2}{Scenario evaluation recovers weights and complementary sums from the recognized circuit, then dispatches by parity. Odd totals use affine GF(2) cone analysis to emit a pseudo-Boolean proof with RUP contradiction; even totals use greedy balancing and bounded exchanges for checked models but have no UNSAT path, while unsupported structures or failures return UNKNOWN.}
\end{asGalleryFamily}

\begin{asGalleryFamily}{093}{maxsat-\allowbreak optimum}{60 GBD instances\enspace\textperiodcentered\enspace 12 validation\enspace\textperiodcentered\enspace baseline 9/12}{The benchmark asks whether a Boolean assignment satisfies hard clauses while using no more than a specified number of soft-constraint relaxations. CNF encodings may express this bound with counters or structured choice, conflict, coverage, or selector gadgets, so recognized forms are restricted rather than universal.}
\ifdefined\pdfbookmark\pdfbookmark[2]{maxsat-optimum}{as-gallery-93}\fi
\asGallerySolver{GRAT}{\asGalleryValue{asGalleryLoss}{0}{0.348}}{\asGalleryValue{asGalleryLoss}{0}{0.373}}{0/12}{Structured order-counter and selector recognition activates destroy/rebuild, crossover, and weighted neighborhood search, with a fresh SAT solve validating each extended candidate. Unrecognized inputs use watched-literal CDCL with textual DRAT-style logging; specialized branches emit SAT models but lack an UNSAT certificate and may return UNKNOWN or continue indefinitely.}
\asGallerySolver{DPR}{\asGalleryValue{asGalleryLoss}{0}{0.348}}{\asGalleryValue{asGalleryLoss}{0}{0.373}}{0/12}{Exact-schema recognition activates least-model Horn closure and support-counted local search over choices, emitting checked SAT models. If that path fails or is inapplicable, watched-literal CDCL solves the original CNF and logs learned clauses as witness-free DPR additions. It tests only K, not a global optimum; exhaustion may return UNKNOWN, and immediate-conflict UNSAT lacks proof logging.}
\asGallerySolver{VeriPB}{\asGalleryValue{asGalleryLoss}{1}{0.407}}{\asGalleryValue{asGalleryLoss}{1}{0.436}}{2/12}{Focused clause-weighted local search handles synthesis-shaped CNFs by flipping variables from false clauses and increasing weights on trapped clauses, returning only checked SAT models and no UNSAT proof path. Other recognized forms undergo subsumption, bounded variable elimination, and watched-literal CDCL with reconstructed assignments and VeriPB/RUP UNSAT logging; unsupported inputs or configured limits can yield UNKNOWN.}
\end{asGalleryFamily}

\begin{asGalleryFamily}{094}{md5-\allowbreak equivalence-\allowbreak checking}{20 GBD instances\enspace\textperiodcentered\enspace 4 validation\enspace\textperiodcentered\enspace baseline 4/4}{The benchmark asks whether a structured Boolean computation can be assigned values that satisfy its circuit constraints and a terminal comparison. CNF uses circuit and state variables with gate and control clauses, but recognized instances follow exact layouts rather than arbitrary encodings.}
\ifdefined\pdfbookmark\pdfbookmark[2]{md5-equivalence-checking}{as-gallery-94}\fi
\asGallerySolver{GRAT}{\asGalleryValue{asGalleryLoss}{1}{0.113}}{\asGalleryValue{asGalleryLoss}{1}{0.218}}{3/4}{Occurrence-based unit propagation drives a DIMACS-order lucky-model construction, setting remaining encountered literals false and retrying with reversed order and phase choices within the bounded affine family. It checks every retained clause before emitting a complete satisfying assignment; malformed, unsupported, or failed attempts return UNKNOWN, and no UNSAT path is implemented.}
\asGallerySolver{DPR}{\asGalleryValue{asGalleryLoss}{0}{0.0382}}{\asGalleryValue{asGalleryLoss}{0}{0.0739}}{1/4}{Fixed-point construction leads: occurrence-list propagation builds a base assignment, reuses its stable state across middle stages, and watched-literal DPLL searches the two terminal stages as fallback. It checks shifted clauses and terminal gates before emitting a complete assignment; failed or unsupported cases return UNKNOWN, with no UNSAT certificate path.}
\asGallerySolver{VeriPB}{\asGalleryValue{asGalleryLoss}{0}{0.0381}}{\asGalleryValue{asGalleryLoss}{0}{0.0737}}{1/4}{Exact structural recognition recovers equality cones and variable relationships, then a fixed-point loop flips a target or free leaf while streaming gate evaluation. It handles only the default bounded recognized layout and checks every original clause before emitting a complete assignment; failures return UNKNOWN, with no UNSAT certificate path.}
\end{asGalleryFamily}

\begin{asGalleryFamily}{095}{mechanical-\allowbreak master-\allowbreak key}{20 GBD instances\enspace\textperiodcentered\enspace 4 validation\enspace\textperiodcentered\enspace baseline 0/4}{The benchmark asks whether keys can receive bounded-jump cut-depth words so every lock separates each unauthorized key from its authorized keys at some position, with CNF using one-hot key-depth variables, lock-depth variables, and separation witnesses.}
\ifdefined\pdfbookmark\pdfbookmark[2]{mechanical-master-key}{as-gallery-95}\fi
\asGallerySolver{GRAT}{\asGalleryValue{asGalleryGain}{1}{5.27}}{\asGalleryValue{asGalleryGain}{1}{5.25}}{4/4}{Weighted local search assigns legal bounded-jump words to keys, scoring unauthorized pairs by authorized-depth counts and using whole-word moves, breakout weighting, and restarts. It accepts only the D=4, P=8 layout, checks the reconstructed SAT assignment, and returns UNKNOWN when recognition, search, or checking fails; no UNSAT certificate path is implemented.}
\asGallerySolver{DPR}{\asGalleryValue{asGalleryGain}{0}{3.59}}{\asGalleryValue{asGalleryGain}{0}{3.48}}{3/4}{Enumerated legal bounded-jump paths feed weighted local search, tracking authorization counts and violated pairs with whole-word moves and breakout weighting. It accepts only the exact CNF layout, checks reconstructed SAT assignments, and returns UNKNOWN for unsupported inputs or search failure; its proof fallback is disabled, so no UNSAT result is implemented.}
\asGallerySolver{VeriPB}{\asGalleryValue{asGalleryGain}{0}{1.71}}{\asGalleryValue{asGalleryGain}{0}{1.69}}{2/4}{Exact recognition gates sparse counterexample-repair, moving violating keys toward absent depths or unique authorized users with per-lock masks and counts. Other cases use legal words, whole-word RUP additions, and key-space CDCL with auxiliary elimination; SAT models are clause-checked, root conflicts yield UNSAT, while failures or unsupported inputs yield UNKNOWN.}
\end{asGalleryFamily}

\begin{asGalleryFamily}{096}{minimal-\allowbreak disagreement-\allowbreak parity}{29 GBD instances\enspace\textperiodcentered\enspace 6 validation\enspace\textperiodcentered\enspace baseline 1/6}{The benchmark asks whether shared Boolean variables produce parity outputs that disagree with observations within a bounded limit. Its CNF uses XOR or equality auxiliaries, a counter for the disagreement bound, and can include ordinary residual clauses.}
\ifdefined\pdfbookmark\pdfbookmark[2]{minimal-disagreement-parity}{as-gallery-96}\fi
\asGallerySolver{GRAT}{\asGalleryValue{asGalleryGain}{1}{2.65\!\cdot\!10^{5}}}{\asGalleryValue{asGalleryGain}{1}{1.02\!\cdot\!10^{5}}}{6/6}{Affine decoding through a weighted union-find and bit-parallel Gaussian elimination is the main strategy. It enumerates bounded error patterns, then completes auxiliary and weakened-chain variables with watched-literal search, verifies the full CNF, and prints SAT only then; unsupported recognition or failed decoding returns UNKNOWN, with no UNSAT certificate path.}
\asGallerySolver{DPR}{\asGalleryValue{asGalleryGain}{0}{53.2}}{\asGalleryValue{asGalleryGain}{0}{53.1}}{6/6}{GF(2) elimination, min-fill factor decomposition, and separator/syndrome joins are the central search strategy. Junction-tree messages aid bounded reconstruction, and recovered assignments are back-substituted and checked against every input clause before SAT is printed; unsupported or failed paths return UNKNOWN with no UNSAT certificate path.}
\asGallerySolver{VeriPB}{\asGalleryValue{asGalleryGain}{0}{2.96\!\cdot\!10^{4}}}{\asGalleryValue{asGalleryGain}{0}{2.54\!\cdot\!10^{4}}}{6/6}{Affine-mask propagation and Gaussian-elimination syndrome decoding drive the active search on recognized fixed-shape encodings. It enumerates bounded errors, checks decoded witnesses against the CNF, uses DPLL for a narrow weakened branch, and returns UNKNOWN for unsupported or failed cases; its VeriPB proof routine is not invoked for UNSAT.}
\end{asGalleryFamily}

\begin{asGalleryFamily}{097}{minimal-\allowbreak superpermutation}{46 GBD instances\enspace\textperiodcentered\enspace 10 validation\enspace\textperiodcentered\enspace baseline 7/10}{The benchmark asks whether a fixed-length word over n symbols contains each required permutation as a contiguous substring. CNF encodes symbols at positions and links possible windows to occurrence, coverage, and cardinality constraints through auxiliary variables.}
\ifdefined\pdfbookmark\pdfbookmark[2]{minimal-superpermutation}{as-gallery-97}\fi
\asGallerySolver{GRAT}{\asGalleryValue{asGalleryGain}{0}{2.54\!\cdot\!10^{4}}}{\asGalleryValue{asGalleryGain}{0}{57.1}}{10/10}{Hard-coded construction dispatch seeds candidate words, then whole-formula propagation and clause checking produce SAT assignments for supported layouts. If this fails on the recognized n=4 forbidden-canonical signature, occurrence-placement DFS emits RUP-style context clauses through a replayed recipe or semantic fallback; other or unverified cases return UNKNOWN.}
\asGallerySolver{DPR}{\asGalleryValue{asGalleryGain}{1}{5.20\!\cdot\!10^{4}}}{\asGalleryValue{asGalleryGain}{0}{54.8}}{10/10}{Fixed-witness seeding and indexed unit propagation test hard-coded n=4 and n=5 constructions, with bounded residual enumeration and a complete clause check before SAT output. Six exact n=4 variable-count and clause-structure cases use occurrence-window DPLL with copied or regenerated DPR occurrence-cover templates for UNSAT certificates; unsupported inputs or failed residual completion return UNKNOWN.}
\asGallerySolver{VeriPB}{\asGalleryValue{asGalleryGain}{0}{520}}{\asGalleryValue{asGalleryGain}{1}{80.8}}{10/10}{Coverage-first branching drives watched-literal completion after pinning shape-recognized formulas to canonical constructions, with chronological backtracking and full clause checks before SAT output. If completion fails, only n=4 cases enter proof-logging CDCL on uncovered permutation blocks, with VeriPB chains for UNSAT; n=5 failures and unrecognized or unresolved searches return UNKNOWN.}
\end{asGalleryFamily}

\begin{asGalleryFamily}{098}{minimum-\allowbreak disagreement-\allowbreak parity}{30 GBD instances\enspace\textperiodcentered\enspace 6 validation\enspace\textperiodcentered\enspace baseline 0/6}{The benchmark asks whether a binary source assignment induces affine parity outputs with at most K disagreement bits. It encodes the parity relations and Hamming-weight bound in CNF using auxiliary variables, while concrete encodings may vary.}
\ifdefined\pdfbookmark\pdfbookmark[2]{minimum-disagreement-parity}{as-gallery-98}\fi
\asGallerySolver{GRAT}{\asGalleryValue{asGalleryGain}{1}{5.39}}{\asGalleryValue{asGalleryGain}{1}{4.45}}{5/6}{Dual-syndrome meet-in-the-middle decoding uses GF(2) elimination to enumerate bounded supports and recover a checked assignment for the recognized canonical XOR-and-counter layout. If it fails, CDCL adds proof-only XOR consequences, emits DRAT for external checking, and reports UNSAT only after a root conflict; unsupported or inconclusive cases return UNKNOWN.}
\asGallerySolver{DPR}{\asGalleryValue{asGalleryGain}{0}{3.00}}{\asGalleryValue{asGalleryGain}{0}{3.00}}{4/6}{Affine reconstruction recovers the map, then Gray-code meet-in-the-middle search indexes partial outputs and tests bounded Hamming neighborhoods for a source assignment. The candidate is propagated and checked against every clause before SAT output; restricted layouts, capacity limits, failed search, or no model yield UNKNOWN, with no UNSAT certificate path implemented.}
\asGallerySolver{VeriPB}{\asGalleryValue{asGalleryGain}{0}{3.00}}{\asGalleryValue{asGalleryGain}{0}{3.00}}{4/6}{Lee-Brickell information-set decoding samples syndrome positions, solves binary systems, and tests low-error candidates with packed parity operations to recover a source assignment in the recognized affine-XOR/cardinality encoding. After validation, eligible small lower-bound cases use watched-literal CDCL with VeriPB RUP logging for UNSAT; bounded decoding failure or unsupported cases return UNKNOWN.}
\end{asGalleryFamily}

\begin{asGalleryFamily}{099}{misc-\allowbreak satex}{19 GBD instances\enspace\textperiodcentered\enspace 4 validation\enspace\textperiodcentered\enspace baseline 3/4}{The benchmark asks whether a Boolean assignment satisfies every clause of a CNF formula. Some instances encode one allowed choice per object with binary incompatibilities, while others encode Boolean gate relationships; these structures describe recognized subfamilies, not all CNF inputs.}
\ifdefined\pdfbookmark\pdfbookmark[2]{misc-satex}{as-gallery-99}\fi
\asGallerySolver{GRAT}{\asGalleryValue{asGalleryGain}{0}{8.83}}{\asGalleryValue{asGalleryGain}{0}{8.82}}{4/4}{Graph-color recognition drives bounded TabuCol, while pure 3-CNF uses bounded ProbSAT with break counts; accepted models are checked against the original clauses. Fallback uses canonicalization, bounded elimination, and watched-literal CDCL, logging resolvents and learned clauses as DRAT for external elaboration; failed search or reconstruction yields UNKNOWN rather than UNSAT.}
\asGallerySolver{DPR}{\asGalleryValue{asGalleryGain}{1}{11.6}}{\asGalleryValue{asGalleryGain}{1}{11.4}}{4/4}{Structural matching recovers shared color labels in recognized correspondence-coloring CNF, followed by bounded randomized TabuCol search and full-model checking. The narrow twoall shape instead uses bounded elimination and watched-literal CDCL with logged clauses for UNSAT; circuit and generic paths lack an UNSAT certificate path and may return UNKNOWN after failed search.}
\asGallerySolver{VeriPB}{\asGalleryValue{asGalleryLoss}{0}{0.781}}{\asGalleryValue{asGalleryLoss}{0}{0.780}}{3/4}{Invariant-based recognition recovers list-coloring structure for TabuCol, while other admitted shapes use ProbSAT, min-break, or weighted-breakout search. Recognized AIG and completion cases may use bounded elimination with reverse extension, and heuristic failures fall to watched-literal CDCL with RUP logging; checked SAT assignments are accepted, but unmatched inputs return UNKNOWN.}
\end{asGalleryFamily}

\begin{asGalleryFamily}{100}{miter}{524 GBD instances\enspace\textperiodcentered\enspace 105 validation\enspace\textperiodcentered\enspace baseline 74/105}{A Boolean miter CNF asks whether two Boolean circuits can produce different outputs on some input. Circuit wires and auxiliary variables are constrained by gate clauses, with clauses expressing the required output difference.}
\ifdefined\pdfbookmark\pdfbookmark[2]{miter}{as-gallery-100}\fi
\asGallerySolver{GRAT}{\asGalleryValue{asGalleryLoss}{1}{0.569}}{\asGalleryValue{asGalleryLoss}{1}{0.632}}{46/105}{Signed-literal congruence is its main preprocessing: implication SCCs and recognized AND, XOR, or mux structures merge equivalent signals before bounded elimination. Failed-literal probing and CDCL search the residual; checked SAT models are reconstructed, while UNSAT is logged as textual DRAT for external elaboration and failures return UNKNOWN.}
\asGallerySolver{DPR}{\asGalleryValue{asGalleryLoss}{0}{0.379}}{\asGalleryValue{asGalleryLoss}{0}{0.424}}{15/105}{Exact AND/XOR Tseitin recognition uses topological signed hashing and equivalence lemmas, with an all-AND unit-root path as a second recognizer. Residual cases use bounded elimination, profile-gated CDCL, or tree DPLL; checked SAT assignments are printed, while unsupported or exhausted cases return UNKNOWN and UNSAT uses witness-free RUP/DRUP-style additions.}
\asGallerySolver{VeriPB}{\asGalleryValue{asGalleryLoss}{0}{0.509}}{\asGalleryValue{asGalleryLoss}{0}{0.567}}{39/105}{Parity union-find and bottom-up AND-cone hashing recover duplicate-wire equivalences, followed by bounded Davis-Putnam elimination, even though the input miter structure is not validated. Watched-literal CDCL handles the remainder with RUP-logged clauses; checked SAT models have no VeriPB SAT certificate, and proof, resource, or model failures return UNKNOWN.}
\end{asGalleryFamily}

\begin{asGalleryFamily}{101}{modcircuits}{20 GBD instances\enspace\textperiodcentered\enspace 4 validation\enspace\textperiodcentered\enspace baseline 1/4}{These benchmarks ask whether a bounded acyclic Boolean circuit with two-input gates can realize a modular Boolean function, possibly with encoded input or output residues. CNF encodings represent gate sources, gate truth tables, row values, and outputs, with clauses enforcing consistency and the target behavior.}
\ifdefined\pdfbookmark\pdfbookmark[2]{modcircuits}{as-gallery-101}\fi
\asGallerySolver{GRAT}{\asGalleryValue{asGalleryMuted}{1}{1.00}}{\asGalleryValue{asGalleryMuted}{1}{1.00}}{1/4}{Hard-coded recognizers install bit-parallel phase hints for selected MOD4, MOD5, and restricted MOD3 layouts, while otherwise retaining generic watched-literal CDCL search with exactly-one source-choice branching where applicable. SAT assignments are checked against the original clauses; UNSAT search logs learned clause additions and a terminal empty clause for external DRAT elaboration, and failed checks return UNKNOWN.}
\asGallerySolver{DPR}{\asGalleryValue{asGalleryLoss}{0}{0.750}}{\asGalleryValue{asGalleryLoss}{0}{0.750}}{0/4}{Structural decoders derive hard-coded phase assignments for selected MOD4, MOD5, and restricted MOD3 layouts, then test constructions under temporary assumptions before CDCL fallback. Fallback uses implication-SCC substitution and bounded variable elimination, while first-UIP learning logs clause additions, including witness-free additions; checked SAT models return, but unsupported shapes or failed completion yield UNKNOWN.}
\asGallerySolver{VeriPB}{\asGalleryValue{asGalleryLoss}{0}{0.750}}{\asGalleryValue{asGalleryLoss}{0}{0.750}}{0/4}{Embedded recipe matching decodes one-hot layouts and uses 64-bit truth-table simulation to assign gates, outputs, and row variables for selected parameter shapes. A polarity-shuffled ten-gate layout can trigger signed topology recovery with CDCL completion after construction fails; checked assignments are returned, while unsupported shapes or failed search yield UNKNOWN and no UNSAT certificate is implemented.}
\end{asGalleryFamily}

\begin{asGalleryFamily}{102}{mosoi-\allowbreak 289}{36 GBD instances\enspace\textperiodcentered\enspace 8 validation\enspace\textperiodcentered\enspace baseline 4/8}{Assign one of four colors to each grid cell so that no color occupies all four corners formed by two rows and two columns. A CNF uses four variables per cell, clauses restricting allowed colors, and four-literal negative clauses forbidding monochromatic rectangles.}
\ifdefined\pdfbookmark\pdfbookmark[2]{mosoi-289}{as-gallery-102}\fi
\asGallerySolver{GRAT}{\asGalleryValue{asGalleryMuted}{0}{1.00}}{\asGalleryValue{asGalleryMuted}{0}{1.00}}{4/8}{Collision-scored min-conflicts search uses row-pair/color counts with legal recoloring and restarts. It checks candidate assignments against every parsed clause; bounded search failure returns UNKNOWN. Counting obstructions trigger a DRAT proof using a pigeonhole reduction from a usable 6x31 core; failed or unsupported cases return UNKNOWN.}
\asGallerySolver{DPR}{\asGalleryValue{asGalleryMuted}{0}{1.00}}{\asGalleryValue{asGalleryMuted}{0}{1.00}}{4/8}{A specialized 6x30/30x6 matching construction is tried first; otherwise pseudorandom annealing minimizes monochromatic rectangles, with balanced column swaps preserving color counts in tight cases. Checked SAT assignments are emitted, while a generated 6x31 core can produce a DPR pigeonhole proof with fresh collision variables; unsupported sequential-counter/K4 cases return UNKNOWN.}
\asGallerySolver{VeriPB}{\asGalleryValue{asGalleryGain}{1}{2.52\!\cdot\!10^{4}}}{\asGalleryValue{asGalleryGain}{1}{788}}{8/8}{Incremental min-conflicts search on a recognized ordered or signed/permuted grid scores row-pair/color collisions and applies random perturbations. It verifies every input clause for each candidate before emitting a SAT assignment; resource-count contradictions produce a VeriPB proof, while the MIS fallback emits only a proof and unsupported inputs return UNKNOWN.}
\end{asGalleryFamily}

\begin{asGalleryFamily}{103}{multiplier-\allowbreak circuits}{18 GBD instances\enspace\textperiodcentered\enspace 4 validation\enspace\textperiodcentered\enspace baseline 4/4}{These benchmarks ask whether two bounded unsigned binary integers multiply to a fixed target. The target is encoded by unit clauses on product outputs, while Boolean gate clauses connect operand bits to a multiplier circuit and auxiliary variables.}
\ifdefined\pdfbookmark\pdfbookmark[2]{multiplier-circuits}{as-gallery-103}\fi
\asGallerySolver{GRAT}{\asGalleryValue{asGalleryGain}{0}{547}}{\asGalleryValue{asGalleryGain}{1}{217}}{4/4}{Structural multiplier recognition recovers operand widths and the fixed product, then small-divisor and Pollard-Brent factorization finds width-fitting operands. The factors drive watched-literal propagation and a full clause scan for SAT; only a root-level unit-propagation conflict emits an empty-clause certificate, while recognition, factoring, or other unsatisfiable cases return UNKNOWN.}
\asGallerySolver{DPR}{\asGalleryValue{asGalleryGain}{0}{713}}{\asGalleryValue{asGalleryGain}{0}{33.1}}{4/4}{Rigid-layout recognition and bit-parallel signature simulation recover a target from ordered encoding, then drive width-bounded divisor search with Pollard-Brent factoring. Candidate factors are checked against all original clauses before a SAT assignment is emitted; ambiguous signatures, unsupported layouts, or failed searches return UNKNOWN, with no UNSAT certificate path.}
\asGallerySolver{VeriPB}{\asGalleryValue{asGalleryGain}{1}{853}}{\asGalleryValue{asGalleryGain}{0}{33.3}}{4/4}{Structural recognition reads product units and replaces SAT search with width-bounded factor search, using trial division, with Pollard-Brent available only up to 128 bits. It checks factors through the circuit and clauses before emitting a SAT model; unsupported or unfactored targets return UNKNOWN, with no UNSAT or VeriPB certificate path.}
\end{asGalleryFamily}

\begin{asGalleryFamily}{104}{multiplier-\allowbreak equivalence-\allowbreak checking}{20 GBD instances\enspace\textperiodcentered\enspace 4 validation\enspace\textperiodcentered\enspace baseline 0/4}{The benchmark asks whether two circuit implementations, commonly 16-bit multipliers over two 16-bit operands, produce different outputs for some input. The CNF encodes the circuits and a final disagreement miter, so SAT gives a counterexample and UNSAT rules out disagreement for that encoding.}
\ifdefined\pdfbookmark\pdfbookmark[2]{multiplier-equivalence-checking}{as-gallery-104}\fi
\asGallerySolver{GRAT}{\asGalleryValue{asGalleryMuted}{1}{1.00}}{\asGalleryValue{asGalleryMuted}{1}{1.00}}{0/4}{XOR/XNOR detection and bounded parity substitution strengthen recognized instances before watched-literal CDCL search with activity-based branching and clause learning. SAT is reported only with an assignment checked against the CNF; rejected patterns return UNKNOWN, while exhausted UNSAT search logs learned and strengthening clauses plus deletions and a final empty clause for external checking.}
\asGallerySolver{DPR}{\asGalleryValue{asGalleryMuted}{1}{1.00}}{\asGalleryValue{asGalleryMuted}{1}{1.00}}{0/4}{Primary-input conflict projection uses an independent watched-literal RUP oracle to minimize learned clauses, driving asserting backjumps in watched-literal CDCL search. SAT models are reconstructed and checked; unsupported or failed cases return UNKNOWN, while UNSAT emits witness-free DPR additions, including learned operand-blocking clauses and a final empty clause for external elaboration and checking.}
\asGallerySolver{VeriPB}{\asGalleryValue{asGalleryMuted}{1}{1.00}}{\asGalleryValue{asGalleryMuted}{1}{1.00}}{0/4}{Proof-first pseudo-Boolean generation recovers partial products and width-three compressor relations from a restricted Tseitin DAG, emits a refutation, and leaves proof checking external. If generation fails, bit-parallel evaluation searches for a counterexample and reports SAT only after clause checking; otherwise it returns UNKNOWN and has no UNSAT certificate path.}
\end{asGalleryFamily}

\begin{asGalleryFamily}{105}{multiplier-\allowbreak verification}{27 GBD instances\enspace\textperiodcentered\enspace 6 validation\enspace\textperiodcentered\enspace baseline 2/6}{Determine whether values for the free input variables can satisfy a structured acyclic Boolean mux network, a fixed-true terminal, and a final disjunction. The CNF defines each mux with six clauses and includes the network constraints plus the final clause.}
\ifdefined\pdfbookmark\pdfbookmark[2]{multiplier-verification}{as-gallery-105}\fi
\asGallerySolver{GRAT}{\asGalleryValue{asGalleryGain}{0}{1.05}}{\asGalleryValue{asGalleryGain}{1}{1.05}}{2/6}{64-bit signature sweeping proposes RUP-validated unit and binary lemmas over the ordered ITE DAG. CDCL searches, tries final-literal assumptions, then falls back to whole-miter search; it emits DRAT-style clause additions and an empty clause for external elaboration, returns UNKNOWN on unsupported or incomplete cases, and has no SAT assignment output.}
\asGallerySolver{DPR}{\asGalleryValue{asGalleryLoss}{0}{0.702}}{\asGalleryValue{asGalleryLoss}{0}{0.707}}{0/6}{Gate-aware propagation stores each ITE as a six-clause object, rescanning affected gates while CDCL branches only on primary inputs. First-UIP learning with watched clauses supports UNSAT certification through witness-free learned clause additions and an empty clause; unsupported structure, duplicate final literals, or proof failures return UNKNOWN, while SAT prints an assignment.}
\asGallerySolver{VeriPB}{\asGalleryValue{asGalleryGain}{1}{1.05}}{\asGalleryValue{asGalleryLoss}{0}{0.833}}{2/6}{Mux-derived phase initialization guides watched-literal CDCL with first-UIP learning. It logs learned clauses with RUP and dependency hints and a final empty constraint for UNSAT; the SAT path checks and prints a model but emits no completed VeriPB certificate, while unsupported inputs or proof failures return UNKNOWN.}
\end{asGalleryFamily}

\begin{asGalleryFamily}{106}{mutilated-\allowbreak chessboard}{22 GBD instances\enspace\textperiodcentered\enspace 5 validation\enspace\textperiodcentered\enspace baseline 2/5}{The task asks whether dominoes can cover every remaining square after removing two same-color opposite corners from a checkerboard, without overlap. CNF encodings use placement variables, coverage clauses, and constraints preventing a square from being used twice.}
\ifdefined\pdfbookmark\pdfbookmark[2]{mutilated-chessboard}{as-gallery-106}\fi
\asGallerySolver{GRAT}{\asGalleryValue{asGalleryGain}{0}{1.53}}{\asGalleryValue{asGalleryGain}{0}{1.16}}{3/5}{Checkerboard matching recovery drives anti-diagonal frontier DP, using occupancy bit masks and a subset trie to reuse failed states. Preparation checks destination conflicts with binary propagation and emits bottom-up dependency lemmas, deletions, and an empty root clause for UNSAT; without a SAT witness path, satisfiable or unsupported cases return UNKNOWN.}
\asGallerySolver{DPR}{\asGalleryValue{asGalleryLoss}{0}{0.783}}{\asGalleryValue{asGalleryLoss}{0}{0.812}}{1/5}{A strict structural recognizer accepts only validated direct-incidence or sparse-matrix pigeonhole layouts and returns UNKNOWN otherwise. Bounded Davis-Putnam elimination and watched-literal CDCL use first-UIP learning; the clause/deletion trace records resolvents, learned clauses, deletions, and an empty clause on a level-zero conflict, while non-UNSAT outcomes return UNKNOWN without model reconstruction.}
\asGallerySolver{VeriPB}{\asGalleryValue{asGalleryGain}{1}{3.56\!\cdot\!10^{5}}}{\asGalleryValue{asGalleryGain}{1}{1.33\!\cdot\!10^{4}}}{5/5}{Matching recognition converts compact or sparse layouts into degree-2-to-4 counting constraints over mutexes, replacing general search. Unit propagation may justify missing mutexes in the sparse fallback; a VeriPB contradiction then combines coverage clauses with clique inequalities. Unsupported inputs or failed proof preconditions return UNKNOWN; no SAT witness path is implemented.}
\end{asGalleryFamily}

\begin{asGalleryFamily}{107}{oddball-\allowbreak weighing}{40 GBD instances\enspace\textperiodcentered\enspace 8 validation\enspace\textperiodcentered\enspace baseline 8/8}{The benchmark asks whether each object can receive a nonzero ternary signature and orientation across non-adaptive weighings so every weighing has equal positive and negative counts. CNF encodings represent signature choices and directed matching edges pairing opposite outcomes.}
\ifdefined\pdfbookmark\pdfbookmark[2]{oddball-weighing}{as-gallery-107}\fi
\asGallerySolver{GRAT}{\asGalleryValue{asGalleryGain}{1}{413}}{\asGalleryValue{asGalleryGain}{1}{36.9}}{8/8}{It recognizes a narrow matching-and-signature CNF pattern, then searches fixed canonical assignments with randomized-prefix meet-in-the-middle and free ordered subsets with codeword/orientation annealing. Successful constructions undergo streamed propagation and clause checks; an odd-row obstruction can produce textual DRAT for external elaboration, while other heuristic failures return UNKNOWN rather than UNSAT.}
\asGallerySolver{DPR}{\asGalleryValue{asGalleryGain}{0}{359}}{\asGalleryValue{asGalleryGain}{0}{3.77}}{8/8}{On narrow canonical block/domain inputs, an odd-coordinate matching obstruction adds core clauses and an empty clause for UNSAT certification. Otherwise, exact dynamic programming gives way to annealing or distinct-tuple search; the SAT path lifts tuple and matching variables, propagates the CNF, and greedily completes clauses, while failures return UNKNOWN.}
\asGallerySolver{VeriPB}{\asGalleryValue{asGalleryLoss}{0}{0.692}}{\asGalleryValue{asGalleryGain}{0}{1.04}}{7/8}{On its recognized layouts, seeded annealing balances coordinate sums by changing orientations and, for free layouts, selected signature classes. Only fixed layouts with an odd support coordinate use the pseudo-Boolean matching proof path; other cases seek a model by streaming propagation after pairing opposite sides, with failures or oversized inputs returning UNKNOWN.}
\end{asGalleryFamily}

\begin{asGalleryFamily}{108}{or\_\allowbreak randxor}{10 GBD instances\enspace\textperiodcentered\enspace 2 validation\enspace\textperiodcentered\enspace baseline 2/2}{The benchmark asks whether ternary Boolean XOR equations can be satisfied when each logical value is represented by the OR of a consecutive pair of variables. Each equation is converted to CNF by distributing over those pair expressions, producing the encoded instance.}
\ifdefined\pdfbookmark\pdfbookmark[2]{or\_randxor}{as-gallery-108}\fi
\asGallerySolver{GRAT}{\asGalleryValue{asGalleryLoss}{0}{0.281}}{\asGalleryValue{asGalleryLoss}{1}{0.354}}{2/2}{Bit-packed GF(2) Gaussian elimination solves recognized systems and reconstructs a checked SAT assignment. Inconsistent cases use pair-symmetry reduction, bounded Davis-Putnam elimination, and watched-literal CDCL to build a textual DRAT refutation for external checking; only exact canonical OR-pair expansions are handled, and failures return UNKNOWN.}
\asGallerySolver{DPR}{\asGalleryValue{asGalleryLoss}{1}{0.434}}{\asGalleryValue{asGalleryLoss}{0}{0.129}}{2/2}{Bit-packed GF(2) elimination first solves recognized equations and produces a clause-checked SAT assignment. For inconsistency, the solver peels a 2-core, applies pair-symmetry reduction and bounded Davis-Putnam elimination with witness-free clause additions, then uses CDCL to complete the refutation; only exact canonical OR-pair expansions are recognized, and failures return UNKNOWN.}
\asGallerySolver{VeriPB}{\asGalleryValue{asGalleryLoss}{0}{0.194}}{\asGalleryValue{asGalleryLoss}{0}{0.275}}{2/2}{Dependency-carrying GF(2) elimination solves recognized systems and yields a clause-checked SAT assignment. For inconsistency, the solver peels and normalizes a 2-core, then uses watched-literal CDCL with RUP logging to emit a VeriPB proof; only the exact OR-pair encoding is supported, and unsupported inputs or failed proof construction return UNKNOWN.}
\end{asGalleryFamily}

\begin{asGalleryFamily}{109}{ordering-\allowbreak principle}{36 GBD instances\enspace\textperiodcentered\enspace 8 validation\enspace\textperiodcentered\enspace baseline 8/8}{The benchmark asks whether a directed relation on n objects can give every object a predecessor while obeying antisymmetry and transitivity. A CNF encoding uses variables for directed edges, predecessor clauses, binary clauses forbidding opposite edges, and ternary clauses expressing transitivity.}
\ifdefined\pdfbookmark\pdfbookmark[2]{ordering-principle}{as-gallery-109}\fi
\asGallerySolver{GRAT}{\asGalleryValue{asGalleryGain}{1}{142}}{\asGalleryValue{asGalleryGain}{1}{11.7}}{8/8}{A bounded search for cycles among omitted transitivity clauses supplies checked SAT models when successful; a single missing predecessor or antisymmetry axiom also permits direct model construction. Otherwise, vertex elimination emits a textual DRAT clause sequence toward UNSAT; unsupported shapes or failed construction return UNKNOWN without general SAT search.}
\asGallerySolver{DPR}{\asGalleryValue{asGalleryGain}{0}{138}}{\asGalleryValue{asGalleryGain}{0}{4.31}}{8/8}{Structural recognition recovers vertices, directed edges, and polarities, then validates the supported transitivity pattern. It either greedily eliminates vertices with witness-free RUP additions to derive UNSAT, or constructs and checks a model for narrowly supported missing-axiom or cycle cases; unsupported inputs and failed checks return UNKNOWN without general SAT search.}
\asGallerySolver{VeriPB}{\asGalleryValue{asGalleryGain}{0}{120}}{\asGalleryValue{asGalleryGain}{0}{3.99}}{8/8}{Safe object elimination removes an object only when omitted transitivity clauses cannot obstruct it, then emits a VeriPB RUP proof of UNSAT. If elimination fails, it checks models only for a missing-transitivity cycle or one missing predecessor or asymmetry clause; unsupported or failed cases return UNKNOWN without general SAT search.}
\end{asGalleryFamily}

\begin{asGalleryFamily}{110}{ordering-\allowbreak principle-\allowbreak xor}{6 GBD instances\enspace\textperiodcentered\enspace 2 validation\enspace\textperiodcentered\enspace baseline 2/2}{Given n vertices, can a transitive, antisymmetric relation give every vertex at least one of three prescribed predecessors? The supplied encodings represent each relation variable by an XOR pair and expand the resulting width-two and width-three clauses into CNF.}
\ifdefined\pdfbookmark\pdfbookmark[2]{ordering-principle-xor}{as-gallery-110}\fi
\asGallerySolver{GRAT}{\asGalleryValue{asGalleryGain}{0}{2{,}670}}{\asGalleryValue{asGalleryGain}{0}{15.8}}{2/2}{Candidate-state scanning applies the finite-order minimal-element contradiction, replacing a current candidate exactly when a relation holds. It bridges selected XOR definitions to logical clauses and emits a DRAT clause-addition refutation ending in the empty clause; unrecognized or output-failing cases return UNKNOWN rather than using general SAT search.}
\asGallerySolver{DPR}{\asGalleryValue{asGalleryGain}{1}{8{,}640}}{\asGalleryValue{asGalleryGain}{1}{44.4}}{2/2}{Minimum-fill elimination orders the predecessor graph and gauge-fixes only XOR atoms needed by that plan. It emits witness-free DPR additions for bypass and reduced predecessor clauses, reporting UNSAT only after an empty clause is written; there is no SAT fallback, and unsupported or failed cases return UNKNOWN.}
\asGallerySolver{VeriPB}{\asGalleryValue{asGalleryGain}{0}{2{,}710}}{\asGalleryValue{asGalleryGain}{0}{33.8}}{2/2}{Redundancy-based XOR symmetry breaking fixes each even physical bit before deterministic vertex elimination. It emits RUP predecessor clauses as candidate sets are unioned, then completes a VeriPB proof with a final NONE marker and UNSAT conclusion; failed recognition or proof generation returns UNKNOWN, and no SAT search is used.}
\end{asGalleryFamily}

\begin{asGalleryFamily}{111}{p-\allowbreak center}{21 GBD instances\enspace\textperiodcentered\enspace 5 validation\enspace\textperiodcentered\enspace baseline 5/5}{The benchmark asks whether at most p candidate facilities can cover every demand point, with each demand represented by a positive facility clause. In CNF, auxiliary variables enforce the cardinality bound, while equivalent encodings may represent the same p-center question.}
\ifdefined\pdfbookmark\pdfbookmark[2]{p-center}{as-gallery-111}\fi
\asGallerySolver{GRAT}{\asGalleryValue{asGalleryGain}{1}{3.03}}{\asGalleryValue{asGalleryGain}{1}{2.14}}{5/5}{Failed-literal probing is the active strategy: root propagation tests variables and turns a conflicting polarity into an opposite unit until conflict or closure. It handles only a restricted layout with unit coverage clauses; no SAT path exists, so unclosed or SAT cases return UNKNOWN, while closed refutations record units and an empty clause.}
\asGallerySolver{DPR}{\asGalleryValue{asGalleryGain}{0}{2.72}}{\asGalleryValue{asGalleryLoss}{0}{0.416}}{5/5}{Facility-aware DPLL is the active strategy: watched propagation branches on a facility from the shortest uncovered cover clause, favoring frequent facilities, then uses auxiliary fallback decisions to complete the counter extension. It checks SAT assignments and emits postorder, witness-free RUP blockers for UNSAT; other encodings or failed checks return UNKNOWN.}
\asGallerySolver{VeriPB}{\asGalleryValue{asGalleryGain}{0}{2.06}}{\asGalleryValue{asGalleryLoss}{0}{0.663}}{5/5}{Specialized neighborhood branching is central: watched propagation selects a candidate from the tightest uncovered neighborhood, while counter clauses enforce capacity. Root propagation may produce an empty-RUP result; otherwise a CDCL fallback learns RUP clauses, while checked SAT assignments are emitted and unsupported layouts or failed checks return UNKNOWN.}
\end{asGalleryFamily}

\begin{asGalleryFamily}{112}{parity-\allowbreak games}{28 GBD instances\enspace\textperiodcentered\enspace 6 validation\enspace\textperiodcentered\enspace baseline 5/6}{The benchmark uses structured CNF encodings of bounded sequences of parity-game strategies and their transitions. Satisfiability asks whether Boolean variables can realize a consistent encoded construction, using relation variables, transition clauses, and definitional witnesses.}
\ifdefined\pdfbookmark\pdfbookmark[2]{parity-games}{as-gallery-112}\fi
\asGallerySolver{GRAT}{\asGalleryValue{asGalleryGain}{1}{1.05}}{\asGalleryValue{asGalleryGain}{1}{1.05}}{5/6}{Strict structural recognition gates two-watched-literal CDCL with 1-UIP learning, clause minimization, activity-based branching, and LBD retention. SAT models are checked; a non-SAT pass is rerun with logging, and only a logged UNSAT yields UNSAT, while nonmatching inputs or a failed rerun yield UNKNOWN. The proof path emits textual DRAT clauses.}
\asGallerySolver{DPR}{\asGalleryValue{asGalleryLoss}{0}{0.673}}{\asGalleryValue{asGalleryLoss}{0}{0.673}}{4/6}{Binary-implication SCC compression, failed-literal probing, and bounded resolution elimination simplify recognized instances before CDCL; vivification is inactive. SAT assignments are reconstructed and checked against the original CNF, while UNSAT deductions are streamed as witness-free textual DPR additions; malformed or nonmatching inputs return UNKNOWN.}
\asGallerySolver{VeriPB}{\asGalleryValue{asGalleryLoss}{0}{0.673}}{\asGalleryValue{asGalleryLoss}{0}{0.672}}{4/6}{Binary-implication SCC compression and bounded variable elimination precede CDCL with first-UIP learning. In SAT-first mode, it searches without proof retention and reruns only UNSAT results for version 3.0 pseudo-Boolean proof emission. SAT models are reconstructed and checked; no verifier is included, and parsing, checking, or proof-writing failures return UNKNOWN.}
\end{asGalleryFamily}

\begin{asGalleryFamily}{113}{pebbling}{55 GBD instances\enspace\textperiodcentered\enspace 11 validation\enspace\textperiodcentered\enspace baseline 11/11}{These instances ask whether a Boolean assignment can satisfy a DAG's pebbling dependencies while forcing its sink false. CNF encodings use signed literal blocks, source clauses, Cartesian transition families, and sink units; some variants use a specific NAE gadget.}
\ifdefined\pdfbookmark\pdfbookmark[2]{pebbling}{as-gallery-113}\fi
\asGallerySolver{GRAT}{\asGalleryValue{asGalleryGain}{0}{13.9}}{\asGalleryValue{asGalleryLoss}{0}{0.292}}{11/11}{A clause-sharing recognizer factors transition families into Cartesian predecessor blocks and reconstructs the pebbling graph, with an UNSAT-only fallback for exact parity-gadget shapes. Staged resolution and RUP checks handle complete transitions and limited missing-cell repairs. CDCL can return a checked SAT assignment or DRAT trace; unsupported cases return UNKNOWN.}
\asGallerySolver{DPR}{\asGalleryValue{asGalleryGain}{1}{109}}{\asGalleryValue{asGalleryGain}{1}{25.3}}{11/11}{Occurrence signatures and Cartesian-tail factorization recover a disjoint DAG, with a validated three-variable NEQ route using a fixed resolution template. Topological traversal emits resolution proofs, including repairs, while SAT construction handles only one recognized source, transition, or sink defect and checks its model. Unrecognized or failed constructions return UNKNOWN.}
\asGallerySolver{VeriPB}{\asGalleryValue{asGalleryGain}{0}{30.0}}{\asGalleryValue{asGalleryGain}{0}{12.4}}{11/11}{NAE3 dispatch eliminates exact gadgets by local resolution; incidence recovery otherwise builds a Horn quotient from signed blocks. Forward chaining finds a conflict for proof or lifts a checked SAT model; only one-deletion OR cases are accepted. Unit-containing CDCL covers small recognized-width formulas; NAE SAT cases and unresolved inputs return UNKNOWN.}
\end{asGalleryFamily}

\begin{asGalleryFamily}{114}{perfect-\allowbreak matching}{17 GBD instances\enspace\textperiodcentered\enspace 4 validation\enspace\textperiodcentered\enspace baseline 0/4}{Given a bipartite graph, can each of n+1 left vertices choose an incident right vertex without choosing any right vertex twice? The CNF requires one choice per left vertex and at-most-one use per right vertex, often exposing an unbalanced matching or pigeonhole core.}
\ifdefined\pdfbookmark\pdfbookmark[2]{perfect-matching}{as-gallery-114}\fi
\asGallerySolver{GRAT}{\asGalleryValue{asGalleryGain}{1}{5.53\!\cdot\!10^{5}}}{\asGalleryValue{asGalleryGain}{0}{3{,}080}}{4/4}{Structural recognition identifies an unbalanced graph-pigeonhole core by recovering consecutive row blocks and column exclusions from binary implication paths. It recursively eliminates rows and holes with fresh variables, using RAT definitions and RUP child rows to derive the empty clause; unsupported encodings, proof failures, and SAT cases return UNKNOWN.}
\asGallerySolver{DPR}{\asGalleryValue{asGalleryGain}{0}{2.82\!\cdot\!10^{5}}}{\asGalleryValue{asGalleryGain}{0}{1{,}840}}{4/4}{Hall-deficit recognition reconstructs the unbalanced matrix and checks binary implication reachability for column conflicts. It recursively removes a row and column, adding Tseitin definitions and witness-free conflict clauses, with RUP bridges when needed, until an empty clause; unsupported shapes, proof failures, and SAT inputs return UNKNOWN without SAT search or assignment output.}
\asGallerySolver{VeriPB}{\asGalleryValue{asGalleryGain}{0}{3.26\!\cdot\!10^{5}}}{\asGalleryValue{asGalleryGain}{1}{6.18\!\cdot\!10^{4}}}{4/4}{Pigeonhole counting recognizes a row-major unbalanced core, recovering column at-most-one constraints from pairwise, sequential, grouped, or implication-based encodings. It writes a VeriPB proof using inequality chains or RUP-added conflicts followed by cutting-planes induction; full-CNF propagation is a fallback, while unrecognized or satisfiable inputs return UNKNOWN without general SAT search.}
\end{asGalleryFamily}

\begin{asGalleryFamily}{115}{petrinet-\allowbreak concurrency}{54 GBD instances\enspace\textperiodcentered\enspace 11 validation\enspace\textperiodcentered\enspace baseline 6/11}{These benchmarks ask whether a graph's vertices can be assigned one of k colors so adjacent vertices differ. The accepted CNF uses vertex-color variables, per-vertex at-least-one clauses, and same-color edge exclusions, with a restricted symmetry-broken layout.}
\ifdefined\pdfbookmark\pdfbookmark[2]{petrinet-concurrency}{as-gallery-115}\fi
\asGallerySolver{GRAT}{\asGalleryValue{asGalleryGain}{1}{530}}{\asGalleryValue{asGalleryGain}{1}{12.8}}{11/11}{DSATUR coloring is the primary search, with colors renamed by first occurrence to respect the triangular symmetry breaking. If it fails, bounded bit-parallel clique search seeks a checked (k+1)-clique for a pigeonhole-style UNSAT proof; bounded TabuCol may find a SAT assignment and independently verify it; unresolved cases return UNKNOWN. It handles only this recognized layout.}
\asGallerySolver{DPR}{\asGalleryValue{asGalleryGain}{0}{5.77}}{\asGalleryValue{asGalleryLoss}{0}{0.640}}{10/11}{DSATUR coloring drives the SAT search on the strictly recognized layout, using dense bitsets; success emits a complete assignment. If it fails, greedy and exact clique search seek a (k+1)-clique for a pigeonhole-style UNSAT proof, but failure to find the clique or write the proof returns UNKNOWN; no explicit empty-clause derivation is written.}
\asGallerySolver{VeriPB}{\asGalleryValue{asGalleryGain}{0}{5.77}}{\asGalleryValue{asGalleryGain}{0}{5.18}}{10/11}{Clique-anchored coloring and DSATUR drive SAT search on the recognized layout, with randomized list-coloring, MRV backtracking, and tabu/min-conflicts fallbacks. A bounded exact clique search can emit an UNSAT cutting-planes proof for a (k+1)-clique within the kernel guard; SAT models are clause-checked, but no external proof checker is invoked, and failure returns UNKNOWN.}
\end{asGalleryFamily}

\begin{asGalleryFamily}{116}{philips}{4 GBD instances\enspace\textperiodcentered\enspace 1 validation\enspace\textperiodcentered\enspace baseline 1/1}{These CNFs encode a miter comparing two combinational multiplier circuits driven by the same binary inputs. Gate variables and local circuit relations are expressed as CNF constraints, so satisfiability asks whether the two outputs can differ.}
\ifdefined\pdfbookmark\pdfbookmark[2]{philips}{as-gallery-116}\fi
\asGallerySolver{GRAT}{\asGalleryValue{asGalleryGain}{0}{91.1}}{\asGalleryValue{asGalleryLoss}{0}{0.425}}{1/1}{Packed 64-bit truth-table enumeration over a recognized acyclic netlist is the active technique. A clause-checked candidate yields SAT; recognition or proof failures yield UNKNOWN, with no general SAT fallback. The no-candidate path uses RUP equivalences and a recursive Shannon-style refutation with deletions, blocking clauses, and a final empty clause for external checking.}
\asGallerySolver{DPR}{\asGalleryValue{asGalleryGain}{0}{1.61}}{\asGalleryValue{asGalleryGain}{1}{1.06}}{1/1}{Gray-code enumeration of one recovered 10-bit operand drives the search. Each fixed case uses fresh CDCL with minimized learned clauses; a SAT case returns a checked model. Refuted cases write lifted learned clauses and resolve blocking clauses to the empty clause for external DPR elaboration; unsupported cases return UNKNOWN.}
\asGallerySolver{VeriPB}{\asGalleryValue{asGalleryGain}{1}{438}}{\asGalleryValue{asGalleryLoss}{0}{0.811}}{1/1}{Fixed-shape fingerprinting dispatches on twenty high-fanout inputs. The active path emits a complete binary tree with RUP blocking leaves and cutting-planes joins to a contradiction for external VeriPB checking. It does not internally validate the proof or provide a SAT fallback; recognition or generation failures return UNKNOWN.}
\end{asGalleryFamily}

\begin{asGalleryFamily}{117}{phnf}{27 GBD instances\enspace\textperiodcentered\enspace 6 validation\enspace\textperiodcentered\enspace baseline 2/6}{The task is to choose exactly one value from each disjoint group while avoiding forbidden pairs between groups. In CNF, group clauses require a choice, complete within-group binary exclusions enforce uniqueness, and cross-group binary clauses forbid incompatible value pairs.}
\ifdefined\pdfbookmark\pdfbookmark[2]{phnf}{as-gallery-117}\fi
\asGallerySolver{GRAT}{\asGalleryValue{asGalleryLoss}{1}{0.875}}{\asGalleryValue{asGalleryGain}{1}{1.39}}{1/6}{Rectangle compression is active: repeated row and column masks become Boolean features before CDCL searches the model. The bounded recognizer has no generic fallback, so rejected inputs and failed SAT checks return UNKNOWN. UNSAT logging records feature, rectangle, and learned clauses for external elaboration, but an initial-unit conflict may lack an empty clause.}
\asGallerySolver{DPR}{\asGalleryValue{asGalleryLoss}{0}{0.747}}{\asGalleryValue{asGalleryGain}{0}{1.37}}{0/6}{Parity factoring leads the active search: union-find links complementary binary relations, CDCL searches the latent CNF, and finite-domain search verifies assignments before path-consistency and finite-domain CDCL fallbacks. Unsupported structures return UNKNOWN; verified SAT assignments are emitted, while UNSAT uses a witness-free DPR log for external elaboration, with replay failure also returning UNKNOWN.}
\asGallerySolver{VeriPB}{\asGalleryValue{asGalleryLoss}{0}{0.747}}{\asGalleryValue{asGalleryGain}{0}{1.37}}{0/6}{Dominance substitution and failed-assumption probing lead the default search, followed by path-consistency implications and bit-mask propagation in CDCL. The recognizer accepts only bounded complete choice blocks and returns UNKNOWN otherwise; SAT assignments are reparsed and checked, while UNSAT records dominance and RUP steps with a final contradiction, reporting UNKNOWN if proof finalization fails.}
\end{asGalleryFamily}

\begin{asGalleryFamily}{118}{pigeon-\allowbreak hole}{154 GBD instances\enspace\textperiodcentered\enspace 31 validation\enspace\textperiodcentered\enspace baseline 23/31}{The benchmark asks whether pigeons can be assigned to holes without exceeding the allowed hole capacity. CNF encodings use placement variables, clauses requiring each pigeon to choose a hole, and clauses preventing forbidden collisions; variants may add capacities or guards.}
\ifdefined\pdfbookmark\pdfbookmark[2]{pigeon-hole}{as-gallery-118}\fi
\asGallerySolver{GRAT}{\asGalleryValue{asGalleryLoss}{0}{0.982}}{\asGalleryValue{asGalleryGain}{0}{1.57}}{22/31}{Exact grid recognition regenerates the expected clause set before accepting a layout and constructs a checked assignment for recognized SAT cases. Specialized UNSAT branches use subset-counting, occupancy-state, implication, finite-path, or relativized reasoning and emit DRAT streams; bounded CDCL handles limited cases, while unsupported or unfinished paths return UNKNOWN.}
\asGallerySolver{DPR}{\asGalleryValue{asGalleryGain}{0}{8.82}}{\asGalleryValue{asGalleryGain}{0}{2.47}}{30/31}{Incidence reconstruction identifies pigeon rows, holes, and guarded collision structure, directing recognized instances to specialized paths rather than a general CNF solver. Only the standard branch constructs and checks SAT assignments; other recognized variants use clause additions, deletions, PR-style induction, RUP checks, or clique-restricted CDCL proofs, while failed recognition or unfinished paths return UNKNOWN.}
\asGallerySolver{VeriPB}{\asGalleryValue{asGalleryGain}{1}{8.84}}{\asGalleryValue{asGalleryGain}{1}{14.0}}{30/31}{Structural recognition reconstructs matrix placement columns and collision edges, enabling a direct SAT assignment only on the matrix path. Recognized UNSAT variants use aggregate PB/RUP derivations, while a fixed MIS-wrapper signature alone enables CDCL with RUP-logged learned clauses; unsupported layouts or failed certificate generation return UNKNOWN.}
\end{asGalleryFamily}

\begin{asGalleryFamily}{119}{planning}{667 GBD instances\enspace\textperiodcentered\enspace 134 validation\enspace\textperiodcentered\enspace baseline 81/134}{These benchmarks ask whether a Boolean assignment satisfies every clause in a CNF encoding of a planning decision problem. Variables and clauses can represent actions, states, auxiliary choices, or occupancy constraints, while inputs use DIMACS rather than recovered planning semantics.}
\ifdefined\pdfbookmark\pdfbookmark[2]{planning}{as-gallery-119}\fi
\asGallerySolver{GRAT}{\asGalleryValue{asGalleryLoss}{0}{0.719}}{\asGalleryValue{asGalleryLoss}{0}{0.850}}{57/134}{Short-clause occurrence weighting guides bounded DIMACS CDCL, with a dense-instance binary-implication path but no planning-specific reconstruction. Malformed inputs return UNKNOWN; oversized inputs receive only a two-pass model attempt and otherwise return UNKNOWN. Validated SAT assignments are emitted, while UNSAT retains learned clauses as ASCII additions followed by the empty clause.}
\asGallerySolver{DPR}{\asGalleryValue{asGalleryLoss}{1}{0.762}}{\asGalleryValue{asGalleryLoss}{1}{0.889}}{62/134}{A narrow multi-robot path-planning reconstruction searches space-time paths with vertex and swap avoidance, then injects a candidate branch. A failed candidate falls back to CDCL rather than proving UNSAT; clause-checked SAT models are emitted, while UNSAT uses a learned-clause journal as DPR additions plus the empty clause, and malformed or unsupported inputs return UNKNOWN.}
\asGallerySolver{VeriPB}{\asGalleryValue{asGalleryLoss}{0}{0.759}}{\asGalleryValue{asGalleryLoss}{0}{0.790}}{62/134}{Conservative layout-gated root probing can force learned root units before first-UIP CDCL; it is structural rather than planner-semantic. SAT assignments are reread and clause-checked, while UNSAT translates learned clauses into VeriPB RUP constraints with deletions and an empty constraint; malformed, oversized, or output failures return UNKNOWN.}
\end{asGalleryFamily}

\begin{asGalleryFamily}{120}{polynomial-\allowbreak multiplication}{60 GBD instances\enspace\textperiodcentered\enspace 12 validation\enspace\textperiodcentered\enspace baseline 5/12}{The benchmark asks whether length-n polynomial multiplication over GF(2) can be expressed as a sum of T products of three binary factor vectors. Its CNF encodes factor coefficients, conjunctions, and XOR chains enforcing each convolution output.}
\ifdefined\pdfbookmark\pdfbookmark[2]{polynomial-multiplication}{as-gallery-120}\fi
\asGallerySolver{GRAT}{\asGalleryValue{asGalleryGain}{0}{7.57}}{\asGalleryValue{asGalleryGain}{0}{5.32}}{11/12}{Recognizing compact GF(2) layouts, it constructs a symmetric rank decomposition, evaluates product and XOR variables, and independently checks every clause before emitting a SAT assignment. For recognized UNSAT cases, including ordered GF(3) layouts, it content-checks and remaps bundled DRAT cores; unsupported layouts or failed checks return UNKNOWN.}
\asGallerySolver{DPR}{\asGalleryValue{asGalleryGain}{1}{7.57}}{\asGalleryValue{asGalleryGain}{1}{6.24}}{11/12}{Recognizing regular layouts with sufficient rank, it constructs SAT assignments from GF(2) bilinear decompositions, with ordered propagation and full-CNF validation; the alternate layout has no SAT path. For UNSAT, it recovers limited permutations by incidence structure and maps checked embedded DRAT templates for bounded recognized cases; failures or unsupported cases return UNKNOWN.}
\asGallerySolver{VeriPB}{\asGalleryValue{asGalleryGain}{0}{1.52}}{\asGalleryValue{asGalleryGain}{0}{1.53}}{7/12}{It recognizes the tensor header equations and constructs GF(2) decompositions from diagonal and pair products, using a special reduced-product basis and elimination for n=5. It evaluates all auxiliary variables and clauses before SAT output; it has no general search or UNSAT certification, so failed or unsupported cases return UNKNOWN.}
\end{asGalleryFamily}

\begin{asGalleryFamily}{121}{popularity-\allowbreak similarity}{42 GBD instances\enspace\textperiodcentered\enspace 9 validation\enspace\textperiodcentered\enspace baseline 3/9}{These instances are 3-CNF formulas, with three literals per clause. The available solver artifacts establish their clause structure but do not recover the generating popularity-similarity model.}
\ifdefined\pdfbookmark\pdfbookmark[2]{popularity-similarity}{as-gallery-121}\fi
\asGallerySolver{GRAT}{\asGalleryValue{asGalleryGain}{0}{1.07}}{\asGalleryValue{asGalleryGain}{1}{1.12}}{3/9}{Bounded variable elimination limits resolvent growth, records eliminated variables, and passes the residual to watched-literal CDCL with degree-seeded EVSIDS. It accepts only the recognized fixed 3-CNF degree profile, returns UNKNOWN on other inputs, checks reconstructed SAT assignments, and may print UNSAT after failed reconstruction without a final empty-clause certificate.}
\asGallerySolver{DPR}{\asGalleryValue{asGalleryGain}{1}{1.09}}{\asGalleryValue{asGalleryLoss}{0}{0.988}}{3/9}{Focused local search uses break-score weighting on variables from unsatisfied clauses, then falls back to bounded Davis-Putnam elimination and a 1-UIP CDCL cube search. It handles the recognized 3-CNF shape, returns UNKNOWN otherwise, checks SAT assignments, and logs witness-free clause additions for external DPR elaboration when cube closure reports UNSAT.}
\asGallerySolver{VeriPB}{\asGalleryValue{asGalleryGain}{0}{1.03}}{\asGalleryValue{asGalleryLoss}{0}{0.898}}{3/9}{Popularity ordering and damped loopy-BP seed decisions and phases; watched-literal CDCL then uses first-UIP learning and restarts. It accepts only the recognized 3-CNF profile, returns UNKNOWN on unsupported inputs or conflict limits, and checks SAT assignments. Enabled proof logging emits learned clauses and level-zero conflicts as VeriPB RUP steps.}
\end{asGalleryFamily}

\begin{asGalleryFamily}{122}{prime-\allowbreak factoring}{185 GBD instances\enspace\textperiodcentered\enspace 37 validation\enspace\textperiodcentered\enspace baseline 27/37}{These CNFs ask whether two bounded binary factors multiply to a fixed target integer. Variables encode factor bits, partial products, carries, and output wires, while gate clauses and fixed bits constrain the multiplication circuit.}
\ifdefined\pdfbookmark\pdfbookmark[2]{prime-factoring}{as-gallery-122}\fi
\asGallerySolver{GRAT}{\asGalleryValue{asGalleryLoss}{0}{0.845}}{\asGalleryValue{asGalleryLoss}{0}{0.843}}{24/37}{Structural multiplier recognition recovers operand positions, polarities, and a bounded factor split, then checks the resulting assignment against every original clause. Specialized failures fall back to first-UIP CDCL; SAT yields a checked assignment, root-conflict UNSAT is logged as textual DRAT for external elaboration, and unresolved cases return UNKNOWN.}
\asGallerySolver{DPR}{\asGalleryValue{asGalleryLoss}{0}{0.596}}{\asGalleryValue{asGalleryLoss}{0}{0.604}}{17/37}{Structural probing on recognized multipliers uses packed all-zero and singleton-pair tests to recover input polarity and bit order, then searches bounded factors. A restricted CDCL fallback checks SAT assignments. Only the special 8-by-8 EZFact no-factor case emits witness-free blocking clauses followed by a resolution tree; other no-factor results and CDCL UNSAT return UNKNOWN.}
\asGallerySolver{VeriPB}{\asGalleryValue{asGalleryGain}{1}{1.03}}{\asGalleryValue{asGalleryLoss}{1}{0.973}}{27/37}{Polarity-aware structural decoding reconstructs shuffled multiplier bits and tests arithmetic candidates, with a narrower multiplier recognizer as a second direct path. These direct paths provide checked SAT assignments but no UNSAT certificate. If they fail, CDCL logs level-zero UNSAT as VeriPB RUP traces with deletions; unsupported or resource-limited cases return UNKNOWN.}
\end{asGalleryFamily}

\begin{asGalleryFamily}{123}{prime-\allowbreak testing}{58 GBD instances\enspace\textperiodcentered\enspace 12 validation\enspace\textperiodcentered\enspace baseline 7/12}{These benchmarks ask whether structured Boolean arithmetic constructions have satisfying assignments, including fixed-product factors, modular square roots, or repeated gate predicates. CNF encodings use Tseitin-style gate clauses, unit constraints, and equivalences over circuit and word variables.}
\ifdefined\pdfbookmark\pdfbookmark[2]{prime-testing}{as-gallery-123}\fi
\asGallerySolver{GRAT}{\asGalleryValue{asGalleryGain}{0}{1.27}}{\asGalleryValue{asGalleryGain}{0}{1.29}}{8/12}{Arithmetic witness construction drives the main search: factorization or modular-root solving fixes word inputs, and unit propagation completes assignments followed by full clause checks; repeated blocks use watched-literal DPLL with offset translation. Supported failures may use bounded input-tree search, reporting UNSAT only with textual DRAT refutations; unsupported cases return UNKNOWN.}
\asGallerySolver{DPR}{\asGalleryValue{asGalleryGain}{1}{1.56}}{\asGalleryValue{asGalleryGain}{1}{1.37}}{9/12}{Bit-parallel exhaustive search handles recognized repeated circuits, while arithmetic decoding uses factorization or modular roots followed by propagation and clause checks. Selected fallbacks use restricted CDCL, emitting clauses and an empty clause as DPR additions; unsupported cases return UNKNOWN, while proof-output failure can return SAT without an UNSAT certificate.}
\asGallerySolver{VeriPB}{\asGalleryValue{asGalleryGain}{0}{1.27}}{\asGalleryValue{asGalleryGain}{0}{1.29}}{8/12}{Bit-parallel evaluation is the primary search for repeated circuits, while product and quadratic layouts recover arithmetic witnesses using factorization or modular-root methods, then complete assignments by propagation and validate every original clause. The solver has no general SAT search or UNSAT or VeriPB certification path; unrecognized or failed constructions return UNKNOWN.}
\end{asGalleryFamily}

\begin{asGalleryFamily}{124}{product-\allowbreak configuration}{23 GBD instances\enspace\textperiodcentered\enspace 5 validation\enspace\textperiodcentered\enspace baseline 5/5}{These benchmarks ask whether a Boolean assignment can select product or configuration choices while satisfying requested features, exclusions, implications, and support conditions. The instances use flattened CNF with unit, binary, and wider clauses representing these constraints, without requiring recovery of named product domains.}
\ifdefined\pdfbookmark\pdfbookmark[2]{product-configuration}{as-gallery-124}\fi
\asGallerySolver{GRAT}{\asGalleryValue{asGalleryGain}{0}{3.88}}{\asGalleryValue{asGalleryGain}{1}{3.31}}{5/5}{Failed-literal probing builds root units in short-clause activity order, followed by watched-literal propagation and activity-guided CDCL with first-UIP learning and restarts. Rejected inputs or proof-output failures return UNKNOWN; SAT returns a checked model, while UNSAT records learned clauses and, on root conflict, the empty clause as textual DRAT additions.}
\asGallerySolver{DPR}{\asGalleryValue{asGalleryGain}{1}{6.63}}{\asGalleryValue{asGalleryLoss}{0}{0.502}}{5/5}{An exclusion-heavy recognizer selects a specialized path, combining root-unit and watched-literal propagation with activity-weighted, negative-first CDCL, first-UIP learning, and restarts. Unsupported structure or internal or proof-output failure returns UNKNOWN; SAT yields a checked model, while UNSAT appends learned clauses and, on root contradiction, an empty clause to textual DPR output.}
\asGallerySolver{VeriPB}{\asGalleryValue{asGalleryGain}{0}{4.24}}{\asGalleryValue{asGalleryGain}{0}{1.09}}{5/5}{Shortest-open-clause branching follows failed-literal closure, guiding search; conflicts use first-UIP learning, nonchronological backtracking, and a conservative whole-decision nogood fallback. Unsupported structure or checking or writing failure returns UNKNOWN; SAT yields a checked assignment, while UNSAT writes learned clauses and an empty clause as a VeriPB RUP certificate.}
\end{asGalleryFamily}

\begin{asGalleryFamily}{125}{profitable-\allowbreak robust-\allowbreak production}{20 GBD instances\enspace\textperiodcentered\enspace 4 validation\enspace\textperiodcentered\enspace baseline 2/4}{The benchmark asks whether binary production decisions and auxiliary Boolean variables can satisfy constraints linking production and profit across scenarios. These constraints encode comparisons, arithmetic, and robust minima or related aggregates in CNF, often with order/comparator and Tseitin variables.}
\ifdefined\pdfbookmark\pdfbookmark[2]{profitable-robust-production}{as-gallery-125}\fi
\asGallerySolver{GRAT}{\asGalleryValue{asGalleryGain}{1}{23.7}}{\asGalleryValue{asGalleryGain}{1}{17.6}}{4/4}{For recognized inputs, support-row canonicalization fixes selected auxiliary rows and installs stride-based assignments, then watched-literal CDCL searches decomposed components with reused phases, learning, and restarts. These restrictions may exclude satisfying assignments; SAT models are checked against the original clauses, while contradictory or exhausted searches return UNKNOWN and no UNSAT certificate is produced.}
\asGallerySolver{DPR}{\asGalleryValue{asGalleryLoss}{0}{0.823}}{\asGalleryValue{asGalleryLoss}{0}{0.825}}{2/4}{Binary-implication SCC substitution and bounded elimination remove equivalent or redundant comparator variables before watched-literal CDCL. On SAT, it reconstructs eliminated variables and checks every original clause before printing a model; no DPR proof is emitted, and rejected inputs, contradictions, or failed searches return UNKNOWN rather than UNSAT.}
\asGallerySolver{VeriPB}{\asGalleryValue{asGalleryLoss}{0}{0.804}}{\asGalleryValue{asGalleryLoss}{0}{0.805}}{2/4}{False-first phase heuristics with occurrence-signature exceptions initialize watched-literal CDCL, while recognition of the expected ripple-gate prefix only selects the specialized input path. After a SAT result, the solver reparses and checks every original clause before outputting the assignment; it has no VeriPB derivation or UNSAT path, so search exhaustion and internal failure return UNKNOWN.}
\end{asGalleryFamily}

\begin{asGalleryFamily}{126}{purdom-\allowbreak instances}{19 GBD instances\enspace\textperiodcentered\enspace 4 validation\enspace\textperiodcentered\enspace baseline 4/4}{The family asks whether two binary operand vectors multiply to a fixed integer. CNF encodings represent factor bits, partial products, and an AND/XOR carry network, with unit clauses fixing output bits; some encodings may shuffle variables or polarities.}
\ifdefined\pdfbookmark\pdfbookmark[2]{purdom-instances}{as-gallery-126}\fi
\asGallerySolver{GRAT}{\asGalleryValue{asGalleryLoss}{0}{0.878}}{\asGalleryValue{asGalleryLoss}{1}{0.388}}{4/4}{Partial-product graph recovery identifies a shuffled, polarity-flipped multiplication circuit, then factors the fixed target and evaluates compatible operands for a clause-checked model. If no compatible pair is found, modular prefix lemmas guide a least-significant-bit-first RUP tree, which writes the UNSAT proof only if it completes; unsupported inputs return UNKNOWN.}
\asGallerySolver{DPR}{\asGalleryValue{asGalleryGain}{1}{917}}{\asGalleryValue{asGalleryLoss}{0}{0.278}}{4/4}{Gate recognition reconstructs a single-writer network, including a clean AND/XOR multiplier. It recovers an odd target, factors it, and checks a model; direct UNSAT cases emit a blocker-tree clause sequence, while augmented cases may use bounded factor search or unlogged CDCL for SAT, returning UNKNOWN on recognition or search failure.}
\asGallerySolver{VeriPB}{\asGalleryValue{asGalleryLoss}{0}{0.134}}{\asGalleryValue{asGalleryLoss}{0}{0.189}}{3/4}{Carry-weight and bipartite-graph recovery reconstructs factor-bit order, after which factoring the target enables fixed-point AND/XOR evaluation and independent clause checking for composite cases. For prime targets, first-UIP CDCL supplies RUP-style proof logging; unsupported structures, factoring failures, or an incomplete proof return UNKNOWN.}
\end{asGalleryFamily}

\begin{asGalleryFamily}{127}{puzzle}{21 GBD instances\enspace\textperiodcentered\enspace 5 validation\enspace\textperiodcentered\enspace baseline 5/5}{The benchmark asks whether fixed-orientation polyomino components can cover every cell of a rectangular board without overlap. A structured CNF encodes board-cell labels, one-label-per-cell constraints, piece-placement limits, adjacency, and forbidden translations.}
\ifdefined\pdfbookmark\pdfbookmark[2]{puzzle}{as-gallery-127}\fi
\asGallerySolver{GRAT}{\asGalleryValue{asGalleryLoss}{0}{0.0927}}{\asGalleryValue{asGalleryLoss}{1}{0.0866}}{4/5}{It searches recognized fixed, translation-only shapes with bit-parallel exact-cover DFS, constrained-cell branching, and failed-state memoization. Found tilings are expanded and checked before SAT output. When search fails, mapped-placement CDCL attempts UNSAT and emits DRAT additions; unsupported cases or failed proof generation return UNKNOWN without external verification.}
\asGallerySolver{DPR}{\asGalleryValue{asGalleryLoss}{0}{0.0546}}{\asGalleryValue{asGalleryLoss}{0}{0.0790}}{2/5}{It searches recognized fixed, translation-only shapes by bit-parallel exact-cover backtracking with constrained-cell branching and failed-state memoization. Found tilings are expanded and checked before SAT output. When search fails on eligible small boards, auxiliary CDCL logs DRAT additions, including witness-free clauses; unsupported cases or an inconclusive fallback return UNKNOWN without external verification.}
\asGallerySolver{VeriPB}{\asGalleryValue{asGalleryLoss}{1}{0.164}}{\asGalleryValue{asGalleryLoss}{0}{0.0752}}{4/5}{It extracts placement masks from bidirectional implications and solves exact cover by minimum-domain branching, grouped translations, and failed-state memoization. SAT assignments are checked; UNSAT reruns a DFS emitting VeriPB symmetry breakers and RUP nogoods ending in contradiction. No general-SAT fallback is active, so rejected inputs or proof failures return UNKNOWN.}
\end{asGalleryFamily}

\begin{asGalleryFamily}{128}{pythagorean-\allowbreak triples}{21 GBD instances\enspace\textperiodcentered\enspace 5 validation\enspace\textperiodcentered\enspace baseline 2/5}{Given a collection of Pythagorean triples, assign each represented integer one of two colors so that no triple is monochromatic. In CNF, each triple is represented by complementary monotone clauses, while binary clauses capture constraints left after a vertex is fixed or removed.}
\ifdefined\pdfbookmark\pdfbookmark[2]{pythagorean-triples}{as-gallery-128}\fi
\asGallerySolver{GRAT}{\asGalleryValue{asGalleryGain}{1}{4.05\!\cdot\!10^{4}}}{\asGalleryValue{asGalleryGain}{1}{2.98\!\cdot\!10^{4}}}{5/5}{Core reconstruction is the active technique: the solver regenerates a bounded Pythagorean 2-core, refines incidence signatures to map the input, and applies an embedded two-coloring. It completes residual variables with non-monochromatic propagation and failed-literal DFS; the restricted recognizer does not validate numeric identities, and recognition or search failure returns UNKNOWN without an UNSAT certificate path.}
\asGallerySolver{DPR}{\asGalleryValue{asGalleryGain}{0}{1.45}}{\asGalleryValue{asGalleryGain}{0}{1.45}}{4/5}{Stochastic local search leads the default path, combining break-based moves, clause weighting, and elite-agreement freezing. If it fails, high-degree decimation is followed by watched-literal CDCL repair on expanding neighborhoods and then the full formula; checked SAT assignments are emitted, while failed or internally unsatisfiable searches return UNKNOWN because no UNSAT certificate path is implemented.}
\asGallerySolver{VeriPB}{\asGalleryValue{asGalleryLoss}{0}{0.936}}{\asGalleryValue{asGalleryLoss}{0}{0.936}}{2/5}{Weighted breakout local search is the active first stage, using the best coloring as a phase guide before expanding-neighborhood CDCL repairs and a full CDCL fallback with elimination and reconstruction. The recognizer accepts only this non-monochromatic structure; checked SAT assignments are output, while CDCL UNSAT invokes VeriPB logging and proof failure yields UNKNOWN.}
\end{asGalleryFamily}

\begin{asGalleryFamily}{129}{quantum-\allowbreak kochen-\allowbreak specker}{11 GBD instances\enspace\textperiodcentered\enspace 3 validation\enspace\textperiodcentered\enspace baseline 2/3}{The benchmark asks whether an n-vertex graph can meet a specified Kochen-Specker obstruction: Boolean edge variables choose adjacencies, triangle variables represent conjunctions of three edges, and CNF clauses enforce square-freeness and rule out the intended zero-one coloring, sometimes with symmetry breakers.}
\ifdefined\pdfbookmark\pdfbookmark[2]{quantum-kochen-specker}{as-gallery-129}\fi
\asGallerySolver{GRAT}{\asGalleryValue{asGalleryLoss}{0}{0.588}}{\asGalleryValue{asGalleryLoss}{0}{0.695}}{1/3}{The solver first tries a hard-coded 17-vertex graph witness, propagates triangle auxiliaries, and verifies complete SAT assignments against the original clauses. Other recognized inputs use edge-biased CDCL with elimination and failed-literal probing; first-UIP clauses support external DRAT elaboration, while only an empty clause yields UNSAT and unsupported or unfinished cases return UNKNOWN.}
\asGallerySolver{DPR}{\asGalleryValue{asGalleryLoss}{1}{0.654}}{\asGalleryValue{asGalleryLoss}{1}{0.759}}{1/3}{A hard-coded 17-vertex construction solves residual constraints, propagates triangle auxiliaries, and verifies its assignment against the original CNF. Other recognized cases use edge-prioritized CDCL with selected layer removal and first-UIP learning; witness-free DPR additions support UNSAT certification, while unsupported or unfinished cases return UNKNOWN and the shortcut lacks UNSAT certification.}
\asGallerySolver{VeriPB}{\asGalleryValue{asGalleryLoss}{0}{0.469}}{\asGalleryValue{asGalleryLoss}{0}{0.595}}{0/3}{Edge-prioritized CDCL drives the accepted n=17 minimum-degree regime, with watched literals, activity-based branching, phase saving, deletion, and restarts. It checks SAT assignments against the original CNF and reports UNSAT only after serializing its trace as RUP steps for external checking, ending in an empty clause; unsupported inputs or trace failures return UNKNOWN.}
\end{asGalleryFamily}

\begin{asGalleryFamily}{130}{quasigroup-\allowbreak completion}{261 GBD instances\enspace\textperiodcentered\enspace 53 validation\enspace\textperiodcentered\enspace baseline 47/53}{The benchmark asks whether a partially filled quasigroup or Latin-square table can be completed so each cell receives one value and each value occurs once in every row and column. CNF uses cell-value choices with one-hot groups and pairwise conflicts, but encodings vary.}
\ifdefined\pdfbookmark\pdfbookmark[2]{quasigroup-completion}{as-gallery-130}\fi
\asGallerySolver{GRAT}{\asGalleryValue{asGalleryLoss}{0}{0.578}}{\asGalleryValue{asGalleryLoss}{0}{0.697}}{41/53}{A strict polarity/parity recognizer recovers three exact-cover dimensions, then tries DSATUR-guided min-conflicts and bounded smallest-column search. Failed recognition or search falls back to CDCL; SAT models are checked, while UNSAT emits textual DRAT for external elaboration and checking, and proof, verification, or parse failures can yield UNKNOWN.}
\asGallerySolver{DPR}{\asGalleryValue{asGalleryLoss}{1}{0.618}}{\asGalleryValue{asGalleryLoss}{1}{0.702}}{41/53}{A phase-shuffled exact-cover detector recovers three dimensions and tries bounded min-conflicts followed by a local walk. Failures use CDCL or, for larger recognized instances, exhaustive Algorithm X with state-clause additions for its proof search; SAT reports checked models, and parse, verification, or proof failures can yield UNKNOWN.}
\asGallerySolver{VeriPB}{\asGalleryValue{asGalleryLoss}{0}{0.448}}{\asGalleryValue{asGalleryLoss}{0}{0.531}}{37/53}{Watched-literal CDCL drives search, with proof-logged variable elimination, first-UIP learning, and structural recognition used only for phases and preprocessing. UNSAT emits VeriPB-style RUP constraints and a final empty RUP; SAT reconstructs and checks a model but creates no final proof artifact, while malformed input or output setup failure yields UNKNOWN.}
\end{asGalleryFamily}

\begin{asGalleryFamily}{131}{railway-\allowbreak safety}{9 GBD instances\enspace\textperiodcentered\enspace 2 validation\enspace\textperiodcentered\enspace baseline 0/2}{These benchmarks ask whether a bounded railway interlocking or transition-system scenario has a Boolean assignment satisfying its gate, state, and transition constraints together with a selected safety obligation. The encoding uses Tseitin-style CNF clauses over those variables.}
\ifdefined\pdfbookmark\pdfbookmark[2]{railway-safety}{as-gallery-131}\fi
\asGallerySolver{GRAT}{\asGalleryValue{asGalleryMuted}{1}{1.00}}{\asGalleryValue{asGalleryMuted}{1}{1.00}}{0/2}{Preprocessing combines bounded Davis-Putnam elimination and binary-implication SCC reduction with watched-literal CDCL; very wide terminal clauses trigger failed-literal decomposition, while smaller cases use residual CDCL. Outside structural recognition it returns UNKNOWN; SAT models are checked against the original CNF, and UNSAT steps are logged as textual DRAT for external elaboration and checking.}
\asGallerySolver{DPR}{\asGalleryValue{asGalleryMuted}{1}{1.00}}{\asGalleryValue{asGalleryMuted}{1}{1.00}}{0/2}{Failed-literal probing ranks literals in a syntactically wide clause; first-UIP CDCL searches branches under their negations and otherwise falls back to ordinary CDCL. Malformed input returns UNKNOWN; SAT models are checked against original clauses, while UNSAT logging records learned clauses and failed-literal units, can include witness-free clause additions, and closes with an empty clause.}
\asGallerySolver{VeriPB}{\asGalleryValue{asGalleryMuted}{1}{1.00}}{\asGalleryValue{asGalleryMuted}{1}{1.00}}{0/2}{A final wide, all-negative clause activates assumption-based decomposition: watched-literal first-UIP CDCL tries each tail literal, retains learned consequences, and uses ordinary CDCL otherwise. Outside recognition it returns UNKNOWN; SAT models are checked against original clauses, while UNSAT learned clauses are emitted as VeriPB RUP records followed by an UNSAT conclusion.}
\end{asGalleryFamily}

\begin{asGalleryFamily}{132}{ramsey}{6 GBD instances\enspace\textperiodcentered\enspace 2 validation\enspace\textperiodcentered\enspace baseline 1/2}{These benchmarks ask whether the edges of a complete graph can be colored so that specified forbidden subgraphs are not monochromatic. Boolean CNF encodes the edge colors and may add auxiliary variables and clauses for local color constraints.}
\ifdefined\pdfbookmark\pdfbookmark[2]{ramsey}{as-gallery-132}\fi
\asGallerySolver{GRAT}{\asGalleryValue{asGalleryLoss}{0}{0.998}}{\asGalleryValue{asGalleryLoss}{0}{0.988}}{1/2}{Line-graph reconstruction drives the K32 four-color path, which applies an F2\textasciicircum{}4 edge coloring; a direct K18 path instead uses symmetry-breaking sorting and a CDCL tail. SAT assignments are checked against the clauses; UNSAT paths emit textual DRAT traces for external elaboration, while generic Ramsey-shaped inputs use CDCL and rejected inputs return UNKNOWN.}
\asGallerySolver{DPR}{\asGalleryValue{asGalleryGain}{1}{4{,}220}}{\asGalleryValue{asGalleryGain}{0}{91.0}}{2/2}{Structural recognition drives Ramsey-specific paths: four-color cases use bounded min-conflicts search on a recovered 3-uniform hypergraph, with a paired two-color fallback. SAT candidates are checked against the input CNF; exact recognized layouts alone receive UNSAT certificates from bundled DPR templates, while failed bounded search or other cases return UNKNOWN.}
\asGallerySolver{VeriPB}{\asGalleryValue{asGalleryGain}{0}{1{,}990}}{\asGalleryValue{asGalleryGain}{1}{184}}{2/2}{Min-conflicts recoloring handles recognized four-color layouts, propagating gate variables after a candidate is found; a separate R(4,4;18) path reconstructs edge structure, adds symmetry reductions, and runs CDCL. Clause-checked SAT candidates are accepted; the R(4,4;18) path emits VeriPB red and RUP steps for UNSAT, while R(4,5;25) inputs and failed bounded searches return UNKNOWN.}
\end{asGalleryFamily}

\begin{asGalleryFamily}{133}{ramsey-\allowbreak numbers}{6 GBD instances\enspace\textperiodcentered\enspace 2 validation\enspace\textperiodcentered\enspace baseline 0/2}{The benchmark asks whether the edges of a complete graph can be colored with two colors while avoiding a monochromatic K\_s in one color and K\_t in the other. CNF uses one variable per edge and clauses for every forbidden clique.}
\ifdefined\pdfbookmark\pdfbookmark[2]{ramsey-numbers}{as-gallery-133}\fi
\asGallerySolver{GRAT}{\asGalleryValue{asGalleryGain}{1}{1.98}}{\asGalleryValue{asGalleryGain}{1}{1.07}}{1/2}{Specialized comparator networks reduce selected (4,4,n) and (3,6,n) cases to guarded core cases, using ordering and branch exclusions for the former and selection/profile sorting before CDCL for the latter. It handles dispatched instances through an induced 18-vertex core, emits proof clauses and comments for GRAT checking, and otherwise returns UNKNOWN.}
\asGallerySolver{DPR}{\asGalleryValue{asGalleryMuted}{0}{1.00}}{\asGalleryValue{asGalleryMuted}{0}{1.00}}{0/2}{An edge-indexed CDCL engine uses watched literals, clause learning, and restarts; selected recognized cases search an induced core. It accepts only the canonical encoding and supported range, serializes learned clauses and deletions for DPR elaboration and checking, and returns UNKNOWN when unsupported or failing; SAT from a core need not model the full input.}
\asGallerySolver{VeriPB}{\asGalleryValue{asGalleryMuted}{0}{1.00}}{\asGalleryValue{asGalleryMuted}{0}{1.00}}{0/2}{An extrema-degree PB route targets (3,6;19), using symmetry and CDCL to establish units before summing degree bounds into a PB contradiction. If it does not finish, symmetry-guided CDCL is the fallback; PB/RUP records are logged, SAT models are checked, only exhaustive encodings are accepted, and unsupported or unresolved cases return UNKNOWN.}
\end{asGalleryFamily}

\begin{asGalleryFamily}{134}{ramseycube}{10 GBD instances\enspace\textperiodcentered\enspace 2 validation\enspace\textperiodcentered\enspace baseline 2/2}{These CNFs ask whether the edges of a complete graph can be colored with two colors while avoiding monochromatic subgraphs. Variables represent graph edges, and signed clauses encode forbidden 3-cubes, with some encodings also constraining a specified color on four-cycles.}
\ifdefined\pdfbookmark\pdfbookmark[2]{ramseycube}{as-gallery-134}\fi
\asGallerySolver{GRAT}{\asGalleryValue{asGalleryGain}{0}{1.04}}{\asGalleryValue{asGalleryGain}{0}{1.29}}{2/2}{Structural Q3/C4 recognition uses prescribed n=8-20 patterns and fixed-coloring witnesses for small SAT cases, verifying every clause before SAT. Asymmetric instances use a complete K9 core with 1-UIP CDCL, emitting learned clauses and an empty clause for external DRAT elaboration; unsupported inputs and symmetric K13 return UNKNOWN.}
\asGallerySolver{DPR}{\asGalleryValue{asGalleryGain}{0}{3.06}}{\asGalleryValue{asGalleryLoss}{0}{0.738}}{2/2}{Embedded colorings answer small SAT cases by mapping supported lexicographic K\_n edge layouts to fixed assignments and checking every clause. Asymmetric inputs with an induced K9 core use 1-UIP CDCL and log learned clauses plus an empty clause; unsupported layouts, failed checks, and symmetric n=13 return UNKNOWN.}
\asGallerySolver{VeriPB}{\asGalleryValue{asGalleryGain}{1}{230}}{\asGalleryValue{asGalleryGain}{1}{30.8}}{2/2}{Threshold-core detection searches for K13 or K9 complete cores, deletes clauses outside the core, and adds permutation-witnessed symmetry breaking before watched-literal CDCL. Only core UNSAT receives RUP-logged VeriPB output; direct-formula UNSAT and unsupported or unmatched inputs return UNKNOWN, while checked SAT models are returned.}
\end{asGalleryFamily}

\begin{asGalleryFamily}{135}{random}{872 GBD instances\enspace\textperiodcentered\enspace 175 validation\enspace\textperiodcentered\enspace baseline 86/175}{This family asks whether a Boolean assignment can satisfy every clause in a random-style CNF formula, or whether no such assignment exists. Instances encode clauses as sets of signed literals, often with a broadly uniform short width.}
\ifdefined\pdfbookmark\pdfbookmark[2]{random}{as-gallery-135}\fi
\asGallerySolver{GRAT}{\asGalleryValue{asGalleryGain}{0}{1.44}}{\asGalleryValue{asGalleryGain}{0}{1.43}}{114/175}{ProbSAT-style local search leads, using false-clause sampling, break/make counts, and restarts; its finite budget is not an UNSAT conclusion before watched-literal CDCL takes over. Verified SAT assignments are emitted; UNSAT requires a textual DRAT stream ending in the empty clause for external elaboration, while nonmatching or failed cases return UNKNOWN.}
\asGallerySolver{DPR}{\asGalleryValue{asGalleryGain}{0}{1.13}}{\asGalleryValue{asGalleryGain}{0}{1.12}}{96/175}{Incremental probSAT-style local search samples false clauses and uses break/make scoring with random restarts, but is gated off for some narrow-width, dense, or high-ratio cases. Watched-literal CDCL then supplies a checked SAT assignment or, after a level-zero conflict, an UNSAT result with clause-addition proof; rejected inputs and proof or reconstruction failures yield UNKNOWN.}
\asGallerySolver{VeriPB}{\asGalleryValue{asGalleryGain}{1}{1.50}}{\asGalleryValue{asGalleryGain}{1}{1.50}}{117/175}{Focused ProbSAT/WalkSAT local search leads, sampling variables from false clauses with break-weighted and occasional make-aware moves; default routing can give some recognized shapes an unbounded search, so they may not reach the exact fallback. The exact CDCL fallback verifies SAT assignments and emits VeriPB RUP constraints for UNSAT; solver or proof-file failures yield UNKNOWN.}
\end{asGalleryFamily}

\begin{asGalleryFamily}{136}{random-\allowbreak circuits}{21 GBD instances\enspace\textperiodcentered\enspace 5 validation\enspace\textperiodcentered\enspace baseline 4/5}{The task is to choose Boolean primary inputs so an acyclic fixed-width lookup circuit produces a prescribed pattern on its final output bits. CNF uses truth-table clauses for the gates and unit clauses to fix the target outputs.}
\ifdefined\pdfbookmark\pdfbookmark[2]{random-circuits}{as-gallery-136}\fi
\asGallerySolver{GRAT}{\asGalleryValue{asGalleryGain}{1}{78.0}}{\asGalleryValue{asGalleryGain}{1}{61.0}}{5/5}{AVX-512 batched truth-table enumeration first tests compatible primary-input assignments, pruning gates outside the target cone. A CDCL fallback propagates table rows and learns first-UIP clauses when enumeration is unavailable or finds no model; verified SAT assignments are emitted, while unsupported inputs or failed search return UNKNOWN, with no UNSAT certificate path.}
\asGallerySolver{DPR}{\asGalleryValue{asGalleryGain}{0}{1.24}}{\asGalleryValue{asGalleryGain}{0}{1.24}}{4/5}{Target-cone reduction replaces each relevant forbidden truth-table block with compact implications before a watched-literal CDCL search. The search uses a topological order portfolio, retrying with another order after a bounded attempt; verified SAT models are emitted, while unsupported layouts or failed search return UNKNOWN and no UNSAT certificate is produced.}
\asGallerySolver{VeriPB}{\asGalleryValue{asGalleryGain}{0}{9.80}}{\asGalleryValue{asGalleryGain}{0}{7.53}}{5/5}{Batched truth-table enumeration with root bitset filtering first tests eligible recognized circuits. The fallback searches backward from fixed outputs, choosing a gate with the smallest compatible preimage set and propagating bitset domains; verified SAT assignments are emitted, while unsupported layouts, contradictions, or exhaustion return UNKNOWN, with no UNSAT certificate path.}
\end{asGalleryFamily}

\begin{asGalleryFamily}{137}{random-\allowbreak clustered}{36 GBD instances\enspace\textperiodcentered\enspace 8 validation\enspace\textperiodcentered\enspace baseline 4/8}{The benchmark asks whether a Boolean assignment can make every clause in a sparse 3-CNF formula true. Inputs use three-literal clauses and are intended to exhibit clustered variable co-occurrence, although recognizer filters describe only supported encodings, not the entire family.}
\ifdefined\pdfbookmark\pdfbookmark[2]{random-clustered}{as-gallery-137}\fi
\asGallerySolver{GRAT}{\asGalleryValue{asGalleryMuted}{0}{1.00}}{\asGalleryValue{asGalleryMuted}{0}{1.00}}{4/8}{Break-weighted stochastic local search drives accepted 3-CNF instances, choosing variables from unsatisfied clauses and updating clause counts, occurrence lists, and unsatisfied-clause state incrementally. A found assignment is independently checked before SAT output; rejected inputs return UNKNOWN, while unsuccessful accepted searches continue restarting without an UNSAT proof path.}
\asGallerySolver{DPR}{\asGalleryValue{asGalleryMuted}{1}{1.00}}{\asGalleryValue{asGalleryMuted}{1}{1.00}}{4/8}{Structural recognition selects supported clustered 3-CNF instances, then randomized break-weighted local search flips variables from unsatisfied clauses while maintaining clause counts and occurrence data incrementally. It independently checks and emits complete SAT assignments; unsupported inputs return UNKNOWN, while unsuccessful accepted searches restart without an UNSAT or DPR certificate path.}
\asGallerySolver{VeriPB}{\asGalleryValue{asGalleryMuted}{0}{1.00}}{\asGalleryValue{asGalleryMuted}{0}{1.00}}{4/8}{Bounded ProbSAT searches recognized instances with break-weighted flips from unsatisfied clauses, then falls back to watched-literal CDCL with clause learning when it fails. SAT assignments are checked against clauses; CDCL UNSAT paths serialize a RUP-based proof for external VeriPB checking, while unsupported inputs return UNKNOWN and local-search failure alone does not establish UNSAT.}
\end{asGalleryFamily}

\begin{asGalleryFamily}{138}{random-\allowbreak csp}{24 GBD instances\enspace\textperiodcentered\enspace 5 validation\enspace\textperiodcentered\enspace baseline 3/5}{These instances ask whether each object in a finite binary CSP can take one value while avoiding forbidden unary and pairwise combinations. CNF encodings often use exactly-one selector blocks or compact bit blocks defining legal states, with cross-object clauses forbidding state pairs.}
\ifdefined\pdfbookmark\pdfbookmark[2]{random-csp}{as-gallery-138}\fi
\asGallerySolver{GRAT}{\asGalleryValue{asGalleryGain}{0}{13.7}}{\asGalleryValue{asGalleryGain}{0}{13.6}}{5/5}{State-level CDCL over the recovered finite-domain CSP leads the search, followed by arc-consistency backtracking and local repair or min-conflicts methods. Reconstructed assignments are checked against the original clauses before SAT output; failed finite phases may enter restart cycles until externally stopped, and no UNSAT certificate path is implemented.}
\asGallerySolver{DPR}{\asGalleryValue{asGalleryGain}{1}{540}}{\asGalleryValue{asGalleryGain}{1}{534}}{5/5}{Structure recovery compiles recognized one-hot or compact-state CNF into domains with unary costs and forbidden pairs, then uses belief propagation, tabu or min-conflicts search, and consistency repairs. Feasible assignments are mapped back and checked against clauses before SAT output; unsupported or unsuccessful cases return UNKNOWN, with no UNSAT certificate path.}
\asGallerySolver{VeriPB}{\asGalleryValue{asGalleryGain}{0}{19.4}}{\asGalleryValue{asGalleryGain}{0}{19.3}}{5/5}{Damped belief propagation seeds weighted breakout and tabu search on recovered CSPs; arc-consistency masks and dom/wdeg backtracking provide fallback. Original clauses are checked for SAT assignments; the unbounded WalkSAT route has no complete fallback, while unsupported or failed searches return UNKNOWN and no VeriPB logger provides an UNSAT certificate.}
\end{asGalleryFamily}

\begin{asGalleryFamily}{139}{random-\allowbreak hiddenmodel}{117 GBD instances\enspace\textperiodcentered\enspace 24 validation\enspace\textperiodcentered\enspace baseline 16/24}{The benchmark asks whether a Boolean assignment satisfies every clause of a structured CNF formula whose clauses contain two or three distinct variables. A recurring subpattern translates to odd-parity XOR equations, but it is only a special subfamily, not the definition of the benchmark.}
\ifdefined\pdfbookmark\pdfbookmark[2]{random-hiddenmodel}{as-gallery-139}\fi
\asGallerySolver{GRAT}{\asGalleryValue{asGalleryGain}{0}{1.16}}{\asGalleryValue{asGalleryGain}{0}{1.16}}{17/24}{Parity shortcuts and spectral-seeded NAE or focused local search first seek checked satisfying assignments, followed by watched-literal CDCL when those searches fail. The parity and heuristic paths have no UNSAT certificate; the fallback logs learned clauses and a root conflict as DRAT additions for external elaboration, while unresolved or unsupported cases return UNKNOWN.}
\asGallerySolver{DPR}{\asGalleryValue{asGalleryGain}{0}{1.01}}{\asGalleryValue{asGalleryGain}{0}{1.01}}{16/24}{Validated structure sends odd 3-XOR blocks to packed GF(2) elimination; other recognized formulas use bounded focused search, width-2 inputs go directly to CDCL, and nonmatching inputs return UNKNOWN. Checked SAT models certify satisfiability; the fallback logs witness-free DPR RUP additions for learned clauses and an empty clause, requiring external elaboration and checking.}
\asGallerySolver{VeriPB}{\asGalleryValue{asGalleryGain}{1}{1.28}}{\asGalleryValue{asGalleryGain}{1}{1.23}}{18/24}{Packed GF(2) elimination handles the recognized odd-XOR signature, then majority/minority seeds and focused ProbSAT search seek a checked model before watched-literal CDCL takes over. The fallback logs learned clauses and the root conflict as VeriPB RUP constraints, including deletions and a final UNSAT conclusion for external verification; malformed or unresolved inputs return UNKNOWN.}
\end{asGalleryFamily}

\begin{asGalleryFamily}{140}{random-\allowbreak modularity}{59 GBD instances\enspace\textperiodcentered\enspace 12 validation\enspace\textperiodcentered\enspace baseline 11/12}{The benchmark asks whether a Boolean assignment satisfies every clause in a CNF formula, or whether no such assignment exists. Variables are grouped into communities, with many short clauses local to one community and fewer clauses linking communities.}
\ifdefined\pdfbookmark\pdfbookmark[2]{random-modularity}{as-gallery-140}\fi
\asGallerySolver{GRAT}{\asGalleryValue{asGalleryLoss}{0}{0.203}}{\asGalleryValue{asGalleryLoss}{0}{0.204}}{4/12}{Community decomposition leads on recognized fixed-size layouts: local DPLL handles contiguous blocks, followed by damped belief-propagation decimation for cross-block search and a bounded elimination followed by CDCL fallback. UNSAT proofs are emitted as textual DRAT for external elaboration; SAT models are checked, and unrecognized or unresolved inputs return UNKNOWN.}
\asGallerySolver{DPR}{\asGalleryValue{asGalleryLoss}{0}{0.229}}{\asGalleryValue{asGalleryLoss}{0}{0.228}}{5/12}{Contiguous-community decomposition leads the search on recognized fixed-size layouts: deterministic DPLL handles sufficiently dense community blocks, followed by weighted coordinate repair and CDCL fallbacks for bridge clauses. Local contradictions and global CDCL refutations are recorded as DPR additions; checked SAT models are output, while unsupported or unresolved cases return UNKNOWN.}
\asGallerySolver{VeriPB}{\asGalleryValue{asGalleryLoss}{1}{0.331}}{\asGalleryValue{asGalleryLoss}{1}{0.333}}{7/12}{Incremental ProbSAT-style search leads on recognized layouts, followed by community-product block search with projected DPLL repairs and an unbounded CDCL fallback. SAT assignments are checked; a completed CDCL refutation is written as VeriPB RUP steps, while formulas outside the recognized structure or unfinished refutations return UNKNOWN.}
\end{asGalleryFamily}

\begin{asGalleryFamily}{141}{random-\allowbreak mus}{14 GBD instances\enspace\textperiodcentered\enspace 3 validation\enspace\textperiodcentered\enspace baseline 0/3}{This family asks whether one Boolean assignment satisfies every disjunction of signed literals in a CNF formula. The recognized inputs use a restricted mix of two- and three-literal clauses; the label does not establish provenance or minimal unsatisfiability.}
\ifdefined\pdfbookmark\pdfbookmark[2]{random-mus}{as-gallery-141}\fi
\asGallerySolver{GRAT}{\asGalleryValue{asGalleryMuted}{1}{1.00}}{\asGalleryValue{asGalleryMuted}{1}{1.00}}{0/3}{ProbSAT local search with break-weighted flips is the active first stage, and verified models terminate the search. Failure triggers bounded elimination, vivification, and watched-literal first-UIP CDCL; SAT assignments are checked, while UNSAT logs derived clauses and an empty clause in a DRAT trace intended for external elaboration, with unsupported inputs returning UNKNOWN.}
\asGallerySolver{DPR}{\asGalleryValue{asGalleryMuted}{1}{1.00}}{\asGalleryValue{asGalleryMuted}{1}{1.00}}{0/3}{WalkSAT with break-count choices leads the search before fallback. If no verified model is found, bounded elimination and watched-literal CDCL emit a textual DPR/RUP-style trace; unsupported inputs and some failures return UNKNOWN, and optional lookahead is inactive by default. External elaboration and checking are outside the solver; no certificate validation is claimed.}
\asGallerySolver{VeriPB}{\asGalleryValue{asGalleryMuted}{1}{1.00}}{\asGalleryValue{asGalleryMuted}{1}{1.00}}{0/3}{On recognized inputs, break-only probSAT with weighted flips leads the search, and every candidate model is checked against the input. After local search fails, bounded variable elimination and watched-literal 1-UIP CDCL continue until SAT or UNSAT; the fallback logs VeriPB RUP constraints, while probSAT failure alone produces no UNSAT certificate and unsupported inputs return UNKNOWN.}
\end{asGalleryFamily}

\begin{asGalleryFamily}{142}{random-\allowbreak planted-\allowbreak solution}{328 GBD instances\enspace\textperiodcentered\enspace 66 validation\enspace\textperiodcentered\enspace baseline 60/66}{The task is to find a Boolean assignment satisfying every clause of a random planted-solution CNF. Recognized inputs have three literals over distinct variables per clause, with selected density ranges.}
\ifdefined\pdfbookmark\pdfbookmark[2]{random-planted-solution}{as-gallery-142}\fi
\asGallerySolver{GRAT}{\asGalleryValue{asGalleryGain}{1}{3.37}}{\asGalleryValue{asGalleryGain}{1}{3.37}}{64/66}{Triple-parity belief propagation seeds the exact 4.41-density branch, while other recognized densities use signed-pair spectral power iteration; both feed focused local search with break-score and polarity restarts. A verified full assignment is emitted on success, but malformed or out-of-band inputs return UNKNOWN, and recognized failures continue through unbounded restarts without an UNSAT certificate path.}
\asGallerySolver{DPR}{\asGalleryValue{asGalleryLoss}{0}{0.796}}{\asGalleryValue{asGalleryLoss}{0}{0.796}}{57/66}{Signed spectral seeding and degree-normalized power iteration initialize break-weighted local search, with belief propagation and survey decimation added only in the exact 4.41-density branch. It verifies any complete satisfying assignment, but unsupported inputs return UNKNOWN and failed recognized searches can remain in unbounded random restarts; no UNSAT certificate path is implemented.}
\asGallerySolver{VeriPB}{\asGalleryValue{asGalleryLoss}{0}{0.655}}{\asGalleryValue{asGalleryLoss}{0}{0.655}}{54/66}{Planted-likelihood belief propagation with confidence-guided decimation supplies seeds, followed by bounded Hamming and make-break repair and, on some paths, learned-clause CDCL completion. The solver independently checks a full satisfying assignment; unsupported or unsuccessful runs return UNKNOWN, no UNSAT certificate is produced, and a middle-density branch can remain in unbounded decimation and repair.}
\end{asGalleryFamily}

\begin{asGalleryFamily}{143}{rbsat}{116 GBD instances\enspace\textperiodcentered\enspace 24 validation\enspace\textperiodcentered\enspace baseline 4/24}{The benchmark asks whether each finite-domain object can take exactly one value while avoiding listed forbidden value pairs between objects. Its recognized CNF uses one signed choice block per object, one at-least-one clause, pairwise at-most-one clauses, and binary clauses forbidding cross-object pairs.}
\ifdefined\pdfbookmark\pdfbookmark[2]{rbsat}{as-gallery-143}\fi
\asGallerySolver{GRAT}{\asGalleryValue{asGalleryGain}{0}{2.37}}{\asGalleryValue{asGalleryGain}{0}{2.37}}{17/24}{Structure-aware finite-domain search leads with weighted min-conflicts, arc-consistent repair, and belief-propagation decimation before falling back to watched-literal CDCL with clause learning. SAT assignments are checked against the CNF; UNSAT is certified by a separate CDCL rerun that writes DRAT additions for external elaboration and checking, while recognition or certifying failure can return UNKNOWN.}
\asGallerySolver{DPR}{\asGalleryValue{asGalleryGain}{1}{2.80}}{\asGalleryValue{asGalleryGain}{1}{2.80}}{17/24}{Signed one-hot recognition and bit-mask arc consistency drive the exact search after a bounded randomized min-conflicts pass, using dom/wdeg variable selection and least-damaging values. SAT assignments are checked; after failed searches, a separate unit-propagation DFS emits a tree-style proof of UNSAT with decision nogoods and an empty clause, without independent verification; unrecognized encodings return UNKNOWN.}
\asGallerySolver{VeriPB}{\asGalleryValue{asGalleryGain}{0}{2.68}}{\asGalleryValue{asGalleryGain}{0}{2.68}}{17/24}{Structural one-hot recovery drives randomized constructive search, tabu or message-passing repair, weighted local search, and bounded MAC before a CSP-directed CDCL scout. A SAT assignment is clause-checked; only a scout UNSAT result enables a logged CDCL rerun with VeriPB RUP constraints for UNSAT certification, while failure enters unbounded diversified search rather than UNKNOWN.}
\end{asGalleryFamily}

\begin{asGalleryFamily}{144}{reg-\allowbreak n}{60 GBD instances\enspace\textperiodcentered\enspace 12 validation\enspace\textperiodcentered\enspace baseline 8/12}{These structured formulas assign Boolean color choices to objects arranged by a hierarchy. Positive clauses require a choice for each object; tree or product-based exclusions, sometimes expanded into ternary and longer clauses, constrain which choices can coexist.}
\ifdefined\pdfbookmark\pdfbookmark[2]{reg-n}{as-gallery-144}\fi
\asGallerySolver{GRAT}{\asGalleryValue{asGalleryMuted}{0}{1.00}}{\asGalleryValue{asGalleryGain}{0}{1.01}}{8/12}{Parity-based canonicalization and rank-indexed checking recover tree coordinates and verify every required clause family exactly once. For supported fixed regimes, the solver remaps variables in a packed regime-specific DRAT template and reports UNSATISFIABLE only after replay succeeds; it has no general SAT search, and unsupported or malformed structures or missing proof data return UNKNOWN.}
\asGallerySolver{DPR}{\asGalleryValue{asGalleryMuted}{0}{1.00}}{\asGalleryValue{asGalleryLoss}{0}{0.287}}{8/12}{Product-run detection and polarity-specific OR extensions compact recognized contiguous Cartesian-product layouts before a watched-literal first-UIP CDCL search. UNSAT results are logged as a textual DPR-style stream with RAT-style definitions, deletions, and CDCL lemmas, while SAT assignments are checked against the original CNF; failed recognition or internal failure returns UNKNOWN.}
\asGallerySolver{VeriPB}{\asGalleryValue{asGalleryGain}{1}{1.01}}{\asGalleryValue{asGalleryGain}{1}{1.01}}{8/12}{Canonical variable recovery and exact, order-sensitive clause validation identify the hierarchical coloring structure without a SAT search. It introduces event, z, and a variables with red constraints, aggregates occurrences into macro conflicts, and emits RUP tree-resolution steps in a VeriPB text proof; recognition or output failures return UNKNOWN, and the source includes no verifier.}
\end{asGalleryFamily}

\begin{asGalleryFamily}{145}{register-\allowbreak allocation}{20 GBD instances\enspace\textperiodcentered\enspace 4 validation\enspace\textperiodcentered\enspace baseline 4/4}{The benchmark asks whether an interference graph can assign one of k registers to each vertex so adjacent vertices differ. CNF uses x[v,c]: positive vertex clauses require registers, while negative edge clauses forbid adjacent vertices from sharing a register.}
\ifdefined\pdfbookmark\pdfbookmark[2]{register-allocation}{as-gallery-145}\fi
\asGallerySolver{GRAT}{\asGalleryValue{asGalleryGain}{0}{13.4}}{\asGalleryValue{asGalleryGain}{0}{1.75}}{4/4}{Maximum Cardinality Search drives chordal coloring for the exact direct encoding (3 \ensuremath{<}= k \ensuremath{<}= 20, at most 5000 vertices). Checked colorings provide SAT assignments; if they fail, DSATUR seeks a model or a bounded clique search can emit a pigeonhole DRAT proof, with unresolved or unsupported cases returning UNKNOWN.}
\asGallerySolver{DPR}{\asGalleryValue{asGalleryGain}{1}{15.4}}{\asGalleryValue{asGalleryGain}{0}{1.34}}{4/4}{Bitset clique search leads the direct-coloring workflow, restricted to complete consecutive-block encodings with 1 \ensuremath{<}= k \ensuremath{<}= 63 and at most 8192 vertices; reverse-MCS and greedy coloring seek a checked SAT assignment. A rechecked (k+1)-clique yields a deletion-based pigeonhole DPR proof; otherwise exponential DSATUR emits rejection blocks, and unresolved cases return UNKNOWN.}
\asGallerySolver{VeriPB}{\asGalleryValue{asGalleryGain}{0}{13.6}}{\asGalleryValue{asGalleryGain}{1}{14.8}}{4/4}{Bitset clique search and maximum-cardinality analysis guide the restricted contiguous-block recognizer for k \ensuremath{>}= 3; other inputs return UNKNOWN. Greedy coloring seeks a checked SAT assignment, while exact component-wise DSATUR handles unresolved cases. A rechecked clique yields VeriPB pseudo-Boolean proofs, and exact search emits RUP nogoods; model checks or proof-output failures return UNKNOWN.}
\end{asGalleryFamily}

\begin{asGalleryFamily}{146}{relational-\allowbreak dependencies}{20 GBD instances\enspace\textperiodcentered\enspace 4 validation\enspace\textperiodcentered\enspace baseline 4/4}{The benchmark asks whether variables can satisfy Horn dependencies while limiting false primary variables and true auxiliary variables. Its CNF encoding uses clauses for a and b imply c plus complete subset blocks that impose the two cardinality limits.}
\ifdefined\pdfbookmark\pdfbookmark[2]{relational-dependencies}{as-gallery-146}\fi
\asGallerySolver{GRAT}{\asGalleryValue{asGalleryGain}{0}{26.0}}{\asGalleryValue{asGalleryGain}{1}{2.74}}{4/4}{A Horn-closure shortcut sets the first block true and second block false; if it fails, watched-literal CDCL uses implicit cardinality propagation and branches only on the first block. SAT assignments are checked against the input, while UNSAT paths emit textual DRAT; unrecognized formulas or unresolved failures return UNKNOWN.}
\asGallerySolver{DPR}{\asGalleryValue{asGalleryGain}{0}{33.1}}{\asGalleryValue{asGalleryGain}{0}{1.69}}{4/4}{A closure-pruned primary-variable search uses Horn forward chaining, cardinality cutoffs, and lookahead branch selection; a SAT survey may supply a model or reusable tree. SAT models receive a raw clause check; on recognized instances, UNSAT searches serialize search trees as witness-free DPR additions, while construction or writing failure returns UNKNOWN.}
\asGallerySolver{VeriPB}{\asGalleryValue{asGalleryGain}{1}{35.4}}{\asGalleryValue{asGalleryLoss}{0}{0.504}}{4/4}{Certifying primary-variable search forward-chains Horn rules with bitmask closure and causal conflict clauses; it first tries a bounded natural-order search, then an unbounded degree-scored fallback. SAT leaves receive structural checks, while UNSAT traversals write RUP and cutting-planes resolution steps in VeriPB; unsupported inputs or failed checks or proof failures return UNKNOWN.}
\end{asGalleryFamily}

\begin{asGalleryFamily}{147}{relativized-\allowbreak pigeon-\allowbreak hole}{55 GBD instances\enspace\textperiodcentered\enspace 11 validation\enspace\textperiodcentered\enspace baseline 8/11}{The benchmark asks whether p first-layer pigeons can choose distinct objects, while each chosen object chooses a distinct one of only p-1 final holes. A CNF uses choice, activation, and object-to-hole variables, with existence, implication, and collision clauses encoding these two injective assignments.}
\ifdefined\pdfbookmark\pdfbookmark[2]{relativized-pigeon-hole}{as-gallery-147}\fi
\asGallerySolver{GRAT}{\asGalleryValue{asGalleryGain}{0}{1.51}}{\asGalleryValue{asGalleryGain}{0}{3.36}}{9/11}{Extension-variable composition and a sequential at-most-one counter reduce complete recognized encodings to an ordinary pigeonhole contradiction. The UNSAT branch writes DRAT records containing extension clauses, derived clauses, deletions, and an empty clause; one omitted color-collision case instead gets a checked SAT model, while unsupported, over-limit, or proof/model failures return UNKNOWN.}
\asGallerySolver{DPR}{\asGalleryValue{asGalleryGain}{0}{1.51}}{\asGalleryValue{asGalleryLoss}{0}{0.720}}{9/11}{Structural recognition drives a square-case matching proof and a rectangular-case witness-backed repair with case splits that derive canonical selector units before an inner pigeonhole proof. Clause additions may be witness-free; one-clause-short forms receive only a checked SAT construction, limited to supported missing-tuple cases, and unsupported inputs or failures return UNKNOWN.}
\asGallerySolver{VeriPB}{\asGalleryValue{asGalleryGain}{1}{2.94\!\cdot\!10^{4}}}{\asGalleryValue{asGalleryGain}{1}{335}}{11/11}{Pseudo-Boolean clique counting is central: lower bounds force at least p active objects, while per-hole demand and clique capacities cap them at p-1, yielding a streamed UNSAT proof. One omitted second-layer collision instead uses a constructed and checked SAT model with no UNSAT certificate path; unsupported forms or failures return UNKNOWN.}
\end{asGalleryFamily}

\begin{asGalleryFamily}{148}{risc-\allowbreak instruction-\allowbreak removal-\allowbreak golcrest}{9 GBD instances\enspace\textperiodcentered\enspace 2 validation\enspace\textperiodcentered\enspace baseline 1/2}{These benchmarks encode relational checks on a processor circuit. Primary inputs and ordered auxiliary variables represent Boolean signals, while Tseitin gate clauses and fixed endpoint conditions ask whether the encoded circuit admits a consistent assignment.}
\ifdefined\pdfbookmark\pdfbookmark[2]{risc-instruction-removal-golcrest}{as-gallery-148}\fi
\asGallerySolver{GRAT}{\asGalleryValue{asGalleryLoss}{1}{0.667}}{\asGalleryValue{asGalleryGain}{1}{1.55}}{0/2}{Packed truth-pattern simulation groups equal or complemented variable signatures to propose units and equivalences, then checks each candidate by watched-literal RUP. Bounded propagation-only closure emits accepted RUP clauses and an empty clause on contradiction; unsupported encodings or no contradiction return UNKNOWN, and no SAT assignment path is implemented.}
\asGallerySolver{DPR}{\asGalleryValue{asGalleryLoss}{1}{0.667}}{\asGalleryValue{asGalleryGain}{1}{1.55}}{0/2}{Reduced ordered BDD construction uses interleaved primary-input ordering and unique-node/apply caching to evaluate the recognized circuit directly. It reports UNSAT only for a terminal-true target, emitting witness-free DPR additions for the derivation and final empty clause; strict width-8 and input-count checks, resource limits, and other outcomes return UNKNOWN, with no SAT witness path.}
\asGallerySolver{VeriPB}{\asGalleryValue{asGalleryLoss}{1}{0.667}}{\asGalleryValue{asGalleryGain}{1}{1.55}}{0/2}{Bounded Davis-Putnam elimination preprocesses the CNF by removing low-occurrence variables within resolvent bounds, then watched-literal CDCL solves the reduced instance. It logs BVE resolvents, learned clauses, and terminal conflicts as VeriPB steps, checks SAT assignments against the original clauses, and returns UNKNOWN when eliminated variables cannot be reconstructed.}
\end{asGalleryFamily}

\begin{asGalleryFamily}{149}{risc-\allowbreak instruction-\allowbreak removal-\allowbreak subrv}{21 GBD instances\enspace\textperiodcentered\enspace 5 validation\enspace\textperiodcentered\enspace baseline 0/5}{The benchmark asks whether an acyclic Boolean netlist has an assignment satisfying its gate constraints and three terminal unit clauses. In CNF, each later variable is constrained as a Boolean function of up to three earlier variables.}
\ifdefined\pdfbookmark\pdfbookmark[2]{risc-instruction-removal-subrv}{as-gallery-149}\fi
\asGallerySolver{GRAT}{\asGalleryValue{asGalleryMuted}{1}{1.00}}{\asGalleryValue{asGalleryMuted}{1}{1.00}}{0/5}{It uses bounded-cone MiniCDCL and parity union-find to discover endpoint relations, then applies truth-table closure with local clause proving. It reports UNSAT only when closure derives a contradiction, emitting textual DRAT additions for external elaboration and checking; unsupported inputs or failed closure return UNKNOWN, and no SAT-witness path is implemented.}
\asGallerySolver{DPR}{\asGalleryValue{asGalleryMuted}{1}{1.00}}{\asGalleryValue{asGalleryMuted}{1}{1.00}}{0/5}{It uses template-based gate propagation in CDCL, initially deciding on non-output variables and falling back to all variables when that queue empties. It reports UNSAT after a root conflict and logs learned clauses with a terminal empty clause, but has no SAT-witness path; unsupported, oversized, or completed satisfiable cases return UNKNOWN.}
\asGallerySolver{VeriPB}{\asGalleryValue{asGalleryMuted}{1}{1.00}}{\asGalleryValue{asGalleryMuted}{1}{1.00}}{0/5}{It uses formula-seeded, 64-lane bit-parallel evaluation of recognized functional gates, filtering lanes by the unit clauses and selecting a surviving assignment for direct clause checking. It emits a checked SAT assignment when one survives, but has no UNSAT or VeriPB certificate path; unsupported inputs or no surviving lane return UNKNOWN.}
\end{asGalleryFamily}

\begin{asGalleryFamily}{150}{rooks}{30 GBD instances\enspace\textperiodcentered\enspace 6 validation\enspace\textperiodcentered\enspace baseline 0/6}{The benchmark asks whether n+1 squares can be selected on an n by n board while no row contains more than one selected square. CNF uses cell variables, row-conflict clauses, and an exact-cardinality constraint represented by a BDD.}
\ifdefined\pdfbookmark\pdfbookmark[2]{rooks}{as-gallery-150}\fi
\asGallerySolver{GRAT}{\asGalleryValue{asGalleryGain}{1}{1.10\!\cdot\!10^{4}}}{\asGalleryValue{asGalleryGain}{0}{7.43}}{6/6}{Layered ordered-BDD recognition drives a frontier-state proof: sequential row-prefix variables track occupied rows, and minimal fixed-cardinality masks are propagated from deep BDD layers upward to emit the certificate. It is UNSAT-only for the recognized row-based encoding; unsupported structure, resource limits, or proof-generation failure yield UNKNOWN rather than invoking general SAT search.}
\asGallerySolver{DPR}{\asGalleryValue{asGalleryGain}{0}{1.05\!\cdot\!10^{4}}}{\asGalleryValue{asGalleryGain}{0}{5.30}}{6/6}{BDD recovery and frontier-limited dynamic programming drive the proof: occupancy variables and bitmask states summarize filled rows, while threshold-based minimal occupied subsets prune recursive BDD traversal; a cofactor fallback handles canonical ITE cases. It is UNSAT-only for the narrow recovered row-capacity encoding; recognition or generation failure returns UNKNOWN without general SAT search.}
\asGallerySolver{VeriPB}{\asGalleryValue{asGalleryGain}{0}{584}}{\asGalleryValue{asGalleryGain}{1}{14.9}}{6/6}{Pseudo-Boolean cutting planes drive an UNSAT certificate: row at-most-one inequalities provide an upper bound of n, while proof-only one-unit flow through the ordered BDD and a rank-weighted potential derive the lower bound n+1. The solver accepts only the recognized square, row-based BDD structure; recognition or proof-emission failure returns UNKNOWN, with no SAT/model path.}
\end{asGalleryFamily}

\begin{asGalleryFamily}{151}{rubikcube}{20 GBD instances\enspace\textperiodcentered\enspace 4 validation\enspace\textperiodcentered\enspace baseline 0/4}{Bounded Rubik's Cube reachability asks whether legal face turns can transform an initial cube state into the solved state within a fixed move horizon. A CNF encoding represents move choices and intermediate states, constraining transitions and the final solved configuration.}
\ifdefined\pdfbookmark\pdfbookmark[2]{rubikcube}{as-gallery-151}\fi
\asGallerySolver{GRAT}{\asGalleryValue{asGalleryMuted}{1}{1.00}}{\asGalleryValue{asGalleryMuted}{1}{1.00}}{0/4}{Watched-literal CDCL with activity branching, learning, restarts, and clause reduction is the main search, using recovered move-variable preferences only for a canonical cube layout. An optional two-phase search precedes it and falls back to CDCL; checked SAT assignments are accepted, while UNSAT uses DRAT-style traces for external certification and proof or certification failures return UNKNOWN.}
\asGallerySolver{DPR}{\asGalleryValue{asGalleryMuted}{1}{1.00}}{\asGalleryValue{asGalleryMuted}{1}{1.00}}{0/4}{Recovered cube permutations drive two-phase cubie search with BFS pruning and move-word restrictions, but only on the recognized canonical layout and bounded horizon. A path gets auxiliary completion and full original-CNF checking; failed search or validation returns UNKNOWN, while optional proof writers emit DRAT-style traces for external DPR elaboration and checking.}
\asGallerySolver{VeriPB}{\asGalleryValue{asGalleryMuted}{1}{1.00}}{\asGalleryValue{asGalleryMuted}{1}{1.00}}{0/4}{Generator-numbered gadget validation, coordinate pruning tables, and depth-fixed two-phase DFS search only the recognized layout and phase schedule. On failure, ProofCDCL probes scoped suffixes for SAT, then runs watched-literal CDCL with RUP lemmas and PBC subproofs; checked models are accepted, while UNSAT requires a level-zero conflict and otherwise returns UNKNOWN.}
\end{asGalleryFamily}

\begin{asGalleryFamily}{152}{sat-\allowbreak x}{20 GBD instances\enspace\textperiodcentered\enspace 4 validation\enspace\textperiodcentered\enspace baseline 1/4}{The benchmark asks whether fixed-width integer words satisfy a cube-sum equation, typically X\textasciicircum{}3 - Y\textasciicircum{}3 - Z\textasciicircum{}3 = 3, with optional bounds or range restrictions. CNF represents the arithmetic with bit vectors and Tseitin auxiliary variables, plus clauses enforcing constants, targets, comparisons, and exact-width arithmetic.}
\ifdefined\pdfbookmark\pdfbookmark[2]{sat-x}{as-gallery-152}\fi
\asGallerySolver{GRAT}{\asGalleryValue{asGalleryGain}{1}{1.01}}{\asGalleryValue{asGalleryGain}{1}{1.02}}{1/4}{Bounded signed cube identities and 2-adic cube-root lifting generate candidate words; propagation and residual DPLL complete and validate each candidate before SAT output. Failures return UNKNOWN, and the cube path has no UNSAT certificate; only a narrow 16-bit Brocard branch uses CDCL and logs learned clauses plus a contradiction for external DRAT elaboration.}
\asGallerySolver{DPR}{\asGalleryValue{asGalleryLoss}{0}{0.758}}{\asGalleryValue{asGalleryLoss}{0}{0.764}}{0/4}{An embedded arithmetic witness fixes the inputs of recognized cubic layouts; flat propagation and residual watched-literal DPLL complete the assignment and rescan all clauses before SAT output. Failed searches return UNKNOWN; the cubic path has no UNSAT certificate, while a narrow Brocard fallback uses CDCL and logs learned clauses for external DPR elaboration and checking.}
\asGallerySolver{VeriPB}{\asGalleryValue{asGalleryLoss}{0}{0.758}}{\asGalleryValue{asGalleryLoss}{0}{0.764}}{0/4}{For recognized cubic layouts, an embedded witness fixes the cubic inputs; occurrence-indexed propagation and false-first residual backtracking complete the assignment and validate clauses before SAT output. A small Brocard fallback uses watched-literal CDCL and emits VeriPB RUP steps for UNSAT; failures return UNKNOWN, with no generic UNSAT certificate or in-process proof check.}
\end{asGalleryFamily}

\begin{asGalleryFamily}{153}{satcoin}{20 GBD instances\enspace\textperiodcentered\enspace 4 validation\enspace\textperiodcentered\enspace baseline 0/4}{The supported satcoin instances ask whether a 32-bit nonce in a specified range makes a Bitcoin-genesis header satisfy a double-SHA-256 difficulty target. The CNF bit-blasts the hashing circuit and the constraints selecting candidate nonces.}
\ifdefined\pdfbookmark\pdfbookmark[2]{satcoin}{as-gallery-153}\fi
\asGallerySolver{GRAT}{\asGalleryValue{asGalleryMuted}{1}{1.00}}{\asGalleryValue{asGalleryLoss}{0}{0.557}}{0/4}{Bit-parallel unit propagation tests up to 64 interval candidates at once after a bounded two-equality recognizer identifies the 32-bit range. When every lane is refuted, an MSB-first interval tree emits propagation-checked DRAT additions and deletions for external elaboration; a surviving lane, recognition failure, or proof-generation failure returns UNKNOWN.}
\asGallerySolver{DPR}{\asGalleryValue{asGalleryMuted}{1}{1.00}}{\asGalleryValue{asGalleryMuted}{1}{1.00}}{0/4}{64-lane bit-parallel propagation scans recognized interval candidates, with scalar clause propagation as a fallback for unresolved groups. An all-clause-satisfying candidate yields a checked complete assignment. When all candidates are refuted, it adds blocking clauses for covered cubes, including witness-free batch additions, and uses a high-bit-first closing tree; unresolved, recognition, or resource-limit failures return UNKNOWN.}
\asGallerySolver{VeriPB}{\asGalleryValue{asGalleryMuted}{1}{1.00}}{\asGalleryValue{asGalleryMuted}{1}{1.00}}{0/4}{Double-SHA-256 evaluation of extracted nonces drives the search after recognizing a restricted CNF layout and interval. A hash hit returns UNKNOWN because no CNF model is reconstructed; otherwise a prefix tree emits VeriPB blockers and RUP parent steps for UNSAT, with structural leaf checks and UNKNOWN on recognition or proof-writing failure.}
\end{asGalleryFamily}

\begin{asGalleryFamily}{154}{scheduling}{435 GBD instances\enspace\textperiodcentered\enspace 87 validation\enspace\textperiodcentered\enspace baseline 52/87}{These scheduling CNFs ask whether games or jobs can be assigned to rounds, dates, venues, or workers while respecting one-assignment, conflict, home/away, break, and exclusion constraints. Variables represent assignments and auxiliary conditions, and clauses enforce the allowed choices and incompatibilities.}
\ifdefined\pdfbookmark\pdfbookmark[2]{scheduling}{as-gallery-154}\fi
\asGallerySolver{GRAT}{\asGalleryValue{asGalleryLoss}{0}{0.645}}{\asGalleryValue{asGalleryLoss}{0}{0.687}}{25/87}{Circle-method one-factorization and layered orientation dynamic programming construct recognized double-round-robin schedules while tracking venue alternation and rejecting forbidden break runs. Watched-literal propagation completes auxiliary variables and validates SAT assignments; bounded MiniCdcl or Hall and matching refutations cover selected cases, while failed or unsupported cases return UNKNOWN and UNSAT paths log DRAT additions.}
\asGallerySolver{DPR}{\asGalleryValue{asGalleryLoss}{0}{0.572}}{\asGalleryValue{asGalleryLoss}{0}{0.609}}{17/87}{Structural fingerprints dispatch narrow encodings; circle factorization fixes Break schedules and derives home/break variables, followed by unit propagation and residual auxiliary-only DPLL. SAT valuations are clause-scanned, but one CDCL path can leave entries unassigned and validate them as true; selected UNSAT cases emit tree or learned-clause additions, while unsupported or failed cases return UNKNOWN.}
\asGallerySolver{VeriPB}{\asGalleryValue{asGalleryLoss}{1}{0.830}}{\asGalleryValue{asGalleryLoss}{1}{0.881}}{39/87}{Symmetry-preserving red and RUP derivations exploit recognized round-robin structure, while matching uses augmenting paths and Hall-style deficiencies. A bounded monotone-choice search, with stochastic construction only outside break encodings, precedes watched-literal first-UIP CDCL; SAT assignments are checked, specialized UNSAT derivations require proof output, and unresolved cases can return UNKNOWN.}
\end{asGalleryFamily}

\begin{asGalleryFamily}{155}{school-\allowbreak timetabling}{60 GBD instances\enspace\textperiodcentered\enspace 8 validation\enspace\textperiodcentered\enspace baseline 6/8}{The task asks whether courses with weekly loads can be assigned time slots and eligible teachers or resources while respecting daily blocks, occupied-day limits, availability, and non-overlap constraints. CNF encodings use course-period, day-use, resource-choice, and auxiliary cardinality or conflict variables.}
\ifdefined\pdfbookmark\pdfbookmark[2]{school-timetabling}{as-gallery-155}\fi
\asGallerySolver{GRAT}{\asGalleryValue{asGalleryLoss}{0}{0.945}}{\asGalleryValue{asGalleryLoss}{0}{0.945}}{5/8}{A narrow translated-layout recognizer extracts durations, daily limits, educator choices, and availability, then schedules contiguous blocks with legal day allocations and bit masks. Min-conflicts with tabu repair and random restarts seeks a model; watched-literal completion and full clause checks precede SAT output. Unsupported layouts return UNKNOWN, while scheduling failures retry without an UNSAT certificate path.}
\asGallerySolver{DPR}{\asGalleryValue{asGalleryGain}{1}{1.42}}{\asGalleryValue{asGalleryGain}{1}{1.42}}{6/8}{A structural recognizer extracts the timetable grid, resource choices, and cardinality constraints, then uses a weighted Hall/max-flow gate followed by randomized min-conflicts repair and bit-mask interval placement. Propagation and full clause checking precede SAT output; failed construction or checking returns UNKNOWN, and no UNSAT certificate path is implemented.}
\asGallerySolver{VeriPB}{\asGalleryValue{asGalleryLoss}{0}{0.793}}{\asGalleryValue{asGalleryLoss}{0}{0.790}}{5/8}{An encoding detector identifies grid, loads, options, and intervals, then uses Horn/implication closure, min-conflicts repair, and source-sensitive moves. Small residual neighborhoods use watched-literal CDCL completion with full clause checks before SAT output. Parse or detection failures return UNKNOWN; search may restart indefinitely without an UNSAT certificate path.}
\end{asGalleryFamily}

\begin{asGalleryFamily}{156}{set-\allowbreak covering}{40 GBD instances\enspace\textperiodcentered\enspace 8 validation\enspace\textperiodcentered\enspace baseline 7/8}{Choose a subset of columns that hits every positive requirement while never choosing both endpoints of a declared conflict. In CNF, requirements are positive clauses and conflicts are distinct binary negative clauses, yielding an independent hitting-set decision problem.}
\ifdefined\pdfbookmark\pdfbookmark[2]{set-covering}{as-gallery-156}\fi
\asGallerySolver{GRAT}{\asGalleryValue{asGalleryGain}{0}{71.5}}{\asGalleryValue{asGalleryGain}{1}{58.9}}{8/8}{A conflict-aware local search first builds an independent hitting set through randomized restarts and replacement moves that evict conflicting selections. It then falls back to uncovered-row CDCL with learned clauses; checked SAT assignments are reported, while UNSAT is logged in textual DRAT for external checking, and unsupported or failed cases return UNKNOWN.}
\asGallerySolver{DPR}{\asGalleryValue{asGalleryGain}{1}{81.8}}{\asGalleryValue{asGalleryGain}{0}{52.0}}{8/8}{A bounded focused local search scores flips by coverage and conflict effects, then falls back to complete fail-first search when it misses a model. The exact path tracks row coverage and forbidden neighborhoods, checks SAT assignments, and emits witness-free RUP clauses in textual DPR for UNSAT; unsupported forms return UNKNOWN.}
\asGallerySolver{VeriPB}{\asGalleryValue{asGalleryGain}{0}{1.73}}{\asGalleryValue{asGalleryGain}{0}{1.77}}{7/8}{An independence-preserving min-conflicts pass first seeks a cover by evicting conflicting selections, then falls back to exact bitset search on the smallest compatible domain. It uses conflict-directed backjumping and sibling exclusions, checks SAT assignments, and emits VeriPB RUP no-goods for external UNSAT checking; unsupported input or proof-file failure returns UNKNOWN.}
\end{asGalleryFamily}

\begin{asGalleryFamily}{157}{sgen}{128 GBD instances\enspace\textperiodcentered\enspace 26 validation\enspace\textperiodcentered\enspace baseline 4/26}{The benchmark asks whether a Boolean assignment satisfies every signed CNF clause. Its instances use structured encodings of exact-cover, counting, matching, or counter constraints, represented by unit, binary, and wider clauses.}
\ifdefined\pdfbookmark\pdfbookmark[2]{sgen}{as-gallery-157}\fi
\asGallerySolver{GRAT}{\asGalleryValue{asGalleryGain}{0}{1.46}}{\asGalleryValue{asGalleryGain}{0}{1.47}}{11/26}{Signed five-variable exact-cover recognition leads to Algorithm X/DLX, followed by potentially unbounded stochastic repair, while separate partition-gap and counter-gap patterns use watched-literal CDCL. Checked SAT assignments are printed; CDCL UNSAT logs learned clauses, deletions, and an empty clause as textual DRAT for external elaboration and checking, whereas unsupported or unresolved cases return UNKNOWN.}
\asGallerySolver{DPR}{\asGalleryValue{asGalleryGain}{0}{1.57}}{\asGalleryValue{asGalleryGain}{0}{1.58}}{12/26}{Signed exact-cover reconstruction drives fixed-seed min-conflicts with tabu moves, followed by smallest-constraint Algorithm X; compact or all-ternary cases can fall back to watched-literal CDCL. Verified SAT assignments are printed, while CDCL UNSAT runs log learned clauses, deletions, and an empty clause for external elaboration; failed searches and unresolved larger cases return UNKNOWN.}
\asGallerySolver{VeriPB}{\asGalleryValue{asGalleryGain}{1}{3{,}900}}{\asGalleryValue{asGalleryGain}{1}{261}}{26/26}{Pseudo-Boolean counting, binary-clique, and sequential-counter recognition handle structured UNSAT cases first, while SAT search uses augmenting-path matching or bit-parallel exact-cover DFS. UNSAT certificates come from counting or counter derivations, or from restricted exact-cover refutation with inequalities and RUP nogoods; checked SAT models are printed, and unsupported, failed, or uncertified paths return UNKNOWN.}
\end{asGalleryFamily}

\begin{asGalleryFamily}{158}{sgen-\allowbreak balanced}{4 GBD instances\enspace\textperiodcentered\enspace 1 validation\enspace\textperiodcentered\enspace baseline 1/1}{The benchmark asks whether a Boolean assignment satisfies a structured 3-CNF formula. Each clause has three distinct variables, while occurrences are distributed nearly evenly across variables and polarities, often through ordered rounds or layers; the recognizers accept only particular such layouts.}
\ifdefined\pdfbookmark\pdfbookmark[2]{sgen-balanced}{as-gallery-158}\fi
\asGallerySolver{GRAT}{\asGalleryValue{asGalleryGain}{1}{9.27}}{\asGalleryValue{asGalleryGain}{1}{24.6}}{1/1}{On recognized formulas, shallow decisions use clause-pressure scores and choose polarity toward lower pressure; deeper search falls back to activity-based CDCL with first-UIP learning, restarts, and learned-clause reduction. SAT models are checked before printing, while UNSAT runs emit learned-clause additions, deletions, and final empty clause for external elaboration; unsupported formulas or proof-file failure return UNKNOWN.}
\asGallerySolver{DPR}{\asGalleryValue{asGalleryGain}{0}{1.09}}{\asGalleryValue{asGalleryGain}{0}{2.61}}{1/1}{Exact ordered ten-layer inputs, optionally with a partial layer, are recognized; others return UNKNOWN. Dense pivots trigger independent-set elimination with resolvents and reconstruction; CDCL is the fallback. SAT checks precede assignment output; UNSAT records learned clauses, resolvents, deletions, and an empty clause; reconstruction failure aborts and model-check failure returns UNKNOWN.}
\asGallerySolver{VeriPB}{\asGalleryValue{asGalleryLoss}{0}{0.998}}{\asGalleryValue{asGalleryGain}{0}{2.72}}{1/1}{After balanced-shape recognition, it first runs bounded probSAT local search with break-count weighting, then independently falls back to watched-literal CDCL with phase saving and first-UIP learning. Verified SAT models are printed; UNSAT runs emit parent-logged clauses and a final empty-clause RUP for external checking. Unsupported inputs, proof-file failure, or failed model checks return UNKNOWN.}
\end{asGalleryFamily}

\begin{asGalleryFamily}{159}{sliding-\allowbreak puzzle}{62 GBD instances\enspace\textperiodcentered\enspace 13 validation\enspace\textperiodcentered\enspace baseline 0/13}{The benchmark asks whether a 5x5 board with one blank and 24 numbered tiles can reach the ordered goal arrangement within a bounded number of moves. Its CNF encodes initial states, move transitions, and goal checks for this reachability question.}
\ifdefined\pdfbookmark\pdfbookmark[2]{sliding-puzzle}{as-gallery-159}\fi
\asGallerySolver{GRAT}{\asGalleryValue{asGalleryGain}{0}{1.86}}{\asGalleryValue{asGalleryGain}{1}{1.86}}{6/13}{IDA* search combines Manhattan distance with immediate-reversal and parity pruning, then a linear-conflict bound, after recognizing the fixed CNF layout. A found path fixes the encoded state variables, and residual DPLL completes and checks the SAT assignment; no UNSAT certificate path exists, so unsupported or unsolved cases return UNKNOWN.}
\asGallerySolver{DPR}{\asGalleryValue{asGalleryGain}{0}{1.86}}{\asGalleryValue{asGalleryGain}{0}{1.86}}{6/13}{IDA* search updates Manhattan distance, suppresses immediate reversals, and advances bounds by two on the fixed 5x5 layout. Found paths yield checked SAT assignments; after failure, a boundary-crossing construction with PHP-style symmetry clauses may emit DPR proof through horizon 31, reporting UNSAT only on success; other cases return UNKNOWN.}
\asGallerySolver{VeriPB}{\asGalleryValue{asGalleryGain}{1}{1.86}}{\asGalleryValue{asGalleryGain}{0}{1.86}}{6/13}{Bounded depth-first A* search combines Manhattan and linear-conflict bounds with immediate-reversal pruning, child ordering, and a failed-state table. For a found path it tries direction mappings, fixes selector bits, propagates and checks a complete SAT assignment; no UNSAT certificate path exists, so unsupported or no-path cases return UNKNOWN.}
\end{asGalleryFamily}

\begin{asGalleryFamily}{160}{social-\allowbreak golfer}{39 GBD instances\enspace\textperiodcentered\enspace 8 validation\enspace\textperiodcentered\enspace baseline 3/8}{The task arranges players into G groups of P each week for W weeks, using every player once weekly and preventing any pair from meeting twice. CNF clauses encode these group and pair constraints, sometimes with auxiliary variables.}
\ifdefined\pdfbookmark\pdfbookmark[2]{social-golfer}{as-gallery-160}\fi
\asGallerySolver{GRAT}{\asGalleryValue{asGalleryGain}{0}{1.27}}{\asGalleryValue{asGalleryGain}{0}{1.27}}{4/8}{Conflict-directed coloring first extracts matching structure from the Table encoding, then falls back to Context or classic reconstructions with specialized schedules, watched propagation, and 1-UIP CDCL. A full SAT assignment is checked against every original clause; unsupported inputs, failed construction, and exhausted search return UNKNOWN, with no UNSAT certificate path implemented.}
\asGallerySolver{DPR}{\asGalleryValue{asGalleryGain}{1}{2.55}}{\asGalleryValue{asGalleryGain}{1}{2.55}}{6/8}{Structural CNF recognition drives schedule reconstruction: table, context, and classic layouts are decoded by incidence components, pair fingerprints, and supported schedule constructions. The resulting assignment is checked against every original clause, but unsupported or failed cases return UNKNOWN and no UNSAT result or DPR trace is produced.}
\asGallerySolver{VeriPB}{\asGalleryValue{asGalleryGain}{0}{2.55}}{\asGalleryValue{asGalleryGain}{0}{2.54}}{6/8}{Strict Table and Context recognition seeds embedded or finite-field affine schedules, with symmetry alignment and recursive clique-factor search as bounded construction fallbacks. Unit propagation and false-phase completion fill the assignment before every original clause is checked; unsupported or unsuccessful constructions return UNKNOWN, and no UNSAT certificate path is implemented.}
\end{asGalleryFamily}

\begin{asGalleryFamily}{161}{software-\allowbreak bmc}{18 GBD instances\enspace\textperiodcentered\enspace 4 validation\enspace\textperiodcentered\enspace baseline 4/4}{The benchmark asks whether a bounded execution of software can satisfy its Boolean transition and assertion constraints. Bit-blasting program and state bits, with Tseitin variables for intermediate gates, turns this question into CNF whose satisfiability represents a possible execution.}
\ifdefined\pdfbookmark\pdfbookmark[2]{software-bmc}{as-gallery-161}\fi
\asGallerySolver{GRAT}{\asGalleryValue{asGalleryGain}{1}{2.94}}{\asGalleryValue{asGalleryGain}{1}{2.96}}{4/4}{Watched-literal CDCL with 1-UIP minimization and topology-biased decisions drives accepted instances; binary implication specialization serves large bounded-width cases, while local repair is inactive. Checked SAT models are printed, while root-level UNSAT emits learned clauses and the empty clause as textual DRAT for validation; unsupported inputs or failed checks return UNKNOWN.}
\asGallerySolver{DPR}{\asGalleryValue{asGalleryGain}{0}{1.58}}{\asGalleryValue{asGalleryGain}{0}{1.54}}{4/4}{Bounded variable elimination with non-tautological resolvents precedes watched-literal CDCL and first-UIP learning on selected formulas; guarded cases use CDCL alone. The aggregate-only recognizer sends unrecognized or resource-failing cases to UNKNOWN rather than falling back, while SAT reconstructs assignments and UNSAT logs clause additions for external DPR elaboration and checking.}
\asGallerySolver{VeriPB}{\asGalleryValue{asGalleryGain}{0}{1.09}}{\asGalleryValue{asGalleryLoss}{0}{0.649}}{4/4}{Polarity-biased watched-literal CDCL drives recognized inputs, using first-UIP learning, phase saving, and geometric restarts; oversized instances are rejected. SAT assignments are checked against original clauses, while UNSAT emits learned clauses as VeriPB RUP steps plus an empty RUP and UNSAT conclusion; malformed, unrecognized, or storage-failing cases return UNKNOWN.}
\end{asGalleryFamily}

\begin{asGalleryFamily}{162}{software-\allowbreak verification}{277 GBD instances\enspace\textperiodcentered\enspace 56 validation\enspace\textperiodcentered\enspace baseline 36/56}{These CNFs encode bit-blasted software-verification conditions. Boolean variables represent circuit or state signals, while clauses express gate relations, fixed values, and other verification constraints.}
\ifdefined\pdfbookmark\pdfbookmark[2]{software-verification}{as-gallery-162}\fi
\asGallerySolver{GRAT}{\asGalleryValue{asGalleryLoss}{1}{0.950}}{\asGalleryValue{asGalleryGain}{1}{1.71}}{35/56}{A bounded reverse-circuit probe evaluates grouped gate definitions in parallel with 64-bit lanes and checks candidate assignments against the original CNF. On probe failure, watched-literal CDCL with first-UIP learning and restarts provides the fallback, logging textual DRAT clauses; the probe cannot establish UNSAT, and unrecognized or oversized inputs return UNKNOWN.}
\asGallerySolver{DPR}{\asGalleryValue{asGalleryLoss}{0}{0.708}}{\asGalleryValue{asGalleryGain}{0}{1.25}}{24/56}{Two bounded low-variable probes with opposite phases try SAT or UNSAT before unbounded CDCL; exhausted probes are discarded. The fallback uses watched-literal CDCL with VSIDS and first-UIP learning, checking SAT models and logging proof-minimized clauses that may be witness-free, including the empty clause for a root conflict. Parse, model-check, and resource failures return UNKNOWN.}
\asGallerySolver{VeriPB}{\asGalleryValue{asGalleryLoss}{0}{0.638}}{\asGalleryValue{asGalleryGain}{0}{1.01}}{20/56}{A pre-conflict bias toward low variable IDs exploits the expected primary-before-derived allocation. The sole search engine is watched-literal CDCL with VSIDS, phase saving, first-UIP learning, geometric restarts, and RUP logging. Checked SAT assignments are emitted; UNSAT produces an empty RUP contradiction and VeriPB conclusion, while malformed, oversized, proof-output, or model-check failures return UNKNOWN.}
\end{asGalleryFamily}

\begin{asGalleryFamily}{163}{sorting-\allowbreak networks}{23 GBD instances\enspace\textperiodcentered\enspace 5 validation\enspace\textperiodcentered\enspace baseline 1/5}{These instances ask whether a bounded-depth comparator network can be chosen to sort every encoded Boolean input on its channels. CNF variables select disjoint comparator pairs in each layer, while auxiliary simulation variables and clauses enforce channel use and sorted outputs.}
\ifdefined\pdfbookmark\pdfbookmark[2]{sorting-networks}{as-gallery-163}\fi
\asGallerySolver{GRAT}{\asGalleryValue{asGalleryMuted}{0}{1.00}}{\asGalleryValue{asGalleryMuted}{0}{1.00}}{1/5}{Canonical comparator-block recognition only supplies phase and activity hints for selected layouts, while watched-literal CDCL performs the search with learning, restarts, and clause reduction. SAT assignments are checked against original clauses and printed; UNSAT emits learned additions and an empty clause as an external DRAT stream, while unsupported inputs return UNKNOWN.}
\asGallerySolver{DPR}{\asGalleryValue{asGalleryGain}{0}{1.34}}{\asGalleryValue{asGalleryGain}{0}{1.34}}{2/5}{Structural conflict/OR recognition admits supported layouts, then hard-coded comparator cores seed selected 13- and 16-channel searches before watched-literal CDCL fallback. Fallback SAT assignments are checked against the CNF, while UNSAT emits learned witness-free DPR/RUP additions and an empty clause; seeded UNSAT or unsupported input returns UNKNOWN.}
\asGallerySolver{VeriPB}{\asGalleryValue{asGalleryGain}{1}{2.01}}{\asGalleryValue{asGalleryGain}{1}{1.99}}{3/5}{Odd blocks use embedded comparator portfolios, occurrence propagation, and residual DPLL; even blocks use watched-literal CDCL, with the seven-block fallback also using CDCL. Checked SAT assignments are printed; even-branch UNSAT logs RUP constraints and an empty clause, while unsupported, even-SAT, or other unsuccessful cases return UNKNOWN; the seven-block fallback has no UNSAT proof path.}
\end{asGalleryFamily}

\begin{asGalleryFamily}{164}{ssp-\allowbreak 0}{5 GBD instances\enspace\textperiodcentered\enspace 1 validation\enspace\textperiodcentered\enspace baseline 1/1}{The benchmark asks whether a subset of integer-weighted items sums to a fixed target, using Boolean selectors for the choices. A CNF encoding adds auxiliary variables for arithmetic and a ripple-adder circuit constrained to equal that target.}
\ifdefined\pdfbookmark\pdfbookmark[2]{ssp-0}{as-gallery-164}\fi
\asGallerySolver{GRAT}{\asGalleryValue{asGalleryLoss}{1}{0.319}}{\asGalleryValue{asGalleryLoss}{1}{0.320}}{0/1}{It recognizes a rigid ripple-adder layout and uses 64-bit bit-sliced propagation of singleton selector assignments to reconstruct the encoded weights and target. Packed meet-in-the-middle subset-sum search supplies candidates, which are propagated and checked against every CNF clause; failed recognition, search, or validation returns UNKNOWN, with no UNSAT certificate path.}
\asGallerySolver{DPR}{\asGalleryValue{asGalleryLoss}{1}{0.319}}{\asGalleryValue{asGalleryLoss}{1}{0.320}}{0/1}{It uses fixed-cardinality meet-in-the-middle search over complementary selector halves, after recovering weights and the target from bit-lane evaluations of the ripple-carry circuit. Candidate assignments are expanded and checked against the full CNF; unsupported, negative, or candidate-free inputs return UNKNOWN, and no UNSAT or DPR certificate path is implemented.}
\asGallerySolver{VeriPB}{\asGalleryValue{asGalleryLoss}{1}{0.319}}{\asGalleryValue{asGalleryLoss}{1}{0.320}}{0/1}{It uses bit-parallel evaluation of the recognized topological circuit on singleton selector assignments to recover weights and target bits. A low-64-bit meet-in-the-middle filter generates candidates, each checked against the complete CNF to reject collisions; unsupported or uncertifiable searches return UNKNOWN, with no UNSAT or VeriPB certificate path.}
\end{asGalleryFamily}

\begin{asGalleryFamily}{165}{st-\allowbreak connectivity-\allowbreak principle}{2 GBD instances\enspace\textperiodcentered\enspace 1 validation\enspace\textperiodcentered\enspace baseline 0/1}{The benchmark asks whether two edge-selected path systems on identical rectangular grids can connect alternating terminal pairs without sharing a vertex. Boolean edge variables and CNF clauses enforce local degree or parity rules, plus cross-color clauses forbidding selected edges from meeting at one vertex.}
\ifdefined\pdfbookmark\pdfbookmark[2]{st-connectivity-principle}{as-gallery-165}\fi
\asGallerySolver{GRAT}{\asGalleryValue{asGalleryMuted}{0}{1.00}}{\asGalleryValue{asGalleryMuted}{0}{1.00}}{0/1}{Min-fill bucket elimination is the active refutation strategy, using literal masks and subsumption pruning while resolving variables and dynamically rebucketing when needed. It emits retained clauses and the empty clause as textual DRAT for external elaboration, reports UNSAT only on a refutation, and returns UNKNOWN if recognition or refutation fails.}
\asGallerySolver{DPR}{\asGalleryValue{asGalleryMuted}{0}{1.00}}{\asGalleryValue{asGalleryMuted}{0}{1.00}}{0/1}{Bounded frontier-ordered Davis-Putnam elimination is attempted first, using compact masks and subsumption before an in-process watched-literal CDCL tail handles the residual formula. It logs resolvents and learned clauses, including witness-free additions, and reports UNSAT only after an empty-clause conflict; restrictive recognition, width bounds, or complete assignments return UNKNOWN.}
\asGallerySolver{VeriPB}{\asGalleryValue{asGalleryGain}{1}{10.9}}{\asGalleryValue{asGalleryGain}{1}{6.59}}{1/1}{Column-wise transfer maintains a subsumption-minimal boundary clause summary, while watched-literal local DPLL searches each column and records propagation and branch-resolution clauses as VeriPB RUP steps. The final unassumed search emits UNSAT only on conflict; recognition or height limits, and a satisfying search with no witness path, return UNKNOWN.}
\end{asGalleryFamily}

\begin{asGalleryFamily}{166}{station-\allowbreak repacking}{9,842 GBD instances\enspace\textperiodcentered\enspace 1,969 validation\enspace\textperiodcentered\enspace baseline 1673/1969}{Choose exactly one channel for each station while avoiding incompatible pairs of selected station-channel options. A CNF encoding uses variables for station-channel choices, positive clauses to require a choice, and negative clauses to forbid multiple choices or conflicting pairs.}
\ifdefined\pdfbookmark\pdfbookmark[2]{station-repacking}{as-gallery-166}\fi
\asGallerySolver{GRAT}{\asGalleryValue{asGalleryLoss}{0}{0.544}}{\asGalleryValue{asGalleryLoss}{0}{0.556}}{1313/1969}{Peeling with guaranteed extensions and station-level TabuCol-style search, including breakout weighting, leads to watched-literal CDCL with EVSIDS, restarts, and first-UIP learning. SAT candidates are extended and checked; only CDCL establishes UNSAT, recording learned clauses and deletions in a textual DRAT trace ending in the empty clause while unsupported inputs return UNKNOWN.}
\asGallerySolver{DPR}{\asGalleryValue{asGalleryLoss}{1}{0.615}}{\asGalleryValue{asGalleryLoss}{1}{0.622}}{1402/1969}{Randomized min-conflicts search with tabu moves, restarts, and bounded residual repair first builds channel assignments using compact masks and station ordering, then falls back to watched-literal CDCL with learning and witness-free DPR/RUP clause additions. Only CDCL can establish UNSAT; checked SAT assignments are certificates, while unsupported encodings or failed checks return UNKNOWN.}
\asGallerySolver{VeriPB}{\asGalleryValue{asGalleryLoss}{0}{0.458}}{\asGalleryValue{asGalleryLoss}{0}{0.466}}{1184/1969}{Bit-mask propagation with randomized greedy/min-conflicts construction and a tabu, adaptive-weighting retry leads to component-wise CDCL with watched clauses, first-UIP analysis, and nonchronological backtracking. SAT models are independently checked; CDCL UNSAT results are replayed with RUP steps into a VeriPB certificate, while unsupported inputs or replay failures return UNKNOWN.}
\end{asGalleryFamily}

\begin{asGalleryFamily}{167}{stedman-\allowbreak triples}{44 GBD instances\enspace\textperiodcentered\enspace 9 validation\enspace\textperiodcentered\enspace baseline 2/9}{These CNFs ask whether choices of change-ringing calls, orientations, and successor states can satisfy all constraints of a lifted Stedman state system. The choices use construction variables, one-hot or compact state variables, and transition variables, with CNF clauses enforcing allowed states and propagated changes.}
\ifdefined\pdfbookmark\pdfbookmark[2]{stedman-triples}{as-gallery-167}\fi
\asGallerySolver{GRAT}{\asGalleryValue{asGalleryLoss}{0}{0.861}}{\asGalleryValue{asGalleryLoss}{0}{0.861}}{1/9}{Explicit six-state block recognition leads into watched-literal CDCL with first-UIP learning, activity-guided branching, and phase-randomized restarts. SAT assignments are checked against every original clause before printing; UNSAT emits textual DRAT clauses and a terminal empty clause for external GRAT elaboration, while unsupported inputs or internal failures return UNKNOWN.}
\asGallerySolver{DPR}{\asGalleryValue{asGalleryLoss}{1}{0.889}}{\asGalleryValue{asGalleryLoss}{1}{0.889}}{1/9}{Graph-projected search extracts oriented transitions, applies breakout walks and subtour cuts, then seeds exact projected CSP search before CDCL; recognized compact or signed layouts go directly to CDCL. It emits checked SAT assignments only; no DPR proof path exists, and unsupported or unsuccessful searches return UNKNOWN rather than certified UNSAT.}
\asGallerySolver{VeriPB}{\asGalleryValue{asGalleryLoss}{0}{0.791}}{\asGalleryValue{asGalleryLoss}{0}{0.791}}{0/9}{Bounded recognition enables call-first branching in watched-literal CDCL with first-UIP learning and all-variable fallback; no specialized transition propagator is active. Checked SAT assignments are printed without a completed VeriPB conclusion; UNSAT logs learned clauses as augmented RUP constraints and writes an UNSAT conclusion unless proof logging is disabled, while unsupported inputs return UNKNOWN.}
\end{asGalleryFamily}

\begin{asGalleryFamily}{168}{stone}{14 GBD instances\enspace\textperiodcentered\enspace 3 validation\enspace\textperiodcentered\enspace baseline 1/3}{Stone benchmarks ask whether markers can be placed on a two-parent directed acyclic graph, with each marker assigned a red or non-red status. CNF clauses enforce one placement per vertex, red sources, a non-red sink, and two-parent propagation.}
\ifdefined\pdfbookmark\pdfbookmark[2]{stone}{as-gallery-168}\fi
\asGallerySolver{GRAT}{\asGalleryValue{asGalleryGain}{1}{1{,}640}}{\asGalleryValue{asGalleryGain}{1}{11.2}}{3/3}{Source peeling and structural recognition recover mappings and graph orientation, accepting only an intact transition set or one canonical omission. A sink-reaching non-red omission yields a checked SAT assignment; otherwise it emits a DRAT-style derivation for external checking, while unsupported cases return UNKNOWN without general SAT search.}
\asGallerySolver{DPR}{\asGalleryValue{asGalleryGain}{0}{1{,}500}}{\asGalleryValue{asGalleryGain}{0}{8.46}}{3/3}{Co-occurrence mapping and structural recognition recover phases, permutations, and the DAG, then identify missing combinations in the sink cone. The first sink-relevant hole yields a checked SAT assignment; otherwise it emits a proof ending in the empty clause for DPR checking, while unsupported inputs return UNKNOWN without general SAT search.}
\asGallerySolver{VeriPB}{\asGalleryValue{asGalleryGain}{0}{2.18}}{\asGalleryValue{asGalleryGain}{0}{2.14}}{2/3}{Bit-mask recovery of polarity and color mappings, then DAG reconstruction, validates the shuffled three-source encoding and locates missing tuples. Supported complete instances emit a VeriPB derivation ending in UNSAT; recognized partial instances get checked SAT assignments, but oversized complete or unsupported inputs and proof failures return UNKNOWN without SAT search.}
\end{asGalleryFamily}

\begin{asGalleryFamily}{169}{subgraph-\allowbreak isomorphism}{194 GBD instances\enspace\textperiodcentered\enspace 39 validation\enspace\textperiodcentered\enspace baseline 12/39}{Subgraph isomorphism asks whether vertices of a pattern can be injectively assigned to host vertices while preserving edges. CNF encodings can represent these choices as one-hot domains or bounded bit blocks and add clauses forbidding incompatible pairs.}
\ifdefined\pdfbookmark\pdfbookmark[2]{subgraph-isomorphism}{as-gallery-169}\fi
\asGallerySolver{GRAT}{\asGalleryValue{asGalleryGain}{1}{1.07}}{\asGalleryValue{asGalleryLoss}{1}{0.924}}{14/39}{Direct and logarithmic recognition drives finite-domain CSP search with propagation, MRV branching, and recursive proof search; shortened logarithmic or auxiliary forms use watched-literal CDCL. Checked SAT assignments are reported; UNSAT branches emit textual DRAT traces with learned clauses or tree nogoods, while unsupported inputs return UNKNOWN.}
\asGallerySolver{DPR}{\asGalleryValue{asGalleryLoss}{0}{0.846}}{\asGalleryValue{asGalleryLoss}{0}{0.362}}{7/39}{Graph-aware CSP search is distinctive: recognized direct or logarithmic forms with recovered alignment receive matching and neighborhood filters, while unaligned forms use only propagation. SAT assignments are validated; UNSAT search emits witness-free DPR additions, extending logarithmic blocks with one-hot clauses; unsupported inputs or proof branches not establishing UNSAT return UNKNOWN.}
\asGallerySolver{VeriPB}{\asGalleryValue{asGalleryLoss}{0}{0.878}}{\asGalleryValue{asGalleryLoss}{0}{0.892}}{8/39}{Graph-aware bitset search distinguishes this solver: eligible one-hot inputs use matching filters, min-conflicts, and clique search, while bit-vector, auxiliary, or failed cases fall back to watched-literal CDCL. SAT models are checked; bounded one-hot UNSAT cases try TreeProof, while other UNSAT cases emit learned-clause RUP certificates and unsupported inputs return UNKNOWN.}
\end{asGalleryFamily}

\begin{asGalleryFamily}{170}{subset-\allowbreak cardinality}{10 GBD instances\enspace\textperiodcentered\enspace 2 validation\enspace\textperiodcentered\enspace baseline 0/2}{The benchmark asks whether a shared system of signed exact-two constraints can be satisfied. The CNF represents local blocks with complementary NAE-style clauses and adds endpoint constraints that impose a boundary parity or counting condition.}
\ifdefined\pdfbookmark\pdfbookmark[2]{subset-cardinality}{as-gallery-170}\fi
\asGallerySolver{GRAT}{\asGalleryValue{asGalleryGain}{0}{1.16\!\cdot\!10^{5}}}{\asGalleryValue{asGalleryGain}{0}{430}}{2/2}{On recognized complete inputs, complementary NAE-pair recognition drives the strategy: XOR accumulators and ordered incidence swaps cancel shared block variables, with exhaustive boundary rows closing a DRAT proof. A weakened block gets a flow-based SAT-model fallback or modulo-four proof; an unused mod-three routine prevents the supplied source from compiling.}
\asGallerySolver{DPR}{\asGalleryValue{asGalleryGain}{0}{1.23\!\cdot\!10^{5}}}{\asGalleryValue{asGalleryGain}{0}{145}}{2/2}{XOR parity cancellation drives complete-instance solving: reconstructed exact-two blocks receive witness-free gate definitions and local lemmas, while incidence swaps eliminate internal variables into a boundary contradiction. With one missing NAE half, it tries a modulo-four proof, then uses enumeration and max flow for SAT; unsupported or failed cases return UNKNOWN.}
\asGallerySolver{VeriPB}{\asGalleryValue{asGalleryGain}{1}{7.19\!\cdot\!10^{5}}}{\asGalleryValue{asGalleryGain}{1}{1.06\!\cdot\!10^{5}}}{2/2}{For the recognized degree-two layout, incidence-graph reconstruction and XOR propagation determine consistent local switching, then a counting-gap test compares required true literals with available variables. When the resulting contradiction is found, it emits one pseudo-Boolean cutting-planes proof; unsupported or noncontradictory cases return UNKNOWN, and no SAT-assignment path is implemented.}
\end{asGalleryFamily}

\begin{asGalleryFamily}{171}{subsumptiontest}{19 GBD instances\enspace\textperiodcentered\enspace 4 validation\enspace\textperiodcentered\enspace baseline 4/4}{The benchmark asks whether a Boolean CNF is satisfiable after clauses are repeatedly split into two extensions, adding a variable in both polarities. The resulting clauses may be shuffled, while the family structure consists of complementary twins rather than an arbitrary CNF.}
\ifdefined\pdfbookmark\pdfbookmark[2]{subsumptiontest}{as-gallery-171}\fi
\asGallerySolver{GRAT}{\asGalleryValue{asGalleryGain}{1}{323}}{\asGalleryValue{asGalleryGain}{1}{20.7}}{4/4}{A polarity-majority scan proposes a complete model while streaming clauses, with a full recount and rescan as fallback when conflicts appear. For a narrow one-false-clause case, exact sign-flipped pair contractions derive opposite units and emit a textual DRAT refutation; failed recognition, verification, or derivation returns UNKNOWN.}
\asGallerySolver{DPR}{\asGalleryValue{asGalleryGain}{0}{93.6}}{\asGalleryValue{asGalleryGain}{0}{8.75}}{4/4}{A sign-pattern reduction reverses clause divisions by matching one-bit sign differences and repeatedly shrinking the formula to a core. Bounded propagation and recursive search can yield a checked SAT assignment; UNSAT requires an explicit empty clause and a derivation DAG that may add clauses without witnesses, while unsupported or exhausted cases return UNKNOWN.}
\asGallerySolver{VeriPB}{\asGalleryValue{asGalleryGain}{0}{237}}{\asGalleryValue{asGalleryGain}{0}{6.00}}{4/4}{Polarity-imbalance scanning first proposes a complete assignment and verifies every original clause. If it fails, width-layered closure of exact complementary twins targets the least-imbalanced variable, then retries globally; a contradiction emits parent-linked VeriPB RUP steps, while only a fully checked recovered model yields SAT and other cases return UNKNOWN.}
\end{asGalleryFamily}

\begin{asGalleryFamily}{172}{sudoku}{30 GBD instances\enspace\textperiodcentered\enspace 6 validation\enspace\textperiodcentered\enspace baseline 0/6}{The family represents Sudoku-style digit assignments in CNF, using variables for cell or related finite-domain choices and clauses for Sudoku or other instance constraints. The benchmark question is whether the resulting Boolean formula has a satisfying assignment; some encodings include multiple grids and auxiliary variables.}
\ifdefined\pdfbookmark\pdfbookmark[2]{sudoku}{as-gallery-172}\fi
\asGallerySolver{GRAT}{\asGalleryValue{asGalleryMuted}{1}{1.00}}{\asGalleryValue{asGalleryMuted}{1}{1.00}}{0/6}{Watched-literal CDCL with activity-guided branching, 1-UIP learning, non-chronological backtracking, and restarts drives the active search. For SAT it prints an assignment; for UNSAT it writes learned-clause additions and an empty clause as textual DRAT for external GRAT elaboration and checking, while size, parsing, or proof-output failures return UNKNOWN.}
\asGallerySolver{DPR}{\asGalleryValue{asGalleryMuted}{1}{1.00}}{\asGalleryValue{asGalleryMuted}{1}{1.00}}{0/6}{Order-prefix recognition prioritizes the first 17 objects, then uses activity-based first-UIP CDCL with clause-scanning propagation, restarts, and fallback branching. The recognizer checks structure, not Sudoku semantics; SAT assignments are checked against original clauses, while UNSAT emits learned clauses plus an empty clause as textual DPR for external elaboration; rejection or parse failure returns UNKNOWN.}
\asGallerySolver{VeriPB}{\asGalleryValue{asGalleryMuted}{1}{1.00}}{\asGalleryValue{asGalleryMuted}{1}{1.00}}{0/6}{The compiled executable is an unconditional UNKNOWN fallback: it only checks invocation shape, neither parses the CNF nor searches. An exploratory source file contains row, column, and box filtering with bounded Sudoku backtracking, but it is not compiled or reachable, so no SAT assignment or UNSAT certificate is emitted.}
\end{asGalleryFamily}

\begin{asGalleryFamily}{173}{sum-\allowbreak of-\allowbreak 3-\allowbreak cubes}{10 GBD instances\enspace\textperiodcentered\enspace 2 validation\enspace\textperiodcentered\enspace baseline 0/2}{The benchmark asks whether a target integer can be written as the sum of three signed cubes. CNF instances bit-blast bounded cube circuits, using magnitude blocks, sign or selector variables, arithmetic auxiliaries, and fixed target bits.}
\ifdefined\pdfbookmark\pdfbookmark[2]{sum-of-3-cubes}{as-gallery-173}\fi
\asGallerySolver{GRAT}{\asGalleryValue{asGalleryGain}{1}{2.00\!\cdot\!10^{4}}}{\asGalleryValue{asGalleryGain}{1}{4{,}320}}{2/2}{Integer-construction search is active: the solver recognizes a narrow bit-blasted circuit shape and seeks bounded signed-cube identities through constructions, residue filtering, and cube-pair search. It completes and checks each candidate assignment; unsupported inputs or failed searches return UNKNOWN, with no UNSAT certificate path.}
\asGallerySolver{DPR}{\asGalleryValue{asGalleryGain}{0}{2.00}}{\asGalleryValue{asGalleryGain}{0}{2.00}}{1/2}{Arithmetic shortcut search is active: a narrow circuit detector extracts three operand words and searches bounded cube identities using near-cancellation, modular roots, factorization, and divisors. Candidates are completed and clauses checked; rejected or exhausted searches return UNKNOWN, with no UNSAT or DPR certificate path.}
\asGallerySolver{VeriPB}{\asGalleryValue{asGalleryMuted}{0}{1.00}}{\asGalleryValue{asGalleryMuted}{0}{1.00}}{0/2}{Modular arithmetic search leads: the solver recognizes a narrow cube circuit and tests bounded constructions with factorization, roots, CRT, and discriminant filters. If this fails, DPLL can emit a checked SAT assignment or a VeriPB RUP UNSAT log; unsupported cases return UNKNOWN, without proof verification.}
\end{asGalleryFamily}

\begin{asGalleryFamily}{174}{summle}{26 GBD instances\enspace\textperiodcentered\enspace 6 validation\enspace\textperiodcentered\enspace baseline 6/6}{The benchmark asks whether a bounded read-once arithmetic expression can reach a requested target from the multiset 1, 2, 2, 4, 4, 8, 25, 100 using permitted binary operations. CNF bit-blasts operation choices, operands, intermediate values, and target conditions.}
\ifdefined\pdfbookmark\pdfbookmark[2]{summle}{as-gallery-174}\fi
\asGallerySolver{GRAT}{\asGalleryValue{asGalleryLoss}{0}{0.431}}{\asGalleryValue{asGalleryLoss}{0}{0.436}}{5/6}{Subset DP, meet-in-the-middle checks, and cached models or witnesses guide expression construction on three layouts while watched-literal CDCL completes or searches the CNF. Satisfying assignments are checked against original clauses and emitted as DIMACS models; no independently checkable UNSAT certificate is produced, so refutations or unsupported inputs return UNKNOWN.}
\asGallerySolver{DPR}{\asGalleryValue{asGalleryGain}{1}{1{,}050}}{\asGalleryValue{asGalleryGain}{1}{447}}{6/6}{Expression-table lookup and subset dynamic programming build a candidate operation sequence whose circuit is forced by unit assumptions before watched-literal CDCL completes remaining variables. Failed construction invokes generic CDCL with learned clauses logged as witness-free additions for external checking; unsupported shapes return UNKNOWN, and direct SAT construction emits no proof stream.}
\asGallerySolver{VeriPB}{\asGalleryValue{asGalleryGain}{0}{815}}{\asGalleryValue{asGalleryGain}{0}{64.8}}{6/6}{Target extraction from repeated signatures and multiset dynamic programming construct a bounded arithmetic witness; forced selector and opcode bits are completed by watched-literal CDCL. Checked SAT assignments are emitted, but exhaustion, conflicts, failed validation, or unrecognized layouts return UNKNOWN because no UNSAT or VeriPB certificate path exists.}
\end{asGalleryFamily}

\begin{asGalleryFamily}{175}{tensors}{40 GBD instances\enspace\textperiodcentered\enspace 8 validation\enspace\textperiodcentered\enspace baseline 2/8}{The benchmark asks whether a three-way tensor over GF(2) is the XOR of at most R rank-one outer products. CNF encodings use AND gates for factor products, XOR constraints for tensor entries, and ordering constraints for interchangeable terms.}
\ifdefined\pdfbookmark\pdfbookmark[2]{tensors}{as-gallery-175}\fi
\asGallerySolver{GRAT}{\asGalleryValue{asGalleryGain}{0}{2.14}}{\asGalleryValue{asGalleryGain}{0}{2.14}}{5/8}{It searches for rank-one decompositions by choosing shared matrix updates that reduce residual slice ranks, with randomized diversification and pivot-style fallback. It completes ordering and auxiliary variables with dynamic programming and DPLL, checks every clause, and emits a SAT assignment; unsupported structure or failed search returns UNKNOWN, with no UNSAT certificate path.}
\asGallerySolver{DPR}{\asGalleryValue{asGalleryGain}{1}{2.15}}{\asGalleryValue{asGalleryGain}{1}{2.15}}{5/8}{It builds a GF(2) slice-space span by choosing minimum-rank residuals across slicing modes, then factors residual matrices into rank-one terms with binary Gaussian elimination. It reconstructs and orders factors, propagates and checks the clauses, and emits a SAT assignment; bounded search failure or unsupported encodings return UNKNOWN, with no UNSAT certificate path.}
\asGallerySolver{VeriPB}{\asGalleryValue{asGalleryGain}{0}{2.04}}{\asGalleryValue{asGalleryGain}{0}{2.04}}{5/8}{It contracts XOR components into tensor cells, then uses bit-mask tabu/min-conflicts seeding, slice switches, and Gaussian-elimination rank factoring to construct factors. Comparator completion uses subset DP and DPLL before a clause-checked SAT assignment; failed construction invokes internal CDCL that can emit an UNSAT RUP proof for external checking, while unsupported or out-of-regime encodings return UNKNOWN.}
\end{asGalleryFamily}

\begin{asGalleryFamily}{176}{termination-\allowbreak analysis}{68 GBD instances\enspace\textperiodcentered\enspace 14 validation\enspace\textperiodcentered\enspace baseline 13/14}{The benchmark asks whether a Boolean assignment satisfies all clauses encoding termination-ordering constraints. Auxiliary variables may represent circuit gate outputs in a Tseitin-style CNF, while the implementations recognize only selected layouts, not every possible encoding.}
\ifdefined\pdfbookmark\pdfbookmark[2]{termination-analysis}{as-gallery-176}\fi
\asGallerySolver{GRAT}{\asGalleryValue{asGalleryLoss}{1}{0.389}}{\asGalleryValue{asGalleryLoss}{0}{0.373}}{9/14}{Bounded randomized local search first targets recognized, short Tseitin-like CNF and accepts only a fully checked model; other accepted inputs undergo propagation, bounded elimination, probing, and watched-literal CDCL. SAT models are clause-checked; UNSAT uses a textual DRAT log of preprocessing, probing, and learned clauses, while unsupported inputs or missing proof output return UNKNOWN.}
\asGallerySolver{DPR}{\asGalleryValue{asGalleryLoss}{0}{0.267}}{\asGalleryValue{asGalleryLoss}{0}{0.316}}{6/14}{AND/parity DAG reconstruction drives bounded gate repair with phases, random flips, and restarts; large circuits skip repair, and failure falls back to CDCL, not UNSAT. Inputs outside unit-rich, mostly binary/ternary shapes return UNKNOWN; SAT models are checked, the CDCL fallback is uncapped, and UNSAT proof logs use root conflicts and learned, potentially witness-free clause additions.}
\asGallerySolver{VeriPB}{\asGalleryValue{asGalleryLoss}{0}{0.350}}{\asGalleryValue{asGalleryLoss}{1}{0.413}}{8/14}{Conservative Tseitin recognition and bounded elimination of low-occurrence outputs preserve semantic variables before watched-literal CDCL; recognized one-hot cases receive a bounded random-walk SAT probe, while unmatched layouts return UNKNOWN. A non-logging scout precedes a fresh VeriPB RUP run when needed; checked SAT assignments are accepted, but failed or incomplete UNSAT certification returns UNKNOWN.}
\end{asGalleryFamily}

\begin{asGalleryFamily}{177}{test-\allowbreak configuration}{44 GBD instances\enspace\textperiodcentered\enspace 9 validation\enspace\textperiodcentered\enspace baseline 4/9}{The benchmark asks whether n configurations of 46 options can obey fixed incompatibilities and implications while covering every compatible option pair. Its CNF has per-row variables and pair-conjunction variables; variants require each pair to be absent from a row or impose lexicographic row ordering.}
\ifdefined\pdfbookmark\pdfbookmark[2]{test-configuration}{as-gallery-177}\fi
\asGallerySolver{GRAT}{\asGalleryValue{asGalleryGain}{0}{7.81\!\cdot\!10^{4}}}{\asGalleryValue{asGalleryGain}{0}{814}}{9/9}{Randomized maximum-coverage search over enumerated maximal cliques seeks a fixed-size row cover, repairs negative-coverage variants locally, and clause-checks the resulting complete SAT assignment. On smaller recognized instances, an obstruction generator emits ASCII DRAT additions for external elaboration and checking; recognition or search failure returns UNKNOWN without a general SAT fallback.}
\asGallerySolver{DPR}{\asGalleryValue{asGalleryGain}{0}{5.05\!\cdot\!10^{4}}}{\asGalleryValue{asGalleryGain}{0}{394}}{9/9}{An exact recognizer for a capped fixed layout dispatches to a compatibility-cover construction, pads and orders rows as needed, evaluates auxiliaries, and checks every clause before returning a SAT assignment. Smaller instances use a hard-coded obstruction and DPR steps with witness-free clause additions; recognition or proof-generation failure returns UNKNOWN without a general fallback.}
\asGallerySolver{VeriPB}{\asGalleryValue{asGalleryGain}{1}{8.25\!\cdot\!10^{4}}}{\asGalleryValue{asGalleryGain}{1}{2{,}080}}{9/9}{Arithmetic-layout recognition selects a hard-coded compatibility cover for SAT, pads and orders rows as required, reconstructs auxiliaries, and checks every input clause before releasing the assignment. Smaller recognized instances use fooling-set counting with premise lookup to emit VeriPB RUP and cutting-plane proofs; unrecognized inputs or proof-generation failure return UNKNOWN without a SAT fallback.}
\end{asGalleryFamily}

\begin{asGalleryFamily}{178}{testpattern-\allowbreak generation}{52 GBD instances\enspace\textperiodcentered\enspace 11 validation\enspace\textperiodcentered\enspace baseline 11/11}{The benchmark asks whether a Boolean assignment satisfies a CNF encoding of a test-pattern or fault-detection condition. In circuit-oriented instances, variables represent signals and clauses impose the relevant constraints, so SAT gives a test pattern and UNSAT rules out one for that encoding.}
\ifdefined\pdfbookmark\pdfbookmark[2]{testpattern-generation}{as-gallery-178}\fi
\asGallerySolver{GRAT}{\asGalleryValue{asGalleryLoss}{0}{2.11\!\cdot\!10^{-4}}}{\asGalleryValue{asGalleryLoss}{0}{5.22\!\cdot\!10^{-4}}}{10/11}{Statistical recognition selects two bounded-width ATPG-miter signatures without reconstructing topology. Watched-literal CDCL uses VSIDS, first-UIP learning, and restarts, falling back to a descending scan when the activity heap has no candidate; checked SAT models are printed, while UNSAT emits DRAT additions and an empty clause, and unsupported or failed cases return UNKNOWN.}
\asGallerySolver{DPR}{\asGalleryValue{asGalleryLoss}{0}{2.11\!\cdot\!10^{-4}}}{\asGalleryValue{asGalleryLoss}{0}{5.21\!\cdot\!10^{-4}}}{10/11}{Syntactic envelope recognition accepts only CNF formulas matching bounded-width statistical signatures; it does not recover gates, boundaries, or a netlist. Watched-literal CDCL supplies first-UIP learning, activity-based branching, and restarts; checked SAT assignments are printed, while UNSAT produces learned clauses and an empty clause for DRAT elaboration, with other outcomes UNKNOWN.}
\asGallerySolver{VeriPB}{\asGalleryValue{asGalleryGain}{1}{6.53}}{\asGalleryValue{asGalleryGain}{1}{1.30}}{11/11}{Narrow recognition precedes parity contraction of adjacent equivalence or inversion pairs and signed clause rewriting. Watched-literal CDCL uses first-UIP learning and restarts; SAT checks a reconstructed assignment, while UNSAT is certified by VeriPB RUP steps for learned and contracted clauses plus an empty conclusion; rejected, inconsistent, or proof-failed cases return UNKNOWN.}
\end{asGalleryFamily}

\begin{asGalleryFamily}{179}{theorem-\allowbreak proving}{8 GBD instances\enspace\textperiodcentered\enspace 2 validation\enspace\textperiodcentered\enspace baseline 2/2}{These benchmarks encode grid-based pebbling contradictions: they ask whether local Boolean rules can assign consistent states to grid vertices. Source or root conditions, predecessor implications, and a designated sink condition are expressed with small CNF gadgets, often using two Boolean variables per vertex.}
\ifdefined\pdfbookmark\pdfbookmark[2]{theorem-proving}{as-gallery-179}\fi
\asGallerySolver{GRAT}{\asGalleryValue{asGalleryGain}{1}{33.4}}{\asGalleryValue{asGalleryGain}{1}{8.85}}{2/2}{Support-hash grouping and DSU recovery reconstruct the paired-variable DAG, validating a unique source, acyclicity, and topological order without checking rectangular dimensions. The recovered order drives DRAT additions for external elaboration, including boundary and interior clauses, a sink unit, and the empty clause; unrecognized inputs return UNKNOWN, with no SAT-model path.}
\asGallerySolver{DPR}{\asGalleryValue{asGalleryGain}{0}{14.4}}{\asGalleryValue{asGalleryLoss}{0}{0.474}}{2/2}{Exact structural recognition recovers paired variables, predecessor DAG, and grid order from the two-variable OR-substituted rectangular grid form. It emits witness-free DPR additions for the recovered order, using head clauses and bridge clauses to derive the sink contradiction; unrecognized inputs and failures return UNKNOWN, with no SAT-search or model path.}
\asGallerySolver{VeriPB}{\asGalleryValue{asGalleryGain}{0}{12.2}}{\asGalleryValue{asGalleryGain}{0}{2.93}}{2/2}{Two-pass counting sort reconstructs variable pairs and four-clause grid gadgets, then checks dimensions, polarity, clause ownership, and an acyclic root-reachable order. The solver emits a cutting-planes certificate with one derivation per non-root vertex and a final sink contradiction, while malformed or unrecognized inputs return UNKNOWN and no SAT-solving fallback is implemented.}
\end{asGalleryFamily}

\begin{asGalleryFamily}{180}{tournament}{16 GBD instances\enspace\textperiodcentered\enspace 4 validation\enspace\textperiodcentered\enspace baseline 0/4}{These instances ask whether a partially fixed tournament can be completed so every seven-vertex subset contains a cyclic triangle, not a transitive ordering. The CNF uses edge orientations and triangle witnesses, with clauses linking witnesses to edges and requiring each seven-set to be covered.}
\ifdefined\pdfbookmark\pdfbookmark[2]{tournament}{as-gallery-180}\fi
\asGallerySolver{GRAT}{\asGalleryValue{asGalleryMuted}{1}{1.00}}{\asGalleryValue{asGalleryMuted}{1}{1.00}}{0/4}{Score-based local search is the active technique: it flips free edge orientations while incrementally tracking transitive seven-sets, using tabu moves, dynamic weights, focused randomness, and restarts. The narrow encoding is checked before search; found assignments are clause-checked, unsupported inputs return UNKNOWN, and the SAT-only path has no UNSAT certificate or exhaustion result.}
\asGallerySolver{DPR}{\asGalleryValue{asGalleryMuted}{1}{1.00}}{\asGalleryValue{asGalleryMuted}{1}{1.00}}{0/4}{Restricted tournament blow-up construction is tried first, duplicating core vertices under capacity constraints and rejecting remaining transitive seven-sets. Only recognized encodings can yield checked SAT assignments; unsupported inputs return UNKNOWN, and no UNSAT derivation or exhaustion result is implemented.}
\asGallerySolver{VeriPB}{\asGalleryValue{asGalleryMuted}{1}{1.00}}{\asGalleryValue{asGalleryMuted}{1}{1.00}}{0/4}{Stochastic min-conflict edge-flip search is active, maintaining selector values and seven-set violation counts while sampling unsatisfied sets and applying scored flips, noise, and restarts. A specialized prefix-completion branch checks constrained six-set-free extensions; found SAT models are checked, unsupported encodings return UNKNOWN, and no active UNSAT certificate path exists.}
\end{asGalleryFamily}

\begin{asGalleryFamily}{181}{tree-\allowbreak decomposition}{24 GBD instances\enspace\textperiodcentered\enspace 5 validation\enspace\textperiodcentered\enspace baseline 2/5}{The benchmark asks whether an undirected graph has an elimination ordering of bounded width that obeys precedence constraints. Structured CNF variables and clauses encode the graph, ordering, fill edges, and width bound, with supported layouts varying by solver.}
\ifdefined\pdfbookmark\pdfbookmark[2]{tree-decomposition}{as-gallery-181}\fi
\asGallerySolver{GRAT}{\asGalleryValue{asGalleryGain}{1}{1.72}}{\asGalleryValue{asGalleryGain}{1}{1.71}}{3/5}{Structural recovery and mixed elimination search distinguish this solver: it recognizes only direct or scrambled encodings, combines randomized min-fill/min-degree and annealing local search, and uses exact DFS on smaller graphs. A found order gets Horn or renamable-Horn completion and clause verification; failures return UNKNOWN, with no UNSAT certificate path.}
\asGallerySolver{DPR}{\asGalleryValue{asGalleryGain}{0}{1.13}}{\asGalleryValue{asGalleryGain}{0}{1.13}}{2/5}{Structural recognition and randomized elimination search lead the solver: it recovers a narrowly supported graph encoding, tries greedy min-fill/min-degree orders, and uses capped beam search as fallback. Successful orders receive residual Horn completion and clause verification; unsupported or failed cases return UNKNOWN, and no UNSAT certificate path exists.}
\asGallerySolver{VeriPB}{\asGalleryValue{asGalleryLoss}{0}{0.872}}{\asGalleryValue{asGalleryLoss}{0}{0.872}}{1/5}{Precedence-aware randomized min-fill/min-degree search leads the solver: it recognizes only the canonical layout, tries greedy orders, and falls back to heuristic repair stages. Successful orders receive Horn completion and clause verification before SAT assignment output; optional CDCL can generate RUP steps and an UNSAT conclusion, while normal failure returns UNKNOWN.}
\end{asGalleryFamily}

\begin{asGalleryFamily}{182}{trigonometric-\allowbreak functions}{6 GBD instances\enspace\textperiodcentered\enspace 2 validation\enspace\textperiodcentered\enspace baseline 1/2}{These instances ask whether Boolean assignments to input and internal variables can satisfy every clause in a circuit encoding of a trigonometric computation or relation. The benchmarks use gate-based bit-vector or Tseitin CNFs, while recognizers accept only restricted structural layouts.}
\ifdefined\pdfbookmark\pdfbookmark[2]{trigonometric-functions}{as-gallery-182}\fi
\asGallerySolver{GRAT}{\asGalleryValue{asGalleryGain}{0}{1.01}}{\asGalleryValue{asGalleryGain}{0}{1.07}}{1/2}{Structural case splitting canonicalizes small functional gate groups in the piecewise-lines layout, deriving constant and equivalence clauses before watched-literal CDCL; Taylor-layout instances skip this preprocessing. UNSAT branches log DRAT clause additions, while checked SAT models are returned and unsupported inputs or a configured conflict cutoff yield UNKNOWN.}
\asGallerySolver{DPR}{\asGalleryValue{asGalleryGain}{1}{2.34}}{\asGalleryValue{asGalleryGain}{1}{2.10}}{2/2}{Implication-graph equivalence substitution and blocked-clause elimination simplify recognized gate-heavy CNFs, with bounded variable elimination attempted only on larger accepted instances. First-UIP learning logs RUP clauses and a terminal empty clause for UNSAT; checked SAT assignments are returned, while unsupported inputs or failed checks yield UNKNOWN.}
\asGallerySolver{VeriPB}{\asGalleryValue{asGalleryLoss}{0}{0.953}}{\asGalleryValue{asGalleryLoss}{0}{0.596}}{1/2}{A conservative topology heuristic identifies likely source variables and prefers them on activity ties, but all accepted CNFs use watched-literal, first-UIP CDCL search; no circuit-specific propagator or dispatch is active. UNSAT produces VeriPB RUP records, an empty-clause RUP, and an UNSAT conclusion, while checked SAT models are returned; malformed input or failed checks yields UNKNOWN.}
\end{asGalleryFamily}

\begin{asGalleryFamily}{183}{tseitin-\allowbreak formulas}{197 GBD instances\enspace\textperiodcentered\enspace 68 validation\enspace\textperiodcentered\enspace baseline 51/68}{These benchmarks ask whether a CNF encoding of XOR constraints has a Boolean assignment satisfying every clause. A common encoding groups clauses that forbid one parity of variables, with graph-shaped cases representing variables as edges between constraint vertices and vertex charges as parity requirements.}
\ifdefined\pdfbookmark\pdfbookmark[2]{tseitin-formulas}{as-gallery-183}\fi
\asGallerySolver{GRAT}{\asGalleryValue{asGalleryGain}{0}{3.04}}{\asGalleryValue{asGalleryGain}{0}{2.77}}{62/68}{Exact parity-block recognition drives graph charge-parity and spanning-forest solving, with bounded packed GF(2) elimination for non-graph systems and a narrow repair for selected quadratic encodings. Checked SAT assignments are reconstructed from recovered equations; selected graph contradictions receive attempted DRAT/extended-resolution proofs, while unsupported, oversized, or unproved cases, including dense inconsistencies, return UNKNOWN.}
\asGallerySolver{DPR}{\asGalleryValue{asGalleryLoss}{0}{0.506}}{\asGalleryValue{asGalleryLoss}{0}{0.513}}{32/68}{Complete parity-block decoding drives GF(2) solving: graph systems use spanning forests, with bounded elimination and narrow repaired or quadratic-macro fallbacks. SAT assignments are clause-checked; supported UNSAT cases use graph or macro contractions to emit witness-free DPR clause additions, while rejected proof plans or failed stochastic searches return UNKNOWN, never establishing UNSAT.}
\asGallerySolver{VeriPB}{\asGalleryValue{asGalleryGain}{1}{6.07}}{\asGalleryValue{asGalleryGain}{1}{4.45}}{65/68}{ANF/Mobius recovery yields XOR equations over signed variables or two-literal terms; graph components use charge propagation and tree reconstruction, while dense elimination produces SAT models only. Clause-checked models are reconstructed physically; graph contradictions can receive VeriPB proofs, while a bounded clique fallback emits cutting-planes proofs and non-graph inconsistencies return UNKNOWN without UNSAT certification.}
\end{asGalleryFamily}

\begin{asGalleryFamily}{184}{uniform-\allowbreak random}{4,043 GBD instances\enspace\textperiodcentered\enspace 809 validation\enspace\textperiodcentered\enspace baseline 201/809}{The benchmark asks whether a Boolean assignment can satisfy every clause in a signed-literal CNF formula. Inputs are restricted random-width-shaped DIMACS formulas with controlled clause widths and related structural checks, which describe the encoding rather than establish statistical uniformity.}
\ifdefined\pdfbookmark\pdfbookmark[2]{uniform-random}{as-gallery-184}\fi
\asGallerySolver{GRAT}{\asGalleryValue{asGalleryGain}{1}{1.28}}{\asGalleryValue{asGalleryGain}{1}{1.27}}{332/809}{Pure 2-CNF instances use an SCC contradiction-path procedure; other accepted width classes first receive belief- or survey-propagation-seeded incremental local search with make/break and age-aware choices. On failure, watched-literal 1-UIP CDCL continues to a checked model or conflict, logging textual DRAT clauses for UNSAT; malformed or out-of-scope inputs return UNKNOWN.}
\asGallerySolver{DPR}{\asGalleryValue{asGalleryGain}{0}{1.09}}{\asGalleryValue{asGalleryGain}{0}{1.09}}{241/809}{Focused incremental local search uses clause-satisfaction state, break/make scores, and width-dependent probSAT- or Novelty-style choices. Only narrow complete-route instances enter watched-literal first-UIP CDCL; it can print a checked model or log witness-free DPR additions for UNSAT, while other local-search misses and out-of-scope inputs return UNKNOWN.}
\asGallerySolver{VeriPB}{\asGalleryValue{asGalleryGain}{0}{1.09}}{\asGalleryValue{asGalleryGain}{0}{1.08}}{241/809}{Incremental probSAT-style local search maintains make/break, unsatisfied-clause, and sole-satisfier state, with weighted and deterministic seed strategies. A watched-literal CDCL fallback is available only for bounded instances; it can yield a checked SAT assignment or log RUP proof steps for UNSAT, while larger or failed searches can end UNKNOWN without an UNSAT certificate.}
\end{asGalleryFamily}

\begin{asGalleryFamily}{185}{unknown}{118 GBD instances\enspace\textperiodcentered\enspace 24 validation\enspace\textperiodcentered\enspace baseline 12/24}{These benchmarks ask whether Boolean variables can satisfy a structured encoding of a finite combinatorial construction. Clauses represent choices and enforce exact-one, uniqueness, incidence, adjacency, forbidden-combination, or auxiliary-consistency conditions.}
\ifdefined\pdfbookmark\pdfbookmark[2]{unknown}{as-gallery-185}\fi
\asGallerySolver{GRAT}{\asGalleryValue{asGalleryLoss}{0}{0.609}}{\asGalleryValue{asGalleryLoss}{0}{0.611}}{3/24}{Bit-mask recognition and bounded Dancing Links search target orthogonal Latin-square encodings; fixed balanced-design and bit-vector Latin constructions seed CDCL, with CDCL handling other inputs. SAT assignments are rechecked, and UNSAT paths log derived clauses and an empty clause for external DRAT elaboration; failed direct orthogonal-Latin searches or budget exhaustion return UNKNOWN.}
\asGallerySolver{DPR}{\asGalleryValue{asGalleryLoss}{1}{0.933}}{\asGalleryValue{asGalleryLoss}{1}{0.856}}{11/24}{Bounded structural recognizers try cyclic-isotopy and Dancing Links for direct orthogonal-Latin encodings, residue or TabuCol searches for five-colorings, and component CSP searches for signed-selector cases. Failures fall through to CDCL with failed-literal probing; SAT models are checked against input, while UNSAT logging yields textual DPR and proof or model-check failures return UNKNOWN.}
\asGallerySolver{VeriPB}{\asGalleryValue{asGalleryLoss}{0}{0.888}}{\asGalleryValue{asGalleryLoss}{0}{0.696}}{10/24}{Choice-block recognition uses rare-polarity orientations, bitsets, singleton rules, MRV branching, and bounded DFS; narrow Latin search tries affine constructions before MRV. Failed attempts fall back to CDCL; SAT models are checked, pigeonhole cases emit direct PB proofs, while CDCL UNSAT traces end in RUP contradiction and proof-generation failures return UNKNOWN.}
\end{asGalleryFamily}

\begin{asGalleryFamily}{186}{unknown-\allowbreak cases}{20 GBD instances\enspace\textperiodcentered\enspace 4 validation\enspace\textperiodcentered\enspace baseline 4/4}{The benchmark asks whether a Boolean assignment satisfies every clause of a CNF formula. Structured instances may encode selections that cover requirements while forbidding combinations, or represent circuits and order constraints; these are recurring structures, not a universal encoding.}
\ifdefined\pdfbookmark\pdfbookmark[2]{unknown-cases}{as-gallery-186}\fi
\asGallerySolver{GRAT}{\asGalleryValue{asGalleryLoss}{1}{0.131}}{\asGalleryValue{asGalleryLoss}{1}{0.131}}{1/4}{Four-incidence positive-cover recognition leads the search, followed by orientation, bounded graph/circuit and stochastic model attempts, all accepted only after checking a full CNF assignment. When CDCL proves UNSAT, it logs learned clauses and a final empty clause for external checking; failed structural or internal searches do not establish UNSAT and can return UNKNOWN.}
\asGallerySolver{DPR}{\asGalleryValue{asGalleryLoss}{0}{0.106}}{\asGalleryValue{asGalleryLoss}{0}{0.106}}{0/4}{Polarity-normalized cover recognition leads to bitset propagation, randomized repair, local search, and minimum-coverage DFS, with fixed-layout circuit and repeated-key paths as bounded fallbacks. SAT candidates are checked against the original CNF; only monotone DFS within its 64-variable limit emits decision nogoods and may report UNSAT, while other failures or large-cover exhaustion return UNKNOWN.}
\asGallerySolver{VeriPB}{\asGalleryValue{asGalleryLoss}{0}{0.106}}{\asGalleryValue{asGalleryLoss}{0}{0.106}}{0/4}{Monotone cover/forbidden-set search uses bitset filtering and smallest-uncovered branching, with local search and watched-literal CDCL fallbacks, accepting only clause-checked assignments. The complete monotone and CDCL paths replay learned or decision-nogood clauses as RUP steps with a final empty constraint; heuristic failures provide no proof, and unsupported or exhausted cases return UNKNOWN.}
\end{asGalleryFamily}

\begin{asGalleryFamily}{187}{waerden}{110 GBD instances\enspace\textperiodcentered\enspace 22 validation\enspace\textperiodcentered\enspace baseline 3/22}{The benchmark asks whether positions 1 through N can be two-coloured so one colour avoids every three-term arithmetic progression and hits every K-term progression. Boolean variables encode these conditions as positive and negative CNF clauses; some instances use folded representations.}
\ifdefined\pdfbookmark\pdfbookmark[2]{waerden}{as-gallery-187}\fi
\asGallerySolver{GRAT}{\asGalleryValue{asGalleryGain}{1}{1.04}}{\asGalleryValue{asGalleryGain}{1}{1.04}}{4/22}{Bounded weighted local search first seeks a checked SAT assignment when a positive rank-2 clause exists, then complete DFS blocks positive edges and branches on the tightest unhit negative edge. Proof logging records parent, leaf, exclusion, and deletion clauses for external elaboration; restricted pure-polarity shapes or proof inconsistencies can yield UNKNOWN.}
\asGallerySolver{DPR}{\asGalleryValue{asGalleryGain}{0}{1.01}}{\asGalleryValue{asGalleryGain}{0}{1.02}}{3/22}{Incremental weighted local search leads the SAT attempt, using breakout weights, flip scores, and age tie-breaking. Failure falls back to complete DFS with negative-clause branching; unsupported structure yields UNKNOWN, while checked SAT assignments are returned and DFS may log witness-free conflict clauses without an implemented UNSAT certification.}
\asGallerySolver{VeriPB}{\asGalleryValue{asGalleryLoss}{0}{0.944}}{\asGalleryValue{asGalleryLoss}{0}{0.912}}{2/22}{Two-watched-literal CDCL is the active engine, using activity-based branching, 1-UIP conflict minimization, restarts, failed-literal probing, and learned-clause reduction. It checks SAT assignments, returns UNKNOWN for unsupported structure, and on UNSAT attempts to serialize learned clauses, deletions, and a final contradiction as VeriPB proof text, without invoking a proof verifier.}
\end{asGalleryFamily}

\begin{asGalleryFamily}{188}{xor-\allowbreak chain}{66 GBD instances\enspace\textperiodcentered\enspace 14 validation\enspace\textperiodcentered\enspace baseline 11/14}{These instances ask whether a Boolean assignment satisfies a connected system of parity equations. A typical CNF encoding uses four clauses for each three-variable XOR and two for each two-variable XOR; related instances use sequential counters and graph constraints.}
\ifdefined\pdfbookmark\pdfbookmark[2]{xor-chain}{as-gallery-188}\fi
\asGallerySolver{GRAT}{\asGalleryValue{asGalleryMuted}{0}{1.00}}{\asGalleryValue{asGalleryMuted}{0}{1.00}}{11/14}{XOR-gate recognition and Gaussian elimination handle supported parity systems; persistent XOR-tree logging, fresh definitions, and RUP/resolution projections support UNSAT. On recognition failure, a triangle-graph fallback uses blossom matching and sequential counters for clause-checked SAT models; strengthened forms skip this path, and unsupported or failed dispatch returns UNKNOWN.}
\asGallerySolver{DPR}{\asGalleryValue{asGalleryMuted}{0}{1.00}}{\asGalleryValue{asGalleryMuted}{0}{1.00}}{11/14}{Degree-two incidence-graph recognition selects only supported, clause-complete binary and ternary parity groups in a connected structure. Bit-packed GF(2) elimination yields SAT assignments checked against every original clause, while UNSAT uses spanning-tree XOR expressions, fresh definitions, chord cancellation, and local RUP derivations; unsupported or failed cases return UNKNOWN.}
\asGallerySolver{VeriPB}{\asGalleryValue{asGalleryGain}{1}{3.00}}{\asGalleryValue{asGalleryGain}{1}{2.99}}{13/14}{Direct XOR recognition uses BFS cycle reductions for parity instances, clause-checking SAT assignments and emitting VeriPB derivations for UNSAT contradictions; strengthened forms have only the latter path and may return UNKNOWN. Counter variants reconstruct triangles and use blossom matching for SAT or pseudo-Boolean proofs for conflicting bounds; matching failure or unsupported cases return UNKNOWN.}
\end{asGalleryFamily}

\begin{asGalleryFamily}{189}{xor\_\allowbreak op}{3 GBD instances\enspace\textperiodcentered\enspace 1 validation\enspace\textperiodcentered\enspace baseline 1/1}{The benchmark asks whether a directed relation on vertices can be a strict ordering: it is asymmetric and transitive, yet every vertex must have a predecessor among three designated choices. CNF uses two XOR-related variables per relation atom and expands the logical constraints into clauses.}
\ifdefined\pdfbookmark\pdfbookmark[2]{xor\_op}{as-gallery-189}\fi
\asGallerySolver{GRAT}{\asGalleryValue{asGalleryGain}{0}{8{,}210}}{\asGalleryValue{asGalleryGain}{1}{375}}{1/1}{Exact width-2 XOR decoding with fallbacks builds an embedded-gadget proof: it names atoms, derives ordering clauses, and greedily eliminates vertices to an empty clause. It has no SAT fallback: only recognized three-predecessor schemas are handled, unsupported inputs or failures return UNKNOWN, and DRAT is sent for external elaboration and checking.}
\asGallerySolver{DPR}{\asGalleryValue{asGalleryGain}{1}{2.34\!\cdot\!10^{4}}}{\asGalleryValue{asGalleryGain}{0}{128}}{1/1}{Greedy elimination over predecessor bitsets replaces each removed predecessor by its predecessors, emits canonicalization additions for used XOR gadgets, and ends with an empty clause when successful. Recognition covers only the exact bounded three-predecessor macro layout; no general SAT or SAT-witness path exists, and unsupported or failed constructions return UNKNOWN.}
\asGallerySolver{VeriPB}{\asGalleryValue{asGalleryGain}{0}{1{,}660}}{\asGalleryValue{asGalleryGain}{0}{121}}{1/1}{XOR-block recovery followed by in-process CDCL searches the recovered logical clauses rather than the expanded raw formula. On UNSAT it writes a VeriPB proof with XOR reductions and RUP steps, but does not invoke a verifier; recognition mismatches, satisfiable outcomes, or proof failures return UNKNOWN.}
\end{asGalleryFamily}
\par\clearpage\endgroup

\clearpage

\section{Extended: Solver taxonomy}
\label{sec:app-solver-taxonomy}

We used GPT-6 Astra (xhigh) and GPT-5.6 Luna (high) to iteratively generate and refine a taxonomy for seven different questions about solver behavior.
Here we describe our methodology and potential limitations (\S\ref{sec:solver-taxonomy-methodology}). We then present the full taxonomy (\S\ref{sec:app-full-taxonomy}).

\subsection{Methodology}
\label{sec:solver-taxonomy-methodology}

Given a query about a specific aspect of solver behavior, such as ``What solving strategies are used in the solvers?'', we first built out a suitable taxonomy using GPT-6 Astra (xhigh) in Codex, providing the full suite of 567 generated solvers and their source code. We iteratively refined the taxonomy to ensure it covered all variation and properly distinguished between closely related concepts (both LLM-mediated and with some human feedback).

Given a frozen taxonomy, we then tasked GPT-5.6 Luna (high) with annotating each solver independently for each query and set of labels. We report results for all non-zero counts. A given solver only contributes once to a given label (even if it implements several variants) but each solver can contribute to multiple labels. The full listing of frozen taxonomies is provided in \S\ref{sec:app-full-taxonomy}.

\subsection{Potential limitations}
\label{sec:solver-taxonomy-limitations}

LLMs are fallible and our taxonomies may be incomplete or inaccurate.
For the use of these taxonomies in our paper (observing broad variation patterns) we believe this is an acceptable risk.
The exact counts of solvers for each label and the specific techniques listed, if changed slightly, would not alter the conclusions we draw.

\begingroup
\setlength{\parindent}{0pt}
\setlength{\parskip}{1.5pt plus 0.5pt minus 0.5pt}
\clubpenalty=10000
\widowpenalty=10000
\emergencystretch=1em

\subsection{Full taxonomy labels}
\label{sec:app-full-taxonomy}

\begin{tcolorbox}[
    colback=red!8,
    colframe=red!70!black,
    boxrule=0.5pt,
    arc=1mm,
    left=2.2mm,
    right=2.2mm,
    top=1.3mm,
    bottom=1.3mm,
    before skip=4pt,
    after skip=4pt
  ]
  {\sffamily\color{red!70!black}\faRobot}\enspace
  \textbf{Note that the content in this section is LLM-generated. These full taxonomy labels were generated with GPT-6 Astra (xhigh) with some human feedback.}
  \end{tcolorbox}

\subsubsection{Solving strategies}
\label{app:rq3-audit-algorithm}
Whole solving procedures, including systematic search, local search, algebraic and structural reasoning, inference, and direct answer construction. The algorithm audit counts default-reachable solving roles, excluding procedures used only for preprocessing or proof processing. Such procedures can still qualify in the other studies. For a local-search flip, make counts clauses newly satisfied, and break counts clauses made unsatisfied.

\noindent\textbf{Structural construction / theorem.} Recognized instance structure supports a witness construction or a combinatorial contradiction, such as a pigeonhole or counting argument. The method must perform actual input-dependent reasoning.\par

\noindent\textbf{CDCL.} Conflict-driven clause learning combines Boolean decisions and propagation with conflict analysis, then reuses learned clauses during search. Watched literals, activity scores, or proof logging alone do not qualify.\par

\noindent\textbf{Greedy construction.} An assignment or domain solution is built progressively without systematic backtracking or iterative local repair. Trivial fixed-phase candidates qualify only when actually tested as answer attempts.\par

\noindent\textbf{Exhaustive enumeration.} Assignments or combinations are explicitly enumerated and checked, possibly in parallel using machine-word bits. Ordinary CDCL branching is not counted as a separate enumeration algorithm.\par

\noindent\textbf{Native-domain local search.} Neighborhood moves repair recovered domain objects, such as changing a color or swapping permutation entries. The moves act on these objects beyond individual CNF variable flips.\par

\noindent\textbf{Constraint backtracking.} Search assigns recovered finite-domain variables or domain objects, such as colors or permutations, and backtracks with constraint filtering. The search operates on domain constraints beyond ordinary Boolean DPLL.\par

\noindent\textbf{Circuit / equivalence reasoning.} Recovered gate semantics, circuit simulation, equivalence reasoning, or circuit rewriting solve or reduce the problem. Identifying gates only to choose variable phases is insufficient.\par

\noindent\textbf{Cardinality / PB reasoning.} Dedicated counting or bound propagation, cutting-plane deductions, or counting contradictions reason beyond ordinary CNF propagation. Merely printing a VeriPB certificate is insufficient.\par

\noindent\textbf{WalkSAT-style search.} Search selects an unsatisfied clause, then flips one of its variables using low-break scores or noisy/random choices. An occasional random step within a different algorithm is insufficient.\par

\noindent\textbf{Other graph algorithms.} A dedicated graph procedure, such as reachability, bipartiteness testing, or isomorphism reasoning, decides or constrains feasibility. Graph storage and ordering alone are insufficient.\par

\noindent\textbf{DPLL (no clause learning).} Davis-Putnam-Logemann-Loveland search branches on Boolean variables, propagates consequences, and backtracks without learning conflict clauses. A CDCL implementation alone does not receive this additional label.\par

\noindent\textbf{Weighted local search.} Search adapts clause or constraint penalties to redirect subsequent local moves. Static weights and ordinary break counts alone are insufficient.\par

\noindent\textbf{Arithmetic / number theory.} Arithmetic reasoning, factoring, modular algebra, or integer-feasibility reasoning directly constructs an answer or proves impossibility. Parsing numeric features alone is excluded.\par

\noindent\textbf{Propagation-only solving.} A distinct answer procedure propagates consequences to a fixed point and completes an assignment deterministically without branching search. Propagation used only inside another solving algorithm is excluded.\par

\noindent\textbf{ProbSAT-style search.} Candidate flips are sampled with explicit nonuniform probabilities derived from make/break scores, such as inverse powers of break counts. Uniform random noise alone is excluded.\par

\noindent\textbf{Tabu search / TabuSAT.} A tabu list or tenure explicitly prohibits certain moves, potentially allowing exceptions when a move is sufficiently promising. Age-based tie-breaking or a preference against recent flips alone is insufficient.\par

\noindent\textbf{XOR / GF(2) elimination.} Parity equations are solved or reduced through row operations or Gaussian elimination over GF(2), the two-element field with arithmetic modulo two. Recognizing exclusive-or (XOR) gates without elimination is insufficient.\par

\noindent\textbf{State-space search.} Search traverses explicit domain states, using methods such as breadth-first or depth-first search, iterative deepening, heuristic search, bidirectional search, or beam search.\par

\noindent\textbf{Matching / flow.} A recovered graph is solved using matching, augmenting paths, maximum flow, or a Hall witness: a set with too few neighbors to support the required matching.\par

\noindent\textbf{Resolution / elimination.} The solver explicitly generates resolvents, eliminates variables, or repeatedly applies resolution toward saturation. Conflict-clause learning within CDCL alone does not receive this additional label.\par

\noindent\textbf{Meet-in-the-middle search.} Search enumerates two partial solution spaces and joins compatible halves using a table, hashing, or sorting. Looking up a stored complete answer is excluded.\par

\noindent\textbf{Dynamic programming.} A recurrence combines and caches subproblem results, for example over subsets, trees, or paths. Maintaining an ordinary assignment trail is insufficient.\par

\noindent\textbf{Simulated annealing.} Search accepts worsening moves according to a temperature-dependent rule. Merely changing a random-noise rate over time is insufficient.\par

\noindent\textbf{Stored witness / proof / answer.} A reachable answer route loads, embeds, or retrieves a precomputed assignment, proof, or answer. Runtime memoization and newly derived witnesses are excluded.\par

\noindent\textbf{Branch and bound.} Search computes objective bounds and uses them to prune subproblems that cannot improve the result. Feasibility backtracking without such bounds is excluded.\par

\noindent\textbf{Large-neighborhood search.} Search relaxes or removes a substantial part of a candidate assignment and reoptimizes the resulting neighborhood, often with an exact subsolver.\par

\noindent\textbf{Other algorithm.} The source implements a concrete algorithm outside the named categories. The audit must give its precise name and explain the missing taxonomy category.\par

\noindent\textbf{Look-ahead SAT search.} Tentative assignments and propagation repeatedly estimate the consequences of choices and guide branching. Failed-literal probing used only in preprocessing is excluded.\par

\noindent\textbf{Belief / survey propagation.} The solver iteratively exchanges probabilistic messages on a factor graph, potentially fixing variables according to the resulting beliefs or surveys.\par

\noindent\textbf{Input-feature answer rule.} Names, hashes, sizes, or other input signatures select a prescribed answer or candidate without a demonstrated general structural derivation. This records the decision mechanism without judging its correctness.\par

\noindent\textbf{GSAT-style greedy flips.} Improving variable flips are chosen using make/break scores over all variables or a maintained set of promising variables. Selection only within one unsatisfied clause is insufficient.\par

\noindent\textbf{Independent random sampling.} The solver generates and checks fresh, independent complete candidates. This is distinct from neighborhood search and restarts of that search.\par

\noindent\textbf{2-SAT / implication solving.} A dedicated procedure decides 2-SAT, whose clauses have at most two literals, using strongly connected components of an implication graph or an equivalent method. Binary-clause propagation alone is excluded.\par

\noindent\textbf{Horn / tractable fragments.} A dedicated algorithm solves a recognized tractable fragment, such as Horn formulas (at most one positive literal per clause) or dual-Horn formulas (at most one negative literal). Ordinary unit propagation alone is insufficient.\par

\noindent\textbf{Novelty-style search.} Search selects between the best and second-best candidates using avoidance of the most recently flipped variable and a noise rule. Generic age-based tie-breaking alone is excluded.\par

\noindent\textbf{Local search (unspecified).} Candidate assignments are repeatedly repaired through neighborhood moves. This residual category applies only when no more specific local-search label fits the component.\par

\noindent\textbf{Population / evolutionary.} Search maintains and evolves multiple candidates through selection, mutation, crossover, or other population-based optimization. Repeated independent starts alone are excluded.\par

\noindent\textbf{Decision diagrams / compilation.} The solver builds or manipulates a compiled Boolean representation, such as a binary decision diagram or zero-suppressed decision diagram, to decide satisfiability or construct a solution.\par

\noindent\textbf{Continuous optimization.} The solver optimizes a real-valued relaxation using gradients or continuous dynamics, then decodes Boolean assignments from the result.\par

\noindent\textbf{Configuration-checking search.} A variable becomes eligible to flip again when a neighboring variable or constraint configuration changes. Timestamps alone do not establish this mechanism.\par

\noindent\textbf{External engine, algorithm opaque.} The solver invokes an external solving engine whose algorithm cannot be established from the supplied implementation. A known engine name alone does not establish CDCL, and proof checkers are excluded.\par

\noindent\textbf{LP / MIP optimization.} An explicit linear-programming or mixed-integer-programming solver or relaxation uses optimization machinery. Native Boolean branching alone is insufficient.\par

\noindent\textbf{Pure random walk.} A distinct search repeatedly makes unguided or clause-focused random variable flips. Random steps within a scored, noisy WalkSAT procedure alone do not receive this additional label.\par

\noindent\textbf{SMT / theory solving.} Satisfiability modulo theories combines Boolean reasoning with a theory-specific decision procedure. Arithmetic preprocessing alone is excluded.\par

\subsubsection{Systems-level optimizations}
\label{app:rq3-audit-systems}
Concrete mechanisms in input/output, CPU execution, memory management, incremental work, and parallelism. Conventional implementations can qualify; novelty and measured benefit are not requirements.

\noindent\textbf{Compiler / ISA flags.} The default build explicitly requests compiler optimization, instruction-set features, link-time optimization, or related tuning. Such flags alone do not establish vectorized execution or a measured speedup.\par

\noindent\textbf{Preallocation / reuse.} Capacity reservation, scratch buffers, free lists, pools, or arenas reduce repeated memory allocation. Ordinary container construction alone is excluded.\par

\noindent\textbf{Incremental state updates.} Derived quantities, such as clause counts, move scores, or circuit values, are updated from local changes instead of recomputed globally. Assignment-trail writes alone are excluded.\par

\noindent\textbf{Flat / compact storage.} Flat arenas, offsets, packed records, separate arrays for record fields, or compressed adjacency organize data more compactly than fragmented or pointer-heavy storage. An ordinary array alone is insufficient.\par

\noindent\textbf{Watched / blocking literals.} Watched literals, cached satisfying literals, binary implication paths, or occurrence lists restrict propagation to potentially affected clauses instead of scanning every clause.\par

\noindent\textbf{Specialized parsing.} A custom byte or integer parser, bulk buffering, or explicit fast-I/O configuration avoids formatted input overhead. Ordinary token splitting, formatted scanning, or stream extraction alone is excluded.\par

\noindent\textbf{Scalar word parallelism.} Multiple logical values are packed into scalar machine words or bitsets and processed together using bitwise operations. Ordinary scalar flags and random-number arithmetic are excluded.\par

\noindent\textbf{Lazy / deferred work.} Generation stamps, lazy invalidation, deferred deletion or compaction, or delayed recomputation avoid recurring bulk work. Simply omitting cleanup is insufficient.\par

\noindent\textbf{Buffered proof output.} Proof bytes or lines are explicitly batched, or a larger output buffer is installed, to reduce output calls. Ordinary proof logging alone is excluded.\par

\noindent\textbf{Bit-count / scan intrinsics.} Reachable computation uses dedicated bit-counting, bit-scanning, or rotation intrinsics or equivalent standard functions. Plain bitwise AND or XOR alone is insufficient.\par

\noindent\textbf{Reduced proof-output work.} A concrete mechanism reuses serialized proof fragments, suppresses redundant emitted steps, or avoids repeated serialization. A mathematically short proof alone does not qualify.\par

\noindent\textbf{Specialized hot loops.} Manual unrolling or dedicated fixed-size or short-clause kernels specialize a recurring computation. A constant loop bound or an inline annotation alone is insufficient.\par

\noindent\textbf{Lookup tables.} Precomputed operation or transition tables replace recurring computation, for example with truth tables or byte population-count tables. Stored complete witnesses and ordinary search memoization are excluded.\par

\noindent\textbf{Memory-mapped input.} The solver maps the input file into memory and accesses it through that mapping. An unused mapping wrapper or header alone is insufficient.\par

\noindent\textbf{Branchless updates.} Explicit masks, arithmetic, or table lookups replace conditional selection in a recurring computation. A conditional expression alone does not establish branchless execution.\par

\noindent\textbf{OS memory hints.} Explicit operating-system advice requests access-pattern, paging, or read-ahead behavior. Memory mapping alone is excluded; these hints are distinct from CPU prefetch instructions.\par

\noindent\textbf{Compact proof encoding.} Proof records use an implemented compact serialization, such as binary records, variable-length integers, deltas, or compression. A proof-format name or ordinary text formatting alone is insufficient.\par

\noindent\textbf{Alignment / cache blocking.} Explicit alignment, padding, tiling, or blocking organizes data or computation for locality. Contiguous storage alone is insufficient.\par

\noindent\textbf{Explicit SIMD.} Packed vector operations execute through intrinsics, vector types, assembly, or a SIMD-specific library, such as SSE, AVX2, AVX-512, or NEON. Compiler flags and scalar bitsets alone are excluded.\par

\noindent\textbf{Branch-prediction hints.} Explicit likelihood annotations or profile hints tell the compiler which branch outcome is expected. Ordinary conditional statements do not qualify.\par

\noindent\textbf{CPU software prefetch.} The solver issues a software prefetch intrinsic or instruction for future memory accesses. Ordinary loads, buffering, capacity reservation, and operating-system advice are separate.\par

\noindent\textbf{GPU / accelerator kernels.} A reachable route launches computation on a GPU or another accelerator. Headers, build support, and unused code alone are excluded.\par

\noindent\textbf{Worker processes.} Multiple solver-worker processes execute concurrently. A single subprocess, an execution wrapper, or process-cleanup code alone is insufficient.\par

\noindent\textbf{Worker threads.} A reachable route starts concurrent threads or a parallel region for solver computation. Thread-library dependencies and sequential restarts alone are excluded.\par

\subsubsection{Preprocessing and inprocessing}
\label{app:rq3-audit-preprocessing}
Transformations of a remaining problem before search or while search is in progress. Before only and During only indicate exclusive observed stages; Before + during means both occur in the same solver. Stage unclear denotes unresolved timing. Ordinary per-decision propagation and general conflict learning alone are excluded.

\noindent\textbf{Root unit simplification.} Root-level unit propagation fixes forced assignments and removes satisfied clauses or literals to simplify the residual problem before search starts or resumes. Ordinary propagation after each decision alone is excluded.\par

\noindent\textbf{Clause cleanup.} Duplicate literals, tautologies, or duplicate clauses are removed, including normalization that enables these reductions. Sorting only to recognize structure is excluded.\par

\noindent\textbf{Cone / core reduction.} A relevant cone, core, or proper subset is extracted and solved or transformed, with a mechanism transferring the result to the original problem. Merely ignoring clauses is insufficient.\par

\noindent\textbf{Structural strengthening.} Implied constraints derived from recovered structure strengthen a residual problem, for example equivalences between duplicate gates or order-relation closure. Ordinary conflict learning alone is excluded.\par

\noindent\textbf{Other preprocessing.} A concrete transformation of the remaining problem falls outside the named categories. The annotation must identify the transformation and explain the taxonomy gap; ordinary search machinery is excluded.\par

\noindent\textbf{Resolution elimination.} A variable is removed by resolving clauses containing its positive and negative literals and retaining a satisfiability-equivalent residual problem, possibly subject to growth limits. Branching and Gaussian elimination are separate.\par

\noindent\textbf{Self-subsuming resolution.} Resolution produces a clause that subsumes a parent and therefore shortens it. Ordinary subsumption and general conflict learning alone are excluded; qualifying resolution-based learned-clause minimization may overlap with the search-tuning category.\par

\noindent\textbf{Gate / circuit reduction.} Recovered gates are constant-folded, substituted, rewritten, or functionally eliminated to reduce a residual search problem. Recognition or standalone circuit simulation alone is insufficient.\par

\noindent\textbf{Domain filtering.} Explicit finite domains or their supported values are pruned using domain-consistency reasoning before search starts or resumes. Ordinary Boolean unit propagation alone is excluded.\par

\noindent\textbf{Failed-literal probing.} Tentative literal assignments and propagation derive forced values or other simplifications. Probing used only to score the next branching decision belongs to search tuning.\par

\noindent\textbf{Cardinality / PB reduction.} Recovered counting or pseudo-Boolean constraints are simplified through bound reasoning, forced values, compression, or rewriting before further search. A direct counting refutation or the proof format alone is excluded.\par

\noindent\textbf{XOR / algebraic reduction.} Parity or algebraic equations are eliminated or substituted to reduce a problem passed to further solving. A complete Gaussian solver without a residual reduction role is excluded.\par

\noindent\textbf{Literal substitution.} Equivalent or complementary literals are merged and substituted to rewrite the residual problem, for example using implication-graph components. Detecting an equivalence only to solve directly is excluded.\par

\noindent\textbf{Problem decomposition.} The original instance is partitioned into independent or explicitly coordinated smaller subproblems for separate solving. Ordinary branches of one search do not qualify.\par

\noindent\textbf{Variable compaction.} Active residual variables or literals are renumbered to shrink solver data structures after or during reduction. Relabeling only for recognition or input permutation is excluded.\par

\noindent\textbf{Symmetry breaking.} Canonical restrictions remove equivalent assignments, or a verified symmetry rewrites the problem, reducing the remaining search space. Recognizing a symmetric family alone is insufficient.\par

\noindent\textbf{Pure-literal elimination.} A variable occurring with only one polarity is assigned or eliminated in a satisfiability-preserving simplification. Merely preferring that polarity during search is excluded.\par

\noindent\textbf{Clause subsumption.} A clause is removed because another clause contains a subset of its literals. Removing only duplicate clauses belongs to clause cleanup.\par

\noindent\textbf{Clause vivification.} Sequential literal assumptions and propagation establish that a clause can be shortened or is redundant. Generic failed-literal probing without clause shrinking is separate.\par

\noindent\textbf{Blocked-clause elimination.} A clause is removed when some literal blocks it: every resolvent on that literal is tautological. Explicit generalized blocked or set-blocked rules also qualify.\par

\noindent\textbf{Implication reduction.} Implied binary edges or clauses are removed by transitive reduction or another demonstrated redundancy test. Building an implication graph alone is insufficient.\par

\subsubsection{Search tuning}
\label{app:rq3-audit-search-tuning}
Policies controlling decisions, learning, restarts, local moves, diversification, and effort. The categories describe choices within search, rather than naming whole solving algorithms.

\noindent\textbf{Static / occurrence ordering.} An explicit fixed variable or value order, occurrence score, or polarity count guides decisions. Ordinary container iteration without a decision policy is excluded.\par

\noindent\textbf{Scheduled restarts.} A search restarts after a fixed budget or according to a predetermined schedule, such as Luby or geometric intervals. Switching to a different algorithm alone is separate.\par

\noindent\textbf{Input-dependent parameters.} Input features determine search budgets, thresholds, heuristic parameters, or modes. Dispatch between unrelated algorithms without tuning a search is separate.\par

\noindent\textbf{Bounded search effort.} An individual search attempt has an explicit step, conflict, flip, node, or time limit. The experiment's overall timeout alone is excluded.\par

\noindent\textbf{Biased / seeded phases.} Preferred Boolean values are initialized or steered using polarity, recovered structure, a candidate assignment, or a configured sign. Reusing past assignments alone belongs to phase saving.\par

\noindent\textbf{Conflict-activity branching.} Conflicts or learned clauses update variable priorities, as in variable-state independent decaying sum (VSIDS) and its exponential variant. Static occurrence scores and clause activity alone are excluded.\par

\noindent\textbf{Phase saving.} Previously assigned Boolean values are remembered and reused in later decisions. A single fixed preferred polarity is insufficient.\par

\noindent\textbf{Domain-aware branching.} Semantic quantities such as remaining-domain size, graph degree, constraint slack, or gate role guide decisions. Generic literal occurrence ordering is separate.\par

\noindent\textbf{Learned-clause retention.} A deliberate policy ranks, retains, or deletes learned clauses using quality, activity, size, recency, or related measures. Proof-format deletion bookkeeping alone is excluded.\par

\noindent\textbf{Diverse search starts.} Search is deliberately retried with distinct seeds, shuffled orders, perturbations, or initial assignments. One random initial state alone is insufficient.\par

\noindent\textbf{Learned-clause minimization.} Redundant literals are removed from a learned conflict clause beyond ordinary conflict analysis. General preprocessing vivification is separate.\par

\noindent\textbf{Random branch / tie choices.} Randomness selects search variables or values or resolves decision ties. Random-number tests, identifiers, and independent initial sampling alone are excluded.\par

\noindent\textbf{Make / break scoring.} Local-search moves are scored by their changes to satisfied or violated clauses or domain constraints. Make counts newly satisfied constraints; break counts newly violated ones.\par

\noindent\textbf{Noisy move selection.} Local moves use explicit noise or a nonuniform probability rule, as in WalkSAT or ProbSAT. Randomized branching in complete search is separate.\par

\noindent\textbf{LBD / glue quality.} The number of distinct decision levels represented in a learned clause, called literal-block distance (LBD) or glue, is computed and used to manage clauses or control search.\par

\noindent\textbf{Look-ahead branch scoring.} Tentative assignments, propagation, or predicted downstream reductions score the next branching decision. Probing solely to derive forced literals is preprocessing.\par

\noindent\textbf{Stagnation control.} Detected lack of progress changes search behavior, perturbs a candidate, or stops an attempt. A fixed stopping budget alone is insufficient.\par

\noindent\textbf{Tabu / recency rules.} Tabu tenure, move age, youngest-variable avoidance, or configuration eligibility constrain or prioritize moves. This broad category does not imply that every annotated solver implements TabuSAT.\par

\noindent\textbf{Adaptive constraint weights.} Penalties on clauses or constraints change during search to redirect local moves. Fixed weights supplied by the problem alone are excluded.\par

\noindent\textbf{Temperature / noise control.} Temperature, noise probability, or another exploration-strength parameter changes over time or with stagnation. A fixed noise probability alone is insufficient.\par

\noindent\textbf{Adaptive restarts.} Observed search behavior, such as clause quality or stagnation, triggers restarts or changes their timing. A fixed restart schedule alone is excluded.\par

\subsubsection{Structure recovered from CNF}
\label{app:rq3-audit-representation}
Semantic objects recovered or exploited beyond generic CNF clauses and literals. Recovering a representation does not by itself imply that it simplifies the problem or tolerates every encoding variant.

\noindent\textbf{Semantic components / cones.} Meaningful components, cones of influence, layers, or decompositions are recovered and exploited. Ordinary recursion or adjacency storage alone is insufficient.\par

\noindent\textbf{Problem graphs.} A native problem graph is reconstructed whose vertices and edges represent domain objects, for example in coloring or reachability. Generic clause-variable incidence and watch graphs are excluded.\par

\noindent\textbf{Gates / Boolean circuits.} Gate semantics or a Boolean circuit are recovered from clauses and used in solving, reduction, or certification. This includes comparison circuits and AND-inverter graphs; suggestive variable names alone are insufficient.\par

\noindent\textbf{Finite-domain variables.} Boolean literals are grouped into multi-valued variables and explicit domain constraints used in native search or propagation. Single Boolean variables or incidental clause groups are excluded.\par

\noindent\textbf{Cardinality / PB constraints.} Counting constraints, weighted Boolean sums, or inequalities are recovered from CNF and used in reasoning. Coefficients appearing only in proof output are insufficient.\par

\noindent\textbf{Permutations / orders.} A permutation, ordering, sequence, or positional combinatorial object is recovered and manipulated directly. Sorting input clauses or variables alone is excluded.\par

\noindent\textbf{Arithmetic / modular structure.} Integer, modular, polynomial, subset-sum, or number-theoretic quantities and equations are recovered beyond generic Boolean counting constraints. Numeric parsing alone is insufficient.\par

\noindent\textbf{Other recovered structure.} A concrete semantic representation falls outside the named categories. The annotation must identify it and explain the taxonomy gap; generic clause storage is excluded.\par

\noindent\textbf{Equivalence / implication.} Semantic equivalence classes, strongly connected implication components, or a tractable implication fragment form a higher-level representation. Generic implication storage for clause learning alone is excluded.\par

\noindent\textbf{Matching / assignment.} Matching, bipartite assignment, permutation-matrix, flow, or Hall-set roles and constraints are recovered. A generic graph representation alone is insufficient.\par

\noindent\textbf{XOR / parity equations.} Recovered parity constraints are represented as equations or an explicit parity system used in reasoning or search. Treating XOR only as a circuit gate or bitwise operation is insufficient.\par

\noindent\textbf{Symmetry / group structure.} A concrete symmetry, orbit, automorphism, or group action is identified and used in reasoning or construction. Generic sorting or a symmetry claim alone is insufficient.\par

\noindent\textbf{State-transition systems.} Automata, transitions, traces, planning states, or temporal steps are recovered and used for reasoning in a state space. A purely combinational circuit alone is insufficient.\par

\noindent\textbf{Geometry / packing.} Coordinates, shapes, intervals, placements, physical-board adjacency, or geometric constraints are recovered for solving. A matrix used only for storage is excluded.\par

\noindent\textbf{Statistical models.} A factor/message, correlation, spectral, or other statistical representation supports inference or candidate generation. Ordinary clause incidence without statistical state is excluded.\par

\subsubsection{Combining solving methods}
\label{app:rq3-audit-coordination}
Ways to select, combine, or exchange information between solving procedures. Outer coordination is distinguished from policies internal to one search loop.

\noindent\textbf{Structural dispatch.} Recognized problem structure or encoding selects a solving procedure. Syntax checking or changing one numerical heuristic alone is insufficient.\par

\noindent\textbf{Sequential methods.} Two or more different solving procedures are attempted in sequence. Repeated seeds, recognition variants, and proof emission after solving alone are excluded.\par

\noindent\textbf{Candidate / phase transfer.} An assignment, incumbent, preferred phase, or partial candidate produced by one method is passed to another. Independent guesses and final result printing are excluded.\par

\noindent\textbf{Constraint / bound transfer.} Constraints, algebraic relations, bounds, or deductions pass between distinct reasoning methods. Ordinary clause sharing within one engine alone is insufficient.\par

\noindent\textbf{Separate SAT / UNSAT routes.} Distinct specialized procedures find satisfying assignments and derive unsatisfiability certificates within one submission. The two outcomes of one generic clause-learning engine alone are insufficient.\par

\noindent\textbf{Size / feature dispatch.} Counts, estimated complexity, or other coarse instance features select different solving procedures. Changing a limit within one unchanged algorithm is insufficient.\par

\noindent\textbf{General SAT fallback.} An unsupported, unsuccessful, or exhausted specialized route can invoke a generic SAT engine. The label records a reachable fallback and does not guarantee completeness.\par

\noindent\textbf{Repeated configurations.} An outer coordinator runs repeated starts or configurations of a method and handles their success or failure. Restarts wholly internal to an unchanged search loop are excluded.\par

\noindent\textbf{Budgeted handoff.} A per-method resource limit triggers another method or an escalation. A final timeout without an alternative, or a restart of the same method, is excluded.\par

\noindent\textbf{Residual subsolver calls.} A higher-level procedure conditionally or repeatedly invokes a distinct solver on residual instances or subproblems and uses its answers. Ordinary recursive branches are excluded.\par

\noindent\textbf{Unlogged / certifying stages.} A search or check runs without proof logging, then is repeated or transfers information into a certifying run. Identical search replay is not required; ordinary proof-format conversion alone is excluded.\par

\noindent\textbf{Refinement loop.} Distinct procedures, or abstraction, candidate, and refutation steps, alternate in a feedback loop. One-way preprocessing followed by a single solve is insufficient.\par

\noindent\textbf{Adaptive effort allocation.} Runtime progress or method outcomes change how effort is allocated among methods. A fixed sequence of predetermined budgets is excluded.\par

\noindent\textbf{Parallel portfolio.} Distinct solving procedures run concurrently and their results are coordinated. Sequential calls and independent starts without different methods are excluded.\par

\subsubsection{Encoding assumptions and safeguards}
\label{app:rq3-audit-assumptions}
Input-representation assumptions, recognition mechanisms, scope guards, and answer checks. Recognition and safeguards can coexist with encoding dependence; these labels are not an empirical test of correctness or shuffle robustness.

\noindent\textbf{Unsupported-case rejection.} A failed semantic or scope check intentionally reports UNKNOWN or rejects the specialized route. Generic timeouts and file-opening errors are excluded.\par

\noindent\textbf{Structural validity guards.} A recovered template, invariant, or encoding condition is checked before specialization, with explicit rejection or fallback on failure. Basic input syntax checking alone is insufficient.\par

\noindent\textbf{Relational recognition.} Semantic roles are inferred from incidence, gate or graph relations, or constraint patterns. This can coexist with variable-ID assumptions and does not establish invariance under input permutation.\par

\noindent\textbf{Exact templates / signatures.} Fixed formula sizes, exact templates, hashes, or names select a semantic route or candidate beyond generic complexity thresholds. Signature matching is distinct from subsequent validation and is not an integrity judgment.\par

\noindent\textbf{Original-CNF model checks.} At least one reachable SAT success route checks a completed candidate against every original clause. The label does not imply that all success routes do so; reduced-only and sampled checks are excluded.\par

\noindent\textbf{Multiple encoding variants.} Multiple encodings, orientations, variable layouts, or clause-pattern variants of the same semantic task are explicitly handled. Supporting different input sizes alone is insufficient.\par

\noindent\textbf{Variable-ID / layout reliance.} Arithmetic on variable identifiers, fixed offsets, or contiguous identifier blocks assigns semantic roles. Generic array indexing, renumbering, and literal encoding alone are excluded.\par

\noindent\textbf{Model reconstruction.} Eliminated variables or a domain solution are lifted to a complete original-variable assignment using dependencies, elimination records, or an explicit mapping. Printing an already complete assignment is excluded.\par

\noindent\textbf{Canonicalized recognition.} Signs, permutations, clause or literal order, or equivalent encodings are canonicalized to recognize structure across variants. Sorting only for deduplication or fast lookup is excluded.\par

\noindent\textbf{Rejected-candidate fallback.} Failure of candidate, domain, or model validation resumes search or invokes another method instead of reporting SAT. Parsing or recognition failure alone is separate.\par

\noindent\textbf{Domain / reduced checks.} A reconstructed domain solution or reduced or partial candidate is explicitly validated beyond ordinary search transitions, potentially before lifting to CNF. This is distinct from checking every original clause.\par

\noindent\textbf{Clause-order reliance.} Input clause positions, adjacency, or order determine semantic roles. Ordinary traversal and sorting into order-independent groups are excluded.\par

\noindent\textbf{Sampled / fingerprint checks.} Bounded simulation, sampling, or fingerprints screen semantic recognition, equivalence, or candidate properties. Later exact validation is separate; generic random search and hash-table equality are excluded.\par

\noindent\textbf{Stored candidates / proofs.} A reachable result route retrieves an embedded or external precomputed model, proof, or answer. Constants for a general construction, local-operation tables, and runtime memoization are excluded.\par

\noindent\textbf{Certificate / RUP checks.} The supplied run path checks its own certificate or replays proof obligations before accepting an UNSAT certificate. This includes reverse-unit-propagation (RUP) obligation checks and does not imply an independent checker of a complete serialized proof.\par

\endgroup

\end{document}